%% file: main.tex
\documentclass[12pt,a4paper]{report}

\counterwithout{footnote}{chapter}
\counterwithout{table}{chapter}

\pdfoutput=1
\usepackage[english]{babel}
\usepackage[latin1]{inputenc}
\usepackage[T1]{fontenc}
\usepackage{amsfonts,amsbsy,bm,euscript,mathrsfs}
\usepackage{amssymb,stmaryrd,faktor,slashed}
\usepackage{color}
\usepackage[tbtags]{amsmath}
\usepackage[bookmarks=true,colorlinks=true,linkcolor=black,citecolor=black,urlcolor=black,bookmarksnumbered,linktoc=all,hyperfootnotes=false]{hyperref}

\usepackage{etoolbox}

\usepackage{graphicx}
\usepackage{chngcntr}
\usepackage{makecell}
\usepackage{mathtools}
\usepackage{afterpage}
\usepackage{enumerate}

\usepackage[a4paper,text={170mm,245mm},centering]{geometry}

\numberwithin{equation}{chapter}

\renewcommand{\arraystretch}{1.3}

\makeatletter
\renewcommand\section{\@startsection {section}{1}{\z@}%
{-3.5ex \@plus -1ex \@minus -.2ex}%
{2.3ex \@plus.2ex}%
{\normalfont\large\bfseries}}
\renewcommand\subsection{\@startsection{subsection}{2}{\z@}%
{-3.25ex\@plus -1ex \@minus -.2ex}%
{1.5ex \@plus .2ex}%
{\normalfont\normalsize\bfseries}}
\makeatother

\expandafter\def\expandafter\bfseries\expandafter{\bfseries\ifmmode\else\boldmath\fi}
\expandafter\def\expandafter\mdseries\expandafter{\mdseries\ifmmode\else\unboldmath\fi}
\expandafter\def\expandafter\normalfont\expandafter{\normalfont\ifmmode\else\unboldmath\fi}

\providecommand{\href}[2]{#2}
\newcommand{\arxivlink}[1]{\href{http://arxiv.org/abs/#1}{[arXiv:#1]}}
\newcommand{\doilink}[2]{\href{http://doi.org/#2}{#1}}

\newcommand{\mathsym}[1]{{}}

\newcommand{\rf}[1]{(\ref{#1})}
\newcommand{\mc}{\mathcal }
\newcommand{\la}{\label}

\makeatletter
\newcommand*{\lodbib@citeorder}{}
\newcommand*{\lodbib@notcited}{}% catch entries that were not cited

\def\citation{%
  \forcsvlist{\citation@i}}

\def\citation@i#1{%
  \ifinlist{#1}{\lodbib@citeorder}
    {}
    {\listxadd{\lodbib@citeorder}{#1}}}

\let\ltxorig@lbibitem\@lbibitem
\let\ltxorig@bibitem\@bibitem

\def\@lbibitem[#1]#2#3{%
  \csdef{lodbib@savedlabel@#2}{#1}%
  \@bibitem{#2}{#3}}

\def\@bibitem#1#2{%
  \xifinlist{#1}{\lodbib@citeorder}
    {}
    {\listadd{\lodbib@notcited}{#1}}%
  \csdef{lodbib@savedentry@#1}{#2}}

\renewenvironment{thebibliography}[1]
     {\settowidth\labelwidth{\@biblabel{#1}}}
     {\def\@noitemerr
       {\@latex@warning{Empty `thebibliography' environment}}%
      \section*{\refname}%
      \@mkboth{\MakeUppercase\refname}{\MakeUppercase\refname}%
      \list{\@biblabel{\@arabic\c@enumiv}}%
           {\leftmargin\labelwidth
            \advance\leftmargin\labelsep
            \@openbib@code
            \usecounter{enumiv}%
            \let\p@enumiv\@empty
            \renewcommand\theenumiv{\@arabic\c@enumiv}}%
      \sloppy
      \clubpenalty4000
      \@clubpenalty \clubpenalty
      \widowpenalty4000%
      \sfcode`\.\@m
      \lodbib@biblistloop
      \endlist}

\def\lodbib@biblistloop{%
  \forlistloop{\lodbib@bibitem}{\lodbib@citeorder}%
  \ifdefvoid{\lodbib@notcited}
    {}
    {\forlistloop{\lodbib@bibitem}{\lodbib@notcited}}}

\def\lodbib@bibitem#1{%
  \ifcsundef{lodbib@savedlabel@#1}
    {\ltxorig@bibitem{#1}}
    {\ltxorig@lbibitem[\csuse{lodbib@savedlabel@#1}]{#1}}%
  \csuse{lodbib@savedentry@#1}}
\makeatother

\def\id{\protect{{1 \kern-.28em{\rm l}}}}
\def\be{\begin{eqnarray}}
\def\ee{\end{eqnarray}}
\def\bi{\bibitem}
\def\tr{{\rm tr}}
\def\ha{\tfrac{1}{2}}
\def\td{\tilde}
\def\ci{\cite}
\def\N{{\mathcal N}}
\def\ww{\Omega}
\def\S{{\mathcal S} }
\def\nn{\nu}
\def\z{\zeta}
\def\dg{\dagger}
\def\a{\alpha}
\def\b{\beta}
\def\ap{\alpha^\prime}
\def\aa{{\a'}}
\def\g{\gamma}
\def\ok{\frac{1}{k}}
\def\jL{{J}}
\def\cL{{\mathcal L}}
\def\cH{{\mathcal H}}
\def\E{{\mathcal E}}
\def\w{\omega}
\def\vep{\varepsilon}
\def\De{{\mathcal D}}
\def\k{\kappa}
\def\four{\tfrac14}
\def\third{\tfrac{1}{3}}
\def\det{\hbox{det}}
\def\tid{\tilde}
\def\vv{{\rm v}}
\def\XX{{\rm X}}
\def\ta{{\tilde \a}}
\def\fo{\frac{1}{4}}
\def\K{{\rm S}}
\def\el{\ell}
\def\Tr{{\rm Tr}}
\def\P{\Phi}
\def\l {\lambda}
\def\bl{{\tilde \l}}
\def\const{{\rm const}}
\def\vn{\vec n}
\def\Prod{\Pi}
\def\O{{\mathcal O}}
\def\m{\mu}
\def\vs{\vec \s}
\def\ie{i.e.}
\def\rS{{\rm S}}
\def\as{{\a}}
\def\e{\epsilon}
\def\foot{\footnote}
\def\ra{\rightarrow}
\def\F{{\cal F}}
\def\cc{\circ}
\def\eqv{\equiv}
\def\ni{\noindent}
\def\bw{{\rm w}}
\def\cT{{\cal T}}
\def\no{\nonumber}
\def\J{{\cal J}}
\def\om{\omega}
\def\l{\lambda}
\def\tk{{\tilde k}}
\def\tl{{\tilde \l}}
\def\sqtl{{\sqrt{\tilde \l}}}
\def\adss{$AdS_5 \times S^5$\ }
\def\D{\Delta}
\def\p{\phi}
\def\r{\rho}
\def\rN{{\rm N}}
\def\tw{{\tilde w}}
\def\vl{ \vec \ell}
\def\varpi{{\rm w}}
\def\bG{\bar \G}
\def\ve{\varepsilon}
\def\I{{\cal I}}
\def\La{{\Lambda}}
\def\R{{\rm R}}
\def\bt{\bar\theta}
\def\Z{{\cal Z}}
\def\pa{\partial}
\def\bea{\be}
\def\eea{\ee}
\def\DD{{\rm D}}
\def\chii{\varepsilon}
\def\th{\theta}
\def\t{\tau}
\def\beq{\be}
\def\eeq{\ee}
\def\beqa{\bea}
\def\eeqa{\eea}
\def\bs{\bigskip}
\def\c{{\rm a}}
\def\del{\partial}
\def\s{\sigma}
\def\eps{{\epsilon}}
\def\n{\nu}
\def\dag{\dagger}
\def\bd{\bar \del}
\def\ed{\end{document}}
\def\iffa{\iffalse}
\def\xp{x_+}
\def\xm{x_-}
\def\te{\textstyle}
\def\rx{{\rm x}}
\def\C {{\rm C}}
\def\hC {{\rm \hat C}}
\def\dd {{\rm d}}
\def\bb{{\rm b}}
\def\sql{{\sqrt{\l}}}
\def\ep{\epsilon}
\def\G{\Gamma}
\def\dx{\dot x}
\def\vp{\varphi}
\def\tx{{\tilde x}}
\def\ad{{\rm ad}}
\def\d{\delta}
\def\de {\del}
\def\L{\mathcal{L} }
\def\pcm {{\rm PCM}}
\def\wz{{\rm WZ}}
\def\pcmq {{\rm PCM}$_q$\ }
\def\ZMq {{\rm ZM}$_q$\ }
\def\ZM {{\rm ZM} }
\def\pp{{\rm p}}
\def\qq{{\rm q}}
\def\rk {{\rm u}}
\def\rrk{{\rm e}}
\def\ss{{\rm s}}
\def\tt{{\rm t}}
\def\uu{{\rm u}}
\def\sm{$\sigma$-model}
\def\sms{$\sigma$-models}
\def\sm{$\sigma$-model\ }
\def\sms{$\sigma$-models\ }
\def \ov {\over}
\def\Ad{\text{Ad}}
\def\lmp{\Lambda}
\def\bl{\bar \lambda}
\def \hb {\hbar}
\def \k {\varkappa} \def \h {{\rm h}}
\def \dt {\dt}
\def\ddt{\frac{d}{dt}}
\def \bb {{\rm b}}
\def \vv {{ v}}
\def \kk {{\rm k}}
\def \ka {{\kappa}}
\def \np { }
\def \cM {\mathcal{M}}
\def\H{{\rm H}}
\def \BB {{\rm B}}
\def \HH {{\rm H}}
\def \GG {{\rm G}}
\def \PCM {{\rm PCM}}
\def \WZ {{\rm WZ}}
\def \Lie {{\rm Lie}}
\def\lam{$\lambda$-model}
\def\etm{$\eta$-model}
\def\bh{\bar h}
\def\hh{{\rm h}}
\def \OO {{\cal O}}
\def \dt {\frac{d}{dt}}
\def \rrr {x} \def \qp {{\rm q}}
\def \qp {{\mathrm h} }
\def \kk { k}
\def \bq {{q'}} \def \barp {x} \def \barq {y}
\def \rk {{\rm k}}
\def\cg{c_{_G}}
\def\cf{c_{_H}}
\def\cff{c_{_\F}}
\def\XX{{X}}
\def\hhh{\widetilde {\hh}}
\def\q{{\rm q}}
\def\RR{{J}}
\def\LL{{K}}
\def\M{{\cal M}}
\def\e{\varepsilon}
\def\tih{{\hat h}}
\def\bl{\bar\lambda}

\def\wt{\widetilde}

\def \GB {{$\GG \BB$}\ }
\def \ccg {c_{_G}}
\def \rh {{\rm h}}
\def \ul {\underline}

\usepackage[onehalfspacing]{setspace}
\usepackage[titletoc]{appendix}

\begin{document}

\thispagestyle{empty}

\setcounter{footnote}{0}

\vspace{2.5cm}

\begin{center}
{\bf{\LARGE \

\

\

\

\noindent Integrability and RG flow in 2d sigma models}}

\

\

\

\

{\large  \noindent Thesis for the degree of \textit{Doctor of Philosophy}  in Physics}

\

\

\

\

\

\

\

\noindent {\bf{ \Large \   Nat Levine \foot{\ Now at {\'E}cole Normale Sup{\'e}rieure de Paris and Institut Philippe Meyer. 

\ \ \   nat.levine@phys.ens.fr}}}

\

\

\

\noindent  {\large \it Blackett Laboratory, Imperial College London, 

\noindent London SW7 2AZ, UK}

\

\

\

\

\

\

\
 
\

\end{center}

\newpage

\renewcommand{\abstractname}{Abstract}

%%%%%%%%%%%%%%%%%%%%%%%%%%%%%%%
\begin{abstract}
\addtocounter{page}{1}
\thispagestyle{plain}

\bigskip
Motivated by the search for solvable string theories, we consider the problem of classifying the integrable bosonic 2d $\s$-models.
We  include  non-conformal $\s$-models, which have historically been a good arena for  discovering integrable models
 that were later generalized to Weyl-invariant ones. General  $\s$-models  feature a quantum RG flow, given by a `generalized Ricci flow' of the target-space geometry.

\medskip
This thesis is based on the conjecture that  \textit{integrable} $\s$-models are \textit{renormalizable}, or \textit{stable} under the  RG flow.
It is widely understood that classically integrable theories are stable at the leading  1-loop order with only a few parameters  running.

\medskip
Here we  address what happens at \textit{higher-loop orders}. We find that integrable $\s$-models generally remain RG-stable at higher-loops provided they receive a particular choice of finite counterterms, or quantum $(\a')$ corrections to the target-space geometry. Thought to be preserving integrability at the quantum level, these corrections are analogous to those required for higher-loop conformal invariance of gauged Wess-Zumino-Witten models.

\medskip
\sloppy
We explicitly construct the quantum corrections restoring higher-loop renormalizability for examples of integrable $\eta$- and $\lambda$-deformed $\s$-models. We then consider the integrable $G\times G$ and $G\times G/H$ models and also construct a new class of integrable 
$G\times G /H$ models with \mbox{abelian $H$}.

\medskip
We also reformulate the $\l$-models as $\s$-models on a ``tripled'' $G\times G\times G$ configuration space, where they become  automatically renormalizable  due to the decoupling of some fields.  In the limit when they become non-abelian dual (NAD) models, this suggests that the corresponding `interpolating models' for NAD are also renormalizable, with 2-loop $\b$-functions matching the  group/symmetric space $\s$-models.

\medskip
We then present a new and different link between integrability and the RG flow in the context of  $\s$-models with `local couplings' depending explicitly on 2d time. Such models are naturally obtained in the light-cone gauge 
in string theory,
  pointing to the possibility of a large, new class of solvable string models.

\end{abstract}
%%%%%%%%%%%%%%%%%%%%%%%%%%%%%%%

\newpage

\renewcommand{\abstractname}{Statement of originality}
\begin{abstract}
\addtocounter{page}{2}
\thispagestyle{plain}
 \bigskip \noindent
I declare that, aside from properly acknowledged or referenced contributions of others, this thesis is my own work.

\bigskip \bigskip \noindent
This thesis is closely based on the following research papers. These papers were written as collaborations with co-authors who contributed equally to the work.

\bigskip \bigskip \noindent
\begin{minipage}[t]{1cm}\cite{HLTS}
\end{minipage}
\begin{minipage}[t]{\textwidth-1cm}
B.~Hoare, N.~Levine and A.~A.~Tseytlin,

\noindent 
``On the massless tree-level S-matrix in 2d sigma models,''

\noindent 
\doilink{J. Phys. A \textbf{52}, no.14, 144005 (2019)}{
doi:10.1088/1751-8121/ab0b79}
\arxivlink{1812.02549}.
\end{minipage}
%16 citations counted in INSPIRE as of 11 Jul 2021

\bigskip \noindent
\begin{minipage}[t]{1cm}\cite{HLT1} 
\end{minipage}
\begin{minipage}[t]{\textwidth-1cm}
B.~Hoare, N.~Levine and A.~A.~Tseytlin,

\noindent ``Integrable 2d sigma models: quantum corrections to geometry from RG flow,''

\noindent 
\doilink{Nucl. Phys. B \textbf{949}, 114798 (2019)}{10.1016/j.nuclphysb.2019.114798}
\arxivlink{1907.04737}.
\end{minipage}

\bigskip \noindent
\begin{minipage}[t]{1cm}\cite{HLT2} 
\end{minipage}
\begin{minipage}[t]{\textwidth-1cm}
B.~Hoare, N.~Levine and A.~A.~Tseytlin,

\noindent 
``Integrable sigma models and 2-loop RG flow,''

\noindent 
\doilink{JHEP {\bf 1912}, 146 (2019)}{doi:10.1007/JHEP12(2019)146}
\arxivlink{1910.00397}.
\end{minipage}

\bigskip \noindent
\begin{minipage}[t]{1cm}\cite{HLT2008}
\end{minipage}
\begin{minipage}[t]{\textwidth-1cm}
B.~Hoare, N.~Levine and A.~A.~Tseytlin,

\noindent 
``Sigma models with local couplings: a new integrability -- RG flow connection,''

\noindent 
\doilink{JHEP \textbf{11}, 020 (2020)}{10.1007/JHEP11(2020)020}
\arxivlink{2008.01112}.
\end{minipage}

\bigskip \noindent
\begin{minipage}[t]{1cm}\cite{LT2103}
\end{minipage}
\begin{minipage}[t]{\textwidth-1cm}
N.~Levine and A.~A.~Tseytlin,

\noindent 
``Integrability vs. RG flow in $G \times G$ and $G \times G /H$ sigma models,''

\noindent 
\doilink{JHEP \textbf{05}, 076 (2021)}{
doi:10.1007/JHEP05(2021)076}
\arxivlink{2103.10513}.
\end{minipage}

\end{abstract}

\renewcommand{\abstractname}{Acknowledgements}
\begin{abstract}
\addtocounter{page}{3}
\thispagestyle{plain}

\bigskip \noindent
I am very grateful to my advisor Arkady Tseytlin, who has taught me so much during my PhD. Thank you for your contagious interest and enthusiasm  for the subject, and for always being happy to share your knowledge. It has been a great  inspiration and  pleasure  working   with you.

\bigskip
Thank you to Ben Hoare for our enjoyable collaborations and  countless interesting discussions and explanations. Your advice and support has been extremely valuable to me.

\bigskip
I would also like to thank Tim Adamo, Lorenz Eberhardt, Falk Hassler, Sylvain Lacroix, Konstantinos Sfetsos, Alessandro Torrielli, Stijn van Tongeren, Beno\^{i}t Vicedo and Linus Wulff for interesting discussions, advice and guidance during my PhD.

\bigskip
I am grateful to EPSRC for supporting my PhD through the grant EP/N509486/1.

\bigskip
I thank my partner, family and friends for many great times during the last four years in London. A special thanks goes to my colleagues and officemates at Imperial for the amusing lunchtime discussions, `PhD meetings' and half-hearted attempts to prove the QCD mass gap. In 2020 we very sadly  lost our friend and  colleague Ed Tasker, and he will will always be dearly missed.

\end{abstract}

\newpage

\renewcommand{\abstractname}{Copyright declaration}
\begin{abstract}
\addtocounter{page}{4}
\thispagestyle{plain}
 \bigskip \noindent

The copyright of this thesis rests with the author. Unless otherwise indicated, its contents are licensed under a Creative Commons Attribution 4.0 International Licence (CC BY).

Under this licence, you may copy and redistribute the material in any medium or format for both commercial and non-commercial purposes. You may also create and distribute modified versions of the work. This on the condition that you credit the author.

When reusing or sharing this work, ensure you make the licence terms clear to others by naming the licence and linking to the licence text. Where a work has been adapted, you should indicate that the work has been changed and describe those changes.

Please seek permission from the copyright holder for uses of this work that are not included in this licence or permitted under UK Copyright Law.

\end{abstract}

\newpage

\addtocounter{page}{5}
\tableofcontents

\ 

\bigskip \bigskip\bigskip  \bigskip\bigskip \bigskip\bigskip\bigskip\bigskip\bigskip \bigskip\bigskip  \bigskip\bigskip
{\bf \large \noindent List of tables}

\bigskip
{\bf \noindent Table 1:} Examples of integrable \sms and their time-dependent generalizations \hfill {\bf \pageref{2008tab1}}

\setcounter{section}{0}

\chapter{Introduction}
\section{Background}

The AdS/CFT correspondence is a remarkable set of dualities between 
string theories in Anti-de Sitter  space AdS$_{d+1}$ 
and
conformal field theories CFT$_d$ in one fewer dimension \cite{Maldacena:1997re,Witten:1998qj,Gubser:1998bc}. In the Poincar\'e coordinates for AdS,
\be
ds^2 = \frac{R^2}{z^2} \big(dz^2 + dx^\m dx_\m \big) \ , \qquad \m = 1,\ldots, d \ , \la{poinc}
\ee
one may picture the CFT as existing at the ``conformal boundary'' $z=0$, where the metric formally diverges but is conformally equivalent to the flat $d$-dimensional one. Gauge-string dualities of this type are often called ``holographic'' as they conjecture a mapping  between the degrees of freedom on the boundary ($z=0$) and  those in the ``bulk'', or  interior of the AdS space. A natural first check  is  the matching of internal symmetries: indeed the isometry group $SO(2,d-1)$ of AdS$_{d+1}$ coincides with the conformal group of $d$-dimensional flat space.

Such a correspondence has several immediately interesting and surprising properties. First, the  $(d+1)$th dimension (the $z$ direction in \rf{poinc}) seems to  emerge  from a $d$-dimensional CFT description. Second, the string theory is gravitational while the CFT is not. Thus it is possible to reformulate a  complicated quantum theory of gravity in terms of a  simpler and better understood CFT in flat space. 

In the context of string theory, the  AdS space must be embedded in a string background. A canonical example discussed in Maldacena's original proposal \cite{Maldacena:1997re} is the type IIB background obtained in the low energy limit from a stack of $N$ coincident D3-branes. The conjecture is that the low energy theory describing open strings ending on the D-branes ---  $SU(N)$ $\mathcal N=4$ super Yang-Mills theory in 4d  --- is dual to the theory of closed strings in the near horizon region --- type IIB strings on \adss. The theories' parameters are related by 
\be
\frac{R^2}{\a'} = \sqrt{g_{\rm YM}^2 N} \ , \qquad g_s = \frac{g_{\rm YM}^2}{4\pi} \ ,
\ee
where $R$ is the AdS radius in \rf{poinc}, $\a'=l_s^2$ the squared string length, $g_{\rm YM}$ the Yang-Mills coupling and $N$ the rank of the Yang-Mills gauge group. 

\sloppy
Among the many examples of the AdS/CFT correspondence, the \adss case is notable because it is \textit{integrable}\foot{Here we give a lightning summary of integrability in the AdS/CFT correspondence, focussing on \mbox{\adss} (and missing many details and important references). For a comprehensive introduction, see the reviews \cite{adscftsol,Arutyunov:2009ga,vanTongeren:2013gva,Bombardelli:2016rwb}.}${}^{,}$\foot{Another well-studied, integrable, holographic string background  is the type IIA background on $AdS_4 \times \mathbb{C}\mathbf{P}^3$, dual to 3d $\mathcal{N}=6$ Chern-Simons theory \cite{ABJM}; see, e.g., the review \cite{Klose:2010ki}.} in the free string ($g_s = 0$) limit, or equivalently in the planar limit ($N=\infty$ with $\l = g_{\rm YM}^2 N$ fixed) of the gauge theory. Here ``integrability'' is understood in the same sense of familiar integrable systems like the KdV or sine-Gordon model,  meaning aspects of the theory can be solved exactly without recourse to perturbation theory. 

On the string side (strong 't Hooft coupling $\l$),  the Green-Schwarz action for strings   in the \adss background takes the Metsaev-Tseytlin form \cite{Metsaev:1998it} equivalent to a \sm on the semi-symmetric space $PSU(2,2 | 4)/[SO(1,4)\times SO(5)]$ with a particular Wess-Zumino term. It is a general fact that such models are classically integrable \cite{Bena:2003wd,Das:2004hy,Young:2005jv,Magro:2008dv,Vicedo:2009sn} in the sense of admitting a flat Lax connection and associated conserved charges (see Section \ref{ci} below). In spite of classic issues preventing a direct integrable quantization due to the non-ultralocality of the classical Poisson structure \cite{maillet,Magro:2008dv}, one may just \textit{assume} the quantum theory to be integrable and bootstrap an exact integrable S-matrix  for the fully gauge-fixed theory based on  symmetries and consistency conditions (e.g. in an expansion around a BMN-type \cite{BMN} geodesic, fixing  the `uniform light-cone' gauge; see the review \cite{Arutyunov:2009ga}). This S-matrix describes scattering on a 2d plane, but the finite size (closed string) spectrum may be extracted by some Bethe ansatz equations, or more properly accounting for the ``wrapping corrections'' by double Wick rotating \cite{ZTBA} and using the Thermodynamic Bethe Ansatz (TBA) \cite{YTBA}. The TBA  (nonlinear integral) equations were reformulated as the simpler Quantum Spectral Curve \cite{Gromov:2013pga}, which is the most modern approach to computing the exact spectrum (see also the review \cite{Levkovich-Maslyuk:2019awk}).

On the gauge side (weak coupling), the structure and origin of integrability is more mysterious. It was observed that the mixing matrix for anomalous dimensions of single-trace local operators coincides with the Hamiltonian of an integrable spin chain \cite{Minahan:2002ve}. This conjures an image  of fields inside the trace corresponding to sites of a spin chain, with their R-charges corresponding to the spins. Including more and more loop orders, the corresponding spin chain was found to remain integrable, but with longer-range interactions at each order \cite{BeisertHL}.  Assuming integrability to all orders, the exact spin-chain S-matrix \cite{BeisertS} and  asymptotic Bethe ansatz equations \cite{BeisertBA} (for long operators or long spin chains) can be obtained, matching  results from the string side. The relation between integrability on the gauge and string sides can be understood qualitatively in terms effective Landau-Lifshitz actions for coherent states, with the string effection action being the direct continuum limit of the spin chain one (see, e.g., \cite{Tseytlin:2004xa} and refs.\ there).

Integrability has been extremely useful	 in understanding the holographic duality in the case of  \adss\unskip. Not only has it allowed the solution of the spectral problem for a non-trivial 4d gauge theory (with qualitative similarities to QCD), but it has also facilitated  non-trivial checks of the AdS/CFT correspondence in this example (see, e.g., \cite{Eden:2012fe}). Motivated by a wish to solve more string theories in curved space, and more 4d gauge theories, one could ask in which ways the \adss background can be \textit{deformed} while preserving integrability. More generally, we would like to know which string theories are integrable. 

The question of classifiying integrable string backgrounds is already very difficult  in the simpler case of \textit{bosonic} strings, which will be our focus. Although focussing on bosonic strings takes a step back from the AdS/CFT correspondence that was our motivation, we note the historical fact that the integrable semi-symmetric space \sm governing the \adss background is a simple generalization of bosonic symmetric space \sms discussed below.
 The motion of a bosonic string is described by a 2d \sm
 coupled to 2d gravity,
 \be \begin{aligned} 
&S  = \frac{1}{4\pi\a'} \int d^2 \xi \, \L \la{csm}\\
&\L= -  \big[ \sqrt{-h} h^{\m\n} \, \GG_{mn}(x) + \e^{\m\n} \,\BB_{mn} (x) \big] \,  \del_\m x ^m \,  \del_\n x^n   - \a' \sqrt{-h}\,  R^{(2)} \p(x)  \ ,\\
&m,n=1,\ldots , D \ , \qquad \quad  \m,\n=1,2 \ .
\end{aligned} \ee
Here the scalar fields $x^m$  may be viewed as embedding coordinates for  the 2d worldsheet (traced out by a string over time) in a $D$-dimensional target space-time. The parameter $\a'$ plays the role of $\hbar$ (counting loops) in the  $\s$-model. In this thesis we will  set $\a'=1$ for convenience, but will sometimes reinstate $\a'$ to emphasize the loop counting. The \sm couplings $(\GG_{mn},\BB_{mn})$ are respectively symmetric and antisymmetric and correspond to  the background values of the target space metric and NSNS 2-form. The auxiliary 2d metric $h_{\m\n}$ is arbitrary and $\e_{\m\n}$ is the antisymmetric 2d Levi-Civita tensor.

The model \rf{csm} is 2d diffeomorphism-invariant and classically 2d Weyl-invariant. The role of the dilaton background $\phi(x)$,  appearing coupled to the 2d curvature $R^{(2)}$ at subleading order in $\a'$, is to compensate quantum breaking of Weyl invariance from the $(\GG,\BB)$ terms \cite{Fradkin:1985ys}. Preservation of the local Weyl invariance, which is necessary for the consistency of the quantum theory, imposes that the background fields satisfy certain `string equations of motion' \cite{Lovelace:1983yv} (generalizing the Einstein equation),
\be \la{se}
\begin{aligned}
&R_{mn} - \tfrac{1}{4} (\HH^2)_{mn} + 2\nabla_m \nabla_n \phi  = \O(\a') \ , \qquad \quad  (\HH_{mnp} \equiv 3 \del_{[m} \BB_{np]})  \\
&-\ha \nabla^p \HH_{mnp} + \HH_{mn}{}^p \,  \nabla_p \p   = \O(\a') \  , \\
&\tfrac{1}{6}(D-26) -\ha \nabla^2 \p + (\nabla \p)^2 - \tfrac{1}{24} \HH^2 = \O(\a') \ ,
\end{aligned}
\ee
which contain corrections at each order in $\a'$ (corresponding to the loop expansion of the $\s$-model). These equations are obtained by varying an effective gravitational action for the massless fields,
\be
S = \frac{1}{2\ka^2} \int d^D x \, e^{-2\p} \big( R - \tfrac{1}{12} \HH^2 + 4 (\del \p)^2 \big) + \O(\a')  \ .
\ee
and it has been checked
to several orders in $\a'$  that this action coincides with the one obtained from the tree-level string S-matrix \cite{mt,Graham:1987ep,Foakes:1987ij,Foakes:1987gg} (see also the review \cite{Tseytlin:1988rr}).
 
 For the purposes of classical integrability, it will be sufficient to fix the `conformal' gauge $\sqrt{-h} h^{\m\n} = \eta^{\m\n}$, obtaining a $\s$-model on flat 2d space,
\unskip\foot{The 2d light-cone derivatives are defined as $\del_\pm = \del_\t \pm \del_\s$, see Appendix \ref{conv} for the conventions.}
\be \la{13}
S = \frac{1}{4\pi\a'} \int d^2 \xi \, \L = \frac{1}{4\pi\a'} \int d^2 \xi \, \big[ \GG_{mn}(x) + \BB_{mn} (x)\big] \,  \del_+ x ^m \del_- x^n 
\ ,
\ee
where we have ignored the quantum dilaton term. At the level of classical integrability, the flat and curved 2d-space $\s$-models are equivalent   since they are related by a redefinition $\xi^\m \to \xi'^\m(\xi)$ (as the $\s$-model is  classically 2d Weyl invariant and all 2d metrics are locally equivalent up to a conformal factor). The Virasoro constraints ($T_{++}=0$, $T_{--}=0$) coming from the variation of the 2d metric $h_{\m\n}$ in \rf{csm} should not destroy classical integrability since these are just constraints on the initial data. 

Our eventual goal is to understand which  string backgrounds $(\GG,\BB)$ (solutions of \rf{se} with some choice of $\p$) correspond to integrable \sms \rf{13}. At the classical level, one criterion for integrability is the 2d equations of motion taking the \textit{zero-curvature} form \cite{kdv,lax,zs,zm}
\be \la{zc}
[ \del_+ + L_+(z) , \del_- + L_-(z) ] = 0 \ , 
\ee
where the ``Lax connection'' $L_\pm=L_\pm(z; x, \del x)$ is a 1-complex-parameter family of connections constructed from the fields. For any given \sm couplings $(\GG,\BB)$, it is very difficult to know whether a zero-curvature representation exists, and moreover there is no general recipe for constructing the Lax connection. This tricky classification problem of integrable \sms is the  motivation for the thesis.

\section{Stability under the RG flow}

In this thesis we will relax the Weyl invariance condtion \rf{se} and consider general \sms on a flat 2d space, again asking the question of which choices of couplings $(\GG,\BB)$ correspond to integrable theories. Although non-Weyl-invariant \sms do not directly correspond to string backgrounds, their study is well-motivated by the historical observation that  non-conformal  integrable \sms such as the PCM and the symmetric space \sms have preceded the discovery of related conformal ones --- the WZW model and the semi-symmetric space \sm with WZ term. In other words, one may discover new examples of Weyl-invariant  integrable theories by first discovering non-Weyl-invariant ones.

Moreover, relaxing Weyl invariance is actually  helpful since it opens up a new `diagnostic tool' for  integrability: the \textit{renormalization group (RG) flow}. The idea is that, in order to be able to identify the integrable theories, one should thoroughly investigate their properties and distinctive features. We shall focus on  one conjectured property in particular: that \textbf{integrable \sms are  \textit{stable} under the quantum RG flow} \cite{Fateev:1992tk,fateev96,lukyanov,Fateev:2019xuq,Fateev:2018yos,Litvinov:2018bou,Valent:2009nv,Demulder:2017zhz,Itsios:2014lca,Appadu:2015nfa}. Let us now explain the meaning of this statement.

\sloppy
General \sms  feature a quantum  RG flow (breaking the scale invariance of the classical theory). The \sm can be quantized  preserving the target space covariance \cite{ecker,friedan} and,  in the simplest case with $\BB_{mn}=0$, the  RG flow is given by a `generalized Ricci flow',\foot{We denote by $t=\log \m$ the 2d RG scale.}
\be \begin{aligned}
&\dt G_{mn}= \b_{mn} + \nabla_{(m} \xi_{n)} \ , \qquad \qquad \qquad (B_{mn}=0) \la{2}\\
&\b_{mn} = \a' \, R_{mn} + \a'^2 \, \ha R_{mpqr} {R_n}^{pqr} + \ldots \ .
\end{aligned} \ee
We have included the 1-loop and 2-loop terms, and there will be further contributions at each loop order. Here $\xi^n(x)$ corresponds to a choice of  renormalizations of the fields $x^n$, i.e.\ target space diffeomorphisms depending on the 2d RG scale. Now including the $\BB$-field, the $\b$-function \rf{2} generalizes, in a particular 2-loop subtraction scheme\foot{See Section \ref{SD} for discussion of scheme dependence.} to \cite{mt} (see also \cite{Braaten:1985is,Hull:1987pc,Ketov})
\be \la{BMT} \begin{aligned}
&\ddt(\GG_{mn} + \BB_{mn}) = \b_{mn} + L_\xi (\GG+\BB)_{mn}  +  \del_{[m}\l_{n]}  \ ,  \\
&\b_{mn}= \a' \, \widehat R_{mn} +  \a'^2 \, \tfrac{1}{2} \Big[\widehat R^{klp}{}_n \widehat R_{mklp} - \tfrac12 \widehat R^{ l p k}{}_n \widehat R_{mklp} 
+\tfrac12 \widehat R_{k mn l} \HH^{k pq} \HH^ {l}_{\ pq} \Big] + \ldots \ ,
\end{aligned} \ee
where $\HH_{kmn} = 3 \del_{[k} \BB_{mn]}$ and $\widehat R$ is the curvature of the generalized connection \mbox{$\widehat{\G}^k{}_{mn} = \G(\GG)^k{}_{mn} - \ha \HH^k{}_{mn}$.} The use of $\widehat{R}$ is simply a convenient notation and one can re-express \rf{BMT} in terms of just $R$, $\HH$ and covariant derivatives: for example, $\widehat{R}_{mn} = R_{mn} - \tfrac{1}{4} \HH_{mpq}\HH_n{}^{pq} - \ha D^k \HH_{kmn}$. Here $L_\xi$ denotes the Lie derivative along a vector $\xi$, generalizing the diffeomorphism term of \rf{2} above. The extra term $\del_{[m}\l_{n]}$ denotes a shift of the $\BB$-field by an exact 2-form depending on the 2d RG scale; this is the usual 2-form gauge freedom, and only modifies the \sm Lagrangian \rf{13} by a total derivative.

The string equations  \rf{se}  correspond to the Weyl-invariant case when the  RG flow \rf{BMT} trivializes, i.e. $\ddt (\GG_{mn}+\BB_{mn}) =0$. The diffeomorphism vector and B-gauge transformation must also take a specific form \cite{Tseytlin:1986ws} $\xi_n = 2 \a' \del_n \phi + \O(\a'^2)$, $\l_n = \O(\a'^2)$ in terms of the dilaton, metric and B-field. The dilaton equation in \rf{se} then follows from the metric and $\BB$-field equations up to a constant term, whose correct value ensures the vanishing of the Weyl anomaly coefficient.

\medskip 
More precisely, we may formulate the conjecture of RG stability as follows. For a family of  integrable  models $\GG_{mn}=\GG_{mn}(g_\a)$, $\BB_{mn}=\BB_{mn}(g_\a)$  parametrized by a few couplings $g_\a$, the claim is that the  RG flow \rf{BMT} of the metric and $\BB$-field consistently truncates to a flow of just  the couplings $g_\a$,
\be
\ddt g_\a = \b_\a(g) \ .
\ee
In other words, integrable \sms are   \textit{renormalizable}, i.e.\ they are  explicit solutions of the generalized Ricci flow equation.  

Let us give an intuitive explanation for this conjecture. As we shall discuss below, classically integrable \sms feature an infinite set of commuting conserved charges, as part of an intricate Poisson structure governed by a so-called Lax operator. In general one may expect this structure to be sufficiently constraining to uniquely determine the theory in question, so that the only deformations preserving it are the coupling constants $g_\a$ of the theory. Now assuming the integrable structure to exist also in the quantum theory, then it should be respected by the quantum RG flow, and hence only the couplings $g_\a$ should run under  RG. Another explanation is that some integrable $\s$-models are understood to have exact S-matrices in the infrared, typically with a small number of parameters (see Section \ref{QIS} below). One may then expect that these parameters should somehow match up with the semiclassical $\s$-model description, and that this matching should be preserved along the RG flow.

While there are no known counterexamples
to the conjecture that  integrable $\s$-models are renormalizable, 
 in most examples this has only been checked to the leading 1-loop order in \rf{BMT},
 and there is no general proof. In this work, we shall ask the question of what happens at \textit{higher loop orders.} We will find that the target space geometry of the classical \sm $\L = (\GG+\BB)_{mn}^{(0)} \, \del_+ x^m \del_- x^n$ must generally be deformed by quantum corrections to remain a solution of the RG equation \rf{BMT} beyond 1-loop,
\begin{align}
(\GG+\BB)_{mn}^{(0)} \to (\GG+\BB)_{mn}^{(0)} + \a' (\GG+\BB)_{mn}^{(1)} + \ldots  \  . \la{qc}
\end{align}
Equivalently, one must add to the $\s$-model a particular choice of finite counterterms.
This is a scheme-independent statement because  the required counterterms will be non-trivial in \textit{arbitrary} covariant subtraction schemes. 

Hence, while the classical Lax integrability corresponds just to the classical (uncorrected) geometry $(\GG+\BB)_{mn}^{(0)}$, 
 it is the quantum-corrected geometry that seems to posses the RG properties associated with integrability at the quantum level.
These counterterms might then be understood as preserving `Ward identites' for hidden integrable symmetries at the quantum level, with the expectation that these hidden symmetries are constraining the RG flow. In this sense, one may hope to interpret the corrected \sm as being the correct choice of bare action consistent with quantum integrability.

\section{Structure of the thesis}

This thesis is structured as follows. The rest of  this introductory chapter is a review of integrable  2d $\s$-models with a focus on classical Lax integrability, examples of integrable  models and their quantum RG flows.

\medskip
In Chapter \ref{QCS} (based on the papers \cite{HLT1,HLT2}) we shall consider the higher-loop RG flow for some of the simplest non-trivial `integrable deformed' models where renormalizability does not immediately follow from manifest global symmetries.
For the $\eta$-deformation of $S^2$ (``sausage model'') or $H^2$, we explicitly construct the counterterms required for RG stability up to 3-loop order and conjecture the all-order form of the counterterms. We also study the closely related $\l$-deformation based on  the $SU(2) \ov U(1)$ coset and the $SU(2)$ group, finding the counterterms needed for 2-loop RG stability. We  point out that these counterterms continue to serve the same function of restoring renormalizability in the   non-abelian dual limit of the $\l$-model. This gives a  prescription in these examples for the leading $\a'$-corrections to the non-abelian duality transformation so that the non-abelian duality commutes with the RG flow, resolving prior confusion in the literature.

\medskip
In Chapter \ref{GGS} (based on \cite{LT2103}) we consider the same question for a different class of integrable $\sigma$-models  on products of group spaces. We    first  study  the integrable $G \times G$   model   derived from the  affine Gaudin construction  (for which the 1-loop $\beta$-functions were found in \cite{today})  and show that its condition of integrability  is  automatically  preserved also  by the 2-loop  RG flow without the need for any quantum corrections.  We  then   investigate the RG flow   in the gauged $G\times G/H$  model, including  the integrable  $T^{1,1}$    model  found in \cite{ABL}. In that case we instead find that the integrability condition relating the couplings requires a certain quantum deformation to preserve renormalizability at 2-loop order. We also  construct   a   new   class of integrable $G\times G/H$ models in the case  when the subgroup  $H$ is abelian. In the  simplest case of $G=SU(2),\ H=U(1)$  this  leads  to an integrable $\sigma$-model on the   $T^{1,q}$  space  (with a particular $\BB$-field). This  model is found to be a special case that is stable under the 2-loop  RG flow without deformation --- and we  relate this   property to its  invariance  under T-duality in an isometric $U(1)$ direction. This  $T^{1,q}$ model   may be interpreted as an integrable  deformation of the GMM model  (of two coupled WZW  theories  with generic levels) away from the conformal point.

\medskip
In Chapter \ref{CT} (based on \cite{HLT2}) we derive an alternate formulation of the $\lambda$-model as a \sm on a ``tripled'' configuration space $G\times G \times G $ by re-writing the 2d gauge field in terms of group-valued scalar fields.
 Our central observation is that the $\l$-model quantized in this formulation is automatically renormalizable to all orders without any correction,  with only the deformation parameter $\lambda$ running. 
 This is in contrast to the standard $\sigma$-model found by integrating out $A_\pm$, which is 2-loop renormalizable only after adding the specific finite local counterterms discussed in Chapter \ref{QCS}.
 We compute the 2-loop $\beta$-function of the $\lambda$-model for general groups
and symmetric spaces.
In the non-abelian dual limit this  yields an alternate formulation
of the interpolating models for 
 non-abelian duality
 that is automatically renormalizable with
 the same 2-loop $\beta$-function as the original PCM/symmetric space $\s$-model.
 
 Section \ref{1910TO} contains unpublished work on the construction of similar ``tripled'' or ``enlarged'' formulations of   (i) a generalization of the  $\l$-model to  product spaces  $G^N$  and (ii)  $G\times G/H$ models.
 
\medskip

 In Chapter \ref{loc} (based on \cite{HLT2008}) we study a new and different link between integrability and the RG flow. We consider several classes of integrable $\sigma$-models (on groups and symmetric spaces, $\eta$-models, $\lambda$-models) with \textit{local couplings} that may depend on the 2d coordinates, e.g.\ on time $\tau$. We observe that (i) starting with a classically integrable 2d $\sigma$-model, (ii) formally promoting its couplings $h_\alpha$ to functions $h_\alpha(\tau)$ of 2d time, and (iii) demanding that the resulting time-dependent model also admits a Lax connection implies that $h_\alpha(\tau)$ must solve the 1-loop RG equations of the original theory with $\tau$ interpreted as RG time. This provides a novel example of an `integrability--RG flow' connection. The existence of a Lax connection suggests that these time-dependent $\sigma$-models may themselves be understood as integrable. We investigate this question by studying the possibility of constructing non-local conserved charges (see also Section 5 of \cite{HLT2008} for attempts to construct local charges). Such $\sigma$-models with $D$-dimensional target space and time-dependent couplings subject to the RG flow naturally appear in string theory upon fixing the light-cone gauge in a $(D+2)$-dimensional conformal $\sigma$-model with a metric admitting a covariantly constant null Killing vector and a dilaton linear in the null coordinate. This suggests the possibility of a large, new class of solvable string models.
 
\medskip

In Chapter \ref{cfd} we  summarize the findings of this work and comment on the open questions, possible applications and future prospects.

 \

\noindent Appendix \ref{conv} is an explanation of the conventions and notation used in this thesis.

\medskip

In Appendix \ref{A}  (based on \cite{HLT1}) we discuss the effect of scheme change on the $\s$-model $\b$-functions, focussing  on the particular case with a 2d target space (relevant for the $SU(2)\ov U(1)$ $\eta$- and $\l$-models considered in Chapter \ref{QCS}).  
 
\medskip
 
In Appendix \ref{2103A} (based on \cite{LT2103}) we consider deriving the classical integrability condition for the couplings of $G^N$ models (discussed in Chapter \ref{GGS}) from first principles. We compare our results with the sufficient conditions for integrability obtained from affine Gaudin models. For general $N$ we find that the dimension of the integrable space of $G^N$ couplings coincides with that of the space of affine Gaudin models, and for the $G\times G$ case we find that the affine Gaudin models are the only integrable ones.

\medskip

In Appendix \ref{2103B}  (based on \cite{LT2103})  we  provide the explicit formulae for the 2-loop
$\b$-functions of the $G\times G$ and $G\times G/H$ models (used in Chapter \ref{GGS}) and explain how they were derived.

\medskip

The remaining appendices, based on \cite{HLT2008}, pertain to the `local coupling' theories discussed in Chapter \ref{loc}. Appendix \ref{2008A} contains details of the derivation of the RG flows from the condition of existence of Lax representations for the local coupling theories. 

Appendix \ref{2008om} is about the construction of non-local charges for time-dependent $\s$-models other than the PCM (which was discussed in Section \ref{2008s4}) --- and in particular the time-dependent $\l$-model based on a symmetric space. 

In Appendix \ref{2008C} we consider 1d harmonic oscillators with time-dependent frequency,  a special case of which was obtained in eq.\ \rf{2008lt} as a linearization of a 1d reduction of the time-dependent PCM.

 In Appendix \ref{2008sing} we explain that the sine-Gordon model also obeys the same pattern as the $\s$-models discussed above (relating its integrability with  local couplings to its RG flow).

\section{Review of key concepts \la{rev}}
In the remainder of this introductory chapter, we shall review some concepts priming the reader for the rest of the thesis.   Let us highlight Section \ref{fv}, which  is a summary of our paper \cite{HLTS} attempting to use the tree-level $\s$-model S-matrix  as  another `diagnostic tool' for integrability.

\subsection{Classical integrability \la{ci}}
Most models in physics cannot be  solved exactly. One example is the Standard Model of particle physics, where there is no hope to make exact predictions, but rather computations are made by expanding perturbatively around a free theory.
A simpler example is a many-body system in Newtonian gravity, where only the 2-body case is solvable and the 3-body and higher cases are chaotic. 
 One could ask: \textit{what goes right} for the 2-body problem? One explanation is that the 2-body problem  is integrable\foot{See also the book \cite{Babelon} for an introduction to classical integrability.} while the others are not. For mechanical 1d systems, integrability means there exist at least as many independent, Poisson-commuting conserved charges as the number of degrees of freedom. In that case it follows that the theory can be completely solved by quadratures. 

For 2d models, and in particular the  \sms that will be our focus, there are infinitely many degrees of freedom. The analogous notion of integrability in 2d is then related to the presence of infinitely many commuting conserved charges. A theory of exactly integrable PDEs in mathematical physics was developed starting from the 1960s \cite{kdv,lax,zs,zm}. The flat Lax connection is a central object, whose zero-curvature representation \rf{zc} expresses the relevant non-linear equation as the consistency condition for an over-determined \textit{linear} system
\be
\del_\pm \psi = \psi \, L_\pm  \ ,
\ee
leading to the construction of explicit  solutions by the ``inverse-scattering'' method.

 If a 2d field theory admits a Lax representation \rf{zc}, then one can immediately build non-local conserved charges using the monodromy of the Lax connection along a constant-time contour from $\s=a$ to $\s=b$,
\be
\M = \mathcal P \exp \int_a^b  L_\s \, d\s  \ . \la{Md}
\ee
The path-ordered exponential in \rf{Md} is defined by the properties
\be
 \M^{-1} \del_b \M = L_\s \big|_{\s=b}  \ ,    \qquad \M \big|_{a=b} =1 \ . \la{px}
\ee
On the support of the equation of motion \rf{zc} of zero-curvature form, it satisfies
\be
\del_\t \M = \M \, L_\t (b) - L_\t(a) \, \M \ . \la{ie}
\ee

Assuming a periodicity condition $L_\t(a)=L_\t(b)$ (e.g., following from periodic boundary conditions on the fields),
 then \rf{ie} becomes  $\del_\t \M  = [\M , L_\t (a)]$ and hence
 \be
 \del_\t  \Tr[ \M^n] = 0   \ , \qquad n \in \mathbb{N} \ ,
 \ee
 i.e.\ the eigenvalues of $\M$ are conserved.

If additionally the boundary value is vanishing, i.e. $L_\t(a)=L_\t(b)=0$  (for example with decaying boundary conditions on the plane  $(a,b)=(-\infty,\infty)$), then \rf{ie} gives  the stronger result that all the entries of $\M$ are conserved, 
\be
\del_\t \M = 0 \ .
\ee

In either case, the number of conserved charges is actually infinite since the entries of $\M$ are  functions of the `spectral parameter' $z$ (assuming the dependence is non-trivial) so each coefficient is individually conserved in any series expansion in $z$. These charges are  `non-local' since the monodromy matrix is a non-local object (it is not an integral of a local density).

We note that a Lax connection valued in the algebra of $G$ is determined up to gauge transformations $L_\pm \to k^{-1} L_\pm k + k^{-1} \del_\pm k$,  \ $k(\xi) \in G$, since this transformation preserves the set of equations encoded in the zero-curvature condition \rf{zc}. The monodromy matrix \rf{Md} rotates covariantly under such transformations so, assuming periodicity of the gauge parameter $k$, the set of non-local  charges is preserved.

\subsection{Principal Chiral Model}
Let us consider the simple example of the Principal Chiral Model (PCM),\foot{We denote the  2d indices by $\m,\n$ and define the 2d light-cone coordinates as $\xi^\pm = \ha (\t \pm \s)$ with derivatives $\del_\pm = \del_\t\pm \del_\s$. The $D$-dimensional target space indices are $m,n=1,\ldots, D$. See Appendix \ref{conv} for conventions.}
\be
\L = -\ha \h \,  \Tr[ J_+ J_- ]  \ ,  \qquad J_\m = g^{-1} \del_\m g \ , \quad g\in G \ . \la{PL}
\ee
This is a special case of the general \sm \rf{13} with the target space taken to be a simple Lie group $G$ with the bi-invariant metric (and vanishing $\BB$-field). The model has a large global $G\times G$ symmetry corresponding to  left- and right-multiplications of the group,
\be
g \to g_L \ g \ g_R \ , \qquad \qquad (g_L,g_R)\in G_L \times G_R \ . \la{gps}
\ee
For this model, the global symmetry \rf{gps} (or isometries of the target space) is sufficient to uniquely fix the form of the Lagrangian \rf{PL} up to the overall constant $\h$. Hence the \sm RG equations \rf{2} are solved with just $\h$ running according to the 2-loop $\b$-function \cite{McKane:1979cm,Hikami:1980hi,friedan},\foot{In this PCM case, since the target space is isotropic, there cannot be any diffeomorphism vector $\xi$ in \rf{2} consistent with the symmetries.}
\be
\ddt \h = \cg+\ha \cg \h^{-1} + \ldots \ .  \la{Prg}
\ee
The constant $\cg$ denotes the dual Coxeter number of the group $G$ (see Appendix \ref{conv} for conventions).

The key to noticing the model's classical integrability is that the current $J_\m$ is conserved on-shell, $ \del^\m J_\m = 0$, and also flat, $dJ + J\wedge J = 0$ (since $J_\m=g^{-1} \del_\m g$). This information can be encoded in the flatness \rf{zc} of the Lax connection \cite{zm},
\be
L_\pm \equiv L_\t \pm L_\s = \ha (1+z^{\pm 1}) \, J_\pm  \ . \la{pl}
\ee
An easy way to evaluate the monodromy matrix and extract conserved charges is by expanding in $z$ around the value $z=-1$ where $L_\s$ vanishes (at which point $\M=I$ is trivial). The expansion of the path-ordered exponential yields first the conserved Noether charge built from $J_\m$ itself (corresponding to the $G_R$ symmetry) and then iterated integrals of $J$ of increasing order,
\begin{align}
&\M =I + \ha (z+1) \int_a^b d\s \, J_\t \la{me} \\
&\qquad\ \ \ \
+ \tfrac{1}{4} (z+1)^2 \Big[ \int_{a<\s_1<\s_2<b} d\s_1 d\s_2 \, J_\t(\s_1)J_\t(\s_2) + \int_a^b d\s \,(J_\t-J_\s) \Big]
+ \O\big((z+1)^3\big) \no
\ .
\end{align}
Each term in the expansion (or just their eigenvalues, depending on the boundary conditions) is individually conserved by the argument above, and their  Poisson brackets satisfy a Yangian algebra \cite{Bernard:1990jw}.

\bigskip
As well as these non-local charges, integrable field theories may posses higher-spin local charges. For the example of the PCM \rf{PL}, 
one can build holomorphic and anti-holomorphic local currents by contracting products of $J_\pm$ with a symmetric invariant tensor  \cite{Evans:1999mj},
\begin{align}
\del_\pm \mathcal{J}_\mp^{(n)} = 0 \ , \qquad \qquad &\mathcal{J}_\pm^{(n)} = 
d_{A_1 \cdots A_n} J_\pm ^{A_1} \cdots J_\pm^{A_n} \ \la{cur} \\
&d_{A_1 \ldots A_n} = d_{(A_1 \ldots A_n)} \ , \qquad\qquad
{f^{A}}_{B(C} d_{A_1 \cdots A_{n-1}) A_n}= 0 \ .\la{e3}
\end{align}
The $n$th currents $\mathcal{J}_\mp^{(n)}$ are of spin $n$  and their (anti-)holomorphic conservation implies the conservation of the associated conserved charges,
\be
\del_\t Q_{\pm}^{(n)} = 0 \ , \qquad \qquad Q_\pm^{(n)} = \int d\s \, \mathcal{J}_\pm^{(n)} \ ,
\ee
which Poisson-commute with the non-local charges above \cite{Evans:1999mj}.

A canonical choice for the invariant
tensor $d_{A_1 \cdots A_n}$ is given by the trace,
$d_{A_1 \cdots A_n} = \Tr[ T_{(A_1} \cdots T_{A_n)} ] $.
The conservation of the currents \rf{cur} follows from the equation of motion and flatness of $J_\m$, written in the 2d light-cone coordinates as
\be
\del_+ J_- + \del_- J_+ = 0 \ , \qquad \del_+ J_- - \del_- J_+ + [J_+, J_-] = 0 \ , \la{pe}
\ee
since these may be re-written as
\be
\del_+ J_- = -\ha [J_+, J_-] \ , \qquad \del_- J_+ = \ha [J_+, J_-] \ .
\ee

\subsection{Symmetric space sigma model \la{sssm}}
Now let us consider instead a \sm on a homogeneous coset space $G/H$, obtained  by gauging a subgroup $H$ of the $G_R$ symmetry of the PCM (here $H\subset G$ are both assumed to be simple Lie groups). The gauging procedure may be understood as (i) promoting the $H$ global symmetry to a local one, $g \to g \, h(\xi)^{-1}$, (ii) writing down the transformation of the PCM Lagrangian \rf{PL}, $\L \to -\ha \hh \,  \Tr[(J_+ - h^{-1}\del_+h)(J_-  - h^{-1}\del_-h)]$, and (iii) promoting the gauge parameter to a gauge connection, $h^{-1} \del_\pm h \to A_\pm$ (no longer fixed to be pure gauge). The resulting $G/H$ theory,
\be
\L = -\ha \h \,  \Tr[ (J_+ - A_+) (J_- - A_-) ]  \ , \qquad J_\pm = g^{-1} \del_\pm g \ , \ \  g\in G  \ , \quad  A_\pm \in \Lie(H) \la{ss1} \ ,
\ee
is invariant under the $H$ gauge transformation
\be
g \to g h \ , \quad A_\pm \to h^{-1} A_\pm h + h^{-1} \del_\pm h \ , \qquad h(\xi) \in H  \ .\la{hg}
\ee

The quadratic gauge field $A_\pm$  in \rf{ss1} is auxilliary (appearing without derivatives) and may be integrated out  to  obtain
\be
\L = -\ha \h \,  \Tr[ J_+\,  P_{G/H} \, J_-]  \ , \la{ss2}
\ee
where $P_{G/H}$ is the projector orthogonal to  $\Lie(H)$ in $\Lie(G)$ with respect to the Killing form. The $H$ gauge transformation \rf{hg} then reduces to
\be
g \to g  \, h \ , \qquad h(\xi) \in H  \ .
\ee
The theory also has a global $G$ symmetry, $g \to g_L \, g$,  \ $g_L \in G_L$ corresponding to the isometry of the coset space  $G/H$. 

If $G/H$ is assumed to be a \textit{symmetric space} (meaning $\Lie(H)$ is the even subalgebra under a $\mathbb{Z}_2$-grading of $\Lie (G)$), then these global and local symmetries uniquely determine the theory \rf{ss1} or \rf{ss2} up to the constant $\h$.
Like the PCM above, this automatically makes the theory renormalizable to all orders with just $\h$ running. The 2-loop $\b$-function following from \rf{2} is (see App.\ B of \cite{HLT2} and particular cases in \cite{Brezin:1975sq,Hikami:1980hi})
\be
\ddt \h = 2 \cg + 4 \cg(\cg-\cf) \h^{-1}  \ . \la{sb}
\ee
In contrast, for non-symmetric cosets $G/H$, the theory \rf{ss2} may not be renormalizable as
it  should be understood as part of a multi-coupling theory $\L = -\ha \Tr[ J_+ \, \big( \sum_{i=1}^n \h_i P_{\mathfrak{p}_i} \big) \, J_-]$ respecting the same symmetries, defined in terms  a decomposition of the Lie algebra
$\Lie(G) = \Lie(H) + \sum_{i=1}^n \mathfrak{p}_i$, with $\mathfrak{p}_i$ being irreducible representations under the adjoint action of $H$, mutually orthogonal under the trace. For general cosets we may have $n>1$ (i.e.\ multi-coupling),  while  symmetric spaces must have $n=1$ (single-coupling) if $G$,$H$  are simple.

The symmetric space condition on $G/H$ is also sufficient for classical integrability of the model \rf{ss2}. The corresponding equations of motion are then of the zero-curvature form \rf{zc} with the Lax connection \cite{Eichenherr:1979ci}
\be
L_\pm =B_\pm  + z^{\pm 1} \, P_\pm \ , \qquad  B \equiv P_H  J , \ \ P \equiv P_{G/H} J \la{sll}
\ee
Let us note that the symmetric space is actually a generalization of the PCM, since the PCM on a group $G$ is equivalent to the \sm on the symmetric space $G\times G/G$ (and is obtained upon fixing the gauge $g^{(2)}=1$, where $(g^{(1)},g^{(2)})\in G\times G$, and replacing $\h \to 2\h$). All of the formulas here for symmetric spaces actually apply in that case, except that $G\times G$ is not simple so the definition of $\cf$ in Appendix \ref{conv} should be suitably modified. Indeed the symmetric space $\b$-function \rf{sb} agrees with the PCM one \rf{Prg} upon replacing $\cf \to \ha \cg$ and $\h \to 2\h$. The symmetric space Lax connection \rf{sl} in that case is related to the PCM one \rf{pl} by a gauge transformation.

Both non-local \cite{lp} and local \cite{Evans:2000qx} charges may be constructed similarly to the PCM case above. For the local charges, one now replaces $J \to P$ in  \rf{cur},
\begin{align}
\del_\pm \mathcal{J}_\mp^{(n)} = 0 \ , \qquad \qquad &\mathcal{J}_\pm^{(n)} =
d_{A_1 \cdots A_n} P_\pm^{A_1} \cdots P_\pm^{A_n}   \la{cur2} \\
&d_{A_1 \ldots A_n} = d_{(A_1 \ldots A_n)} \ , \qquad\qquad
{f^{A}}_{B(C} d_{A_1 \cdots A_{n-1}) A_n}= 0 \ .\la{e32}
\end{align}
The holomorphic conservation law in \rf{cur2} follows again from the equation of motion and flatness of $J$  \cite{Evans:2000qx} (here $D^{B}_\pm \equiv \del_\pm + \ad_{B_\pm}$),
\begin{align}
 \qquad &D^B_+ \, P_- + D^B_- \, P_+ = 0 \ , \qquad \del_+ J_- - \del_- J_+ + [J_+, J_-]  = 0 \ ,
\end{align}
since they now  imply that
\begin{align}
\qquad &\del_\pm \  P_\mp = [P_\mp, B_\pm ]  \ . 
\end{align}

\subsection{Constructing  Lax connections for other models}
The integrable models that follow will admit the  \textit{same} flat Lax connections of the form \rf{pl} or \rf{sll} as the PCM or symmetric space (for models respectively built on groups 
 or  symmetric spaces).\foot{Of course, not all integrable $\s$-models will have Lax connections of this form. For example the bi-Yang-Baxter model  \cite{Klimcik:2008eq,Ki} built on a group does not possess a flat, conserved current $\mc{A}_\pm$ so its Lax connection is not of the form \rf{gl}.} The difference is that, in each case, $J_\pm$ will be replaced with some other model-dependent quantity in terms of the fields,
 \begin{alignat}{3}
&G \ : \qquad\quad & &L_\pm = \ha (1+z^{\pm 1}) \mc{A}_\pm \la{gl} \ ,\qquad \qquad & & \mc{A}_\pm \in \Lie{(G)} \\
&G/H \ : \qquad\quad & &L_\pm = \mc{B}_\pm + z^{\pm 1} \mc{P}_\pm \la{sl} \ ,\qquad \qquad & & \mc{B}_\pm \in \Lie{(H)} \ , \ \ \mc{P}_\pm \in \Lie{(G)}/\Lie{(H)}
\end{alignat}
 We note that admitting Lax connections of the same form does not imply  classical equivalence between these different theories (not being related by a canonical transformation); e.g. for group-based models, it simply means they possess a flat, conserved current $\mc A_\pm$.

Non-local conserved charges are then obtained from the monodromy matrix as discussed above for the PCM. Local conserved charges are also constructed similarly to the cases discussed above, with the general form of the (anti-)holomorphic currents given by
\begin{align}
\del_\pm \mathcal{J}_\mp^{(n)} = 0 \ , \qquad \qquad &\mathcal{J}_\pm^{(n)} = \begin{cases}
d_{a_1 \cdots a_n} \mc{A}_\pm ^{a_1} \cdots \mc{A}_\pm^{a_n} \ & \text{(group} \ G) \ \la{curG}\\
d_{a_1 \cdots a_n} \mc{P}_\pm ^{a_1} \cdots \mc{P}_\pm^{a_n} \ & \text{(symmetric space}\ G/H) \
\end{cases} \\
&d_{a_1 \ldots a_n} = d_{(a_1 \ldots a_n)} \ , \qquad\qquad
{f^{a}}_{b(c} d_{a_1 \cdots a_{n-1}) a}= 0 \ .\la{e3G}
\end{align}
The conservation of the currents \rf{curG} follows from the equations of motion
\begin{align}
G \ : \qquad &\del_+ \mc{A}_- + \del_- \mc{A}_+ = 0 \ , \qquad F_{+-}(\mc{A}) = 0 \ , \\
G/H \ : \qquad &D^\mc{B}_+ \mc{P}_- + D^\mc{B}_- \mc{P}_+ = 0 \ , \qquad F_{+-}(\mc{B} + \mc{P}) = 0 \ ,
\end{align}
since these may be re-written as
\begin{align}
G \ : \qquad &\del_\pm \mc{A}_\mp = \ha [A_\mp, A_\pm] \ , \la{ni}\\
G/H \ : \qquad &\del_\pm \mc{P}_\mp = [P_\mp, B_\pm ] \ , \ \ \ \qquad \ \ F_{+-}(\mc{B}) + [\mc{P}_+, \mc{P}_-] = 0 \ . \la{ni2}
\end{align}

\subsection{T-duality and non-abelian duality \la{TS}}
For \sms with global symmetries, or isometries of the target space, one may ``dualize'' the symmetry to obtain a dual \sm\unskip, with certain observables being related between the original and  dual theories (see the review \cite{Alvarez:1994dn} and refs.\ therein). As we shall review below, the duality procedure may be carried out at the level of the path integral and applies to the quantum theory \cite{Buscher:1987qj}; in particular duality is understood to commute with the quantum RG flow, and to map Weyl-invariant  theories (string solutions) to Weyl-invariant theories. However, we stress that the procedures as discussed below are subject to quantum $\a'$ corrections, with the integrals producing quantum determinants.

First we shall consider \textit{abelian T-duality} for a model with a $U(1)$ isometry, which we take for simplicity to be
\be
\L = g_{ij}(x)  \, \del x^i \del x^j + f(x)  \, (\del y)^2 \la{or}
\ee
 Here the isometry corresponds to constant shifts of $y$, and the $x$ coordinates are ``spectators''. Gauging the isometry $\del_\m y \to \del_\m y -  p_\m$ and imposing the condition  $\epsilon^{\m\n} \, \del_\m p_\n = 0$ using a Lagrange multiplier field $\tilde{y}$ gives a gauge-invariant theory (under $y \to y + u$, $p_\m \to p_\m + \del_\m u$),
\be
\L_{\rm int} = g_{ij}(x)  \, \del x^i \del x^j + f(x) \,  (\del y -  p)^2 + \tilde{y} \, \epsilon^{\m\n}\, \del_\m p_\n  \ . 
\ee
Fixing the gauge $y=0$ gives the ``interpolating model'',
\be
\L_{\rm int} = g_{ij}(x)  \, \del x^i \del x^j + f(x) \,  p^2 + \tilde{y} \, \epsilon^{\m\n}\, \del_\m p_\n  \ . \la{int}
\ee
On one hand, integrating out $\tilde{y}$ imposes that $p_\m$ is pure gauge and locally given by $p_\m = \del_\m y$, giving back the original model \rf{or}. On the other hand, we may instead integrate out the non-dynamical quadratic field $p$ (after integrating by parts $\tilde{y} \, \epsilon^{\m\n} \,  \del_\m p_\n \to -\del_\m \tilde{y} \, \epsilon^{\m\n} \,  p_\n)$) to get the dual \sm with the $(x,\tilde y)$ directions
\be
\tilde{\L} = g_{ij}(x)  \, \del x^i \del x^j + \frac{1}{f(x)}  \, (\del \tilde y)^2  \ .
\ee
The dual theory still has a $U(1)$ isometry, corresponding now to shifts of $\tilde{y}$. In summary, the dual pair of models are related by $f \to \tfrac{1}{f}$, $y \to \tilde{y}$, and they were obtained by integrating out different fields from the interpolating theory \rf{int}. This procedure is easily generalized to cases with off-diagonal metric terms and $\BB$-fields, with the resulting transformation being known as the `Buscher rules' \cite{Buscher:1987qj}.

Now let us consider \textit{non-abelian duality} (NAD), where the above procedure is generalized to  non-abelian isometries \cite{Fridling:1983ha,Fradkin:1984ai,delaOssa:1992vci}. We shall focus on the examples of the PCM and symmetric space \sm (considered above), and the procedure is the same for other theories with non-abelian global symmetries. Starting from one of these  theories
\be
\L = -\ha h \, \Tr[ J_+ \, P \, J_- ]  \ , \qquad \quad P=\begin{cases}1  \qquad & \text{(PCM)} \ ,  \\ P_{G/H} \ \qquad & \text{(symmetric space)} \ , \end{cases}  \la{on}
\ee
we shall dualize the $G_L$ symmetry $g \to k g$. As above, we begin by gauging the symmetry and imposing that the gauge field is pure gauge with a Lagrange multiplier field $v$ (here $F_{+-}(A) \equiv \del_+ A_- - \del_- A_+ + [A_+, A_-]$)
\be
\L = -\ha \h \, \Tr[ (J_+ + g^{-1} A_+ g) \, P \, (J_- + g^{-1} A_- g)  + v \, F_{+-}(A)] \ .
\ee
This resulting theory is invariant under the gauge transformation $g \to k g$, $A_\pm \to k A_\pm k^{-1} - \del_\pm k k^{-1}$. Fixing the gauge $g=1$, we obtain the interpolating theory
\be
\L_{\rm int} =  -\ha \h \, \Tr[  A_+  \, P \,  A_- - v \, F_{+-}(A)]  \ . \la{NADi}
\ee
On one hand, integrating out $v$ imposes that $A_\pm$ is flat and  locally of the form $A_\pm = g^{-1} \del_\pm g$, giving back the original theory \rf{on}. On the other hand, after integrating by parts, $A_\pm$ appears quadratically without derivatives; integrating it out yields the NAD \sm depending on $v$,
\be
\tilde{\L} = -\ha \h \, \Tr[  \del_+ v \, \frac{1}{P+ \ad_{v}} \, \del_- v] \ . \la{NADt}
\ee
Unlike abelian T-duality, non-abelian duality destroys the original symmetry, meaning the original $G_L$ symmetry is not present in the NAD theory \rf{NADt}. Thus, at least within the framework of non-abelian duality, one cannot ``dualize back'' from the NAD model to the original one.\foot{In fact the $G_L$ symmetry becomes a Poisson-Lie symmetry of the NAD model \rf{NADt} (see eq.\ \rf{PLs} below) and, starting from the NAD theory, one can dualize back to original theory by Poisson-Lie duality.}

The path-integral derivations of  abelian and non-abelian duality apply in particular at the level of the  classical equations of motion (they may be formulated as canonical transformations; see, e.g., \cite{Sfetsos:1996xj} and refs.\ therein). Hence these dualities preserve the property of classical integrability. For example the NAD of the PCM (\rf{NADt} with $P=1$) inherits a flat Lax connection of the same form \rf{gl} as the PCM with $\mc A_\pm=A_\pm(v)$ taking its value obtained in the quadratic integral in \rf{NADi},
\be
L_\pm = \ha (1+z^{\pm 1}) A_\pm(v) \ , \qquad A_\pm(v) = \frac{1}{\ad_v \mp 1} \del_\pm v \ .
\ee

\subsection{PCM with Wess-Zumino term \la{PkS}}
The PCM \rf{PL} is not the most general $G$-valued \sm with   $G_L \times G_R$ global symmetry \rf{gps}. In fact, there is also a $\BB$-field coupling respecting this symmetry called the \textit{Wess-Zumino} (WZ) term, defined by \cite{Wess:1971yu,novikov,Witten:1983ar}
\begin{align}
\H\equiv d\BB =  \tfrac{1}{12} k \, \Tr[ J \wedge J \wedge J ] \ , \qquad  \BB \equiv \ha \BB_{mn} dx^m \wedge dx^n \ .
\end{align}
The free coefficient $k$ is called the `WZ level'.

One can check that $d\H = 0$, implying the local existence of a $\BB$-field satisfying $\H=d\BB$. Although $\BB$ is not globally defined (the expressions for $\BB$ in different coordinate charts may be related by gauge transformations $\BB \to \BB+ d\l$), the \sm action \rf{13} is well-defined up to an additive constant. Let us take the 2d space to be the Euclidean plane $\mathbb{C}$ (after analytic continuation), which we compactify to the Riemann sphere $S^2=\mathbb{C}\cup \{\infty\}$. Then a coordinate-invariant way to write down the WZ term's contribution to the action is\foot{The factor of $i$ in \rf{iWZ} results from the Wick rotation of the antisymmetric Levi-Civita tensor $\e^{\m\n}$ in the $\s$-model \rf{csm}.}
\be
S_{\WZ} =  \frac{i}{2\pi } \int_{B_3} \H \ , \la{iWZ}
\ee
where $B_3$ is the solid ball whose boundary is $S^2$. One is instructed to continue the fields in an arbitrary way from the boundary into the interior of $B_3$ (this is always possible since Lie groups have trivial 2nd homotopy group $\pi_2(G)=0$). Since $\H$ is closed then its integral \rf{iWZ} is independent of the choice of continuation of the fields, up to an overall topological `winding number' constant. For compact simple groups\foot{For non-compact groups there may be no ambiguity at all and $k$ may be any real number.} this gives an ambiguous contribution $\Delta S_{\WZ} = 2 \pi i n k$ with $n$ an arbitrary integer in the homotopy group $\pi_3(G)=\mathbb{Z}$, which can be shown to classify the configurations. The Euclidean path integral  $Z = \int Dx \, e^{-S[x]}$ is then well-defined as long as the  level $k$ is quantized to be an integer.

We shall denote by `PCM$_k$' the PCM with WZ term with level $k$ (with $\L_{\WZ}$ being a local density for $S_{\WZ}$),
\be
\L = -\ha \h \, \Tr[ J_+ J_- ] + k \, \L_{\WZ}(g) \ . \la{pk}
\ee
The PCM$_k$ is classically integrable for arbitrary values of the couplings $\h$ and $k$, with the Lax connection taking the same form \rf{gl} as the PCM in terms of a modified flat, conserved current \cite{Piette:1987bt}
\be
L_\pm = \ha (1+z^{\pm 1}) \mc{A}_\pm \ , \qquad \qquad \mc A_\pm = (1\pm \tfrac{k}{h}) J_\pm \ . 
\ee

It is a general fact that WZ levels do not run under the RG flow (even for non-compact groups where they are not quantized). This is because the antisymmetric $\BB$-field terms in the \sm $\b$-functions \rf{BMT} are all globally defined,  so cannot be proportional to  WZ terms  whose $\BB$-fields are not globally defined. The WZ term does, however, modify the RG flow of the coupling $\h$, and the PCM$_k$ has 2-loop $\b$-function
\be
\ddt \h = (1-\tfrac{k^2}{\h^2}) \big( \cg + \ha \cg^2 (1-3\tfrac{k^2}{\h^2}) + \ldots \big) \ , \qquad \ddt k = 0 \ . \la{wzwb}
\ee

\subsection{Wess-Zumino-Witten model \la{WS}}
The PCM$_k$ $\b$-function \rf{wzwb} admits a fixed point $\h= k$ (with the other fixed point $\h=-k$ related by 2d parity or $g\to g^{-1}$), which is in fact an  exactly conformal theory called the  \textit{Wess Zumino Witten} (WZW) model,
\be
k \,  \L_G(g) \equiv    k \big( -\ha  \, \Tr[J_+ J_-]  + \L_{\WZ}(g) \big) \ .
\ee
 For this special value of the coupling, the $G_L\times G_R$ symmetry enhances to a ``chiral gauge'' symmetry \cite{Witten:1983ar},\foot{In Minkowski signature $\xi^\pm$ in \rf{cgs} are the same 2d light-cone coordinates used above. In Euclidean signature these become complex-conjugate cooordinates $z$, $\bar z$ on the complex plane.}
\be
g \to g_L(\xi^-) \, g \, g_R(\xi^+) \ , \qquad  g_L(\xi^-) \in G \ , \quad g_R(\xi^+) \in G \ . \la{cgs}
\ee
At the WZW fixed point $h=k$, the equation of motion becomes the statement of holomorphic and anti-holomorphic conservation of the Noether currents corresponding to the chiral gauge symmetries \rf{cgs},
\be
\del_- J_+ = 0  \iff  \del_+ K_- = 0 \ , \qquad J_+ \equiv g^{-1} \del_+ g \ , \quad K_- \equiv  \del_- g g^{-1} \la{WZWe} \ .
\ee
This equation is explicitly solvable in terms of arbitrary $G$-valued (anti-)holomorpic functions,
\be
g = g_1(\xi^-) \,  g_2(\xi^+) \ ,
\ee
with the currents $J_+ = J_+(\xi^+)$ and $K_-  = K_-(\xi^-)$ being holomorphic and anti-holomorphic on-shell. Computing the currents' Poisson brackets and canonically quantizing, one finds that they satisfy the affine Kac-Moody algebra of $G$ with level $k$, with the corresponding OPE,
\be\begin{aligned}
J_+^a(z) J_+^b(z')  \sim {}&{} \frac{1}{k}\Big[\frac{\ka^{ab}}{(z-z')^2} + \frac{i f^{ab}{}_c J_+^c(z')}{z-z'} \Big] \qquad  (+\text{ regular terms})
\ . \la{ca}
\end{aligned}
\ee
One can show that the affine Kac-Moody algebra contains the Virasoro algebra, i.e.\ the current algebra OPEs \rf{ca} imply that a certain  local operator $T(z) \equiv \frac{k^2}{2(k+\cg)}\Tr[  : \!J_+ J_+ \! : (z) ]$ satisfies the Virasoro OPE,
\be
T(z) T(z') \sim \frac{c}{2(z-z')^4} + \frac{2T(z')}{(z-z')^2} + \frac{\del T(z')}{z-z'} \ .
\ee
The antiholomorphic currents $K_-$ satisfy the same Kac-Moody algebra and generate another copy of the Virasoro algebra. Hence the WZW model is exactly conformal, being invariant under the 2d conformal algebra (two copies of the Virasoro algebra). Correspondingly the $\b$-function \rf{wzwb} will vanish to all orders in a particular subtraction scheme (see the discussion around eq.\ \rf{2103Pkb} below). The corresponding conformal field theory, and particularly its correlation functions, were studied in \cite{Knizhnik:1984nr}.

\subsection{Gauged WZW model \la{gWZW}}
There is a closely related family of 2d CFTs based on coset spaces $G/H$ \cite{Goddard:1984vk} described by the \textit{gauged Wess Zumino Witten} (gWZW) Lagrangian \cite{Frishman:1992mr,Karabali},
\be \begin{aligned}
k \, \L_{G/H}(g,A) ={}&{} k \big( \L_G(g) + \Tr[ J_+ A_- - K_- A_+ + g^{-1} A_+ g A_- - A_+ A_-  ] \big) \ ,  \la{gws}\\
&J = g^{-1} d g \ , \quad  K = dg g^{-1} \ , \qquad g \in G \ , \quad A_\pm \in \Lie(H) \ .
\end{aligned} \ee
Like the symmetric space \sm  \rf{ss1}, the gWZW model depends on a quadratic 2d gauge field (that may be integrated out to obtain a \sm\unskip) and is gauge-invariant under an action of the subgroup $H$, 
\be
g \to h^{-1} g h \ , \quad A_\pm \to h^{-1} A_\pm h + h^{-1} \del_\pm h \ . \la{ggs}
\ee

The classical gWZW  $\s$-model \rf{gws} receives an infinite tower of $\a'$ corrections in order to be exactly conformal \cite{Dijkgraaf:1991ba,Bars:1992sr,Tseytlin:1992ri,bs,Tseytlin:1993my}. See the $SU(1,1)/U(1)$ example in eq.\ \rf{6} and the review in Section \ref{revg} below.

\subsection{Integrable \texorpdfstring{$\eta$}{eta}-deformation \la{etas}}
\sloppy
Starting from an integrable \sm like the PCM, one may consider squashing or deforming the target space geometry in a way that preserves the property of classical integrability. One such integrable deformation of the PCM, called the \textit{Yang-Baxter deformation}  or \textit{$\eta$-deformation}, is given by \cite{Klimcik:2002zj}
\be
\L = -\ha \h \, \Tr[ J_+ \, \frac{1}{1-\k R} J_- ]  \ , \la{ep}
\ee
where $\k$ is the deformation parameter. The deformation is built from a linear  operator \mbox{$R:\Lie(G) \to \Lie(G)$}, antisymmetric with respect to the Killing form (i.e. \mbox{$\Tr[x R(y)] = -\Tr[R(x) y]$}) and solving the modified classical Yang-Baxter equation,
\be
[Rx,Ry]-R([Rx,y]+[x,Ry]) + c^2 [x,y] = 0 \ , \qquad x,y\in \Lie(G) \ . \la{ybe}
\ee
With $R$ satisfying these conditions, the $\eta$-deformed PCM \rf{ep} is classically integrable with the Lax connection again taking the the same form \rf{gl} as the PCM, now built from a different flat, conserved current \cite{Klimcik:2008eq}
\be
L_\pm = \ha (1+z^{\pm 1}) \mc{A}_\pm \ , \qquad \qquad \mc A_\pm = -(1+\k^2) \Ad_g\frac{1}{1\pm \k R} J_\pm \ .  \la{lae}
\ee

One can check that the integrability of the model (i.e.\ the equivalence between the flatness of \rf{lae} and the equations of motion of \rf{ep}) relies on $R$ being a solution of the Yang-Baxter equation \rf{ybe}. In general there may be various solutions of the Yang-Baxter equation and their classification is unknown.  We shall focus on a particular class of canonical solutions
 in the ``non-split inhomogeneous'' case, i.e.\  taking $c^2>0$ in \rf{ybe}.\foot{We shall then rescale $R$ to set $c^2=1$. There are also other types of solutions with $c^2>0$.  The ``split inhomogeneous'' ($c^2<0$) solutions can only exist for non-compact groups and include  solutions of the same canonical type as \rf{DJR}; the  ``homogeneous'' ($c=0$) solutions include so-called ``abelian'' and ``Jordanian'' solutions. \la{ybs}} Given a choice of Cartan-Weyl basis $\{h_i, e_\a, e_{-\a}\}$  for $\Lie(G)$, where $h_i$ are the Cartan generators, $e_\a$ the positive roots and $e_{-\a}$ the negative roots, we shall assume a canonical choice of $R$-matrix (see Section 2.2 of \cite{Klimcik:2008eq} and refs.\ there) solving \rf{ybe},
\be
R(h_i) = 0 \ , \qquad R(e_\a) = - i e_\a \ , \qquad R(e_{-\a}) = i e_{-\a} \ . \la{DJR}
\ee
This $R$-matrix, and the associated $\eta$-deformed model \rf{ep}, depend on the choice of Cartan-Weyl basis. However, if one restricts to the case of compact simple groups, then there is only choice of Cartan-Weyl basis (up to equivalence by inner automorphisms).

The associated $\eta$-deformation in  \rf{ep}  breaks the manifest  global symmetry $G_L \times G_R$ of the PCM to the subgroup $G_L \times C$ where $C=U(1)^{r}$ is the Cartan subgroup of $G$ commuting with the $R$-matrix. The $G_R$ isometry of the target space is actually deformed to a  \textit{Poisson-Lie (PL) symmetry} \cite{Klimcik:2002zj}, with the associated Killing vectors $\{v^m_a \, \del_m \}_a$ being replaced by vectors satisfying	
\be
L_{v_c} (G+B)_{mn} =  - f^{ab}{}_c \, v_a^p  \, v_b^q \,  (G+B)_{mp} \, (G+B)_{qn}  \ . \la{PLs}
\ee
PL symmetry is thought to be related to a semi-classical limit of  quantum group symmetry \cite{Delduc:2016ihq}, and indeed the PL symmetries of the $\eta$-deformed models correspond to \textit{non-local} conserved charges  (that can be extracted from the monodromy matrix), the analogs of the leading Noether charges appearing in the expansion \rf{me} for the undeformed models. These non-local charges are found to have a classical Poisson bracket algebra obtained from a semi-classical limit of a quantum group algebra \cite{Delduc:2013fga,Delduc:2016ihq}. For this reason it is assumed that the full quantum $\eta$-deformed theories have quantum group symmetry, and this is the symmetry that has been assumed when bootstrapping their exact integrable S-matrices \cite{qds,Arutyunov:2015qva,ses}.

One can also simultaneously deform both the $G_L$ and $G_R$ parts of the global symmetry  to obtain the integrable `bi-Yang-Baxter' $\s$-model \cite{Klimcik:2008eq,Ki}.  The deformations can be generalized \cite{yosh,DMV}  to include WZ terms; additionally including `TsT transformations' \cite{betadef} gives a  4-parameter integrable deformation of the PCM \cite{Delduc:2017fib,Klimcik:2019kkf}.

Theories with  PL symmetry can be dualized in a process called \textit{Poisson-Lie duality} \cite{poissonlie},  generalizing non-abelian duality and abelian T-duality (see Section \ref{TS}). Under PL duality, the $\eta$-models \rf{ep},\rf{ess} are mapped  \cite{Hoare:2015gda,Klimcik:2015gba,Vicedo:2015pna,Sfetsos:2015nya,Hoare:2017ukq} to another class of integrable deformed models  called  $\l$-models, which we shall discuss in the next section.

A similar integrable $\eta$-deformation exists \cite{Delduc:2013fga} for the symmetric space \sm\unskip,
\be
\L = -\ha  \h  \, \Tr[ J_+ \, P_{G/H} \,  \frac{1}{1-\k \, R_g  P_{G/H}} \, J_-] \ , \qquad  \qquad R_g \equiv \Ad_g^{-1} R \, \Ad_g  \ . \la{ess}
\ee
The global $G_L$ symmetry of the symmetric space is thus deformed to a PL symmetry, with only a Cartan subgroup remaining as a manifest isometry (and the $H_R$ gauge invariance  remaining preserved).
The model is classically integrable with Lax connection of the same form \rf{sl} as the symmetric space in terms of different quantities $\mc B, \mc P$
\be \begin{aligned}
L_\pm = \mc{B}_\pm + z^{\pm 1} \mc{P}_\pm  \ , \qquad &\mc{B}_\pm = P_H A_\pm \ , \quad \mc{P}_\pm  = \sqrt{1+\k^2} P_{G/H} A_\pm  \\
&A_\pm = \frac{1}{1\pm \k R_g P_{G/H}} J_\pm 
\end{aligned} \ee

In agreement with the conjecture that integrable \sms are stable under the RG flow, the $\eta$-deformed models are known to be renormalizable at the leading 1-loop order with just the couplings $h$ and $\k$ running \cite{oneloopeta,Valent:2009nv,Squellari:2014jfa}\foot{The renormalizability and $\b$-functions for the coset case \rf{cos} have only been directly studied in special cases, e.g.\ in \cite{Sfetsos:2015nya}. However, Poisson-Lie duality has been shown to be a symmetry of the 1-loop $\b$-functions \cite{oneloopeta} and, given that the $\eta$-models are understood to be related to the $\l$-models by PL duality \cite{Hoare:2015gda,Klimcik:2015gba,Vicedo:2015pna,Sfetsos:2015nya,Hoare:2017ukq}, the expression \rf{cos} may be derived from the corresponding $\l$-model one.}
\begin{align}
&G \, : \qquad   && \ddt \h = \cg(1+\k^2)^2 \ , \qquad && \ddt \k = \cg \h^{-1} \k(1+\k^2)^2 \ , \la{gro}\\
&G/H \, : \qquad  && \ddt \h  = 2 \cg (1+\k^2) \ , \qquad && \ddt \k = 2\cg \h^{-1} \k(1+\k^2) \la{cos} \ .
\end{align}
This is a non-trivial example of the integrability--RG flow relationship because the $\eta$-deformed models are not \textit{automatically} RG-stable due to manifest symmetries.  Their stability is instead thought to be due to their hidden integrable symmetries.

\subsection{Integrable \texorpdfstring{$\l$}{lambda}-deformation \la{las}}
In \cite{Tseytlin:1993hm} Tseytlin considered a ``universal class'' of models related to gWZW models $\L_{G/F}$ by deformations parametrized by a constant linear  operator $Q:\Lie(G)\to \Lie(G)$,
\be
\L_{\rm Tseytlin} = k \big( \L_{G/F}(g,A) + \Tr[ A_+ (Q-1) A_- ] \big) \ , \qquad g\in G \ , \quad A_\pm \in \Lie(F) \ . \la{uc}
\ee
The 1-loop conformal invariance conditions were computed, taking the form of polynomial constraints on 
  $Q$.

Forgetting about conformal invariance, one may instead ask which choices of $Q$ correspond to integrable theories. A particular class of integrable examples of \rf{uc} was later discovered \cite{Sfetsos:2013wia,Hollowood:2014rla}, called the \textit{$\l$-model} based  on a either group $G$ or a symmetric space $G/H$. The model is defined by the choice $F=G$ and $Q=1-(\l^{-1}-1)P$ in \rf{uc}, or 
\be
\L_\l= k \big( \L_{G/G}(g,A) - (\l^{-1} -1) \Tr[ A_+ \, P \, A_- ] \big) \ . \la{lam}
\ee
Here $k$ is a WZ level and the coupling $\l$ is a deformation parameter (we will sometimes use a redefined coupling $\ka \equiv \frac{1-\l}{1+\l}$ for convenience). The two cases of $G$ or $G/H$ are encoded in the choice of the projector $P$,
\be
P = \begin{cases} 1 \qquad \qquad  &(G\text{ case}) \ ,\\ P_{G/H} \qquad \qquad   &(G/H\text{ case}) \ , \end{cases} 
\ee
so that  the coset case has an $H$ gauge invariance (see eq.\ \rf{ggs}), while the group case has no gauge symmetry at all.

In the group case there is a global $G$ symmetry, $g \to k g k^{-1}$, $A_\pm \to k A_\pm k^{-1}$, $k\in G$; whereas the coset case does not generally have any continuous global symmetries.
In both cases, the $\l$-model \rf{lam} is invariant under a 
 $\mathbb{Z}_2$ transformation of the fields and coupling constants \cite{Itsios:2014lca,Hoare:2015gda} (see also \cite{Kutasov:1989aw,Tseytlin:1993hm}),
\unskip\foot{Such a symmetry was also discussed  for the models \rf{uc} in \cite{Tseytlin:1993hm} (see footnotes 3  and 6 there).}
\be  \begin{aligned}
& k \to - k \ , \qquad\qquad \l \to \l^{-1} \ , %\qquad {\rm i.e. } \quad \ka \to -\ka \ ,
 \la{z2sym}
\\ & g\to g^{-1} \ , \qquad A_+ \to A_+ - (1-\l^{-1}) P A_+ \ , \qquad A_- \to g A_- g^{-1} - K_- \ . 
\end{aligned} \ee
As we shall find below in Chapter \ref{CT}, the preservation of this symmetry at the quantum level may require a particular choice of
regularization scheme.

One may integrate out the  2d gauge field in \rf{lam} to obtain a \sm depending on the scalar $g\in G$ (here $\L_G$ is the WZW Lagrangian),
\be \begin{aligned}
\L_\l  = k \big( \L_G(g) + \Tr[ J_+ M(g)^{-1} K_- ] \big) \ , \qquad &J \equiv g^{-1} dg \ , \ \ K \equiv dg g^{-1} \ ,  \la{lami}\\
&M(g) \equiv Ad_g - 1  + (1-\l^{-1})P \ .
\end{aligned} \ee
The $\l$-model is an integrable deformation of a  WZW model on $G$ or a gauged WZW model on $G/H$, obtained in the limit $\l=0$  when the projected parts $P A_\pm$ decouple and are set to zero by their equations.

In another interesting limit, 
\be \begin{aligned} 
&k \to \infty \ , \qquad \l= 1 - \tfrac{\h}{2k}  \ , \qquad g=e^{-\tfrac{h}{2k}v} \ ,  \\
&v \in \Lie(G) \text{ and } \h  \text{ fixed} \ ,
\end{aligned} \ee
the $\l$-model becomes the non-abelian dual theory \rf{NADt} corresponding to either the PCM or symmetric space \sm\unskip. The theory \rf{lam} with unintegrated gauge field becomes the corresponding ``interpolating model'' \rf{NADi}.

The $\l$-models \rf{lami} (after integrating out $A_\pm$) admit flat Lax connections of the same form \rf{gl},\rf{sl} as the PCM/symmetric space, but with modified currents,
\begin{align}
&G \, : \qquad && L_\pm = \ha(1+z^{\pm 1}) \mc A_\pm \ , \qquad &&  \mc A_\pm = \frac{2}{1+\l} A_\pm  \ ,  \\
&G/H \, : \qquad && L_\pm = \mc B_\pm + z^{\pm 1} \mc P_\pm \ , \qquad &&   \mc B_\pm = P_H A_\pm \ , \ \ P_\pm = \tfrac{1}{\sqrt{\l}} P_{G/H} A_\pm  \ ,
\end{align}
where $A_\pm = A_\pm(g)$ are taken to be their on-shell values.

In line with the conjecture, the $\l$-model is known to be renormalizable at  1-loop  with only $\l$ running\foot{The 1-loop $\b$-functions \rf{la1},\rf{la2}
can also be extracted from \cite{Tseytlin:1993hm}. Note that the WZ level is not running due to general arguments.} \cite{Sfetsos:2014jfa,Appadu:2015nfa}
\begin{align}
&G \, :  && \ddt k = 0 \ , \quad \ddt \l = -  \frac{2 \cg}{k} \frac{\l^2}{(1+\l)^2}  \ ,  \la{la1} \\
&G/H \, :  && \ddt k = 0 \ , \quad \ddt \l = -  \frac{\cg \l}{k}  \ . \la{la2}
\end{align}
These $\b$-functions  respect the $\mathbb{Z}_2$ transformation \rf{z2sym}, being invariant under the transformation $(k,\l) \to (-k,\l^{-1})$ of the couplings.

\subsection{Integrable-deformed string backgrounds}
Although we have focussed on the simplest case of bosonic theories, and relaxed the condition of Weyl invariance (i.e.\ that the target space is a string background), the models discussed above do admit Weyl-invariant superstring generalizations. The Green-Schwarz action for type IIB strings on \adss with RR flux  is a \sm on the $\mathbb{Z}_4$ ``semi-symmetric'' coset $PSU(2,2 | 4)/[SO(1,4) \times SO(5)]$ with a particular WZ-type $\BB$-field term (here $P^{(i)}$ is the projector onto the $i$th graded part of the algebra) \cite{Metsaev:1998it}
\be\begin{aligned} \la{string}
&S =  \tfrac{1}{4\pi \a'} \int \, d^2 \xi \L \ , \qquad \L = -\ha R^2\ {\rm STr} \big[\sqrt{-h} h^{\m\n} J_\m^{(2)} J_\n^{(2)} + \eps^{\m\n} J_\m^{(1)} J_\n^{(3)} \big] \ ,\\
&J^{(i)} = P^{(i)} J \ , \qquad J = g^{-1} dg \ , \qquad g\in PSU(2,2|4) \ ,
\end{aligned}\ee
 This is a generalization of the symmetric space model \rf{ss2} above (replacing the $\mathbb{Z}_2$-automorphism by $\mathbb{Z}_4$) and indeed the bosonic part of the model (setting fermions to zero) is simply the symmetric space \sm on \adss $= {SO(2,4) \ov SO(1,4)}\times {SO(6) \ov  SO(5)}$, with $R$ being the radius of the $AdS_5$ and $S^5$ spaces.
 
 The $\eta$-deformation  \cite{Delduc:2013qra,Delduc:2014kha} and $\l$-deformation \cite{Hollowood:2014qma} can both be  generalized to integrable deformations of semi-symmetric spaces. 
 The $\eta$-deformed \adss model was problematically found to be scale- but not Weyl-invariant, with the background solving only the generalized supergravity equations  \cite{Arutyunov:2015qva,Arutyunov:2015mqj,Wulff:2016tju}, and the perturbative light-cone S-matrix did not directly match the conjectured exact S-matrix \cite{Arutyunov:2015qva}. It was then explained \cite{hsei} that, to maintain Weyl invariance, one must in fact use the \textit{purely fermionic} Dynkin diagram for the superalgebra $\mathfrak{psu}(2,2 | 4)$ to construct a \textit{unimodular} $R$-matrix. In the recent paper \cite{seibold}, an exact S-matrix was conjectured that is different from the previous attempt, but related by a so-called `twist';  this S-matrix indeed matches tree-level perturbative results.

Much remains to be
understood about these integrable-deformed string backgrounds.
 For example one significant challenge is to understand the spectrum of the deformed theories
  (a direction in which some progress was made for the $\eta$-deformation \cite{etaspec} through the TBA and  Quantum Spectral Curve).
 Examining the  properties of the
bosonic $\eta$-model and $\lambda$-model
can provide
valuable insights into their superstring counterparts. For example, the
relation between the $\eta$- and $\l$-models by Poisson-Lie duality was first understood in the bosonic case \cite{Hoare:2015gda,Klimcik:2015gba,Vicedo:2015pna,Sfetsos:2015nya,Hoare:2017ukq}.
It is thus natural to first explore the question of quantum corrections and the higher-loop RG equations by
studying the bosonic models. 
In Section \ref{stric} below, we shall comment on the possible applications of our approach to
 the integrable-deformed string backgrounds.

It is also worth emphasizing that the bosonic models are of interest in their
own right in the general context of investigating  integrable 2d theories.
The $\eta$-model has played an important role in generalizing the duality \cite{Fateev:1992tk,fateev96,lukyanov}
between the deformed $O(3)$ and $O(4)$ sigma models and massive integrable QFTs
 to higher-rank groups and  cosets \cite{Fateev:2019xuq,Fateev:2018yos,Litvinov:2018bou} and cosets of supergroups \cite{Alfimov:2020jpy}. While the
dual theories are quantum-exact, the \sm side of the duality is only
understood so far to leading order in the loop expansion. Thus, after
finding quantum corrections to integrable \sms consistent with
renormalizability, it may then be interesting to study their compatibility with this duality.

There are also many other known examples of integrable string backgrounds, including those based on
lower-dimensional AdS spaces \cite{Zarembo:2010sg,Wulff:2014kja} and their integrable deformations \cite{sugra,Borsato:2016zcf,Lunin,hsei,Seibold:2019dvf};
$\beta$-deformations (or TsT-deformations) \cite{betadef,Osten:2016dvf}; and more general homogeneous Yang-Baxter deformations \cite{homog,homog2,sugra}.

\subsection{Quantum integrable S-matrices \la{QIS}}
The presence of higher conserved charges places strict constraints  on the possible scattering behaviour of a theory \cite{Zamolodchikov:1978xm,Parke:1980ki} (see also the review \cite{Dorey:1996gd}). Particle production is prohibited, i.e. the number of incoming  and outgoing particles are equal; the set of initial momenta equals the set of final momenta; and the S-matrix must factorize into a product of $2\to 2$ S-matrices. The consistency of this factorization implies that the $2\to 2$ S-matrix satisfies the Yang-Baxter equation.

This special scattering behaviour is  expected for integrable theories, and has been proven assuming the presence of at least two higher-spin conserved charges satisfying certain technical conditions \cite{Parke:1980ki}. S-matrices of this type may be seen as ``almost free'', and their existence is a special feature of 2-dimensions since in higher dimensions the presence of higher-spin charges implies triviality of the S-matrix (i.e.\ a free theory) by the Coleman-Mandula theorem.

For certain theories there are various arguments for the preservation of integrability at the quantum level. For example, in the case of certain principal chiral models, it has been  argued \cite{Goldschmidt:1980wq} using consistency with global symmetry that the higher-spin conservation laws \rf{cur} cannot be modifed at the quantum level by anything other than a divergence, i.e.\ a quantum-corrected form of the current must still conserved to all orders in perturbation theory (see also \cite{Polyakov:1977vm} in the context of the $O(n)$-model, i.e.\ the $\s$-model on a sphere $S^{n-1}={O(n)\ov O(n-1)}$.). The non-local charges of the $O(n)$ model were also argued to persist at the quantum level \cite{Luscher:1977uq} and to imply factorization of the S-matrix with no particle production.

In the important paper \cite{Zamolodchikov:1978xm}, Zamolodchikov and Zamolodchikov started by assuming quantum integrability and the dynamical generation of a mass-gap for the  $O(n)$-model. Using also the model's symmetries and assuming a `minimal' solution for the phase ambiguity (with minimal number of poles on the physical sheet), they uniquely determined the exact S-matrix for non-perturbative quantum particles.

The $O(4)$-model, equivalent to the $S^3$ $\s$-model or $SU(2)$ PCM, was later  solved  explicitly \cite{Polyakov:1983tt} by proving equivalence to a Thirring-type fermionic model with an infinite number of flavours, finding agreement with the Zamolodchikov-Zamolodchikov S-matrix. This exact solution was later generalized   to the PCM on $SU(N)$ \cite{Wiegmann:1984pw} and with arbitrary WZ term (PCM$_k$) \cite{Polyakov:1984et}.\foot{See also \cite{Fateev:1994ai}, where the exact ground state energy was computed in the large-$N$ limit.}

The Zamolodchikov-Zamolodchikov `rational' S-matrices (with rational dependence on rapidities) are known to admit `trigonometric' deformations \cite{Zamolodchikov:1980ku,Bazhanov:1984gu}. The 2d ``sausage model'' was then found \cite{Fateev:1992tk} as a semi-classical $\s$-model description of the trigonometric $O(3)$ S-matrix, and this  model was later understood \cite{Hoare:2014pna} to be the $SU(2) \ov U(1)$ example of the $\eta$-model. More generally, similar trigonometric S-matrices are understood to be associated to inhomogeneous Yang-Baxter deformations and the associated $q$-deformed symmetries \cite{qds} (see also recent papers \cite{ses} on the light-cone S-matrix for $\eta$-deformed string backgrounds and refs.\ therein).

\subsection{Factorized scattering vs.\ tree-level sigma model S-matrix \la{fv}}
Given that integrable S-matrices are so constrained, one might  hope to classify the integrable $\s$-models by (i) computing the S-matrix for a general $\s$-model in terms of the couplings $(\GG,\BB)$ (ii) imposing factorization and no particle production and (iii) systematically reading off the resulting constraints on $(\GG,\BB)$  from integrability. 
Like the RG flow (which is the focus of the thesis), the S-matrix could thus serve as another `diagnostic tool' for integrability.

Here we shall summarize the paper \cite{HLTS} where we considered such a procedure for the leading \textit{tree-level} S-matrix, 
  whose factorization should correspond to \textit{classical} integrability. The factorization of the loop-corrected S-matrix should then correspond to preservation of integrability at the quantum level. For example, for the complex sine-Gordon model, we recall that the tree-level 2-body S-matrix satisfies the Yang-Baxter equation but the 1-loop-corrected S-matrix only satisfies it if the theory is supplemented by certain quantum corrections \cite{Hoare:2010fb}. These corrections may be traced to the leading quantum corrections to a gauged WZW model (see eq.\ \rf{6} below) --- of which the complex sine-Gordon model is a massive integrable deformation. The corrections are thus expected to correspond to a particular choice of finite counterterms consistent with quantum integrability, similar to those discussed below in the context of the RG flow.
  
\sloppy
To compute an  S-matrix in perturbation theory, one must first decide on a choice of vacuum. In \cite{HLTS} we considered the `trivial' vacuum $x^m=0$ (with all fields vanishing) in an expansion of the metric and $\BB$-field  around flat space, \mbox{$\GG_{mn}+\BB_{mn} = \delta_{mn} +\h_{mn,p} \, x^p + \hh_{mn,pq} \, x^p \,  x^q + \ldots$\ .} This is a natural choice since it is universal to any target space geometry. However, the  fields $x^m$ expanded  around this vacuum are massless 2d scalars, which are known to suffer from IR divergences at the quantum level \cite{Coleman:1973ci}.  This is related to the fact that  massless particles (moving at the speed of light) on a 1d spatial line  do not separate asymptotically, so an S-matrix cannot have the usual physical interpretation. Still, massless 2d S-matrices were formally discussed for integrable theories in the context of a finite-density
TBA \cite{Zamolodchikov:1992zr}. The S-matrix there retains the interpretation as the relative phase when one particle is moved past
another.

One might still hope to define a formal \textit{tree-level} S-matrix for such massless particles (e.g., from the classical action evaluated on a solution with special scattering boundary conditions), but we found  in  \cite{HLTS} that such an object still has IR singularities. The singularities are not true divergences but instead  ambiguous ``0/0'' contributions due to a vertex vanishing and an internal propagator  blowing up at the same time when the scattering amplitude is taken on-shell. We classified the ambiguities into two types and considered some simple approaches to regularize them (the usual $i \e$ prescription, or an alternative ``massive regularization'' where the particles are given a small mass). One can compute finite scattering amplitudes by regularizing the ambiguities, taking the amplitude on-shell, and finally taking the regulator to zero. However, we found that the resulting amplitudes do not have the usual connection to integrability:  in various well-known integrable examples the amplitudes exhibit particle production. These findings were in agreement with earlier attempts to compute scattering amplitudes for massless 2d scalars in integrable theories \cite{Nappi:1979ig}.

Opposite conclusions have been found for massless 2d S-matrices in non-renormalizable (non scale-invariant)
Nambu-like models \cite{Dubovsky}. For Nambu theory with a flat target space (but not for a curved one), each scalar appears
in the action only through its derivative and thus ambiguities related to IR poles appear to be
absent, not only at the tree level but also at the loop level. Here the (naively expected) relation
between factorization of the massless S-matrix and integrability does appear to hold (with the
Nambu action being integrable beyond tree level only in special critical dimensions). 
Moreover, the flat-space Nambu theory is an example of a $T\bar{T}$-deformation \cite{TTbar} of a free theory, for which integrability apparently constrains the  RG flow even in non-renormalizable cases, or rather  dictates the fixing of the counterterms  to be  consistent with the integrable S-matrix   \cite{Rosenhaus:2019utc}.

A similar S-matrix approach has been used \cite{Wulff} to rule out integrability of strings on certain (warped) AdS supergravity backgrounds  by expanding around non-trivial classical solutions (BMN geodesic \cite{BMN}, GKP spinning string \cite{Gubser:2002tv} or `null cusp' \cite{Kruczenski} solutions) and fixing a light-cone gauge. These vacua give a mass to at least some fields and one may then consider particular worlsheet scattering amplitudes that do not contain any ambiguous contributions from massless fields. While this approach is not easily applied to generic backgrounds (where one does not have a good understanding of classical solutions), it would be interesting in future to consider a more general class of backgrounds than AdS ones, but still assuming  enough isometries to allow a similar choice of vacuum.

\subsection{Affine Gaudin models and local charges \la{agm}}
There is a proposal to systematize integrable 2d theories by viewing them as realizations of affine Gaudin models \cite{Lacroix:2017isl,Vicedo:2017cge} (see also \cite{Lacroix:2018njs}).  It so happens that all of the integrable $\s$-models described above are examples of such theories. This is indeed suggestive they may describe a large portion of the integrable theories.

The affine Gaudin-type integrable models have flat Lax connections \rf{zc}. They also have a systematic construction of local, higher-spin, classically conserved charges, generalizing those constructed in \rf{curG} for the PCM/symmetric space and  models with Lax connections of those same form. The classical Poisson brackets of the spatial component of the Lax connection $\L \equiv L_\s=\ha (L_+ - L_-) \in \Lie(G)$  take the  Maillet $r/s$-form \cite{maillet} (here $z_1,z_2$ are spectral parameters and $\s_1,\s_2$ are spatial coordinates),
\be\begin{aligned} \la{mai}
\big\{ \L_{\ul 1}(z_1; \s_1) ,\L_{\ul 2}(z_2; \s_2) \big\} = {}&{} \Big( [r_{\ul{12}}(z_1,z_2), \L_{\ul 1}(z_1; \s_1) + \L_{\ul 2}(z_2; \s_1)] \\
&+ [s_{\ul{12}}(z_1,z_2), \L_{\ul 1}(z_1; \s_1)- \L_{\ul 2}(z_2; \s_1)] \Big)  \d_{\s_1 \s_2} - 2 s_{\ul{12}}(z_1,z_2)\, \delta_{\s_1\s_2}' \ .
\end{aligned}\ee
Here the underlined $\ul{1},\ul{2}$ indices in \rf{mai} denote the first and second factors in a tensor product $\Lie(G) \otimes \Lie(G)$, e.g. $\L_{\ul 1} \equiv \L \otimes 1$, $\L_{\ul 2} \equiv 1\otimes \L$. The Poisson bracket \rf{mai} is called `non-ultralocal' because of the presence of the $\delta'$ term. The constant $r$- and $s$-matrices $r_{\ul{12}}(z_1,z_2) = - r_{\ul{21}}(z_2,z_1)$ and $s_{\ul{12}}(z_1,z_2) =  s_{\ul{21}}(z_2,z_1)$ are respectively antisymmetric and symmetric, and they take the special ``twist'' form,\foot{The form of \rf{twist} is slightly modified is so-called `cyclotomic' examples involving twisting by a $Z_T$ automorphism, e.g. the symmetric space $\s$-model with $T=2$; here we ignore such cases.}
\be
r_{\ul{12}}(z_1,z_2) + s_{\ul{12}}(z_1,z_2) = \frac{1}{(z_2-z_1) \phi(z_2)} \,  T_A  \otimes T^A \ , \la{twist}
\ee
in terms of a model-dependent meromorphic  \textit{twist function} $\p(z)$ of the spectral parameter. Here the `split Casimir' $ T_A  \otimes T^A$, built from any set of generators $T_A$  for $G$ and the dual generators $T^A$ (with respect to $\ka_{AB} \equiv \Tr[T_A T_B]$), is invariant under  change of basis.

At each zero $z_i$ of the twist function $\phi(z)$ (or more properly zeros of the 1-form $\phi(z) dz$)\foot{In particular the poles of $\L$ will be zeros of the twist function.}, one can automatically construct from \rf{mai} an infinite tower of local, higher spin charges. Assuming for simplicity the Lax connection to be valued in a $B$, $C$ or $D$ type algebra (the $A$ type case is slightly more complicated), the following  charges $\{Q^{(2n)}(z_i)\}_{n,i}$  are all in involution \cite{Lacroix:2017isl} (assuming a well-defined limit of $\phi(z)\L(z)$ at $z_i$),
\be
\big\{ Q^{(2n)}(z_i), Q^{(2m)}(z_j) \big\} = 0 \ , \qquad Q^{(2n)}(z_i) = \int d\s \, \Tr\big[ \big(\phi(z) \L(z)\big)^{2n}\big] \Big|_{z=z_i} \ . \la{trr}
\ee
For realizations of affine Gaudin models, the Hamiltonian is a linear combination of some of these charges, so all of the charges are conserved (commuting with the Hamiltonian).

For the group-based models above (PCM, PCM$_k$, group $\eta$-model, group $\l$-model) with Lax connections of the form \rf{gl} in terms of a flat, conserved current $\mc A_\pm$, let us re-define the spectral parameter $z =\tfrac{1+w}{1-w}$ for simplicity so that the Lax matrix $\L=L_\s$ has simple poles at finite values  $w=\pm 1$. These models' twist functions all have simple zeros at these points $\pm 1$, and take the form $\phi(w) = \tfrac{(1-w)(1+w)}{p(w)}$ where $p$ is a polynomial not vanishing at $\pm 1$. The resulting charges at $w_1 =1$ and $w_2 =-1$ are proportional to the  local charges obtained from the (anti-)holomorphic currents in \rf{curG} above,\foot{We seem to have produced only half as many charges in \rf{chag} as in \rf{curG}, but in fact all of the odd currents $J^{(2n+1)}$ vanish for $B$, $C$ and $D$ type groups \cite{Evans:1999mj}. Formally this procedure applies to the complexifed Lie group and one specifies a real form by some reality conditions. This does not affect the charges in this case since the zeros $w=\pm 1$ of the twist function are real and so the charges \rf{chag} are real.}
\begin{align} \la{chag}
\ddt Q^{(2n)}(\pm 1) = 0 \ , \qquad Q^{(2n)}(\pm 1) \sim \int d\s \, \mc J^{(2n)}_\pm \ , \qquad \mc J^{(2n)}_\pm = \Tr[ \mc A_\pm^{2n} ] \ .
\end{align}
Here we have used the trace but, as in \rf{curG}\rf{e3G} above, we could have instead  used any symmetric, invariant tensor in \rf{trr}.

Thus one learns from \rf{trr} that the higher-spin charges \rf{chag} are all in involution.
It also follows from the Poisson bracket \rf{mai} that these charges are in involution with the monodromy matrix and its non-local conserved charges \cite{Lacroix:2017isl} as constructed in Section \ref{ci}. A similar construction of local, conserved charges applies  to models based on symmetric spaces (or other $Z_T$-cosets).

As well as re-interpreting known models, the affine Gaudin approach has the power to generate new integrable theories, essentially by picking new twist functions $\phi(z)$. For example, the $\eta$-deformations of  symmetric space \cite{Delduc:2013fga} and semi-symmetric space \cite{Delduc:2013qra} $\s$-models  were discovered in this way (based on the previously known group space case); more recent examples include the integrable $G^N$ models  \cite{DLMV}, deformed $G^N$ models \cite{Bassi:2019aaf} and $G\times G/H$ models \cite{ABL} (see Chapter \ref{GGS}).

The affine Gaudin construction systematizes certain reformulations of these integrable $\s$-models.  For a certain class of twist functions, the affine-Gaudin-type models are equivalent \cite{Lacroix:2020flf}  to ``E-models'' \cite{poissonlie} (a first order formulation with manifest Poisson-Lie symmetry), and can also be obtained  \cite{Vicedo:2019dej,Delduc:2019whp} from 4d Chern-Simons theory \cite{Costello:2019tri}.

Finally we remark that the affine Gaudin formalism could lead to a more systematic understanding of the RG flow of integrable $\sigma$-models, or perhaps even a first-principles proof of their renormalizability. Along these lines, a recent paper \cite{today} noted a pattern in the 1-loop RG flow for several integrable affine-Gaudin-type $\s$-models: in each case it may be written in the same universal way as a flow of the twist function. It would be interesting to try to derive this statement, or to check it beyond the leading 1-loop order. One complication is that, beyond the  1-loop order, we should expect the integrable theories to require a quantum deformation to remain RG-stable (see below); it is not yet clear how these quantum deformations would be encoded in the affine Gaudin language.

\subsection{On scheme dependence \la{SD}}
Under covariant scheme transformations, or redefinitions of the \sm couplings $(\GG,\BB)$ (in terms of constants $c,d,e$),
\be
\GG_{mn} \to \GG_{mn} + \a' (c \, R_{mn} + d \,  \HH_{mpq}\HH_n{}^{pq}) + \ldots  \ , \quad \BB_{mn} \to \BB_{mn} + \a' \, e \, \nabla^p \HH_{pmn} + \ldots \ , \la{csc}
\ee
the $\b$-function $ \b_{mn} = \ddt (\GG+\BB)_{mn}$ transforms as a contravariant vector in the space of couplings. Simple power counting implies that the leading 1-loop $\b$-function is invariant while, in general, scheme dependence will begin at the 2-loop order.\foot{For the special case of vanishing $\BB=0$, the \sm may be viewed as a ``single-coupling'' theory and its 2-loop $\b$-function \rf{2} is also invariant (see Appendix \ref{A}).}

In this thesis, we will generally use the 
 2-loop $\beta$-functions given in \rf{BMT},
 which were computed  in a  special 2-loop   scheme 
that effectively  treats 
$\GG_{mn}$ and $\BB_{mn}$  as symmetrically as possible  \cite{Bos:1987mw,mt} (see also \cite{Hull:1987pc,Ketov}).
Indeed,  the  
$\beta$-functions   for $\GG$ and $\BB$ in \rf{BMT} are given by the symmetric and antisymmetric parts of a single tensor expression.  
  We shall refer to this $\GG$-$\BB$ symmetric scheme   as  the
  ``\GB scheme'' (see the introduction to Chapter \ref{GGS} for discussion of the properties of this scheme).
  
It is important to emphasize, however, that our conclusions will be scheme-independent. In particular, the statement that `a model is not renormalizable' is understood to mean that it is not renormalizable in any covariant subtraction scheme. We shall find certain finite counterterms that restore higher-loop renormalizability and, while the form of these counterterms will change in different subtraction schemes, they cannot be eliminated by covariant scheme changes such as \rf{csc}.

\chapter{Quantum corrections to geometry  from  RG flow \la{QCS}}

As discussed in the Introduction, it was observed for some years that  integrable \sms appear to be stable under the leading 1-loop RG flow, and our purpose here is to investigate what happens at higher-loop orders. The general finding is that the integrable theories remain stable under the higher-loop RG flow provided they are supplemented with particular finite counterterms, i.e.\ the classical target space geometry $(\GG+\BB)_{mn}^{(0)}$ receives certain quantum corrections,
\begin{align}
(\GG+\BB)_{mn} = (\GG+\BB)_{mn}^{(0)} + \a' (\GG+\BB)_{mn}^{(1)} + \ldots  \  . \la{qc2}
\end{align}

One might first try to study the integrability--RG flow relationship  for a canonical integrable example such as the PCM,
\be
\L_{\PCM} = -\ha \h  \Tr[ J_+ J_- ]    \ . \la{PCMk}
\ee
However, the question trivializes for this simple example, because the renormalizability of the PCM follows immediately from its   large $G \times G$ global symmetry (and parity invariance)
\be
g \to g_L \,  g  \, g_R \ , \qquad (g_L, g_R )  \in G_L  \times  G_R \ . \la{PCMs}
\ee
Only the curvature radius $\sqrt{h}$ can run because the theory \rf{PCMk} does not admit any deformations preserving these symmetries.
 This conclusion is consistent with the conjecture that integrable models are renormalizable, but it is not good evidence for it: we  only used manifest global symmetry, without relying on the  hidden integrable symmetries.

In order to really probe the role of integrability in constraining the RG flow, we shall consider integrable theories with less global symmetry. As we shall observe,  integrability appears to be sufficient for renormalizability,  even in cases without any manifest global symmetries.
In this chapter we shall focus on  the integrable $\eta$- and $\l$-deformations introduced in Sections \ref{etas} and \ref{las}, which
are classically integrable but have reduced global symmetry groups.  In particular, these models are not fully constrained by their global symmetries, and renormalizability does not follow from global symmetries alone.

First in Section \ref{Seta}, we shall consider the simplest $\eta$-deformed model 
 based on the coset $SU(2) \ov U(1)$.  Then in secton \ref{Slam}, we shall consider $\l$-deformed models 
 based on the coset $SU(2) \ov U(1)$ and the group $SU(2)$. As we reviewed above, these models are already known to be stable under the leading 1-loop RG flow; we shall determine the quantum corrections
required to solve the 2- and higher-loop RG equation \rf{BMT}. 

\

\noindent This chapter is based on the papers \cite{HLT1,HLT2} and closely follows the presentation there.

\section{On quantum corrections in sigma models \la{qcs}}
Before proceeding, let us provide  justification for the  presence of quantum corrections \rf{qc2}  in $\s$-models
by  recalling some familiar examples where such corrections are known to occur.

One class of  examples are
 the special integrable \sms  \rf{gws} corresponding to gauged WZW conformal field theories.
The classical $\s$-models \rf{gws} are indeed scale-invariant, i.e.\ fixed-point solutions of the RG equation \rf{BMT}, at the leading 1-loop order.
However, to remain conformally invariant beyond 1-loop, the target space metric and $\BB$-field
 require non-trivial $\a'$ corrections at each order,
which may be derived exactly 
the underlying quantum gWZW theory
(for discussion and examples see \cite{Tseytlin:1993df,Horowitz:1994ei,Tseytlin:1995fh}).
The simplest example is provided by the exact counterpart \cite{Dijkgraaf:1991ba} of the
familiar classical $SU(1,1)/U(1)$ gWZW metric (or 2d Euclidean black hole) \cite{Bardacki:1990wj,Witten:1991yr}
\begin{align}
ds^2 = k(dr^2 + \tanh^2 r \, dy^2) \ \ \to \ \
ds^2 = (k-2) \Big[ dr^2 + \big( \coth^2 r - \tfrac{2 }{k}\big)^{-1} dy^2\Big] \la{6} ~.
\end{align}
Here the inverse level $k^{-1}$ (and later the inverse coupling $h^{-1}$) plays the role of the loop-counting parameter $\a'$  in \rf{13}

As was checked directly in \cite{Tseytlin:1991ht}, the leading $k^{-1}$ correction in \rf{6}
is  required to solve the 2-loop scale invariance equation following from \rf{2}
(see also \cite{Jack:1992mk} for a 4-loop test of \rf{6}).\foot{In this Weyl-invariant case
the diffeomorphism  vector $\xi_n$ can be written as $ 2 \del_n \Phi$ where the exact dilaton field $\Phi$ is given by
$e^{-2 \Phi}= \sinh 2 r \ \big( \coth^2 r - \tfrac{2 }{k}\big)^{1/2}$. \la{f4}} 
The integrable cases that follow will be a direct generalization of this principle: instead of ensuring complete
scale-invariance, the quantum corrections will ensure invariance under the RG flow with just a few couplings running.

It is also well-known that the classical Buscher rules for abelian T-duality are subject to quantum corrections \cite{Tseytlin:1991wr,Panvel:1992he,Kaloper:1997ux,Parsons:1999ze}, so geometries resulting from T-duality will contain non-trivial $\a'$ corrections. For example,
given the metric
$ds^2 = \hh [dx^i dx^i + \cM(x) dy^2] $  with an isometry direction $y$, the standard T-duality rule $\cM \to \wt \cM= \cM^{-1}$ is known to be
modified at the 2-loop level \cite{Tseytlin:1991wr} (see also \cite{Haagensen:1997er,Haagensen:1997bh})
\be
\wt \cM = \cM^{-1} \big( 1 + \tfrac{1}{2} \hh^{-1} \, \del_i \log \cM\, \del^i \log \cM \big) \ . \la{5}\ee
In an example below in Section \ref{Slam}, this quantum correction to the T-duality transformation will be reproduced by special limits of the corrections for the $\l$-model. We shall also produce similar corrections to the transformation rules of non-abelian duality in some special examples.

%%%%%%%%%%%%%%%%%%%%%%%%%%%%%%%
\section{\texorpdfstring{$\eta$}{eta}-model \la{Seta}}
%%%%%%%%%%%%%%%%%%%%%%%%%%%%%%%

The metric of the 2d ``sausage'' model\foot{We note that $\hh$ in this chapter is equivalent to $4h$ in \cite{HLT1}.}
\be
ds^2 = \tfrac{1}{4} \hh \Big[ \frac{dr^2}{(1-r^2)(1+\k^2 r^2)} + \frac{1-r^2}{1+\k^2 r^2} d\p^2 \Big] \ , \la{21}
\ee
was originally written down in \cite{Fateev:1992tk} as the leading semi-classical approximation to the \sm corresponding to the massive integrable trigonometric S-matrix of \cite{Zamolodchikov:1980ku}. It was discovered as a solution to the 1-loop RG equation (the leading Ricci flow term in \rf{2}) and conjectured to be classically integrable. Its classical integrability was later shown in \cite{lukyanov}. In \cite{Hoare:2014pna} this model was identified as the $SU(2) \ov U(1)$ example of the $\eta$-deformation \cite{Klimcik:2002zj,Delduc:2013fga}, introduced in Section \ref{etas}.\foot{It is interesting to note that the metric \rf{21} is formally self T-dual, i.e. invariant under
$\phi\to \td \phi $ and $\td \phi \to i \k^{-1} \phi$,\ \ $r \to i \k^{-1} r^{-1}$. \la{fstd}}

The metric \rf{21} is an integrable deformation of the round $S^2$ ($\k=0$) of radius $\sqrt{\h/4}$, with 
$\k$ being the deformation parameter.
The usual $\eta$-deformation  correspnds to the case when $\varkappa$ is real, however the authors of  \cite{Fateev:1992tk} considered another regime of interest with  $\varkappa$  imaginary, in which case the \sm is UV stable with a free UV fixed point at $\k^2 = -1$.
Alternatively, we can send $\k^2\to -1$ while simultaneously expanding around $r^2 = 1$.
Taking the limit in this way, the metric \rf{21} reduces to the classical metric of the $\frac{SU(1,1)}{U(1)}$ gWZW model (cf \rf{6}),
\be r^2 = 1 - (1+\k^2) \sinh^2 \rrr \ , \ \ \qquad \k^2 \to -1 \la{22}\ , \qquad
ds^2\to \tfrac{1}{4}\hh( d\rrr^2 + \tanh^2 \rrr\, d\p^2) \ , \ee
with $\h/4$ becoming the level.
Another useful limit is the maximal deformation limit $\k \to \infty$, $\hh\to \infty$ with $\hh' \equiv \tfrac{\hh}{\k^2}$ fixed, which yields 
\cite{Hoare:2014pna} the undeformed hyperbolic space $H^2$  with radius $\sqrt{\hh'/4}$ (here we set $r=\tfrac{1}{\cosh{ \theta}}$)
\be
ds^2 = \frac{\hh'}{4 r^2} \Big[ \frac{dr^2}{1-r^2} + (1-r^2) d\p^2 \Big] = \tfrac{1}{4} \hh' \Big[ d \theta^2 + \sinh^2 \theta \, d\p^2 \Big] \ . \la{222}
\ee

The metric \rf{21} solves the 1-loop RG equation \rf{2} with $\hh$ and $\k$ running as (eq.\ \rf{cos} with $c_{_{SU(2)}}=2$)
\be \la{23}
\dt \hh = 4 (1+\k^2) + \OO(\hh^{-1}) \ , \qquad \qquad
\dt \k = 4 \hh^{-1} \k (1+\k^2) + \OO(\hh^{-2}) \ . \ee
There is an RG-invariant combination of couplings
\be
\dt \nu =0\ , \qquad \qquad \la{244}
\nu \equiv {\hh \ov 4 \k} + \O(\hh^0) \ . \ee
In agreement with the above comments,
$\k^2=-1$ is a UV fixed point (corresponding to either a free theory or a gWZW model depending how the limit is taken).

However,  including also the 2-loop term in the RG equations \rf{2}, the $\s$-model on the classical sausage geometry \rf{21} appears not to be renormalizable: it does not solve the 2-loop RG equations with only $\hh$ and $\k$ running. This is true in any covariant subtraction scheme\foot{As discussed in Appendix \ref{A}, for 2d target spaces, the scheme dependence does not begin until 3-loops.} and with any diffeomorphism vector (here one only needs to consider diffeomorphism vectors  of the form $\xi^m=\xi^m(r)$ since $\phi$ is an isometry direction).

The relation to the gWZW model suggests that,
for \rf{21} to remain a solution to the RG equation at higher loops with only $\hh$ and $\k$ running,
it should similarly be modified by quantum (i.e. $1/\hh$) corrections.
Inspired by the analogy with the form of the gWZW metric in \rf{6} we propose
the following conjecture for the exact generalization of \rf{21}
\begin{align}
ds^2 = (\tfrac{1}{4} \hh- 1+\k^2 ) \Big[ \frac{dr^2}{(1-r^2)(1+\k^2r^2)} + \big(
\frac{1+ \k^2r^2}{1- r^2} + \frac{8\k^2}{\hh} \big)^{-1}
d\p^2 \Big] \ . \la{24}
\end{align}
The metric \rf{24} is expected to solve the RG equations to all loop orders
(in a particular scheme and modulo coordinate redefinitions) with $\hh$ and $\k$ running
according to a generalization of \rf{23}.
The conjecture passes some obvious tests:
\begin{enumerate}[(i)]
\item  The metric \rf{24} reduces to $S^2$ when setting $\k=0$, now with
shifted radius-squared coupling 
 ${\tfrac{1}{4}\hh-1}$, which is simply related to the classical one $\tfrac{1}{4}{\hh}$  by a finite coupling redefinition.
Alternatively, one may shift $\h\to \h+4(1-\k^2)$ in \rf{24} to obtain
$ds^2 = \tfrac{1}{4} \h \Big[ \frac{dr^2}{(1-r^2)(1+\k^2r^2)} + \big(
\frac{1+ \k^2r^2}{1- r^2} + \frac{2\k^2}{\tfrac{1}{4} \h+1-\k^2} \big)^{-1}
d\p^2 \Big] ,$
so that the radius-squared of $S^2$ remains $\tfrac{1}{4}\hh$ in the $\k=0$ limit.
\item The metric remains flat for $\k^2=-1$.
\item In the non-trivial $\k^2 \to -1$ limit \rf{22} it reduces to the exact gWZW metric in \rf{6} with level $k=\h/4$.
\item In the maximal deformation limit \rf{222} it reduces to $H^2$, now with shifted radius-squared coupling  $\h'/4+1$.
\end{enumerate}

More importantly,
one can directly check that \rf{24} with the leading $h^{-1}$ correction
included,
\be
\begin{aligned}  ds^2 &= \h \Big[ \frac{dr^2}{(1-r^2)(1+\k^2r^2)} \Big(\frac{1}{4}-{{1-\k^2}\ov{\h}}\Big) \\
&\qquad \qquad\qquad \quad + \frac{1- r^2}{1+ \k^2 r^2} \Big(\frac{1}{4} - {{1-\k^2}\ov{\h}} -{ 2 \k^2 \ov {\h} } \frac{1- r^2}{1+ \k^2 r^2} + \ldots \Big) d\p^2 \Big] \ , \la{277}
\end{aligned}\ee
indeed solves the 2-loop RG equation
\rf{2}, which becomes in this case with 2d target space,\foot{In 2 dimensions the curvature tensor can be written as
 $R_{mnkl} = \ha R ( \GG_{mk} \GG_{nl} - \GG_{nk} \GG_{ml})$. Note that any $\BB$-field on a 2d space is trivial (with zero field strength $\H=d\BB$) so we use the $\b$-function \rf{2} for the case of vanishing $\BB$-field (also taking $\l_m=0$).}
\be \la{27}
\dt \GG_{mn} = (\ha R + \tfrac{1}{4} R^2 ) \GG_{mn} + \ldots + \nabla_{(m} \xi_{n)} \ ,
\ee
for a particular diffeomorphism vector $\xi_n$ with components,\foot{We note that the diffeomorphism vector \rf{2777} begins at 2-loop order; the 1-loop diffeomorphism vanishes in these coordinates.}
\be \la{2777}
\xi_\p = 0 \ , \qquad \quad \xi_r = \frac{8\k^2 (1+\k^2) r}{\h (1+\k^2 r^2)^2} \ .
\ee
The $\beta$-functions \rf{23} receive the following 2-loop corrections
\begin{align}
\la{28}
\dt \h = (1+\k^2)&\big[4 + 16 \, \h^{-1} (1-\k^2) + \OO(\h^{-2})\big] \ , \\
\la{288} \dt \k = \h^{-1} \k (1+\k^2)&\big[4 + 16\,  \h^{-1} (1-\k^2) + \OO(\h^{-2}) \big] \ .
\end{align}
The RG-invariant quantity in \rf{244} remains RG-invariant without 2-loop corrections,
\be
\dt \nu =0\ , \qquad \qquad
\nu \equiv {\h \ov 4 \k} + 0 + \mathcal{O}(\h^{-1}) \ . \la{2888}\ee
Furthermore, the ansatz \rf{24} also solves the 3-loop RG equation \rf{2} with
\be \la{29}
\b_{mn} = \Big[ \ha R + \tfrac{1}{4} R^2 + c_1 R^3 + c_2 (\nabla R)^2 + c_3 R \nabla^2 R \Big] \GG_{mn}
+ c_4 \nabla_m R \nabla_n R \ , \ee
in a particular ``natural'' renormalization scheme corresponding to
\be \qquad c_1 = 0 \ , \qquad c_2 = \tfrac{1}{8} \ , \qquad c_3 = -\tfrac{1}{4} \ , \qquad c_4=- \tfrac{1}{16}\ ,
\la{31}
\ee
in which the 3-loop $\b$-functions for $\h$ and $\k$ take the form
\begin{align}
\la{32}
\dt { \h} =(1+\k^2)&\Big[4 + 16 \, \h^{-1} (1-\k^2) + 64\,  \h^{-2} (1-\k^2)^2 + \O(\h^{-3}) \Big] \ , \\
\la{33} \dt { \k} = \h^{-1} \k (1+\k^2)&\Big[ 4 + 16\,  \h^{-1} (1-\k^2) + 64\,  \h^{-2} (1-\k^2)^2 +\O(\h^{-3})\Big] \ .
\end{align}
The diffeomorphism vector in \rf{2777} receives the following 3-loop correction
\be \la{300}
\xi_\p = 0 \ , \qquad \quad \xi_r = \frac{8\k^2 (1+\k^2) r}{\h (1+\k^2 r^2)^2} \Big[ 1 + \frac{8\k^2 (r^2-1)}{\h(1+\k^2 r^2)} \Big] \ .
\ee
This scheme \rf{31} is related to the minimal subtraction scheme in \cite{Graham:1987ep,Foakes:1987ij,Foakes:1987gg},
\be \la{30}
\qquad c_1 = \tfrac{5}{32} \ , \qquad c_2 = \tfrac{1}{16} \ , \qquad c_3 = 0\ , \qquad c_4=-\tfrac{1}{16}\ ,
\ee
by the covariant coupling redefinition,
\begin{align}
\GG_{mn}^{(\rm nat)}= \big[ \GG_{mn} + \tfrac{5}{8} \, \a'^2 ({R}^2)_{mn} + \ha \, \a'^2 \nabla^2 R_{mn} \big] ^{(\rm min)}\ ,
\end{align}
where $\GG_{mn}^{(\rm nat)} \equiv \GG_{mn}$ is the metric \rf{24} in the ``natural'' scheme and $\GG_{mn}^{(\rm min)}$ is the corresponding metric in the minimal scheme (see Appendix \ref{A}).

We call the scheme \rf{31} ``natural'' because the RG-invariant quantity in \rf{2888} remains invariant with no 3-loop corrections:
\be
\dt \nu =0\ , \qquad \qquad
\nu \equiv {\h \ov 4\k} + 0 + 0 + \mathcal{O}(\h^{-2}) \ . \ee
This prompts us to conjecture that, in a natural choice of
subtraction scheme at each loop order, the  quantity $\nu\equiv \tfrac{\h}{\k}$ is an exact RG-invariant.
This suggests that $\nu$ should be the parameter that appears in the exact quantum trigonometric S-matrix
that generalizes the non-perturbative massive S-matrix of the $O(3)$-invariant $S^2$ \sm
\cite{Zamolodchikov:1978xm}.

Relatedly, we extrapolate from the obvious pattern in \rf{32},\rf{33} to conjecture that, in the same natural scheme, the all-loop $\b$-functions of $h$ and $\k$ are
\begin{align}
\dt {\h} &=\frac{1+\k^2}{\tfrac{1}{4} \h-(1-\k^2) } \, \h \ , \la{exactb1} \\
\dt { \k} &=\frac{1+\k^2}{\tfrac{1}{4} \h- (1-\k^2)} \, \k \ . \la{exactb2}
\end{align}
Here $\k^2=-1$
remains a fixed point, as it should to all orders since it corresponds both to flat space and the gWZW limit.
For $\k=0$ we get $\dt { R^2_{S^2}} =1+R^{-2}_{S^2}$ for the $S^2$ radius $R_{S^2}=\sqrt{\tfrac{1}{4}\h-1}$, which agrees with the 1-loop and 2-loop coefficients in the $\b$-function of the
$S^2$ model. In the maximal deformation limit \rf{222} we get $\dt { R^2_{H^2}} =-1+R^{-2}_{H^2}$ for the $H^2$ radius $R_{H^2} = \sqrt{\tfrac{1}{4}\h'+1} = \sqrt{\tfrac{1}{4}\h\k^{-2}+1}$. The fact that this is correctly related to the $S^2$ $\b$-function by the analytic continuation $R^2_{H^2} = - R^2_{S^2}$ is a
 check of the consistency of the conjectured exact metric \rf{24} and exact $\b$-functions \rf{exactb1},\rf{exactb2}. Moreover, 
 if there is a natural scheme where \rf{exactb1},\rf{exactb2} are exact, 
  then the $S^2$ and $H^2$ $\b$-functions are 2-loop exact in this scheme
   (this may be possible in a special non-minimal scheme  since the 3- and higher-loop  $\b$-function coefficients are scheme-dependent).

Let us mention that
in the case of the
(1,1) supersymmetric generalization of the \sm \rf{13}, the first potential correction to the 1-loop $\b$-function
appears
at 4 loops ($\sim \z(3) R^4$ in the minimal scheme \cite{Grisaru:1986px}, with this particular invariant actually vanishing in the case of 2d target space).
There is also
no deformation of the super-gWZW metric (to all orders) and it is thus natural to expect that the same may apply to the
supersymmetric $\s$-model on the sausage geometry \rf{21} (i.e.\ only $\h$ and $\k$  running without adding corrections to the metric).

One may also repeat the above discussion for the
$\eta$-deformation of hyperbolic space $H^2= \frac{SU(1,1)}{U(1)}$
(or the Lorentzian signature symmetric spaces $dS_2$ and $AdS_2$).
The corresponding metric is related to \rf{21} by the
formal analytic continuation,
\be \la{344}
r\to i r\ , \qquad \ \ \p\to i \p\ , \ \qquad \ \k \to i \k\ , \qquad \ \ \h \to - \h \ .\ee
i.e. we find\foot{\la{ftd}Let us note that the metric \rf{34} is self T-dual, i.e. invariant under
$\phi\to \td \phi $ and $\td \phi \to \k^{-1} \phi$,\ \ $r \to \k^{-1} r^{-1}$.}
\be\la{34}
ds^2 = \tfrac{1}{4} \h \Big[ \frac{dr^2}{(1+r^2)(1+\k^2r^2)} + \frac{1+r^2}{1+\k^2 r^2} d\p^2 \Big] \ . \ee
This inhomogeneous  Yang-Baxter deformation of $H^2$ is based on the \textit{split} $R$-matrix of $\Lie(SU(1,1))$ (see footnote \ref{ybs}).
On the other hand, if we had not continued $\k\to i \k$ in \rf{344}, we would have found Yang-Baxter deformation based on the non-split $R$-matrix.
For $S^2$ only the latter case exists as a Yang-Baxter deformation in a strict sense since there is no split $R$-matrix of $\Lie(SU(2))$ (see, e.g., \cite{BDCGR}).

For
$\k=0$ \rf{34} becomes the $H^2$ metric, while for $\k^2=1$ it is flat.
The limit analogous to \rf{22}, i.e.
$\k \to 1$ with $ r^2\to -1 + (1-\k^2) \sinh^2 t$, is now a formal limit
and gives the metric
$ ds^2 = \tfrac{1}{4}\h( - dt^2 + \tanh^2 t\, d\p^2)$, which is the classical metric of the
$\frac{SU(1,1)}{\mathbb{R}}$ gWZW model.

In
conformally-flat coordinates the metric \rf{34} may be written as
\be \la{35}
ds^2 = {\h \ov 4(p^2 + \k^2 q^2)} ( dp^2 + dq^2) \ , \qquad \qquad
r = \frac{q}{p} ~, \quad \phi = \ha \log({p^2+q^2})\ ,
\ee
where the scaling symmetry $(p, q) \to \l (p, q)$ is the counterpart of
$\p$-shift isometry of \rf{34}.

The analog of the exact metric \rf{24} is found by the analytic continuation \rf{344},
\be \la{36}
ds^2 = (\tfrac{1}{4}\h+1+\k^2) \Big[ \frac{dr^2}{(1+r^2)(1+\k^2r^2)} + \big(
\frac{1+ \k^2r^2}{1+ r^2} + \frac{8\k^2}{\h} \big)^{-1}
d\p^2 \Big]\ . \ee
Again, formally taking $ r^2 \to -1 + (1-\k^2) \sinh^2 t$, the limit
$\k^2 \to 1$ is the exact metric of the $\frac{SU(1,1)}{\mathbb{R}}$ gWZW model.
 
 \sloppy
 Interestingly the exact metric \rf{36} may have different properties compared to the classical metric \rf{34}.
For example, analytically continuing further \mbox{($\k \to  i \k, \     r \to  \rho, \    \phi \to  i t $)} to the non-split $\eta$-deformation of AdS$_2$,
  there is a singularity at 
$\rho^2 = {\h - 8 \k^2 \over  \k^2 (\h+8)}$ that is always  present in the classical limit $\h \gg 1$ 
but   absent  for $\h <  8 \k^2$.

Written in the conformally-flat coordinates as in \rf{35}, the quantum-corrected metric \rf{36} is
\be \la{37}
ds^2 = (\tfrac{1}{4} \h+ 1+\k^2 ) \Big[ { dp^2 + dq^2 \ov p^2 + \k^2 q^2} - \frac{\k^2}{2 \h} \, \frac{ \big[ d (p^2+q^2)\big]^2 }{(p^2+\k^2 q^2)^2( \tfrac{1}{4} + {2\k^2 \ov \h} { p^2+ q^2\ov p^2 + \k^2 q^2})} \Big]\ .
\ee
Expanding the metric \rf{37} to first subleading order in small $\h^{-1} $ one obtains
\be \la{38}
ds^2 = \tfrac{1}{4} \h\, { dp^2 + dq^2 \ov p^2 + \k^2 q^2} + (1+\k^2){ dp^2 + dq^2 \ov p^2 + \k^2 q^2} - \frac{\k^2}{2} \, \frac{ \big[ d (p^2+q^2)\big]^2 }{
( p^2 + \k^2 q^2)^2 } + \OO(\h^{-1})
\ . \ee

%%%%%%%%%%%%%%%%%%%%%%%%%%%%%%%
\section{\texorpdfstring{$\l$}{lambda}-model \la{Slam}}
%%%%%%%%%%%%%%%%%%%%%%%%%%%%%%%

Let us now turn to the  $\l$-deformed \sms\unskip.
As we reviewed in Section \ref{las}, the $\l$-model may be
constructed by starting with the $G/G$ gauged WZW action and adding
a deformation  quadratic in the gauge field, breaking some or all of the gauge symmetry,
\be \begin{aligned} \la{40}
&\L = \kk\, \Big( \L_G(g)  + \Tr\big[ g^{-1} \partial_+ g A_- -  A_+ \partial_- g g^{-1} \\
&\qquad \qquad \qquad  \qquad \qquad + g^{-1} A_+ g A_-  - A_+ A_- + (\l^{-1} -1) A_+ P A_- \big]\Big) \ .
\end{aligned} \ee
One obtains a \sm by integrating out the  gauge field $A_\pm$ (here $\L_G$ is the WZW Lagrangian, see Section \ref{WS})
\begin{equation}\la{lamc}
\L = k \, \Big( \L_G (g) + \Tr[ J_+\, M^{-1}\, K_- ] \Big) \ ,\quad \qquad M = \operatorname{Ad}_g - I +(1-\lambda^{-1}) P \ .
\end{equation}
Here the integration over the gauge field  was done classically so the resulting \sm \rf{lamc} should only be understood as a classical approximation that will be subject to  quantum corrections (cf.\ the classical gWZW geometry in eq.\ \rf{6}).
 In what follows we shall construct such corrections in a manner consistent with 2-loop renormalizability  in the simplest examples based on $SU(2) \ov U(1)$ and $SU(2)$.

\subsection{\texorpdfstring{$SU(2)/U(1)$}{SU(2)/U(1)} example}
First let us consider the $\l$-model with 2d target space  based on the symmetric space ${G\ov H} = {SU(2) \ov U(1)}$.
We fix the $U(1)$ gauge symmetry (generated by $\s_2$) by choosing the following parametrization of the coset element
\be \la{42}
\te g = \exp(i \alpha \s_3) \exp({i} \beta \s_2) \ , \ \ \ \ \ \ \cos  \alpha = \sqrt{p^2+q^2}\ ,\ \ \ \ \ \tan \beta = \frac{p}{q} \ . \ee
The  \lam\   \rf{lamc} then becomes 
\be \la{43}
{\L} = \frac{\kk}{1-p^2-q^2}\big(\ka\, \partial_+p \partial_-p + \ka^{-1}\partial_+ q\partial_-q\big)\ , \qquad \qquad
\ka \equiv \frac{1-\l}{1+\l}
\ . \ee

\sloppy
The model \rf{43} again admits several useful limits. 
In general, in a ``boost'' limit that infinitely rescales the Cartan directions, the $\l$-model gives the T-dual of the \etm\ on an analytically continued coset \cite{Hoare:2015gda} (equivalent to the Poisson-Lie duality relating the $\eta$- and $\l$-models discussed in  Section \ref{etas}).
In this special 2d case, the $\eta$-deformed $S^2$ model is self T-dual (see footnote \ref{fstd}) so the T-duality may be skipped:
infinitely rescaling the coordinates,\foot{This limit takes the coordinates $(p,q)$ out of their proper coordinate range $p^2+q^2<1$ in \rf{43}. This may be avoided by instead considering the $\l$-model on the analytically continued coset $SU(1,1) \ov U(1)$.}
\be (p,q) \to \g(p, \kappa q) \ , \qquad \g \to \infty \la{resc} \ee
the metric becomes equivalent \cite{Hoare:2015gda} to the \etm \ metric \rf{35} on the analytically continued coset $SU(1,1) \ov U(1)$.
Changing coordinates $(p,q) \to (r,\p) = (q/p, \ha \log (p^2+q^2))$ 
 and analytically continuing $(r,\p) \to (ir, i\p)$, we obtain again the $\eta$-model \rf{21} on $SU(2) \ov U(1)$, with the couplings related by
\be
\h = 4 k \ka \ , \qquad \k = i \ka \ . \la{cre}
\ee

The limit $\ka=1$ (or $\l=0$) corresponds to the classical metric of the $SU(2)\ov U(1)$ gWZW model:
\be \la{45}
ds^2 = \frac{k}{1-p^2- q^2 }( { dp^2 + dq^2 }) = \kk(d\a^2 + \cot^2 \alpha\, d \b^2) \ , \ \ \quad (p,q) = \cos\a (\sin \b, \cos \b)
\ . \ee
As discussed in Section \ref{qcs}, the gWZW metric  \rf{45} is known to be subject to quantum corrections at each order. Thus it is already clear that the $\l$-deformation \rf{42} of this model should be subject to  non-trivial quantum corrections.

The \lam\ is known to be 1-loop renormalizable with only $\ka$ running \cite{Itsios:2014lca,Appadu:2015nfa} (eq.\ \rf{la2} with $c_{_{SU(2)}}=2$ and $\ka\equiv \tfrac{1-\l}{1+\l}$):
\begin{equation}
{\dt }\kk = 0+ \mathcal{O}(\kk^{-1}) \ , \qquad \qquad
\dt{\ka} = - \kk^{-1}(1-\ka^2) + \mathcal{O}(\kk^{-2}) \ .\la{310}
\end{equation}
The main observation here is that one may ensure the 2-loop renormalizability of this model by modifying the classical action
\rf{43} with a particular quantum correction.
A natural guess is that the necessary counterterms  originate from a determinant of integration over the 2d gauge field, which appears quadratically in \rf{40}.
Such integrals are not well-defined \cite{Schwarz:1992te}, with the resulting  local terms depending on a choice of regularization. One particular prescription is to write the  gauge field in terms of two scalars, $A= d u + d*v$ and integrate over $u,v$ with measures $M$ \cite{Tseytlin:1991wr,Schwarz:1992te} (see also \cite{Gerasimov:1990fi}), obtaining the result 
\be \la{7}
\int [dA] \, e^{ i \int d^2 \xi\, \sqrt{-h}
\, (-M \, A^\m A_\m)} = \exp \Big[\frac{i}{4 \pi} \int d^2 \xi  \, \sqrt{-h} \, \big(
\ha (\del \log M)^2 + \ha R^{(2)} \log M \big) \Big]\ .
\ee
 For generality we assumed a curved 2d background with curvature $R^{(2)}$, and we have ignored a trivial $M$-independent factor in \rf{7}.
The $R^{(2)} \log M$ term in \rf{7} is related to the familiar dilaton shift found in
\cite{Buscher:1987qj} (see also \cite{DeJaegher:1998pp}), and vanishes on a flat 2d background.

However, the $(\del \log M)^2$ term remains non-trivial on a flat 2d background.
In this case  $M=M(g)$ is the matrix that appears in the part of the $\l$-model action \rf{40} that is quadratic in the gauge field, i.e. $\L = \ldots + \Tr A_+ M(g)A_-$. It generally depends on the group element $g$, taking the the form $M \propto 1 + (\l^{-1} -1) P - \Ad_g$ (written as an operator $\Lie(G) \to \Lie(G)$). Hence the prescription \rf{7} 
suggests the
 following
 correction to the classical Lagrangian \rf{43} of the ${SU(2)}\ov {U(1)}$ \lam,\foot{Recall that the Lagrangian $\L = \GG_{mn} \del_+ x^m \del_- x^n = - \GG_{mn} \del_\m x^m \del^\m x^n$ has a minus sign relative to the target space metric when written in covariant 2d notation. This explains the relative minus sign between the two lines in \rf{47}.}
\be\begin{aligned} \la{47}
&\Delta \L =  \ha ( \del \log \det M )^2  =   \ha \big[ \del \log (1-p^2 - q^2)\big]^2 \ ,\\
&\Delta(ds^2) = -\ha \big[ d \log (1-p^2 - q^2)\big]^2 \ .
\end{aligned} \ee
Somewhat surprisingly, in this case the 1-loop correction \rf{47} happens to be independent of the deformation parameter $\l$.

This leads to a 1-loop corrected metric for the \lam\  \rf{43}  given by (cf. \rf{38})
\be \la{48}
ds^2 = \kk\, {\ka\, dp^2 + \ka^{-1}d q^2 \ov 1-p^2-q^2} - \frac{1}{2} \Big[\frac{d ( 1- p^2 - q^2 )}{1-p^2 - q^2 }\Big]^2
+ \OO ( k^{-1}) \ .
\ee
One can then check directly that the metric \rf{48} indeed solves the 2-loop RG equations %\rf{2},
\rf{27}
with  (recall that $\ka\equiv \tfrac{1-\l}{1+\l}$
)
\be \la{49}
{\dt }\kk = 0 + \kk^{-1}(\ka^{-1}-\ka)^2 + \mathcal{O}(\kk^{-2}) \ , \qquad \qquad
\dt{\ka} =  \kk^{-1}(1-\ka^2) + 0 + \mathcal{O}(\kk^{-3}) \ ,
\ee
and the components of the
diffeomorphism vector $\xi_m$ given by
\begin{align}
\xi_p &= \frac{2p}{1-p^2-q^2}\big[1 + {1 \ov \kk} \frac{\kappa (p^2-1) -\kappa^{-1} p^2 }{1- p^2-q^2} \big] \ ,
\no \\
\xi_q &= \frac{2q}{1-p^2-q^2}\big[1 + {1 \ov \kk} \frac{\kappa^{-1} (q^2-1) - \kappa q^2)}{1- p^2-q^2} \big] \ . \la{50}
\end{align}
The $\b$-function for the deformation parameter $\ka$
does not receive a 2-loop correction.
Surprisingly, the parameter $k$ (which has the interpretation of the level in the gWZW model) starts running
at 2-loop order.
Still, there is an RG invariant (cf. \rf{2888})
\be \la{51}
\dt \rk =0 \ ,\ \ \ \ \qquad \rk \equiv \kk + ({\kappa}^{-1} + {\kappa}) + \mathcal{O}(\kk^{-1}) \ .
\ee
This suggests that it is $\rk$ rather than $k$ that should be identified with level of the gWZW model in \rf{40}.
However, in this 2d case  the WZ term vanishes upon gauge fixing (any $\BB$-field is a total derivative in 2d) so the question of  which is the WZ level becomes somewhat meaningless.

These results are consistent with the limits of the \lam\ discussed above.
Setting (cf. \rf{42},\rf{45})
\be \la{317}
(p,q) = (1 + {\kk}^{-1}) \cos\a (\sin \b, \cos \b)
\ee
in \rf{48} and taking the gWZW limit, $\kappa =1$, we find
\be \la{52}
ds^2 = \kk(d\alpha^2 + \cot^2 \alpha\, d\beta^2)
+ 2 (d\alpha^2 + {\cot^2\alpha\ov \sin^{2}\alpha}\, d \beta^2) + \mathcal{O}(\kk^{-1}) \ , \ee
which precisely matches the large $\kk$ expansion of the exact gWZW metric (cf. \rf{6}), i.e.
\be \la{522}
ds^2 = (\kk+2)\Big(d\alpha^2 + \frac{\cot^2 \alpha}{1-\frac{2}{\kk}\cot^2\alpha} \, d\beta^2\Big) \ . \ee
Note that in this limit the RG invariant in \rf{51} becomes $\rk =\kk+2$, i.e. the
shifted level of the $SU(2)/U(1)$ gWZW model.\foot{In the $\ka = 1$ limit
the diffeomorphism vector \rf{50} becomes a gradient of the exact dilaton of the gWZW model:
$\xi_n = 2 \del_n \Phi$, $ \Phi = - \log (1-p^2 - q^2 ) - { 1 \ov 2 k} { 1 \ov 1-p^2 - q^2 } + \OO ( k^{-2}) $ (cf.
footnote \ref{f4}).}

In the \etm\ limit, the  corrected \lam\ metric \rf{48} reduces to the leading correction in the \etm\ metric  \rf{24}  if one modifies the scaling limit \rf{resc} and coupling relation \rf{cre} by certain quantum corrections,
\begin{align}
& p\to \g p \ ,\qquad \qquad q \to -i\g \k \big[1- 4 \h^{-1} ({1+\k^2}) \big]\, q \ , \qquad \qquad \g\to\infty \ , \la{54}\\
& \ka = -i \k \big[1-8\h^{-1} (1+\k^2)\big] \ , \qquad \qquad \tfrac{1}{4}\h-1+\k^2= -i\kk \k \ ,\la{53}
\end{align}
followed by again making the same change of variables $(r,\p) = (q/p, \ha \log (p^2+q^2))$ and  continuing $(r,\p) \to (ir, i\p)$.
 In this limit the RG equations \rf{49} reproduce the corresponding ones for the \etm\ couplings $\k$ and $\h$ in \rf{28},\rf{288} 
with the RG invariant \rf{51} being mapped to an analytic continuation of the $\eta$-model RG invariant \rf{2888},
\be
\rk = \kk + ({\ka}^{-1} + {\ka}) + \OO(\kk^{-1}) = i \frac{\h}{4\k} + \OO(\h^{-1}) =i \nu +\OO(\nu^{-1}) \ . \la{56}
\ee
This may be viewed as a hint that a quantum deformation parameter
that appears
in the corresponding exact S-matrix should be ${\rm q}= e^{- { i  \ov \rk} } = e^{-{1\ov \n}} $ (cf. \cite{Hoare:2015gda}),
which becomes  ${\rm q}= e^{ - { i  \ov \kk+2} }  $ at the gWZW point $\ka=1$.

\renewcommand{\la}[1]{\label{1910#1}} 
\renewcommand{\rf}[1]{(\ref{1910#1})} 
\renewcommand{\kk}{{\rm k}}
\renewcommand{\sm}{$\s$-model}
\renewcommand{\sm}{$\s$-model}
\renewcommand{\lam}{$\l$-model}
\renewcommand{\etm}{$\eta$-model}

\iffa  
 %%%%%%%%%%%%%%%%%%%%%%%%%%%%%%%%%%%%%%%%
\section{Renormalization of \texorpdfstring{$\lambda$}{lambda}-model: standard
configuration space}\la{standard}
%%%%%%%%%%%%%%%%%%%%%%%%%%%%%%%%%%%%%%%%

In section \ref{1910extended} we demonstrated the 2-loop renormalizability of the $\lambda$-model \rf{1} for general groups $G$ and symmetric spaces $G/\F$.
It is then natural to ask what
this implies
for the model on the standard or
physical configuration space, i.e. the \sm{} found by integrating
out $A_\pm$ in \rf{1}.
Doing so classically gives the following Lagrangian
\begin{equation}\la{lamphys}
\L = k \, \Big( {\PCM}(g) + {\WZ}(g) + \Tr[ J_+\, M^{-1}\, K_- ] \Big) \ ,\quad \qquad M = \operatorname{Ad}_g - I +(1-\lambda^{-1}) P \ .
\end{equation}
Similarly, for the NAD model \rf{14} we find
\begin{equation}\la{32}
\L = -\tfrac12 \hh\,\Tr\big[ \partial_+ v\, \M^{-1}
\, \partial_- v \big] \ , \quad \qquad \M
= \ad_v + P \ .
\end{equation}
The integration over $A_\pm$ may also give rise to quantum counterterms required to preserve the renormalizability of \rf{lamc} at 2 loops \cite{HLT1}.
It is natural to expect that, since the term quadratic in $A_\pm$ in the
Lagrangian \rf{1} has the form $\Tr[ A_+ M A_-]$, these corrections may depend on the matrix $M$ in \rf{lamc},
but determining their form in general
appears to be
non-trivial.
Here we will focus on the
examples of the \lam{} for the $SU(2) \ov U(1)$ symmetric space and $SU(2)$ group space.

%%%%%%%%%%%%%%%%%%%%%%%%%%%%%%%%%%%%%%%%
\subsection{\texorpdfstring{$SU(2)/U(1)$}{SU(2) \ov U(1)}}
%%%%%%%%%%%%%%%%%%%%%%%%%%%%%%%%%%%%%%%%

The $\lambda$-model for $SU(2) \ov U(1)$ is related by analytic continuation to that of $SU(1,1)/U(1)$, which was studied in detail in \cite{HLT1}.
Here we briefly summarize certain key points of the discussion there.
Fixing the $U(1)$ gauge symmetry by choosing the following parametrization of the coset element
\begin{equation}
g = \exp(i\alpha\sigma_3)\exp(i \beta \sigma_2) \ , \qquad \cos \a = \sqrt{p^2 + q^2} \ , \qquad \tan \b = \frac{p}{q} \ , 
\end{equation}
the \sm{} \rf{lamc} yields the following classical metric
(the $\BB$-field is trivial in 2d target space and $\ka$ is defined in \rf{133})
\begin{equation}\la{g0coset}
\GG_0 = \frac{k }{1-p^2-q^2}(\ka\, dp^2 + \ka^{-1} dq^2) \ .
\end{equation}
The observation in \cite{HLT1} was that this metric should be modified by a particular quantum correction
from the determinant \cite{Schwarz:1992te} resulting from integrating over $A_\pm$
\begin{equation}
\delta \GG = -\frac{1}{2}\big(d \log \det M\big)^2 = - \frac{1}{2} \big[d \log (1-p^2-q^2)\big]^2 \ . \la{su2u1counter}
\end{equation}
The 1-loop corrected background $\GG=\GG_0 + \delta \GG$
then solves the 2-loop RG equation \rf{rg} with
\begin{equation}\begin{split}\la{0oldcoset}
\frac{d}{dt} k & = 0 + \frac{1}{k} \frac{(1-\ka^2)^2}{ \ka^2} \ , \qquad\qquad
\frac{d}{dt} \ka = \frac{1}{k} (1-\ka^2) \ ,
\end{split}\end{equation}
and
\begin{equation}
\xi^p = -\frac{p}{k\ka}\Big[1-\frac{\ka}{k}+\frac{\ka^{-1}p^2 + \ka q^2}{k(1-p^2-q^2)}\Big]
\ , \quad
\xi^q = -\frac{q\ka}{k}\Big[1-\frac{1}{k\ka}+\frac{\ka^{-1}p^2 + \ka q^2}{k(1-p^2-q^2)}\Big]
\ , \quad \l_{p,q} = 0.
\end{equation}
In this analysis the symmetry \rf{z2sym}
of the 1-loop RG equation in \rf{0oldcoset}
survives at the 2-loop level.
Indeed, while the leading 1-loop terms in \rf{0oldcoset} agree with the RG equations
\rf{6d2loopbetacoset} found from the analysis on the extended configuration space,
they deviate at the 2-loop order.
Since the 2-loop terms in a two-coupling theory are generally scheme-dependent,
we can, in fact, match the two $\b$-function expressions by
redefining the parameters in \rf{g0coset} as follows
\unskip\foot{\ The most general redefinition achieving this is ($C_1$ and $C_2$ are free constants)
$ k \to k - \tfrac{(1 -\ka)^2}{\ka} + 2 C_1$ and

$\quad \ka \to \ka + \frac{4}{k} \big[(1-\ka)\big(1-(1+\ka)C_2\big) - 2 C_1(1-\ka^2)\operatorname{arctanh}\ka\big]$.
}
\begin{equation}\la{38}
k \to k - \frac{(1 -\ka)^2}{\ka} \ , \qquad \qquad \ka \to \ka + \frac{4(1-\ka)}{k} \ .
\end{equation}
Note that in the $\ka \to 1$ limit the level $k$ remains unmodified, in agreement with this limit corresponding to the $SU(2) \ov U(1)$ gWZW model.

\fi

%%%%%%%%%%%%%%%%%%%%%%%%%%%%%%%%%%%%%%%%
\subsection{\texorpdfstring{$SU(2)$ example}{SU(2) example}}
%%%%%%%%%%%%%%%%%%%%%%%%%%%%%%%%%%%%%%%%

Now let us turn to the \lam{} for the group $SU(2)$
\cite{Balog:1993es,Sfetsos:2013wia}.
Parametrizing the group element as
\be
g = \exp{\big[ -i \arcsin\a \, \big(\cos{\b} \, \s_2 + \sin{\b}(\cos{\g} \, \s_3 - \sin{\g}\, \s_1)\big)\big]} \ ,
\ee
we obtain from \eqref{40} the following 3-dimensional classical \sm{} metric and $\BB$-field \cite{Sfetsos:2013wia}
\unskip\foot{Note that, in these coordinates, WZ term  $k \L_{\WZ}(g)$ in \eqref{40} contributes $k(\arcsin\a - \a \sqrt{1-\a^2}) \sin{\b} \, d\b \wedge d\g$ to the $\BB$-field.}
\begin{align}
(ds^2)^{(0)} & = k \big[\frac{d\a^2}{\ka(1-\a^2)} + \frac{\kappa \a^2}{\Delta} ( d\beta^2 + \sin^2\beta d\gamma^2)\big] \ , \quad\qquad
\Delta(\a) \equiv \ka^2 + (1-\ka^2)\a^2 \ , \no
\\
\BB^{(0)} & = k\big(\arcsin\a - \frac{\ka^2\a\sqrt{1-\a^2}}{\Delta} \big)\sin\beta\, d\beta\wedge d\gamma \ , \la{b1}
\\
\HH^{(0)} & =d\BB^{(0)}= \frac{k\a^2}{\sqrt{1-\a^2}\Delta^2}\big[2\ka^2+(1-\ka^2)\Delta\big] \sin\b\, d\a \wedge d\b \wedge d\g \ . \no
\end{align}
As discussed in Section \ref{las},   $\l$-models are known to be 1-loop renormalizable but, again, one can check that the classical background \rf{b1} does not solve the 2-loop RG equation \eqref{BMT} (with only $k,\ka$ running) in \textit{any} covariant subtraction scheme and with \textit{any} choice of diffeomorphism vector. (Here due to the $SU(2)$ global symmetry acting on $\b,\g$, the diffeomorphism vector must be of the form $\xi^\a=\xi^\a(\a)$, $\xi^{\b,\g}=0$ and the $\BB$-gauge transformation must be trivial, $\del_{[m}\l_{n]}=0$.)

However, we observe that it is possible to restore the 2-loop renormalizability of the model
by adding  special
quantum $(1/k)$ counterterms to this classical background,
\begin{equation}\begin{aligned}\la{b2}
ds^2  & = (ds^2)^{(0)} + \frac{2(1-\ka^2)^2 \a^4}{\ka^2 (1-\a^2)\Delta^2} d\a^2 \ ,
\\
\BB & = \frac{\kk}{k} \BB^{(0)} -2 \big( \arctan\frac{\a}{\ka\sqrt{1-\a^2}}
- \frac{\ka \a \sqrt{1-\a^2}}{\Delta}\big)\sin\beta\, d\beta\wedge d\gamma \ ,
\\
\H & = \frac{\kk}{k} \H^{(0)} - \frac{4\ka\a^2}{\sqrt{1-\a^2}{\Delta^2}} \sin\b\, d\a \wedge d\b \wedge d\g \ ,
\qquad \qquad \kk \equiv k + \frac{4+(1+\ka^2)^2}{4\ka} \ .
\end{aligned}\end{equation}
This corrected background \rf{b2} indeed solves the 2-loop RG equation \eqref{BMT} with\foot{The 1-loop terms in \rf{0} match the general result \eqref{la2} with $c_{_{SU(2)}}=2$ and $\ka \equiv \frac{1-\l}{1+\l}$.}
\begin{equation}\la{0}
\frac{d}{dt}k = 0 + \frac{(1-\ka^2)^3(5+3\ka^2)}{8k\ka^2} \ , \qquad
\qquad \frac{d}{dt} \ka = \frac{(1-\ka^2)^2}{2k} \Big[1 - \frac{(1-\ka^2)^2}{k\ka} \Big] \ .
\end{equation}
The only non-zero component of the diffeomorphism vector $\xi^m$ is
\begin{equation}\begin{split}\la{diffv}
\xi^\a & = \frac{\a(1-\a^2)(1-\ka^2)}{k\Delta}\Big[\ka + \frac{2\a^2\ka^2(1-\ka^2) - (3-2\ka^2+\ka^4)\Delta^2}{k\Delta^2}\Big] \ ,
\end{split}\end{equation}
and the B-gauge transformation in \eqref{BMT} vanishes, $\l_m=0$.

Note that the coupling $\kk$ defined in \rf{b2} does not run at 2-loop order,
\be \frac{d}{dt} \kk = 0 + \frac{1}{\kk}\times 0 + \ldots \la{knr} \ . \ee
This RG invariant $\kk$ is the coefficient of $\arcsin\a$ (present in $\BB_0$ in \rf{b1}) in the 1-loop corrected background \rf{b2}. Choosing it
to be integer-valued removes the global ambiguities arising from the $\arcsin\a$ term.
Furthermore, given that under a large transformation,
\unskip\foot{Note that $\arctan\frac{\a}{\ka\sqrt{1-\a^2}} = \operatorname{sign} \ka \, \arcsin\frac{\a}{\sqrt{\Delta}}$ and $\alpha = \frac{\alpha}{\sqrt{\Delta}}$ for $\alpha = 0,\pm1$.}
\begin{equation}
\delta (\arcsin \a) = \operatorname{sign} \ka \ \delta (\arctan\frac{\a}{\ka\sqrt{1-\a^2}})\ ,
\end{equation}
and that the coefficient of $\arctan\frac{\a}{\ka\sqrt{1-\a^2}}$ in \rf{b2} is integer-valued, the $\arctan$ term does not lead to any additional ambiguities.
The quantization of the flux,
\be
\frac{1}{4\pi^2} \int \H^{(0)} = k \ , \qquad\qquad \frac{1}{4\pi^2} \int \H = \kk -2 \ ,
\ee
also supports the identification of $\kk$ as integer-valued.

In the case of the $SU(2) \ov U(1)$ \lam{}
it was possible to write the 1-loop corrections 
 \eqref{47}
in terms of a simple 
counterterm 
 $\Delta \L = -\ha (\del \log \det M)^2 = -\ha (\Tr [M^{-1} \del M ])^2$ in terms of the matrix $M$ as defined in \eqref{lamc}.
For this $SU(2)$ case, we find however that this particular counterterm is not sufficient to restore 2-loop renormalizability (in any covariant subtraction scheme) --- i.e.\ the required corrections \rf{b2} cannot be written in this form even after a covariant scheme change.
 Instead,
let us also include
   similar counterterms (with coefficients $c_i$)
built out of the quantity $M^{-1} \del M$.
In addition, we may also include counterterms (with coefficients $d_i$)
proportional to the terms present in the classical Lagrangian \eqref{lamc} (and its image under parity). As a result, we are led to the following ansatz\foot{In \rf{qccc} and below, $\L_\PCM (a) \equiv -\ha \Tr[a^{-1} \del_+ a \, a^{-1} \del_- a]$ and $\L_{\WZ}$ denote the principal chiral model and Wess-Zumino Lagrangian terms (see Section \ref{PkS}).}
\begin{equation}\begin{aligned}\la{qccc}
\Delta \L &= c_1 (\del_+ \log \det M) ( \del_- \log \det M)
+ c_2 \L_{\PCM}(M) + c_3 \L_{\WZ}(M)
\\ &\quad + d_1 \L_{\PCM}(g) + d_2 \L_{\WZ}(g) + d_3 \Tr[J_+ M^{-1} K_-] + d_4 \Tr[J_- M^{-1} K_+] \ .
\end{aligned}\end{equation}
We find that this matches the required 1-loop corrected background
\rf{b2} provided the constants $c_i$ and $d_i$ take the following values
\begin{align}\la{origin}
&c_1 =-\tfrac{3+\ka}{2(1-\ka)} \ , \qquad c_2 = - \tfrac{2(1+\ka)}{1-\ka} \ , \qquad c_3 = 1 \ ,
\\
&d_1 = 0 \ , \quad d_2 = {\rm k} - k - 2 \ ,\quad
\ d_3 = \tfrac{1}{2}\big[{\rm k} - k + (1+\kappa^{-1})\big] \ , \quad d_4 = -\tfrac{1}{2}\big[{\rm k}-k - 3 (1+\kappa^{-1})\big] \ .\no
\end{align}
Combining the quantum counterterms \rf{qccc} with the classical Lagrangian \eqref{lamc} allows us to represent the \sm{} corresponding to
the 1-loop corrected geometry \rf{b2} in the form
\begin{equation}\begin{aligned}\la{originlag}
&\L  = k\, \L_{\PCM}(g) + ({\rm k} -2) \L_{\WZ}(g)
\\ &\quad  \qquad
+ \tfrac12 \Tr\Big[ \big({\rm k} + k + (1+\ka^{-1})\big) J_+ M^{-1} K_-
- \big({\rm k}-k - 3 (1+\ka^{-1})\big) J_- M^{-1} K_+ \Big]
\\
&\quad \qquad - \tfrac{3+\ka}{2(1-\ka)} (\del_+ \log \det M)(\del_- \log \det M) - \tfrac{2(1+\ka)}{1-\ka} {\PCM}(M) + {\WZ}(M) \ \\
&M(g) = \big( \tfrac{1-\ka}{1+\ka}\big)^2 - \Ad_g \ .
\end{aligned}\end{equation}
We note that coefficients of the WZ terms in \rf{originlag} are RG invariants, $\kk-2$ (see eq.\ \rf{knr}) and $1$. This is consistent with usual expectations and suggests that the representation \rf{originlag} is a natural one.

In the WZW limit $\ka \to 1$ when the RG invariant $\kk$ in \rf{b2} reduces to the usual shifted WZ level
\begin{equation}
\kk\big|_{\ka = 1} = k+2 \ ,
\end{equation}
the corrections to the metric and $\BB$-field in \rf{b2} vanish, so that
the expression in \rf{originlag}  reduces to the classical WZW Lagrangian
$k \big[\L_{\PCM}(g) + \L_{\WZ}(g) \big]$. Indeed, even though the coefficients $ \tfrac{3+\ka}{2(1-\ka)}$ and $\tfrac{2(1+\ka)}{1-\ka}$ in \rf{b2} blow up in the limit $\ka \to 1$, the corresponding $M$-dependent expressions vanish faster and all $M$-dependence  disappears in the $\ka=1$ limit.

%%%%%%%%%%%%%%%%%%%%%%%%%%%%%%%%%%%%%%%%
\subsubsection{\texorpdfstring{$\eta$}{Eta}-model limit}
%%%%%%%%%%%%%%%%%%%%%%%%%%%%%%%%%%%%%%%%

Like the $SU(2) \ov U(1)$ example discussed above, the $SU(2)$ \lam\ \rf{b1} admits a formal ``boost'' scaling limit,
\begin{equation}
\alpha \to \sin(\a + i \zeta) \ , \qquad \qquad \zeta \to \infty \ , \la{fl}
\end{equation}
when it becomes (dropping a closed 2-form contribution to the $\BB$-field)
\begin{equation}\begin{split}\la{class}
\widetilde{ds^2}  & = k \Big[ \frac{1}{\kappa} d\alpha^2 + \frac{ \kappa}{1-\kappa^2} ( d\beta^2 + \sin^2\beta d\gamma^2)\Big] \ ,
\qquad 
\widetilde{\BB} = k\cos\beta  \, d \alpha\wedge d\gamma \ .
\end{split}\end{equation}
This geometry is related by abelian T-duality (in the $\a$ direction) to the squashed 3-sphere \cite{cherednik},
\begin{equation}\la{323}
{ ds^2} = k\Big[ \kappa (d\tilde\alpha - \cos\beta d\gamma)^2 + \frac{ \ka}{1-\ka^2}(d\beta^2 + \sin^2\beta d\gamma^2)\Big] \ , \qquad  \BB=0 \ , 
\end{equation}
which in turn has the interpretation as the $\eta$-deformation \eqref{ep} of the $SU(2)$ PCM (cf. \cite{Kawaguchi:2010jg}). The squashed 3-sphere geometry \rf{323} is an anomalous case of an $\eta$-model that 
 is automatically renormalizable at 2-loops without needing quantum corrections (this statement was checked in App.\ B of \cite{HLT2} --- and more generally for ``squashed'' models interpolating between a PCM and a symmetric space). As we shall now demonstrate, the quantum corrections derived above for the  $SU(2)$ $\l$-model are consistent with this fact.

Starting  from the 1-loop corrected \lam\ background \rf{b2}
 and taking the same scaling limit \rf{fl}, we get a corrected version of \rf{class},
\be\begin{aligned}
\la{simpler}
\widetilde{ds^2}  & = \big(\frac{k}{\kappa} + \frac{2}{\ka^2}\big) d\alpha^2 + \frac{k \kappa}{1-\kappa^2} ( d\beta^2 + \sin^2\beta d\gamma^2) \ , \\
\widetilde{\BB} & = \big[k + \frac{4+(1+\ka^2)^2}{4\ka}\big]  \cos\beta d\a \wedge d\gamma \ .
\end{aligned}\ee
The background \rf{simpler} solves the 2-loop RG equations \eqref{BMT} with the same $\b$-functions  \rf{0} as the $\l$-model and vanishing diffeomorphism vector  and $\BB$-gauge transformation  ($\xi_m=\l_m=0$). 
To extend the T-duality relation between \rf{class} and \rf{323} to the quantum level,
we  use the 1-loop corrections to the T-duality transformation laws
of \cite{Kaloper:1997ux}
(see also \cite{Parsons:1999ze}), generalizing the quantum correction in \eqref{5}.\foot{We note that, in order to directly apply the results of \cite{Kaloper:1997ux}, one first has to perform a scheme change, $(\GG+\BB)_{\mu\nu} \to (\GG+\BB)_{\mu\nu} + \frac14 \H^2_{\mu\nu}$ .}
As a result, the T-dual of \rf{simpler} is found to be
\begin{equation}\la{2loopdual} \begin{aligned}
ds^2 {}&{}= \big[k \ka +\tfrac{1}{2} (1+\ka^2)^2\big]\big(d\tilde\alpha - \cos\beta d\gamma\big)^2 + \big[\frac{k \kappa}{1-\kappa^2} + \tfrac{1}{2} (1-\ka^2)\big] \big( d\beta^2 + \sin^2\beta d\gamma^2\big) \ , \\
 \BB{}&{}=0 \ ,
 \end{aligned}
\end{equation}
where we have rescaled $\tilde \alpha$.
Indeed this just the squashed 3-sphere geometry \rf{323} with redefined parameters,
\begin{equation}
(k, \ka) \to (\hat k, \hat \ka) \ , \qquad \quad  k = \hat{k} - \frac{3+\hat{\ka}^4}{2\hat{\ka}} \ , \qquad  \ka = \hat{\ka}+\frac{1-\hat{\ka}^2}{\hat{k}} \ .
\end{equation}
Thus the fact that the quantum corrected $\l$-model \rf{b2} is stable under the 2-loop RG flow is perfectly consistent, in the $\eta$-model or ``boost'' limit, with the fact that the squashed 3-sphere is automatically stable without  quantum deformation.

We see in this example that
the known quantum corrections to the T-duality transformation rules
are naturally consistent with the 
quantum deformations of integrable models for higher-loop renormalizability.
As we shall discuss in the next section, similar quantum corrections are expected for non-abelian duality.

\renewcommand{\la}{\label}
\renewcommand{\rf}[1]{(\ref{#1})} 
\def \kk {k}

%%%%%%%%%%%%%%%%%%%%%%%%%%%%%%%%%%%%%%%%
\subsection{Non-abelian dual limit \la{NADqS}}
%%%%%%%%%%%%%%%%%%%%%%%%%%%%%%%%%%%%%%%%

In a particular limit, 
\be \begin{aligned} 
&k \to \infty \ , \qquad \l= 1 - \tfrac{\h}{2k}  \ , \qquad g=e^{-\tfrac{h}{2k}v} \ ,  \\
&v \in \Lie(G) \text{ and } \h  \text{ fixed} \la{limr}
\end{aligned} \ee
 the \lam\ \rf{40} becomes the non-abelian dual (NAD)  of the PCM or symmetric space (see Section \ref{TS})
\be
\L = -\ha \h \, \Tr[ \del_+ v \, \frac{1}{P + \ad_v} \del_- v ] \ . \la{NADrep}
\ee
It was noted earlier in the literature  \cite{n1,n2,n3,Bonneau:2001za} that the NAD theories \rf{NADrep} appear not to be renormalizable beyond the 1-loop order, prompting doubt about their inequivalence with the PCM and  symmetric space \sms\unskip. In this section, we shall resolve this confusion by observing that the classical NAD models \rf{NADrep} must be supplemented with particular finite counterterms to maintain renormalizability beyond 1-loop order.
We shall produce the necessary counterterms in the $SU(2) \ov U(1)$ and $SU(2)$ examples by taking the NAD limit of the corresponding  $\l$-model corrections  constructed above.

\subsubsection{NAD of \texorpdfstring{$S^2$}{S^2} sigma model}

For the $SU(2) \ov U(1)$ $\l$-model in the coordinates \rf{43}, the NAD limit amounts to
\be \la{455}
p= \ka \barp\ , \qquad q= 1 - \ka^2 \barq \ , \qquad
\kk \to \infty\ , \quad \ka\to 0 
\ , \qquad \h\equiv 4 k\ka ={\rm fixed}
 \ . \ee
The resulting theory \rf{NADrep} is the NAD of the \sm\ on $S^2$, with metric
\be \la{46}
ds^2 =\frac{\h}{4}\, {dx^2 + d y^2 \ov  2 y-x^2}\ , \ee
and trivial $\BB$-field since the target space is 2d. In the NAD limit \rf{455}, the leading quantum correction to the $\l$-model in \rf{48} suggests a particular correction to the NAD metric \rf{46},
\be \la{57}
ds^2 =\frac{\h}{4}\, {dx^2 + d y^2 \ov  2 y-x^2} - \frac{1}{2} \Big[\frac{d ( 2 y-x^2 )}{2 y-x^2}\Big]^2 + \OO(h^{-1})  
 \ ,
\ee
which indeed solves the 2-loop RG equation \rf{27} with only $\h$ running,
\be \la{58}
\dt \h =  4 + \frac{16}{\h} + \mathcal{O}(h^{-2}) \ , \ee
and matching the 2-loop $\b$-function the dual undeformed $S^2={SU(2) \ov U(1)}$ $\s$-model (eq.\ \eqref{sb} with $c_{_{SU(2)}} =2$, $c_{_{U(1)}}=0$). Note that, since the coordinate redefinition in the NAD limit \rf{455} depends on $\ka=\ka(t)$, 
the diffeomorphism vector gets an additional contribution compared to \rf{50} and 
 its
 components are given by
\be\begin{aligned}
&\xi_x = { 4x \ov 2 y-x^2} - {1\ov \h} {8 x(1 + 3 x^2 - 4 y)\ov ( 2 y-x^2)^2} + \OO(\h^{-2}) \ , \\
&\xi_y =  { 2 (  2 y-1) \ov  2 y-x^2} + {1\ov \h} { 8(1 + 2 x^2 - 2 y)\ov ( 2 y-x^2)^2} + \OO(\h^{-2}) \ .
\end{aligned} \ee

If one takes a further scaling limit,
\begin{equation}\la{addsca} \begin{aligned}
&x \to \gamma \cos\chi \ , \qquad y \to \tfrac12  \gamma^2 + \gamma y \ , \qquad \gamma \to \infty \ ,
% \\
%&ds^2 \to \frac{h}{1 - x^2 }(dx^2 + dy^2) \  .
\end{aligned}
\end{equation}
the NAD metric  \rf{46} degenerates to an abelian T-dual of $S^2$ (in the $y$ direction),\foot{The T-dual metric \rf{h2dual} can also be found by directly taking the limit $\kk\to \infty$, $\ka \to 0$ with
$p = \cos \chi$, $q = \ka y$ and $\h\equiv 4 k\ka$ fixed in the $\l$-model metric \rf{43}.}
\be\la{h2dual}
ds^2 = \tfrac{1}{4}\h \, ( d \chi^2 +\frac{1}{\sin^2\chi} d y^2) \ .
\ee
Taking a particular 1-loop modification of the limit \rf{addsca} (cf. \rf{317}),
\be\la{329}
x \to \gamma (1+ 4\h^{-1}) \cos{\chi} \ , \qquad y \to \tfrac12  \gamma^2 + \gamma (1+4 \h^{-1}) y \ , 
\qquad \gamma \to \infty \ ,
\ee
and shifting the coupling $\h \to \h -8$ in  the quantum-corrected NAD metric \rf{57}, we obtain a quantum-corrected T-dual metric \rf{h2dual},
\be  \la{3277}
ds^2 =\tfrac{1}{4} \h  \, \big[d\chi^2 +\frac{1}{\sin^2 \chi} \big(1+8 \h^{-1}\cot^2\chi\big)dy^2\big] + \OO(\h^{-1}) \ ,
\ee 
which coincides with the  known corrections  \rf{5} to the T-duality transformation rule \cite{Tseytlin:1991wr}
(with ${\cal M} = \sin^2{\chi}$ and $\h \to \tfrac{1}{4}\h$ so that \rf{5} gives $\tilde {\cal M} =\frac{1}{\sin^2 \chi} \big(1+2 (\h/4)^{-1}\cot^2\chi\big)$ in this case).

\def \kk {\rm k}
\renewcommand{\la}[1]{\label{1910#1}} 
\renewcommand{\rf}[1]{(\ref{1910#1})} 

\subsubsection{NAD of \texorpdfstring{$SU(2)$}{SU(2)} principal chiral model}

In the case of the $SU(2)$ $\lambda$-model in the coordinates \rf{b1}, the NAD limit \eqref{limr} amounts to
\begin{equation}\la{nadlimit2}
\a \to \ka \, \alpha\ , \ \ \ \qquad
\kk \to \infty\ , \quad \ka\to 0 
\ , \qquad \h\equiv 4 k\ka ={\rm fixed}
\end{equation}
where $\hh$ and the new coordinate $\a$ are fixed.
Taking this limit in the 1-loop-corrected $\l$-model background \rf{b2} suggests a particular quantum correction to the $SU(2)$ NAD model,
\begin{align}
ds^2 & = \Big[\frac{\hh}{4} + \frac{2\a^4}{(1+\a^2)^2} \Big] d\alpha^2 + \frac{h}{4}\frac{\a^2}{(1+\a^2)} (d\beta^2 + \sin^2\beta d\gamma^2) \ , \no \\
\BB & = \Big[\frac{\hh+5}{4}\frac{\a^3 }{(1+\a^2)} + \frac{2\a}{1+\a^2} - 2\arctan{\a}\Big] \sin \b d\b\wedge d\g \ , \la{f2zero} \\
\H & = \Big[\frac{\hh+5}{4}\frac{\a^2(3+\a^2) }{4(1+\a^2)^2} - \frac{4\a^2}{(1+\a^2)^2}\Big] \sin \b d\a\wedge d\b\wedge d\g \ .  \no
\end{align}
Indeed it solves the 2-loop RG equations \eqref{BMT} with
$\b$-function,
\begin{align}
\frac{d}{dt} \hh = 2 + {2}{\hh}^{-1} \ , \la{333}
\end{align}
matching that of the $SU(2)$ PCM (i.e. \eqref{Prg} with $c_{_{SU(2)}}=2$) and diffeomorphism vector,
\begin{align}
\la{xcorr2}
\xi^\a = \frac{2\a}{\hh(1+\a^2)}\Big[(1 -\a^2) - \frac{4(5+5\a^2+3\a^4+\a^6)}{h(1+\a^2)^2}\Big] \ , \qquad \xi^{\b,\g} = \l_{\a,\b,\g} = 0 \ ,
\end{align}
equalling the $\l$-model diffeomorphism vector \rf{diffv}  in the NAD limit \rf{nadlimit2} plus an extra contribution due to the RG-dependent rescaling of $\a$.

We can also take the NAD limit \eqref{limr} in the corrected Lagrangian written in the form \rf{originlag},
 getting
\begin{equation}\begin{aligned}\la{originlagnad}
&\L  = - (\tfrac{1}{2} \hh + \tfrac{9}{4}) \Tr\big[\partial_+ v\, M^{-1} \partial_- v \big]
- \tfrac{7}{4} \Tr\big[\partial_- v\, M^{-1} \partial_+ v \big] \qquad \qquad 
\\
&\qquad \quad - \tfrac{3}{2} (\del_+ \log \det M)(\del_- \log \det M) - 2 \, \L_{\rm PCM}(M) + \L_{\rm WZ}(M) \ ,\\
&M \equiv 1 +\ad_v \ .
\end{aligned}\end{equation}
This indeed matches with \rf{f2zero} upon choosing the coordinates,
\begin{equation}
v = -\tfrac{i}{2} \a \big[\cos{\b} \, \s_2 + \sin{\b}(\cos{\g} \, \s_3 - \sin{\g}\, \s_1)\big] \ .
\end{equation}

In a further limit of  infinite shift of $\a$,
\begin{equation}\la{limit}
\alpha \to \alpha + \ell \ , \qquad \ \ \ \ell \to \infty \ ,
\end{equation}
the NAD  background  \rf{f2zero} degenerates to an abelian T-dual of the PCM on $SU(2)$ (in the $\a$ direction). The corrected NAD background then leads to a corrected T-dual background in this limit (dropping a trivial closed 2-form contribution to the $\BB$-field),
\begin{align}
\wt{ds^2} = \tfrac{\hh+8}{4}\,d\alpha^2 + \tfrac{\hh}{4}\big(d\beta^2 + \sin^2\beta d\gamma^2\big) \ ,\qquad \ \ \
\BB = \tfrac{\hh+5}{4} \, \cos \b\, d\a \wedge d\g \la{f2zerolim1}
\ .
\end{align}
This background ($\mathbb{R} \times S^2$ supported by a $\BB$-field) is indeed correctly related to the  $SU(2)$ PCM,
\begin{equation}\begin{split}
{ds^2}%& = \tfrac{\hh+2}{4}\big(d\b^2 + d\tilde{\a}^2 + d\g^2 - 2\cos\b\, d\tilde\a d \g\big)
= \hh \big( d \theta ^2+ \sin^2 \theta\, d \psi^2 + \cos^2 \theta \, d \chi^2 \big) \ ,
\end{split}\end{equation}
by (i) the 1-loop corrected T-duality transformation $\a \to \tilde \a$ of \cite{Kaloper:1997ux}, (ii) shifting the coupling $\hh \to \hh-2$, and (iii) changing coordinates  $\tilde \a = \psi + \chi$, $\g = \psi - \chi$ and $\b = 2\theta$.

\renewcommand{\la}{\label}
\renewcommand{\rf}[1]{(\ref{#1})} 
\def \kk {k}

\subsubsection{Summary}

These findings extend the previous conclusions \cite{Fridling:1983ha,Fradkin:1984ai} about the 1-loop quantum
equivalence of the models related by the non-abelian duality to the 2-loop level. As for abelian T-duality, NAD (properly modified by
quantum $\a'$ corrections) should also be a symmetry of the \sm\  $\beta$-functions or the string effective action to all orders in $\a'$.

Resolving earlier problems checking NAD at 2-loop level \cite{n1,n2,n3,Bonneau:2001za}, 
we found that the preservation of quantum equivalence requires a non-trivial 1-loop correction to the
classical NAD model metric.
In the $SU(2) \ov U(1)$ and $SU(2)$ cases, we explicitly constructed the corrections restoring 
 2-loop renormalizability, with the 2-loop $\b$-functions matching the associated symmetric space and group models. This is a non-trivial result since these 2-loop $\b$-functions are  invariant under scheme changes (redefinitions of $h$) since the PCM/symmetric space and their NADs are \textit{single-coupling} theories.

In further abelian T-dual limits, we obtained corrected versions of T-dual models matching the known quantum corrections to the T-duality rules (the simple case of \rf{5} and the more general case of \cite{Kaloper:1997ux}). In these special cases, this approach  serves as an alternate derivation of the leading correction to the T-duality transformation.

%%%%%%%%%%%%%%%%%%%%%%%%%%%%%%%
\section{Discussion}
%%%%%%%%%%%%%%%%%%%%%%%%%%%%%%%

We have demonstrated that, just as for gauged WZW models,
the invariance of integrable $\s$-models under the
two-loop (and higher) RG flow requires a specific quantum
deformation of the classical Lagrangian (i.e. of the target space metric and $B$-field).

In particular, we proposed an exact metric for the $\eta$-deformation of the $S^2$ (and $H^2$) $\s$-model
that solves the 3-loop RG flow equations and is consistent with the gWZW limit.
We also found the leading-order deformation of the \lam\
for $SU(2) \ov U(1)$, which solves the
2-loop RG equations and is consistent with the gWZW and $\eta$-model limits.
For this example we identified the origin of the deformation as a finite counterterm
resulting from the determinant of integrating over the auxiliary 2d gauge field
(the same determinant that leads to the correction of the dilaton term on a curved 2d background). 

We also found the leading-order deformation of the \lam\ for $SU(2)$ that solves the 2-loop RG equations, 
but in that case the necessary counterterm does not take the same simple and identifiable form.
We checked the consistency of the corrections with the WZW model limit and the limit T-dual to the squashed 3-sphere ($\eta$-model).

In the non-abelian dual limit, the corrections for the $\l$-models generated predictions 
for leading corrections to the $SU(2) \ov U(1)$ and $SU(2)$ NAD models.
These corrections were indeed found to restore the NAD models' 2-loop renormalizability, 
resolving  the earlier problems  \cite{n1,n2,n3,Bonneau:2001za} in checking the consistency of non-abelian duality with the 2-loop RG flow.

%%%%%%%%%%%%%%%%%%%%%%%%%%%%%%%%%%%%%%%%%%%%%%%%%%
 \

The classical $\l$-model's $\mathbb{Z}_2$ transformation  \rf{z2sym}
is respected by the leading corrections found in the above examples.
In explicit coordinates, this is seen as invariance of the counterterm in \rf{48} (in the $SU(2) \ov U(1)$ case) under $k \to -k$, $\ka\to-\ka$, $p \to -p$; and invariance of the counterterms in \rf{1910b2} (in the $SU(2)$ case) under $k\to -k$, $\ka \to -\ka$, $\a \to -\a$.
The resulting 2-loop $\b$-functions \rf{49} and \rf{19100} are thus invariant under $k \to -k$, $\ka \to -\ka$. As we shall see later in Chapter \ref{CT} (see eq.\ \rf{19102777}), the preservation of this $\mathbb{Z}_2$ invariance requires a particular formulation of the quantum theory, and  in generic subtraction schemes it would require a certain coupling redefinition to become manifest.

In the $SU(2) \ov U(1)$ case, the $\l$-model \rf{43} is invariant under an additional $\mathbb{Z}_2$ transformation: $p \leftrightarrow q$, $\ka \to \ka^{-1}$ (i.e. $\l \to -\l$), associated with the subgroup $H=U(1)$ being abelian.\foot{I thank B. Hoare for a discussion of this point.}  Indeed this $\mathbb{Z}_2$ is also respected by the counterterm in \rf{48}, and the resulting 2-loop $\b$-functions \rf{49} are then invariant under $\ka \to \ka^{-1}$.

\ 

In summary, while some models (e.g., on groups or symmetric spaces) have sufficient \textit{manifest} global symmetry to ensure renormalizability to all orders, general integrable models may not have any manifest symmetries. However, the \textit{hidden} integrable symmetries still seem to constrain the RG flow. This pattern is found to persist at higher-loop orders, and in such cases the hidden symmetries seem to require a particular choice of finite counterterms for their preservation at the quantum level.  

A general open question is how to systematically construct the necessary finite counterterms. One natural approach would be to try to derive them by  imposing  ``Ward identities'' for the  hidden integrable symmetries
 at the quantum level, i.e. requiring the  choice of finite counterterms  to be consistent with quantum integrability, and  assuming that such counterterms would also be consistent with renormalizability.
 Another approach, assuming a UV fixed point and a known underlying quantum S-matrix,
 would be to try to systematically reconstruct the corrected \sm\ 
 from a dual massive model, as considered in \cite{fateev96,Litvinov:2018bou,Alfimov:2020jpy}.

\

An interesting open problem would be to construct an
exact generalization of the 1-loop-corrected metric \rf{48} for the $SU(2) \ov U(1)$ \lam\  reducing in the appropriate limits to the exact gWZW
metric \rf{522} and the proposed exact  \etm\ metric \rf{36},\rf{37}. This is non-trivial given the lack of isometries in the \lam\ metric
and that the effective action approach used for the gWZW model \cite{Tseytlin:1993my} does not appear to apply directly to  non-conformal theories.

It would be interesting to see whether the exact metric \rf{24} of the $SU(2)\ov U(1)$ $\eta$-model,
translated to the Hamiltonian framework, would have a natural interpretation in terms of  a deformation of
the classical integrable structure.

A case that was not yet considered is the higher-loop RG flow of the $\eta$-deformation of the PCM (rather than the symmetric space). For the simplest $SU(2)$ case, the $\eta$-model happens to coincide \cite{Klimcik:2002zj} with the $\sigma$-model on a `squashed 3-sphere' \cite{cherednik} and is automatically stable due to manifest symmetries. This is because the $\eta$-deformation breaks the PCM's symmetry to $G_L \times U(1)^r$, i.e. $SU(2)_L\times U(1)_R$  in the $SU(2)$ case, with $U(1)_R$ happening to be a subgroup of $SU(2)$ such that $SU(2) \ov U(1)$ is a symmetric space. This implies that the $SU(2)$ $\eta$-model must coincide with the ``squashed'' model related to  $SU(2) \ov U(1)$ --- part of a general family  built on symmetric spaces $G/H$ considered in App. B of \cite{HLT2}. (Note that general examples from that family are renormalizable due to manifest $G_L\times H_R$ symmetry but \textit{not} classically integrable: the $SU(2) \ov U(1)$ example is atypically integrable.) For $\eta$-models based on general groups, this coincidence does not happen, so the theory is not  protected by manifest symmetries and one may expect to require
 additional  finite local counterterms to get higher-loop renormalizability.

 Finally,
 the \etm\  and the \lam\  are, in general, related \cite{Vicedo:2015pna,Hoare:2015gda,Sfetsos:2015nya,Klimcik:2015gba,Hoare:2017ukq}
by  Poisson-Lie (PL) duality \cite{Klimcik:1995dy} (and analytic continuation).
Since the scaling limits between the $\l$-models and (T-duals of) $\eta$-models revealed that our 2-loop results for the two classes of models are consistent,
then the same observations made above for NAD are likely to apply also to PL duality.
Namely, PL duality (with suitable quantum corrections) should be a symmetry not only at 1-loop order \cite{Valent:2009nv,Sfetsos:2009dj}, but also at higher loops. 
This expectation 
 was recently confirmed by several papers \cite{BW}
 proposing 1-loop corrections to the classical PL duality rules (including NAD as a special case\foot{See also \cite{Borsato:2019oip} on the  leading quantum corrections to homogeneous Yang-Baxter deformations (equivalent to particular NAD transformations \cite{homog2,homog3}) in order for the deformation to preserve 2-loop conformal invariance.}). 
 It would be interesting to check those results against the  corrections \rf{57},\rf{1910originlagnad} found here  for particular NAD models.
 In a related approach, it has recently been proposed \cite{hassler} that there is a particular 2-loop ``subtraction scheme''\foot{The PL-symmetric subtraction scheme of \cite{hassler} is not a covariant
  subtraction scheme  in the usual sense as it is related to the \GB scheme \rf{BMT} by a Lorentz-non-covariant transformation involving the spin connection.}  in which PL symmetry becomes manifest. The $\l$- and $\eta$-models were claimed to be automatically renormalizable in that PL-scheme, suggesting the possibility to derive the corresponding quantum corrections in the standard scheme by reversing the scheme transformation.

%%%%%%%%%%%%%%%%%%%%%%%%%%%%%%%
\renewcommand{\la}[1]{\label{2103#1}} 
\renewcommand{\rf}[1]{(\ref{2103#1})} 
\input{2103.tex}

\renewcommand{\la}{\label}
\renewcommand{\rf}[1]{(\ref{#1})} 
%%%%%%%%%%%%%%%%%%%%%%%%%%%%%%%

%%%%%%%%%%%%%%%%%%%%%%%%%%%%%%%
\renewcommand{\la}[1]{\label{1910#1}} 
\renewcommand{\rf}[1]{(\ref{1910#1})} 
\input{1910v2.tex}

\renewcommand{\la}{\label}
\renewcommand{\rf}[1]{(\ref{#1})} 
%%%%%%%%%%%%%%%%%%%%%%%%%%%%%%%

%%%%%%%%%%%%%%%%%%%%%%%%%%%%%%%
\renewcommand{\la}[1]{\label{2008#1}} 
\renewcommand{\rf}[1]{(\ref{2008#1})} 
\input{2008.tex}

\renewcommand{\la}{\label}
\renewcommand{\rf}[1]{(\ref{#1})} 
%%%%%%%%%%%%%%%%%%%%%%%%%%%%%%%%%%%%%%%%

\chapter{Conclusions and future directions \la{cfd}}

\section{Integrability vs. RG flow}
This thesis is based on the interesting observation that integrable 2d $\s$-models appear to be stable under the leading 1-loop RG flow, i.e. they are solutions of the  (generalized) Ricci flow equation \rf{BMT} with a few couplings running.
Our purpose was to investigate this property beyond the leading 1-loop order. We found in Chapter \ref{QCS} for integrable $\eta$- and $\l$-deformed models, and in Chapter \ref{GGS} for integrable $G\times G/H$ models,  that generic examples will require quantum deformation of the target space geometry to remain stable under the 2-loop and higher RG flow. Thus, in each case, there is a particular finite counterterm  --- or particular choice of bare action --- that restores renormalizability beyond 1-loop.  Moreover, this is a scheme-independent statement since these counterterms cannot be eliminated by a covariant scheme transformation.

In Section \ref{Seta}, we explicitly constructed these counterterms for the simplest $\eta$-deformed $S^2$ (or ``sausage'') model. In that case, by analogy with the gauged WZW model (whose corrections are well-known, and which arises as a particular limit of the sausage model) we conjectured the exact form \rf{24} of the quantum deformation to all orders in $\a'$ (or the $\s$-model loop expansion). We checked explicitly up to 3 loops that these corrections indeed restore renormalizability  in a particular subtraction scheme.

In Section \ref{Slam}, we constructed the leading corrections restoring 2-loop renormalizability to the $\l$-models based on $SU(2) \ov U(1)$ and $SU(2)$. In the simpler $SU(2) \ov U(1)$ case, the corrections were understood as a familiar $(\del \log \det M)^2$ determinant \rf{47} from the quadratic integral over the $\l$-model's 2d gauge field in \rf{lam}. However, such 2d integrals are known to be ill-defined \cite{Schwarz:1992te} with a local term ambiguity depending on a choice of prescription or regularization. Indeed we find that this simple formula is not sufficient to restore renormalizability in the $SU(2)$ case, where the leading corrections take the more complicated form \rf{1910originlag}, or \rf{1910b2} in explicit coordinates.

In Chapter \ref{CT} we proved in general that the $\l$-model \rf{lam} in the formulation with unintegrated gauge field is renormalizable to all orders, by re-writing the gauge field in terms of group valued scalars and obtaining a $\s$-model with a ``tripled'' $G\times G \times G$ configuration space. This means that the  corrections required for renormalizability of the $\l$-model in  ``standard'' configuration space may all be understood as finite counterterms arising from the ambiguous 2d gauge integral. Clearly the simple $(\del \log \det M)^2$ prescription is not generally consistent with renormalizability (since it breaks down already for the $SU(2)$ case).

An interesting open problem is to find a general prescription for the  finite  local counterterms produced by integrals of this type, $\int dA \exp \big( i \int A_+ M A_- \big)$, consistent with the RG flow. This would give a general formula for the quantum corrections needed to restore higher-loop renormalizability for all $\l$-models, generalizing the $SU(2) \ov U(1)$ and $SU(2)$ results found here. One possible approach based on the work of \cite{hassler} would be to try to construct the necessary counterterms from the scheme transformation between our standard 2-loop scheme \rf{BMT} and one that is non-Lorentz-covariant but manifestly respects Poisson-Lie symmetry,  in which the $\l$- and $\eta$-models are claimed to be automatically renormalizable.

For the  $G\times G/H$ models in Section \ref{2103gu}, the manifest (global and local) symmetries alone restrict the theory to a finite set of couplings \rf{2103couH},\rf{2103bb1} that could run under the RG flow. The integrable theories then correspond to a ``surface'' in this space of couplings, specified by polynomial equations \rf{2103condH}. Hence we observed that the necessary counterterms to restore 2-loop stability of the integrable surface under the RG flow can be interpreted as a quantum correction to the shape of the surface. This is like a simple, finite-dimensional analog of the general case where the space of \sm couplings is essentially infinite (e.g., for the integrable deformed models above).

We also constructed  in Chapter \ref{GGS} a new ``branch'' of $H$-gauge-invariant $G\times G/H$ theories in the case when $H$  is abelian, with independent WZ levels on the two copies of the group $G$. We found that a particular subfamily of this new branch is classically integrable and, in the $SU(2)\times SU(2)/U(1)$ case, these correspond to a family of integrable $\s$-models with target space $T^{1,q}$ supported by a particular $\BB$-field. These models are a generalization of the integrable $T^{1,1}$ models of \cite{ABL}, with the extra parameter $q$ being related to the ratio of the two WZ levels.

\

In integrable examples with enough \textit{manifest} symmetries, such as the undeformed PCM or symmetric space, renormalizability can be an automatic result of the symmetries. However, generic integrable $\s$-models may have no manifest global symmetries at all; rather, the integrable structure is related to an intricate structure of \textit{hidden} (non-local or higher spin) symmetries. The effect of such symmetries on the RG flow is far less obvious since, e.g., it is not clear in which way they should constrain the Riemann tensor and $\H$-field that enter the RG equations \rf{BMT}. Remarkably, it appears that hidden integrable symmetries impose drastic constraints on the RG flow, restricting the 1-loop RG flow to just a flow of the couplings of the classical theory. We have observed that this property persists also at higher loop orders provided the target space geometry is suitably deformed by quantum ($\a'$) corrections. Indeed, such a deformation is  natural given the the well-known examples of  conformal $\s$-models where $\a'$ corrections are needed to restore higher-loop conformal invariance.

One interesting open  question is to understand how and why integrability  implies renormalizability, and perhaps to prove this statement, starting at the leading 1-loop order.
As we discussed in the Introduction, the rationale behind the conjectured integrability--RG flow relationship is that the stability of integrable theories under the RG flow simply corresponds to preserving their integrable structure at the quantum level. In fact, it is natural that such hidden integrable symmetries should require a particular choice of finite local counterterms to be manifest in the quantum theory, just like for standard global symmetries. 
It would be interesting to verify this interpretation by checking explicitly that the finite counterterms required for renormalizability are actually restoring some quantum `Ward identities' for hidden integrable symmetries. This would give an interesting semi-classical picture of quantum integrability.
Such computations could be carried out, e.g., for the integrable-deformed models considered in Chapter \ref{QCS}, in relation to either the higher-spin charges or the non-local charges. For the $\eta$-deformed models,  studying the quantum non-local charges would also be an interesting check of the conjectured quantum group symmetry of the quantum theory \cite{qds,Delduc:2013fga,Arutyunov:2015qva,ses}.

 \ 
 
The integrable $G\times G$ models studied in Section \ref{21032} are an anomaly to the discussion above. These theories are \textit{automatically} stable under the 2-loop RG flow, even though (to our best knowledge) the integrability condition is not protected by any manifest symmetries. Generically we would have expected the integrable ``surface'' in coupling space to require some quantum deformation (like the $G\times G/H$ models) in order to remain stable beyond 1-loop. An interesting open problem is to explain this fact, which surely should not be a coincidence.

 \

The $G\times G$ and $G\times G/H$ models demonstrate that, while integrability seems to imply renormalizability, the converse statement is not true. These models are renormalizable with a finite set of couplings but most of them are not integrable (only ones with special values of the couplings solving polynomial equations).

Classical integrability may also  become anomalous at the quantum level, and we note that renormalizability seems too coarse a measure to pick up on this.
For example the bosonic $\mathbb{C}\mathbf{P}^N$ model is well known to suffer such an anomaly \cite{Abdalla:1980jt} (see also \cite{ey1}).
However, it is still 2-loop renormalizable \cite{Valent:1984rj}.\foot{For the $\mathbb{C}\mathbf{P}^N$ model it has been conjectured that quantum integrability
can be restored by including an additional free field \cite{cpn1} in the classical limit (related models have also appeared in \cite{bbr}).
An alternative is to consider the supersymmetric $\mathbb{C}\mathbf{P}^N$ \sm in which there is no anomaly \cite{Gomes:1982qh}.}

\ 

We recall that our original motivation for studying stability under the RG flow was that it might help to \textit{identify} integrable $\s$-models. In a certain sense, this is what we have been doing: starting from classically integrable theories, we have been constructing particular formulations of the quantum theories (with particular finite counterterms) that are RG-stable at higher-loop orders --- with the belief that these formulations are the ones consistent with integrability at the quantum level.

It would be interesting in future to apply the integrability--RG flow connection to discover new examples of integrable $\s$-models. First at the level of classical integrability, this may be carried out by imposing stability under the leading 1-loop RG flow. One possible example would be for non-symmetric $G/H$ cosets, which generally have several couplings,
\be
\L = -\ha \, \Tr[ J_+ \, \big( \sum_{i=1}^n h_i P_{\mathfrak{p_i}} \big) \, J_- ] \ ,
\ee
corresponding to a decomposition of the tangent space into irreducible representations $\mathfrak{p_i}$ under the adjoint action of $H$ (see Section \ref{sssm}). It may be possible in certain cases to discover integrable ``surfaces'' in this coupling space by demanding stability under the RG flow. Indeed, renormalizability of certain non-symmetric coset spaces was previously considered in \cite{bonneau}.

\section{``Tripled'' formulation}
The ``tripled'' formulation of the $\l$-model in Chapter \ref{CT} may be viewed as an alternate formulation of the quantum theory where some fields decouple, likely as a manifestation of integrability. The remaining ``truncated'' theory is automatically stable under the RG flow to all orders due to  manifest global symmetries, without needing any quantum deformation.

The general idea of moving to an enlarged configuration space to make manifest extra symmetries (in this case integrability) is reminiscent of T-duality-invariant formulations of string theory \cite{Tseytlin:1990va}. Such approaches 
 are essentially 1st order Hamiltonian formulations of the $\s$-model, with manifest Lorentz invariance being sacrificed in favour of manifest T-duality invariance. 
A related 1st order formulation of Poisson-Lie symmetric $\s$-models known as the ``E-model'' makes the more general Poisson-Lie duality manifest \cite{poissonlie}.

In Section \ref{1910TO}, we  considered  similar reformulations on enlarged configuration spaces of (i) a variant of the $\l$-model based on $G^N$ and (ii) $G \times G/H$ models. In each case the resulting model was of product type, i.e.\ a particular class of coupled models on some space $G^M$; and in the $G\times G/H$ case there was some ``squashing'', or asymmetry between the $H$ and $G/H$ directions. Re-writing these models in this way is interesting as the corresponding enlarged models will have identical classical integrability and RG flow properties to the original ones. For the $G^N$ $\l$-model we find an identification between its RG flow and classical integrability conditions and those of an apparently simpler undeformed $G^{2N}$ model. We also learn that it is automatically renormalizable in the tripled formulation  due to its manifest symmetries (before imposing the integrability conditions of \cite{Bassi:2019aaf}).  For the $G\times G/H$ models we essentially discover a new class of classically integrable and RG-stable squashed $G\times G \times H$ models. It would be interesting to study the higher-loop RG flows of these theories to see if the  enlarged formulations render the integrability conditions automatically stable  at higher-loops (as for the $\l$-model).

Another interesting class of $\s$-models are those with complex homogeneous target spaces that are equivalent  \cite{byk2,byk3}  to bosonic Gross-Neveu (GN) models \cite{Gross:1974jv} (see also the reviews \cite{byk4}). The reformulation of these $\s$-models as GN models is essentially equivalent to the introduction of a 2d gauge field, so should be closely related to the ``tripled'' or ``extended'' formulations discussed here. It would be interesting to check stability under the 2-loop RG flow in that formulation (which was previously only checked   at 1-loop \cite{byk2}).

An interesting future direction would be to try to construct a similar  tripled formulation of the $\eta$-deformed models. It seems plausible that such a formulation should exist since the $\eta$-models and $\l$-models are related (up to analytic continuation) by Poisson-Lie duality a \cite{Hoare:2015gda,Klimcik:2015gba,Vicedo:2015pna,Hoare:2017ukq}, or by limits and  T-duality \cite{Hoare:2015gda,Hoare:2017ukq}.

\section{Non-abelian duality}

In the non-abelian dual limit of the $\l$-model, the general 2-loop $\b$-functions obtained from the tripled formulation match those of the original PCM/symmetric space $\s$-models. This agreement is a non-trivial check of our $\l$-model $\b$-functions since the 2-loop terms are also scheme-invariant under coupling redefinitions in that single-coupling limit. It is thus natural to expect that the  interpolating models for non-abelian duality are renormalizable with the same 2-loop $\b$-functions as the original models --- justified also by their relation to the original models by just an integral over a linear Lagrange multipler imposing the flatness of the gauge field.

Hence, like the $\l$-model, the non-trivial corrections found for the NAD models based on $SU(2)/U(1)$ and $SU(2)$ in Section \ref{NADqS} should be seen as  resulting from the integral over the 2d gauge field.  
In certain limits these NAD models degenerate to abelian T-duals and in those cases our corrections matched with the known corrections to the T-duality transformation. 

It would be interesting to consider more general models with non-abelian symmetries, which may have spectator coordinates (here $x$ or $u,v$), for example
\be
\L = \del_+ x \, \del_- x + f(x) \, \Tr[ J_+ J_- ]  , \qquad \qquad (J \equiv g^{-1} dg , \ \ g\in G)
\ee
or perhaps the conformal $\s$-model considered in \rf{2008nPCM},
\be
\L = -2\, \del_+ u \, \del_- v + u \, \Tr[ J_+ J_- ] \ .
\ee
Starting from such a theory that is integrable and renormalizable (say, for a particular choice of $f(x)$) the interpolating model for NAD is obtained by replacing $J_\pm \to A_\pm$ and adding a Lagrange multiplier constraint $\Tr[v \, F_{+-} (A)]$. One can then perform the same trick of going to a tripled formulation by re-writing the gauge field in terms of group valued scalars. It would be interesting to study the RG flows for these interpolating models in the tripled formulation and to see if they still automatically match the original models (also at 2-loops) in the presence of spectator fields.

\section{Quantum corrections and string sigma models \la{stric}}
An important application of our results would be to the related $\l$- and $\eta$-deformations of superstring $\s$-models. For example, focussing on the simplest case of \adss\unskip, the $\l$-deformed background has been extracted  from a $\s$-model formulation \cite{Hollowood:2014qma} and checked to be a type IIB supergravity solution \cite{sugra}. Equivalently, the string \sm is 1-loop Weyl-invariant (1-loop scale-invariance was also checked in  \cite{Appadu:2015nfa}). The $\eta$-deformed \adss background of \cite{hsei} was also extracted and checked to be a solution of the supergravity equations.

In general, solutions of the $d=10$ type II supergravity equations should receive $\a'$ corrections to become  exact string solutions, with the first corrections appearing at order $\a'^3$ \cite{Gross:1986iv}. In contrast with the undeformed \adss background where corrections are prohibited by manifest symmetries,\foot{Exact finiteness of  the undeformed \adss superstring $\s$-model \rf{string} follows because the WZ-type  $J^{(1)} J^{(3)}$  term is not renormalized due to general arguments and the $J^{(2)} J^{(2)}$ term may only be renormalized by an overall factor, but this cannot happen since the two terms are related by $\ka$-symmetry (see footnote 4 of \cite{Roiban:2007jf} and refs.\ there).} it is not clear whether the deformed backgrounds will require non-trivial corrections. It may be that, due to the models'  hidden integrable symmetries, the corrections can be made to vanish in some particular subtraction scheme. On the other hand, this did not happen for the bosonic analogs above, which are also integrable.

It would be interesting to study this question and  to understand the structure of the $\a'$ corrections to these backgrounds. For the $\l$-deformation, a systematic way to do this may come from  adapting the tripled formulation to the superstring case. This seems feasible since the $\l$-deformed \adss superstring was formulated in \cite{Hollowood:2014qma} in the language of supergroups with a 2d gauge field. In such a tripled formulation,  the $\a'$ corrections may be expected to  vanish due to
manifest symmetries after a decoupling of some fields
 (like the bosonic case), and  the exact  finiteness  of the model may become manifest.

\section{Sigma models with local couplings}

In Chapter \ref{loc} we discussed  a new and apparently different connection between integrability and the RG flow: upon promoting the couplings of an integrable $\s$-model to functions of 2d time, the resulting ``local coupling'' theory remains classically integrable (with a natural ansatz for its Lax connection) only when the coupling functions solve the 1-loop RG flow, with 2d time playing the role of RG time $t=\log\m$. This curious pattern was observed in a broad class of integrable $\s$-models on groups and symmetric spaces, including integrable $\eta$- and $\l$-deformations. It would be good to check this claim for more examples of integrable $\s$-models, like the $G\times G$ and $G\times G/H$ models. We also found the same result for the  sine-Gordon model, and it would be interesting to check more examples of  integrable theories with potentials, e.g. Toda models.

These results are surprising because it seems that the 1-loop quantum $\b$-functions are derivable in this way from a purely classical procedure (from demanding that the local-coupling theory admits a flat Lax connection). We also noted that theories with explicit time-dependence governed by the RG flow are naturally obtained in string theory upon fixing the light-cone gauge $u=\t$, starting from a class of Weyl-invariant $\s$-models \cite{Tseytlin:1992pq,Tseytlin:1992ee} with a covariantly constant null Killing vector and a dilaton linear in the isometry direction,
\be
\L = -2  \, \del_+ u \,  \del_- v  + \GG_{ij}(u;x) \del_+ x^i \del_- x^j \ , \quad \del_u \GG_{ij} = \b_{ij}(x) \ , \qquad (\p = v + \p(u,x)) \ . \la{ssi}
\ee 
Thus we find that  --- assuming the ``transverse'' ($u=\text{const}$) theory to be integrable --- the light-cone ($u=\t$) theory is apparently classically integrable only when the string $\s$-model \rf{ssi} is 1-loop Weyl invariant. This  points to a  deeper meaning or explanation of our results, and to the possibility of a large, new class of solvable string backgrounds.

\

What remains to be clarified is whether these time-dependent (light-cone) theories may really be viewed as solvable. Although the monodromy matrix is formally conserved, it is difficult to explicitly extract the associated conserved charges because of problems making a good expansion in the spectral parameter. We also made attempts to study higher-spin local charges in the paper \cite{HLT2008}, but our results so far were  inconclusive. One possible way to demonstrate solvability would be to explicitly  construct some  classical solutions;  indeed this has already been done for the time-dependent PCM example \cite{Belinsky:1971nt} 
 by adapting the inverse scattering approach of \cite{zs,zm}.

An interesting extension would be to consider the  $(2,2)$ supersymmetric $\s$-models on certain  K\"{a}hler symmetric spaces whose $\b$-functions are known to be 1-loop exact \cite{Novikov}. Supersymmetric $\s$-models on symmetric spaces are  classically integrable in general  \cite{Abdalla:1985nm}. Hence the pattern found above for bosonic theories may generalize to the supersymmetric ones. In that case the corresponding string backgrounds built from the 1-loop RG flow (which will be \textit{exact} string solutions) may be completely solvable.

A further question is whether the local-coupling $\s$-models  admitting Lax representations can be obtained from  4d Chern-Simons theory \cite{Costello:2019tri} (like all of the regular integrable $\s$-models considered in this thesis \cite{Vicedo:2019dej,Delduc:2019whp}). This is a natural expectation, since the 4d perspective places the 2d space-time and the complex spectral parameter on an equal footing, while the Lax connections for the local coupling theories are related to the original ones by a space-time-dependent redefinition of the spectral parameter, $z \to z(w; \t,\s)$.

%%%%%%%%%%%%%%%%%%%%%%%%%%%%%%%
\begin{appendices}
%%%%%%%%%%%%%%%%%%%%%%%%%%%%%%%

\chapter{Conventions \la{conv}}
Unless otherwise stated, the 2d space is assumed to be flat Minkowski space with signature $(-+)$, i.e. $\eta_{00}=-\eta_{11}=-1$ in terms of the 2d coordinates  $\xi^\m = (\xi^0,\xi^1) = (\t,\s)$ with 2d incides $\m,\n=1,2$. The antisymmetric Levi-Civita symbol $\e_{\m\n}$ is defined by $\e^{01}=-\e^{10}=1$ and $\e_{01}=-\e_{10}=-1$ (with indices raised/lowered by $\eta_{\m\n}$). We use 2d light-cone coordinates defined as $\xi^\pm = \ha(\t\pm\s)$ and light-cone derivatives $\del_\pm = \del_\t \pm \del_\s$.

\bigskip
\noindent The $\s$-model action is normalized as $S= \tfrac{1}{4\pi \a'} \int d^2 \xi \,  \L$ with the usual flat-space volume form $d^2 \xi = d\xi^0 \wedge d \xi^1$.
The Lagrangian is 
\be
\L = - \big[ \eta^{\m\n} \, \GG_{mn}(x) + \e^{\m\n}\, \BB_{mn} (x)  \big] \,  \del_\m x ^m \, \del_\n x^n  = \big[ \GG_{mn}(x) + \BB_{mn} (x)\big] \,  \del_+ x ^m \del_- x^n \ .
\ee
The target space metric is $ds^2 = G_{mn} dx^m dx^n$. The $\BB$-field written as a 2-form is $\BB=\ha dx^m \wedge dx^n$ and its field strength is $\H=d\BB$.

\bigskip
\noindent The components of $n$-forms are normalized in the standard way, $H_{(n)} = \tfrac{1}{n!} H_{i_1 \cdots i_n} dx^{i_1} \wedge \cdots \wedge dx^{i_n}$.

\bigskip
\noindent We take $G$ to be a simple, real Lie group with generators  $T_A$ ($A=1,\ldots,\dim(G)$) satisfying $[T_A, T_B] = i {f^C}_{AB} T_C$.
 
 \bigskip
\noindent We use a non-degenerate bilinear form defined as
\be
\Tr = \tfrac{1}{2 \, \chi_{_{G,R}}} \, \tr_{_R} \ ,
\ee
where $\tr_{_R}$ is the matrix trace in some faithful representation $R$ of $G$. The constant $\chi_{_{G,R}}$ denotes the \textit{index} of the representation $R$, defined by
\be
\tfrac{1}{2 \, \chi_{_{G,R}}} \, \tr_{_R}(T_A T_B) = -\tfrac{1}{2\, \cgg} \, f^C{}_{DA} f^D{}_{CB} \ .
\ee
Here $\cgg$ denotes the dual Coxeter number of $G$: for the classical groups these are $c_{_{SU(N)}} = N$, $c_{_{Sp(N)}} = N+1$  and $c_{_{SO(N)}}=N-2$.

\sloppy
The indices of the adjoint representations coincide with the dual Coxeter numbers, \mbox{$\chi_{_{G,adj}} =c_{_G}$}, and those of the fundamental representations of the classical groups are \mbox{$\chi_{_{SU(n),f}} = \chi_{_{Sp(n),f}} = \ha$} and  \mbox{$\chi_{_{SO(n),f}} = 1$}.

The structure constants then satisfy 
\be
{f^A}_{BC} {f^B}_{AD} = -2 \, \cgg \Tr[T_C T_D] \ , \qquad {f^D}_{EA} {f^E}_{GB} {f^G}_{DC} = \cgg \, f_{ABC} \ ,
\ee
where we have defined $f_{ABC} = \Tr[T_C T_D] {f^D}_{AB}$.

 \bigskip
\noindent  We shall sometimes consider a Lie subgroup $H \subset G$, which is also assumed to be simple. Then we split the algebra directions as $A=(\a,a)$ where $T^\a$ are the generators of the $H$ and the remaining generators $T^a$ are orthogonal to $T^\a$ with respect to the Killing form.

The structure constants ${f^\a}_{\b\g}$ of the subgroup $H$ then satisfy 
\be
{f^\a}_{\b\g} {f^\b}_{\a\d} = -2 \, \cf \, \Tr[T_\g T_\d] \, \qquad {f^\d}_{\e\a} {f^\e}_{\ka \b} {f^\ka }_{\d\g} = \cf \, f_{\a\b\g} \ ,
\ee
where $\cf$ denotes a \textit{rescaled} version of the dual Coxeter number $\tfrac{\chi_{_{H,R}}}{\chi_{_{G,R}}} \cf$ of $H$.

%%%%%%%%%%%%%%%%%%%%%%%%%%%%%%%
\chapter{3-loop \texorpdfstring{$\b$}{beta}-function in different renormalization schemes} \label{A}
%%%%%%%%%%%%%%%%%%%%%%%%%%%%%%%

In general, in quantum field theories
changes of a renormalization scheme are equivalent to local redefinitions of the coupling constants,\foot{Here we reinstate the loop-counting parameter $\a'=\hbar$.}
\be
g^i \to \hat{g}^i (g ) = g^i+ \a' \, c^i_{jk} g^j g^k + \ldots \ . \la{redef}
\ee
Under \rf{redef} the $\b$-functions $\b^i \equiv \dt g^i$ transform as a contravariant vector in the space of couplings:
\be \la{deltab}
\delta \b^i \equiv\hat{\beta}^i - \beta^i = \frac{\del \, \delta g^i}{\del g^j} \b^j - \frac{\del \b^i }{\del g^j} \delta g^j + \ldots \ , \qquad \qquad \delta g^i \equiv \hat{g}^i - g^i \ ,
\ee
where the higher order corrections are non-linear in $\delta g$. Assuming $\b^i$ and $\delta g^i$ both start at 1-loop order, i.e.
are both $\O(\a')$, then $\delta \b$ is $\O(\a'^2)$, so the 1-loop $\b$-function is scheme-invariant.
In single-coupling theories, one can further show that the 2-loop $\b$-function is also scheme-invariant.

Now let us turn to the particular case of 2d $\s$-models, for which the relevant coupling redefinitions \rf{redef} are local redefinitions of the target space geometry.
In order to preserve the manifestly covariant structure of the action \eqref{13} we restrict to covariant redefinitions.
Restricting further to include only redefinitions with a natural interpretation in terms of changing the subtraction scheme,\foot{While the term $\a'^2 \nabla^2 R_{mn}$ in \rf{metricredef} has no natural interpretation as a change of subtraction scheme starting from the minimal subtraction scheme, it is required in order that the transformation
\rf{2dredef} be closed under inversion in the case of the \sm with a 2d target space,
i.e. in order to be able to move back to the minimal subtraction scheme starting from another scheme.} we consider $\GG_{mn} \to
\hat{\GG}_{mn}$ with
\begin{align} \la{metricredef}
&
\hat{\GG}_{mn} = \GG_{mn} + c \, \a' R_{mn} + d \, \a'^2 (R^2)_{mn} + e \, \a'^2 \nabla^2 R_{mn} + \O(\a'^3) \ , \\
&(R^2)_{mn} \equiv R_{mabc} {R_n}^{abc} \ , \no
\end{align}
where $c, d, e$ are arbitrary coefficients (for simplicity we consider the special case with vanishing $\BB$-field; for a general discussion see \cite{mt}).

At the leading $\O(\a')$ order, both $\beta_{mn} = \a' R_{mn} + \ldots$ and $\delta \GG_{mn} = c\, \a' R_{mn}+\ldots $ only contain a single term proportional to $R_{mn}$.
Thus at the leading $\O(\a'^2)$ order in \rf{deltab}, their contributions cancel and the 2-loop $\b$-function is invariant.\foot{Note that if one also included redefinitions of the form $\a' R \GG_{mn}$ then this argument would not go through in general dimensions. However, for the special case of a 2d target space, the identity $R_{mn} = \ha R \GG_{mn}$ means these terms are proportional and the 2-loop $\b$-function remains invariant.}

Specializing further to the case of a 2d target space (relevant for the simplest $SU(2)\ov U(1)$ $\eta$- and $\l$-models considered in Chapter \ref{QCS}), the redefinition \rf{metricredef} becomes
\begin{align} \la{2dredef}
&
\hat{\GG}_{mn} = \GG_{mn} \Big[ 1 + \ha c \, \a' R + \ha d \, \a'^2 R^2 + \ha e \, \a'^2 \nabla^2 R \Big] \ .
\end{align}
Possible scheme dependence starts at 3-loop order, 
 with \rf{metricredef} or \rf{2dredef} being the relevant redefinition at this order.
The most general form of a covariant expression for a 3-loop $\b$-function
is\foot{In \rf{3loopbeta} we have dropped terms of the form $\nabla_{(m} V_{n)}$ since these may be absorbed into the diffeomorphism vector term in \rf{2}. We have excluded the term $\a'^2 \nabla^2 \nabla^2 R \GG_{mn}$ in \rf{3loopbeta} since it does not arise naturally through changes of subtraction scheme starting from the minimal subtraction scheme.}
\be \la{3loopbeta}
\b_{mn} = \Big[ \ha \a' R + \tfrac{1}{4} \a'^2 R^2 + c_1 \, \a'^3 R^3 + c_2 \, \a'^3 (\nabla R)^2 + c_3 \, \a'^3 R \nabla^2 R \Big] \GG_{mn}
+ c_4 \, \a'^3 \nabla_m R \nabla_n R \ . \ee
Not all of the coefficients $c_1,\ldots ,c_4$ can be scheme-dependent, since the space of legitimate schemes should be parametrized by the 3-dimensional space of coupling redefinitions \rf{metricredef}.
To find the values of $c_i$ that are allowed to appear in the $\b$-function in a particular scheme
let us start from the expression for the $\b$-function \rf{3loopbeta} in the
minimal subtraction scheme where \cite{Graham:1987ep,Foakes:1987ij,Foakes:1987gg}
\be
\qquad c_1 = \tfrac{5}{32} \ , \qquad c_2 = \tfrac{1}{16} \ , \qquad c_3 = 0\ , \qquad c_4=-\tfrac{1}{16}\ .
\ee
Implementing the metric redefinition \rf{metricredef} we find from \rf{deltab} that the coefficients in the
transformed $\b$-function are given by
\begin{align}
&\hat{c}_1 = \tfrac{5}{32} +\tfrac{1}{8}(c - 2d) \ , && \hat{c}_2 = \tfrac{1}{16} -\tfrac{1}{4}(c - 2d)-\tfrac{1}{8}(4e+c^2)\ , \no \\
& \hat{c}_3 = -\tfrac{1}{8}(4e+c^2)\ , && \hat{c}_4=-\tfrac{1}{16}\ . \la{genscheme}
\end{align}
We note that since \rf{genscheme} only depends on the redefinition parameters $c,d,e$ through the combinations $c-2d$ and $4e+c^2$, the choice of $c,d,e$ to give any particular values of $c_1,\ldots,c_4$ is not unique, but rather there is one free parameter. In the case of the \etm\ metric studied in Section \ref{Seta}, the ``natural'' scheme \rf{31} is given by \rf{genscheme} with, e.g., $(c,d,e)=(0, \tfrac{5}{8}, \ha)$; i.e. it is related to the minimal subtraction scheme by the redefinition
\begin{align}
\GG_{mn}^{(\rm nat)}= \big[ \GG_{mn} + \tfrac{5}{8} \, \a'^2 ({R}^2)_{mn} + \ha \, \a'^2 \nabla^2 R_{mn} \big] ^{(\rm min)}\ .
\end{align}

\renewcommand{\la}[1]{\label{2103#1}} 
\renewcommand{\rf}[1]{(\ref{2103#1})} 
%%%%%%%%%%%%%%%%%%%%%%%%%%%%%%%%%%%%%%%%
\chapter{Deriving the  integrability conditions for the \texorpdfstring{$G^N$}{G**N} model  \la{A}}
%%%%%%%%%%%%%%%%%%%%%%%%%%%%%%%%%%%%%%%%
It was shown in 
 \cite{DLMV}
 that the coupled $G^N$ model \rf{cou} is integrable for particular choices of the couplings $(\rho_{ij}, k_i)$ corresponding to realisations of  the affine Gaudin models. 
Here  we shall  try to demonstrate 
 the converse statement: these affine Gaudin models are the \textit{only} integrable  cases of the  coupled models \rf{cou}.

We will assume a natural ansatz \rf{Lan} for the Lax connection, valued in $\Lie(G)$
(here we  explicitly  indicate  the summation over $i=1, ..., N$)\foot{While \rf{a1} is the natural ansatz for the Lax connection arising from affine Gaudin models, it does degenerate at certain points in coupling space. For example, taking $\rho_{ij}$ to be diagonal (i.e.\ decoupled PCM$_k$ models), one instead requires a Lax connection valued in $\Lie(G)^N$. Thus it would also be interesting to consider other ansatze for the Lax connection.} 
\be\la{a1}
L_+ = \sum_i \a_{i}(z) \, J_+^{(i)} \ , \qquad\qquad  L_- = \sum_i \b_{i}(z) \, J_-^{(i)} \ ,
\ee
where $z$ is the spectral parameter. 
The curvature of this Lax connection takes the form
\be
F_{+-}(L) = \sum_i \Big( \b_{i}(1-\a_{i}) \  \del_+ J_-^{(i)} - \a_{i}(1-\b_{i}) \  \del_- J_+^{(i)} \Big) + \sum_{i\neq j} \a_{i} \b_{j} \ [J_+^{(i)},J_-^{(j)}] \ . \la{cur}
\ee
The equations of motion of the $G^N$  model \rf{cou}   
(for arbitrary $N$)
\be \begin{aligned}
E_i \quad  \equiv {}&{}\quad \sum_j \Big(  (\rho_{ij} - \delta_{ij} k_j) \ \del_+ J_-^{(j)} + (\rho_{ij} + \delta_{ij} k_j) \ \del_- J_+^{(j)} \la{eom} \\
&\qquad \qquad \qquad \qquad \qquad +  \rho_{ij} \ [J_+^{(i)},J_-^{(j)}] + \rho_{ji} \ [J_-^{(i)},J_+^{(j)}] \Big) \quad  =\quad 0 
\end{aligned} \ee
We note that \rf{cur} and \rf{eom} are the \textit{unique} ways to write these expressions without any terms of the form $[J_+^{(i)},J_-^{(i)}]$, which have been eliminated using the identity $F_{+-}(J^{(i)}) = 0$. 

If  the  model \rf{cou}  is integrable then,\foot{Here we are assuming integrability and deriving necessary conditions on the couplings. Thus we do not need to worry about whether the $v^i(z)$ in \rf{a4} are independent functions (which would be relevant for the converse question).} 
 for some 
  $v^i(z)$,
  we have
\be
F_{+-}(L) = \sum_i v^i(z) \, E_i \ , \la{a4} \ee
which implies that 
\be \la{eqs} \begin{aligned}
&\b_{j}(1-\a_{j}) = \sum_{i} v^i (\rho_{ij} - \delta_{ij} k_j) \ ,\qquad \qquad    \a_{j}(1-\b_{j}) = \sum_{i} v^i (- \rho_{ji} - \delta_{ij} k_j) \ , \\ &\qquad   \a_i \b_j  =  (v^i - v^j) \rho_{ij}  \ , \ \ \ i\neq j\  \  (\text{no summation}) \ . 
\end{aligned}
\ee
This is a system of $N + N + (N^2-N) = N^2+N$ equations. Fixing the freedom to redefine the spectral parameter by setting $v^1 = z$, there are $N+N+(N-1) =3N -1$ ``artificial'' variables, $\a_i, \b_i, 
v^{i\neq 1}$. 
 After solving for these, there are $(N^2+N)-(3N-1)=N^2-2N+1$ remaining equations to be solved for 
 the $N^2+N$ variables $\rho_{ij}, k_i$. After solving all the equations, this leaves $(N^2+N)-(N^2-2N+1)=3N-1$ free parameters for the integrable theory 
 (including the WZ levels, which may be continuous for non-compact groups).\foot{One might worry that the integrability constraints on the couplings $\rho_{ij}, k_i$  resulting from \rf{a4} might depend on the spectral parameter $v^1 = z$. However, this will not happen because there is a rescaling ambiguity  $E_i \to c_i E_i$ in the definition of the equations of motion \rf{eom}. One may thus rescale $E_1$ to effectively set $v^1=z=1$ in \rf{a4}. Since the constraints on the couplings from \rf{a4} must be invariant under such rescalings, then they must not depend on $z$.}

 Thus the space of integrable models is $(3N-1)$-dimensional,
 which 
 coincides with the number of free parameters 
  following from the affine Gaudin  construction  \cite{DLMV}.

Specializing to the $N=2$ case of $G\times G$, this counting suggests a $5$-dimensional space of integrable models.  Then  the 6   free parameters  $(s,t,u,b ,k_1,k_2)$ in \rf{par}
should   be subject to  only one  relation to ensure  integrability. 
 Solving the equations \rf{eqs} in this case, one indeed obtains the condition \rf{cond}  
 originally found  from the affine Gaudin  construction. 

To summarize,  for general $N$,  the space of integrable models has the same dimension as the space of affine Gaudin models. It remains to understand if there may still be  extra branches 
of integrable theories  
 not corresponding to the affine Gaudin models (cf.\ the $G\times G/H$ models,  where this seems to be the case for abelian $H$, see Section \ref{2103sq}).  For the $N=2$ case of $G\times G$ models, we found  exact matching between the space of  integrable  models \rf{eqs} and the space of affine Gaudin models satisfying  the condition \rf{cond}.

%%%%%%%%%%%%%%%%%%%%%%%%%%%%%%%%%%%%%%%%
\chapter{Explicit 2-loop \texorpdfstring{$\b$}{beta}-functions for \texorpdfstring{$G\times G$}{GxG}  and \texorpdfstring{$G\times G/H$}{GxG/H}  models \la{B}} 
%%%%%%%%%%%%%%%%%%%%%%%%%%%%%%%%%%%%%%%%

Here  we shall provide  the explicit formulae for the  2-loop $\beta$-functions  of the  general 
$ G\times G$ and $G\times G/H$   models 
that were
used in 
 the Chapter \ref{GGS}.\foot{The formulae derived in this Appendix are also available in the Mathematica file attached to the arXiv submission of the paper \cite{LT2103}.} We will also briefly explain how they were derived.

\section{\texorpdfstring{$G\times G$}{GxG} model \la{GGapp}}
For the $G\times G$ model \rf{par}, let us use the notation
\be \la{not}
\rho_{ij} = \rh_{(ij)} + b_{[ij]}
= \begin{pmatrix}
s & t \\ t& u
\end{pmatrix}
+ \begin{pmatrix}
0& b\\ -b& 0
\end{pmatrix} \ .
\ee
Let the  $2\times 2$ matrix $n_{ij}$ be the 
 ``square root'' of $\rh_{ij}=\rh_{(ij)}$, and let $m_{ij}$ be its inverse,
\be
n_{ik}\, n_{jk} = \rh_{ij} \ , \qquad m_{ik} \, n_{kj} = \delta_{ij} \ , \qquad \qquad n_{ij} = n_{(ij)}, \quad \ m_{ij} = m_{(ij)} \ . \la{not2}
\ee
The target space metric of the \sm \rf{par} is ``diagonalized'' by the vielbein
1-form,
\unskip\foot{See Appendix \ref{conv} for the group theory conventions.
The generators $T_A$ satisfy $[T_A, T_B] = i {f^C}_{AB} T_C$ and we define $f_{ABC} = \Tr[T_C T_D] {f^D}_{AB}$. For simple groups $G$,  the structure constants satisfy 
 ${f^A}_{BC} {f^B}_{AD} = -2 \cgg \Tr[T_C T_D]$ 
 and ${f^D}_{EA} {f^E}_{GB} {f^G}_{DC} = \cgg f_{ABC}$, where $\cgg$ is the dual Coxeter number of $G$.}
\be \begin{aligned}
&E^{Ai} = n_{ik} \, J^{(k)A} 
\ , \qquad A=1,\ldots,\dim{G} \ , \ \ i = 1,2 \ ,  \la{viel} \\
&J^{(k)} \equiv T_A J^{(k)A}=   \big( g^{(k)} \big)^{-1}  d g^{(k)} \ . 
\end{aligned} \ee
Then  the coefficients of  the metric $ds^2 = \GG_{Ai, Bj} E^{Ai} E^{Bj}$ and the   3-form   $\H = d \BB =  \tfrac{1}{6} \H_{Ai, Bj, Ck} E^{Ai} \wedge E^{Bj} \wedge E^{Ck}$ are given by\foot{The overall factor of $i$ in  \rf{Hf} simply reflects the fact  that the vielbein \rf{viel} is imaginary. This makes no difference and could be eliminated by just multiplying $E^{Ai} \to i E^{Ai}$.}
\begin{align}
&\GG_{Ai, Bj} = -\ha \, \Tr[T_A T_B] \, \delta_{ij} \  , \la{Gf}\\
&\H_{Ai, Bj, Ck} = \tfrac{i}{2} f_{ABC} \big[ k_l  \, m_{il} \, m_{jl} \, m_{kl} + b_{lp}(  m_{ip} \, m_{jl} \, m_{kl} +  m_{il} \, m_{jp} \, m_{kl} +  m_{il} \, m_{jl} \, m_{kp}) \big]  \ . \la{Hf}
\end{align}
From  Cartan's structure equation $d E^{Ai} + \widehat{\w}^{Ai}{}_{Bj} \wedge  E^{Bj} = T^{Ai}$ with torsion $T^{Ai} = \ha  {\H^{Ai}}_{Bj, Ck} E^{Bj} \wedge E^{Ck}$, we obtain the torsionful spin connection,
\begin{align}
&\widehat{\w}^{Ai}{}_{Bj} = \tfrac{i}{2} {f^A}_{BC} M_{ijk} E^{Ck}\ , \\
&M_{ijk} \equiv  m_{il}m_{jl}n_{kl} -  m_{il}n_{jl}m_{kl} -  n_{il}m_{jl}m_{kl}  \la{me}\\
&\qquad\quad  +  k_l  \, m_{il} \, m_{jl} \, m_{kl} + b_{lp}(  m_{ip} \, m_{jl} \, m_{kl} +  m_{il} \, m_{jp} \, m_{kl} +  m_{il} \, m_{jl} \, m_{kp})  \ . \no
\end{align} 
The torsionful Riemann curvature $\widehat{R}^{Ai}{}_{Bj} \equiv \ha \widehat{R}^{Ai}{}_{Bj,Ck,Dl}  E^{Ck} \wedge E^{Dl} = d \widehat{\w}^{Ai}{}_{Bj} + \widehat{\w}^{Ai}{}_{Ck} \wedge \widehat{\w}^{Ck}{}_{Bj} $ is then found in terms of $M_{ijk}$ to be
\be \begin{aligned}
\widehat{R}^{Ai}{}_{Bj,Ck,Dl} ={}& \tfrac{1}{4} \big[  2{f^A}_{BE} {f^E}_{CD} M_{ijp} n_{pq} m_{kq} m_{lq}  \\
&\qquad \qquad + {f^A}_{CE} {f^E}_{BD} M_{ipk}M_{pjl} - {f^A}_{DE} {f^E}_{BC} M_{ipl}M_{pjk} \big] \ . \la{Ri}
\end{aligned}\ee

It is then straightforward to substitute \rf{Hf},\rf{me},\rf{Ri} into the 2-loop $\beta$-functions in the \GB scheme \eqref{BMT}, obtaining  explicit formulae   for 
 the RG equations    $\ddt \rho_{ij} = \b_{ij}(n_{11}, n_{12}, n_{22}, b, k_1, k_2)$ depending on the components of $n_{ij}$. Using a computer symbolic algebra package (e.g.\ Mathematica) it is easy to rewrite these expressions in terms of the components $s,t,u$ of the ``squared'' coupling  $\rh_{ij} = n_{ik} n_{jk}$ in \rf{not}, with all  the square roots cancelling out
  as the Riemann tensor and the $\H$-field must clearly be rational functions of $\rho_{ij}$. We thus obtain the $\beta$-functions
   in the  form  given  in \rf{2c},
\be \begin{aligned}
 & \qquad\qquad \ddt \rho_{ij}  = \a' \, \b_{ij}^{(1)} + \a'^2\,  \b_{ij}^{(2)} + \ldots \ , \\
 \b_{ij}^{(1)} = {}& \ccg (su-t^2)^{-2} \, F^{(4)}_{ij}(s,t,u,b ,k_1,k_2) \ ,  \qquad  
  \b_{ij}^{(2)} = {\rm c}_{_G}^2 (su-t^2)^{-5} \, F^{(9)}_{ij}(s,t,u,b ,k_1,k_2) \ , 
\end{aligned} \ee
where   the explicit form of the homogeneous polynomials  $F^{(4)}_{ij}$   and $F^{(9)}_{ij}$ is:
\ \\

\bigskip
\hspace{-1.26cm}\includegraphics[scale=0.37]{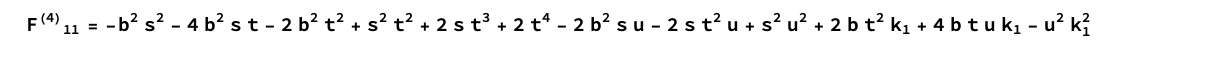}
\medskip

\hspace{-1.26cm}\includegraphics[scale=0.37]{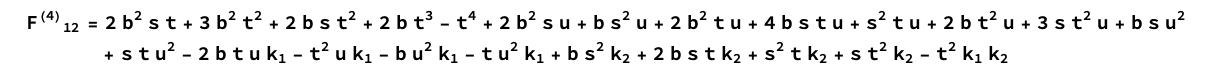}
\medskip

\hspace{-1.26cm}\includegraphics[scale=0.37]{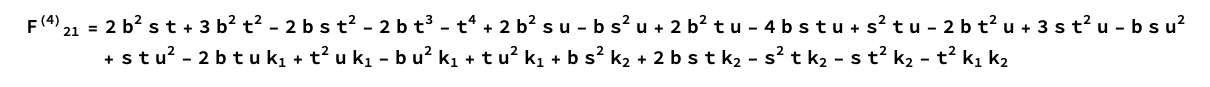}
\medskip

\hspace{-1.26cm}\includegraphics[scale=0.37]{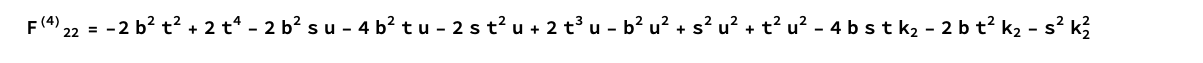}
\medskip

\hspace{-1.3cm}\includegraphics[scale=0.39]{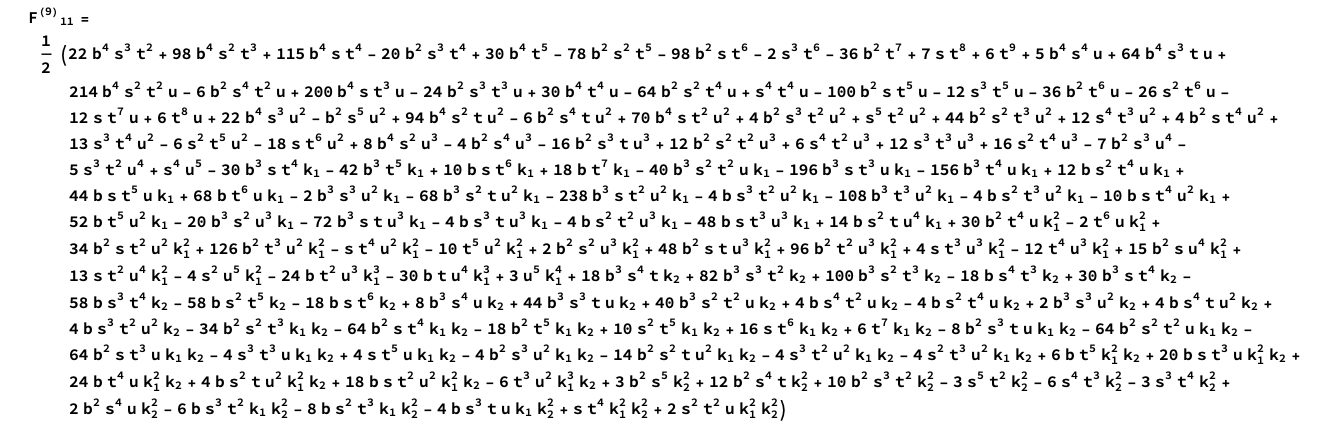}
\medskip

\hspace{-1.3cm}\includegraphics[scale=0.39]{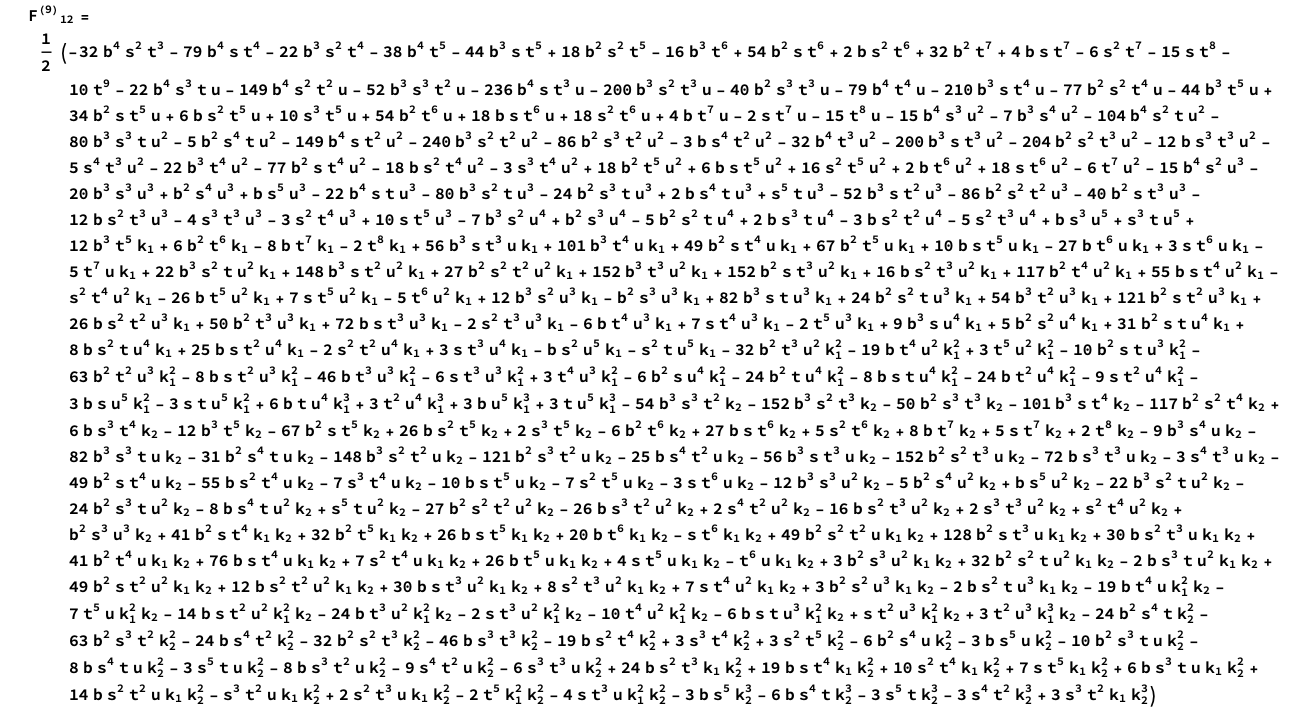}
\medskip

\hspace{-1.3cm}\includegraphics[scale=0.39]{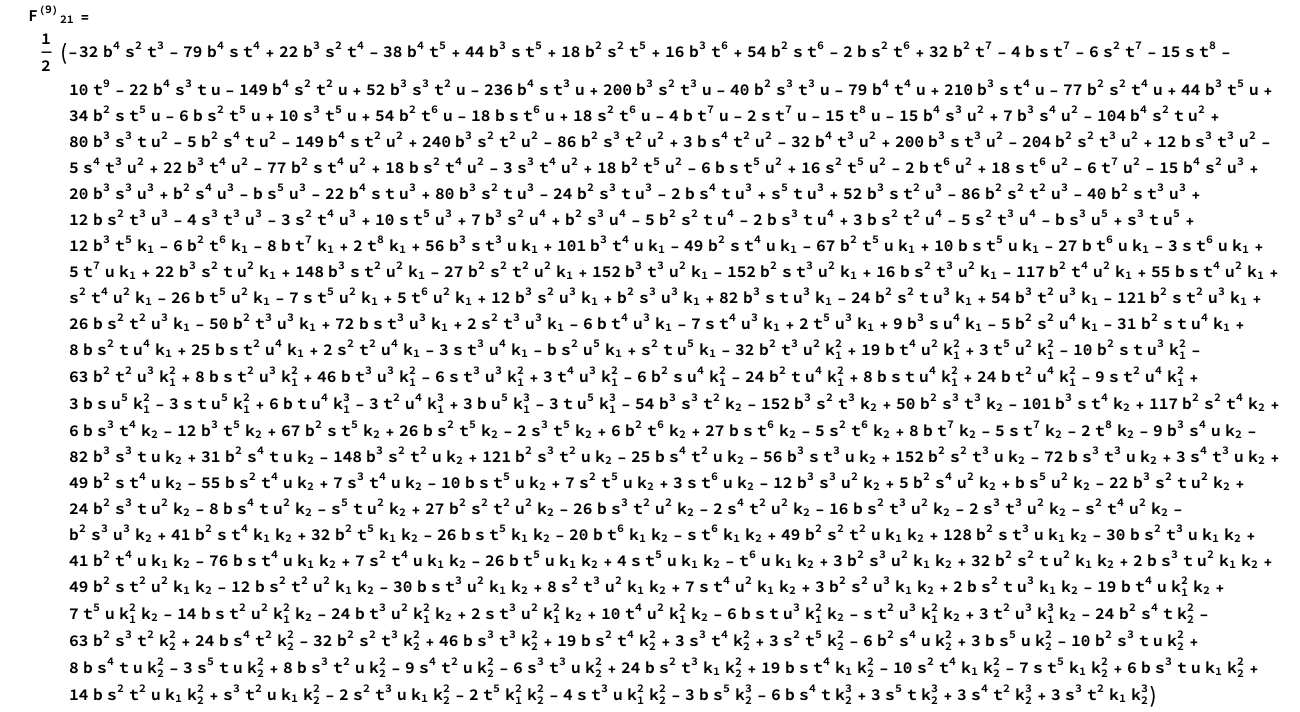}
\medskip

\hspace{-1.3cm}\includegraphics[scale=0.39]{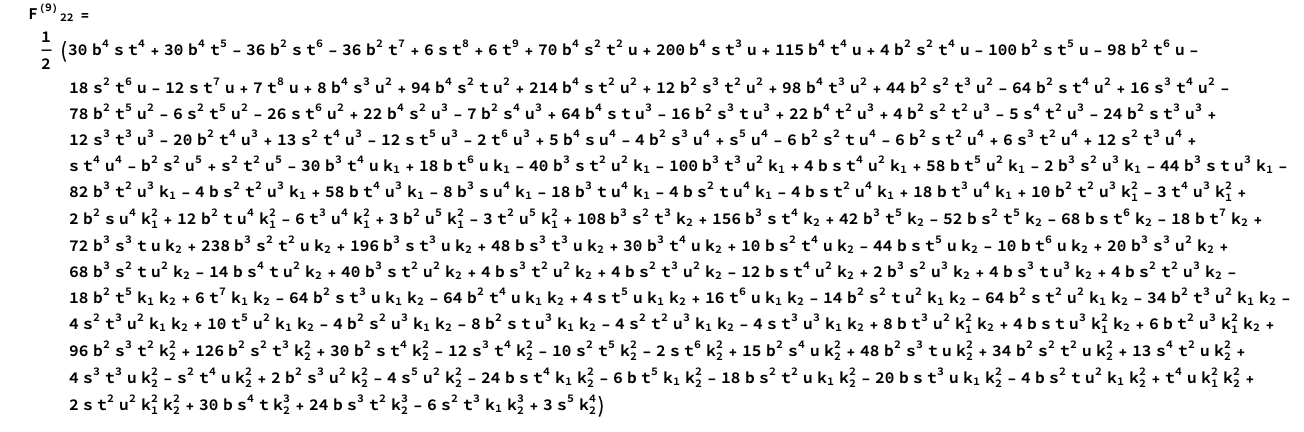}

\section{\texorpdfstring{$G\times G/H$}{GxG/H}  model \la{GGHapp}}
The computation of the $\beta$-functions for the gauge invariant $G\times G/H$ model \rf{couH},\rf{bb1} is similar to the $G\times G$  case  above, except that one has to correctly handle the gauge invariance.

We shall again use the notation \rf{not} and \rf{not2},
  with  the symmetric ``square root'' of 
$\rh_{ij} = \rho_{(ij)}$
 being  $n_{ij}$, and its inverse being  $m_{ij}$.
In the computation below, we shall denote certain combinations of $n_{ij}$ and $m_{ij}$ by 
\be \begin{aligned}
&\hij  = m_{1i}m_{1j} + m_{2i}m_{2j} \ , \quad & &\nu_{ij} = n_{1i}m_{1j} - n_{2i} m_{2j}  \ , \\
&\Hij = \tfrac{k}{2} \hij  + \tfrac{b}{2}(m_{1i} m_{2j} + m_{2i} m_{1j}) \ , \quad &  &\l_{ij} =  m_{1i}m_{1j} - m_{2i}m_{2j} \ .
\end{aligned} \ee
We shall split  up the generators  $T_A$ of $G$ into $T_\a \in \Lie(H)$ and $T_a\in \Lie(G)/\Lie(H)$ (which are orthogonal with respect to the Killing form). 

\renewcommand{\ol}{\overline}
Assuming from the beginning that the matrix $r_{ij}$ satisfies the gauge invariance condition \rf{bb1},\foot{Alternatively, one could obtain the same results by starting with $r_{ij}$ unconstrained, i.e.\ without gauge invariance imposed. One could  first  compute the torsionful Riemann tensor for the target space geometry \rf{couH} with general $r_{ij}$, $\rho_{ij}$. The gauge invariance condition \rf{bb1}  would 
 then  be  imposed and the resulting Riemann tensor   projected onto the non-degenerate directions of the metric $\GG_{MN}$.}  the target space metric of the \sm \rf{couH} is diagonalized by the
  vielbein,\foot{The index  $M$   denotes  all tangent space directions.
 In the $G\times G$ case  in \rf{viel}  we had  $M= (Ai)$,   while here 
$
M = (\a, \bar \a, a i)$. Both $G$ and $H$ are assumed to be simple.
}
\be \begin{aligned}
&E^M = (e^\a, e^{\ol \a} , e^{ai}) = \Big(\sqrt{r}(\B^{(1)\a} -\B^{(2)\a}) \, ,\  \B^{(1)\a} +\B^{(2)\a}\, ,\  n_{ik} P^{(k)a} \Big) \ , \\
& \B^{(k)} \equiv  T_\a \B^{(k)\a} = P_H \big[ \big(g^{(k)}\big)^{-1} d g^{(k)} \big] \ , \qquad P^{(k)} \equiv  T_a P^{(k)a} = P_{G/H} \big[ \big(g^{(k)}\big)^{-1} d g^{(k)} \big]  \ .
\end{aligned}\ee
In this frame, the metric $ds^2 = \GG_{MN} E^{M} E^{N}$ and  the 3-form   $\H = \tfrac{1}{6} \H_{MNP} E^{M} \wedge E^{N} \wedge E^{P}$ have the following non-zero components
\begin{alignat}{3}
&\GG_{\a\b} = -\ha \, \Tr[T_\a T_\b ] \ , \qquad &&\GG_{ai, bj} = -\ha \, \Tr[T_a T_b] \  ,\\
&\H_{\a\b\g} = \tfrac{i}{2} k r^{-3/2} f_{\a\b\g} \ , \qquad &&\H_{\a,bi,cj} = i r^{-1/2} \Hij  f_{\a bc} \ . \la{HH}
\end{alignat}

The $H$ gauge invariance is reflected in the vanishing of all $\ol \a$ components of $\GG_{MN}$ and $\H_{MNP}$, and,  in particular,  the fact that $\GG_{MN}$ is degenerate as a result. One could explicitly fix a gauge, eliminating some target space directions and removing this degeneracy. Instead, we find it more convenient to lift the degeneracy with a small parameter $\eps$ acting as a regulator,\foot{\sloppy
The use of the ``regulator'' $\eps$ is a short-cut for the following gauge-fixing procedure. Fixing an ``axial'' gauge ${i X^\m (\B^{(1)}_\m + \B^{(2)}_\m) = y(\xi) \in \Lie{H}}$ ($\m=1,2$  is the 2d index), the path integral  should be 
 independent of the choice of the constant 2d vector $X^\m$ and the algebra-valued  function  $y(\xi)$  of the 2d coordinates. Inserting the $\delta$-function  of the gauge fixing condition into the path integral   and then 
   integrating over $X^\m$ and $y$ with a Gaussian measure, i.e. 
 $ e ^{-\ha  X^\m X_\m  - \ha \eps \int d^2 \xi  \, \Tr[y y ] }$,  the result  should be  independent of $\eps$ (here we assume 
  Euclidean 2d signature but the  same  is true  also in Minkowski signature after an  analytic continuation). 
  Integrating  first over $y$ we get 
  $ \int d^2 X \exp \big[ {- \ha X^\m X^\n \big( \d_{\m\n} -  \eps \,   \Tr[(\B^{(1)}_\m + \B^{(2)}_\m) (\B^{(1)}_\n + \B^{(2)}_\n)] \big)}\big]$. 
  Integrating over $X_\m$ restores the 2d Euclidean invariance  and  the  result  to leading order in the $\e\to 0$ limit 
  is equivalent to simply adding the regulator term $\Delta \L = -\ha \eps \, \Tr[ (\B^{(1)}_\m + \B^{(2)}_\m)^2]$  corresponding to  \rf{reg}. \la{fgs}}
\be
\GG_{\ol \a \ol \b} = -\ha \eps \, \Tr[T_\a T_\b ] \ . \la{reg}
\ee
Computing the torsionful Riemann tensor as in  Section 
 \ref{2103GGapp}, 
 one finds that it has a finite $\eps\to 0$ limit.
  This means that the resulting Riemann tensor for $\eps=0$ is unambiguous (since there are no divergences that could create finite-term ambiguities). Finally,  we project out the $\ol \a$ directions to obtain the non-zero components
\be \la{RR}
\begin{aligned}
&\widehat{R}^\a{}_{\b\d\e} = -\tfrac{1}{4r} f^\a{}_{\b\g} f^\g{}_{\d\e} + \tfrac{k^2}{4r^3} ( f^\a{}_{\d\g} f^\g{}_{\b\e}  - f^\a{}_{\e\g} f^\g{}_{\b\d}  ) \ , \\
&\widehat{R}^\a{}_{\b,dk,el} = (\tfrac{k}{2r} \l_{kl} -\ha p_{kl}) f^\a{}_{\b\g} f^\g{}_{de} + \tfrac{1}{r} A_{kj}A_{lj}( f^\a{}_{dc} f^c{}_{\b e}  - f^\a{}_{ec} f^c{}_{\b d}  ) \ ,\\
&\widehat{R}^\a{}_{b i,\d,el} = -\tfrac{1}{2r} A_{ji} g_{jl} f^\a{}_{bc} f^c{}_{\d e} - \tfrac{k}{2r^2} A_{li} f^\a{}_{\d \g} f^\g{}_{b e}  + \tfrac{1}{r} A_{lj}C_{ji} f^\a{}_{ec} f^c{}_{b \d}  \ , \\
&\widehat{R}^{ai}{}_{bj,dk,el} = (C_{ij} \l_{kl} -\ha \d_{ij} p_{kl}) f^a{}_{b\g} f^\g{}_{de} + \tfrac{1}{r} A_{ki} A_{lj} f^a{}_{d\g} f^\g{}_{b e}  -  \tfrac{1}{r} A_{li} A_{kj} f^a{}_{e\g} f^\g{}_{b d}    \ , \\
&\widehat{R}^{ai}{}_{bk,\d,\e} = - \tfrac{1}{4r} \d_{ik} f^a{}_{b\g} f^\g{}_{\d\e} + \tfrac{1}{r} C_{ij} C_{jk} (f^a{}_{\d c} f^c{}_{b \e}  -  f^a{}_{\e c} f^c{}_{b \d}  )  \ , \\
&\qquad A_{ij} \equiv \tfrac{1}{4}(\nu _{ji}-\nu _{ij}) - \chi_{ij} + \tfrac{r}{2} \l_{ij} \ ,  \qquad C_{ij} \equiv  -\tfrac{1}{4}(\nu_{ji}+\nu_{ij}) + \chi_{ij} + \tfrac{r}{2} \l_{ij} \ . \\
\end{aligned}
\ee

All that remains is to substitute \rf{HH},\rf{RR} into the 2-loop $\beta$-functions \eqref{BMT} in the \GB scheme. The resulting expression is hard to evaluate as it contains hundreds of terms, each proportional to a contraction of the form $(f f f f)_{\a\b}$ or 
$(f f f f)_{ab}$ 
where each $f$ denotes a component $f_{\g\d\e}$ or $f_{\g d e}$ of the structure constants, and indices are contracted using the Killing form $\Tr[T_A T_B]$. One can show, however, that there are only 11 independent  such contractions after accounting for the antisymmetry of the structure constants.
This  allows for 
the efficient evaluation of the resulting $\beta$-functions for $r$ and $\rho_{ij}$. These are first obtained depending on $n_{11}, n_{12}, n_{22}$ but, as discussed in subsection \ref{2103GGapp}, they may be rewritten in terms of $s,t,u$ with all square roots cancelling.  As a result, we obtain the 2-loop generalization of the 1-loop $\beta$-functions in \rf{1H}--\rf{fH},
\be
\ddt \lh_p  = \a' \b_{\lh_p}^{(1)}+ \a'^2 \b_{\lh_p}^{(2)} \ , \qquad \qquad \lh_p\equiv (r,s,t,b,u) \ ,
\ee
where $\b_{\lh_p}^{(2)}$ are given by the following expressions:
\ \\

\bigskip
\hspace{-1.3cm}\includegraphics[scale=0.39]{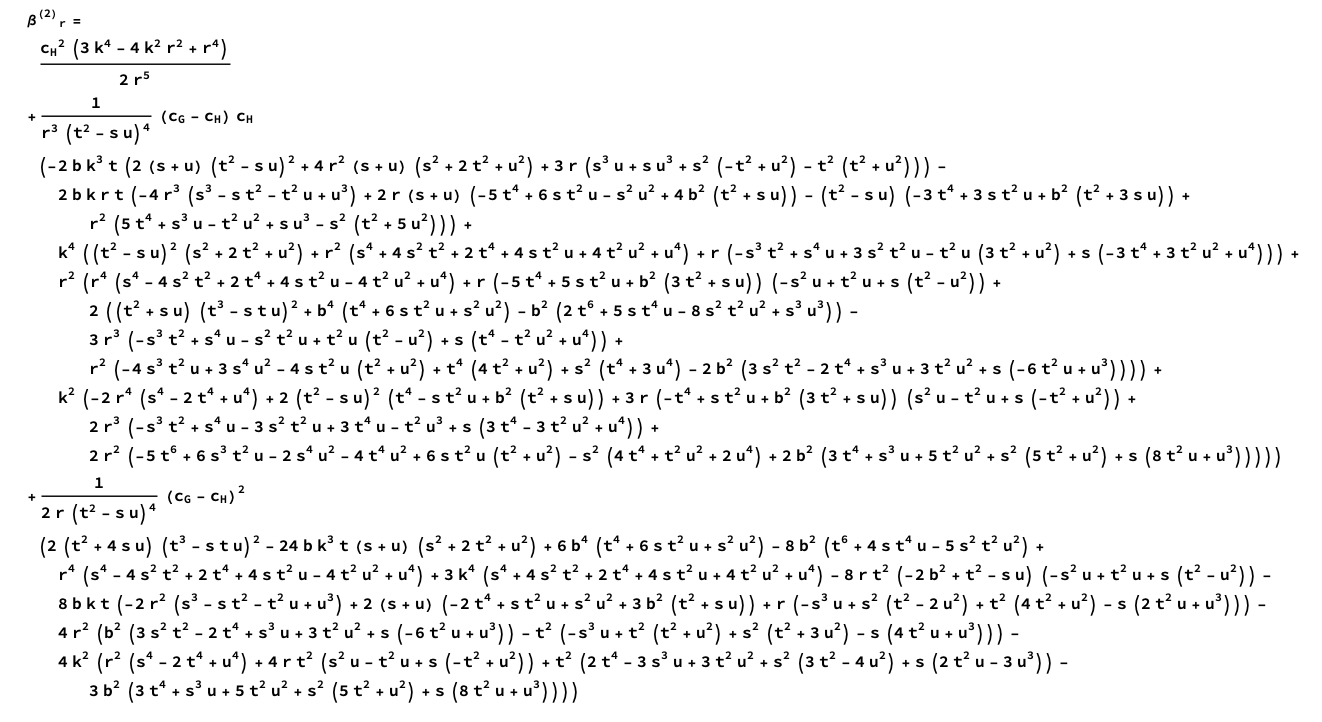}
\medskip

\hspace{-1.3cm}\includegraphics[scale=0.39]{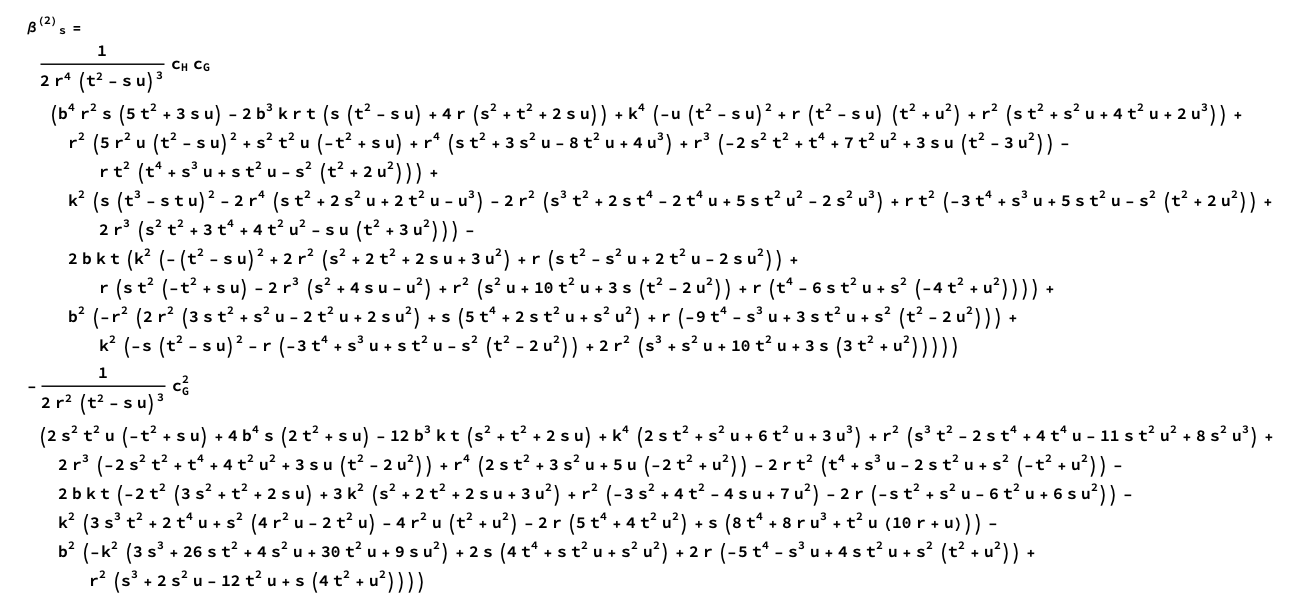}
\medskip

\hspace{-1.3cm}\includegraphics[scale=0.39]{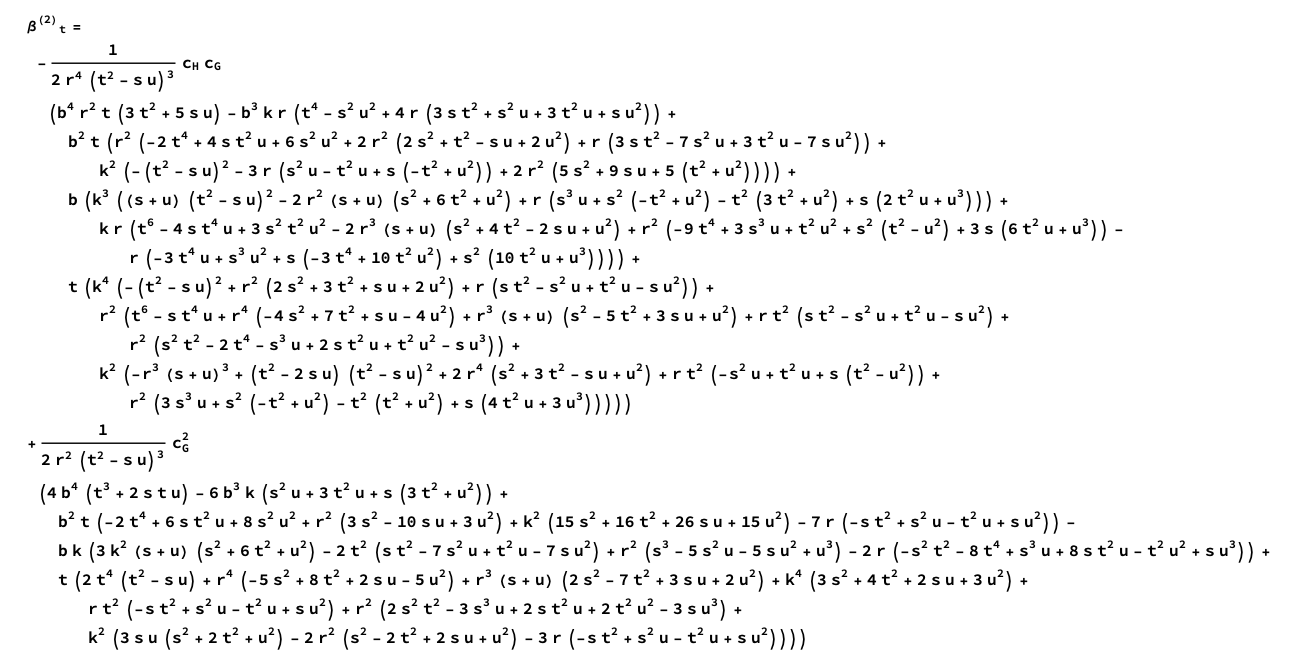}
\medskip

\hspace{-1.3cm}\includegraphics[scale=0.39]{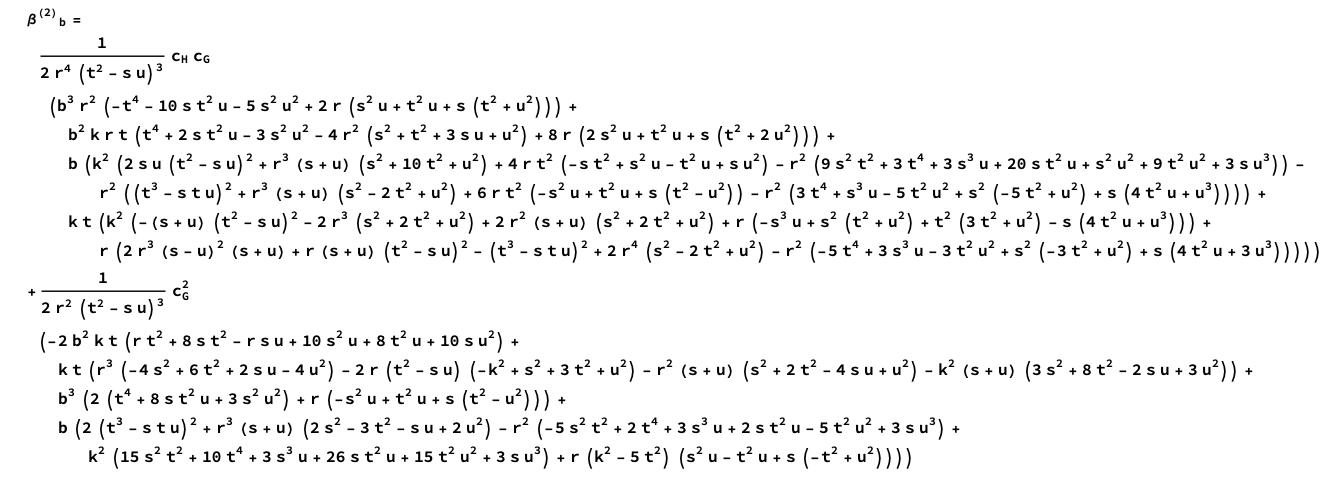}
\medskip

\hspace{-1.3cm}\includegraphics[scale=0.39]{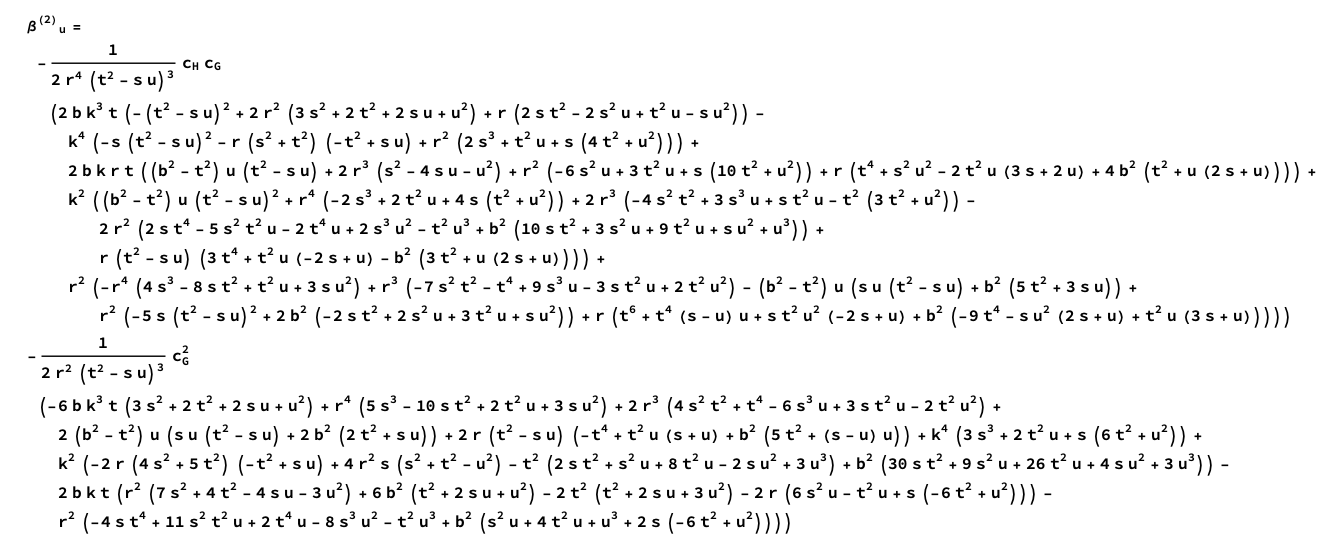}
\medskip

\renewcommand{\la}{\label}
\renewcommand{\rf}[1]{(\ref{#1})}

\renewcommand{\la}[1]{\label{2008#1}} 
\renewcommand{\rf}[1]{(\ref{2008#1})}

%%%%%%%%%%%%%%%%%%%%%%%%%%%%%%%%%%%%%%%%
\chapter{Derivation of RG flow from existence of Lax connection for time-dependent sigma models}\la{A}
%%%%%%%%%%%%%%%%%%%%%%%%%%%%%%%%%%%%%%%%

Here we shall provide some further details of the derivation of the RG flow
from the existence of a Lax representation for the $(\t,\s)$-dependent models in Section \ref{loc}.

\section{Derivation of the equations \rf{alg},\rf{als}}\la{det}

To derive \rf{alg},\rf{als}, we shall ignore terms proportional to the derivatives of the couplings $\del_\pm h_\a$ and match the other terms between the flatness condition of the Lax connection and the equation of motion. The algebraic equations \rf{alg},\rf{als} follow essentially because these `non-derivative'
terms do not change upon the introduction of space-time dependence of the couplings.

Let us consider
the cases of models associated to a group and to a symmetric space
separately.
For the group space case, the original equations of motion following from the flatness of Lax in \rf{gl} are the flatness
and the conservation of the current $\mc{A}_\pm$.
In the $(\t,\s)$-dependent model \rf{sth}, these equations may only be modified by $\O(\del h)$ terms
\be
F_{+-}(\mc{A})= \O(\del h) \ , \qquad\qquad \del_+ \mc{A}_- + \del_- \mc{A}_+ = \O(\del h) \ . \la{gc1}
\ee
They should follow from the flatness of the Lax connection ansatz \rf{glt},
\be
F_{+-} (\widehat{L}) = p_+ p_- F_{+-} (\mc{A}) + \big[p_-(1 - p_+) \del_+ \mc{A}_- - p_+(1-p_-) \del_- \mc{A}_+\big]+ \O(\del p) \ . \la{gc2}
\ee
Comparing first
the terms with $\del_\pm \mc{A}$ and $\mc{A}^2$, i.e.\ neglecting the $\O(\del h)$ and $\O(\del p)$ terms in \rf{gc1},\rf{gc2},
the matching of \rf{gc1} and \rf{gc2} then implies
\be
p_-(1 - p_+) = - p_+(1-p_-) \ .
\ee
This leads to the algebraic equation \rf{alg}.

For the symmetric space case, the original equations of motion following from the vanishing of the curvature of \rf{sl} are the flatness condition for $(\mc{B}+\mc{P})_\pm$ and the equations $D^{\mc{B}}_\pm \mc{P}_\mp = 0$, where $D^{\mc{B}}_\m$ is the covariant derivative with respect to the connection $\mc{B}_\m \in \Lie (H)$.
In the $(\t,\s)$-dependent model \rf{sth} these equations are modified, as in \rf{gc1},
\be
F_{+-}(\mc{B+P})= \O(\del h) \ , \qquad D^\mc{B}_\pm \mc{P}_\mp = \O(\del h) \ . \la{sc1}
\ee
These should follow from the flatness of the ansatz for the Lax pair in \rf{slt},
\begin{align}
F_{+-} (\widehat{L}) ={} &q_+ q_- F_{+-}(\mc{B+P}) + (r_- - q_+ q_-) D^\mc{B}_+ \mc{P}_- - (r_+ - q_+ q_-) D_-^\mc{B} \mc{P}_+
\no \\
& \ + q_-(1-q_+)\del_+ \mc{B}_- - q_+(1-q_-)\del_- \mc{B}_+ + (r_+ r_- - q_+ q_-)[\mc{P}_+, \mc{P}_-] \no \\
& \ + r_- (q_+ -1)[ \mc{B}_+, \mc{P}_-] - r_+ (q_- -1) [ \mc{B}_-, \mc{P}_+]
+ \O(\del q,\del r) \ . \la{sc2}
\end{align}
Comparing the terms in \rf{sc1} and \rf{sc2} that contain
$\del \mc{B}$, $\del \mc{P}$, $\mc{B}^2$, $\mc{P}^2$ and $\mc{BP}$, i.e.\ neglecting the $\O(\del h)$ terms in \rf{sc1} and $\O(\del q, \del r)$ terms in \rf{sc2},
we conclude in particular that the coefficients of the extra $\del\mc{B}$, $[\mc{B,P}]$ and $[\mc{P,P}]$ terms in $\rf{sc2}$ must vanish. Assuming $q_\pm,r_\pm$ are not all zero (so that the Lax connection \rf{slt} is not identically zero) the only solution is $q_+ = q_- = 1$ and $r_+ r_- = 1$, i.e.\ the conditions in \rf{als}.

\section{RG flow in PCM case}\la{fdr}

The matching of the $\O(\del h)$ terms then forces the coupling functions to solve the RG flow equations. The general structure of this argument is explained in eqs.\ \rf{pd1}--\rf{RGs2}, but here we shall run through it explicitly for the simplest PCM example.

Together, \rf{glt} and \rf{alg} lead to the following ansatz for the Lax connection (cf.\ \rf{lpl})
\be
\widehat{L}_\pm = \ha \Big(1+[z(\t,\s)]^{\pm 1}\Big) J_\pm \ , \la{laxh}
\ee
whose curvature is ($F_{+-}(J)=0$ since $J=g^{-1} dg$)
\begin{align}
F_{+-}(\widehat{L}) &= \tfrac{1-{z}^2}{	4{z}} (\del_+ J_- + \del_- J_+) - \tfrac{\del_+ z}{2z^2} J_- - \tfrac{\del_- z}{2} J_+
\ . \la{lah}
\end{align}
The equation of motion for the PCM with coupling $h(\t,\s)$ is (generalizing \rf{pe1})
\be
h (\del_+ J_- + \del_- J_+) + \del_+ h J_- + \del_- h J_+ = 0 \ . \la{eoh}
\ee
For this to be equivalent to the vanishing of the curvature \rf{lah}
the ratios
of the different
coefficients should match.
This leads to the equations of the form \rf{pd1}, i.e.
\be\la{a9}
\del_+ z = - \tfrac{z}{2} (1-z)^2 \,\tfrac{\del_+ h}{h} \ , \qquad \del_- z = - \tfrac{1}{2z} (1-z)^2 \,\tfrac{\del_- h}{h}\ .
\ee
The consistency condition $\del_+(\del_- z) - \del_- (\del_+ z) = 0$ gives
\be
\tfrac{(1-z^2)^2}{2h z} \ \del_+ \del_- h = 0 \ .
\ee
It is remarkable that the $z$ dependence has totally factored out
(a term proportional to $\del_+ h \del_- h$ is absent here due to a special cancellation).
Excluding the trivial cases $z(\t,\s) =\pm 1 $, which would not encode the correct equation of motion, we find that the Lax connection \rf{laxh} only matches the correct equation of motion if $\del_+ \del_- h =0$, i.e.\ if
$h = f^+ (\xi^+) + f^-(\xi^-)$.
Any such solution is related to the PCM's 1-loop RG flow, $h=c\, \t$, by a 2d conformal transformation.

\section{RG flow for theories with multiple couplings}\la{mr}

In Section \ref{2008s3} we explained the derivation of the RG flow
focussing on the case with only one coupling (see eqs.\ \rf{pd1}--\rf{RGs2}).
The same conclusion also holds for the group space $\eta$-model with two couplings $h$ and $\k$ (see Table \ref{2008tab1}), and more generally is expected to be true for multi-coupling theories.

There are multiple independent structures in the equation of motion (for the $\eta$-model these involve different
powers of the $R$-matrix) and correspondingly in the flatness of the Lax connection. Matching
the coefficients of these structures in the $\eta$-model case yields \textit{two} pairs of equations (cf.\ \rf{pd1}),
\begin{align}
&\del_\t z = U_\t(z;h,\k) \ , \qquad \del_\s z = U_\s(z;h,\k) \ , \la{sy1}\\
&\del_\t z = V_\t(z;h,\k) \ , \qquad \del_\s z = V_\s(z;h,\k) \ . \la{sy2}
\end{align}
In general, for an $N$-coupling theory we would expect to find $N$ pairs of equations.

As a system of equations for $z$, \rf{sy1},\rf{sy2} is clearly overdetermined. In two combinations of these equations the $z$ dependence cancels to give relations between $h(\t,\s)$ and $\k(\t,\s)$
\begin{align} \la{a13}
\del_\t( \k^{-1} h ) = \del_\s ( \k^{-1} h ) = 0 \ .
\end{align}
Eq.\ \rf{a13}
implies that $\nu \equiv  \frac{h}{4\k} $ is a constant; this coincides precisely with the 1-loop RG invariant
of the $\eta$-model (see Table \ref{2008tab1}).
For an $N$-coupling theory, we may expect to obtain $(N-1)$ RG invariants $\nu_r$ in this way.

Then
the system \rf{sy1},\rf{sy2} reduces to just two equations --- effectively returning to the single-coupling case of equations \rf{pd1}. Again, the consistency condition for the two remaining equations takes the remarkable form \rf{2w}, where the $\b$-function $\b(h) \equiv \beta^h (h, \nu )$ is understood as a function of the coupling $h$ and the RG invariant. As in eqs.\ \rf{cfs},\rf{RGs2} it then follows (modulo a conformal transformation) that $h(\t,\s)$ depends only on $\t$ and follows the 1-loop RG flow.
The same should generalize to the $N$-coupling case with
$(N-1)$ independent RG invariants $\nu_r$ that can be chosen as constants.

%%%%%%%%%%%%%%%%%%%%%%%%%%%%%%%%%%%%%%%%
\chapter{On non-local charges in time-dependent sigma models
}\la{om}
%%%%%%%%%%%%%%%%%%%%%%%%%%%%%%%%%%%%%%%%

In Section \ref{2008s4} we discussed the construction of non-local charges in the time-dependent PCM.
Here we shall comment on the other time-dependent models in Table \ref{2008tab1},
and, in particular, on the symmetric space $\l$-model.

The construction of the conserved monodromy matrix satisfying \rf{lp} works similarly, although for the models built on symmetric spaces (symmetric space $\s$-model and symmetric space $\l$-model), it is only the eigenvalues of the monodromy matrix that are conserved (because the boundary value of $L_\t$ does not vanish). A sufficient boundary condition in all cases is that $g(\t,\s) \to g_0$ at spatial infinity and that $(g-g_0)$ decays sufficiently fast so
that the monodromy converges at spatial infinity. As in the PCM example it is hard to evaluate the conserved charges explicitly and thus to verify that they are infinite in number (i.e.\ depend non-trivially on the spectral parameter $w$).

Except for the symmetric space $\l$-model, all the other models have global symmetries.
As for the PCM in \rf{noet}, the associated
charges can be obtained by expanding the monodromy around $w=\infty$.\foot{This applies also for the symmetric space $\s$-model after applying the gauge transformation $L_\pm \to g L_\pm g^{-1} - \del_\pm g g^{-1}$ to the Lax connection $L = J^H_\pm + z^{\pm 1} J_\pm^{G/H}$, obtaining the alternate Lax connection $L_\pm = \ha (1-z^{\pm 1}) (-2 g J_\pm^{G/H} g^{-1} )$ of the `group space' form \rf{gl}, instead of \rf{sl} (modulo sign reversal of $z$).}
In the remainder of this appendix, we shall consider
the trivial reduction of the time-dependent symmetric space $\l$-model, which has no manifest global symmetries.
In general, the eigenvalues of the monodromy matrix would be conserved on an infinite spatial line;
however, in the ``trivial'' reduction where the fields depend only on time (see Section \ref{2008s4}), the monodromy matrix does not converge at spatial infinity since $L_\s$ does not vanish there.
Below we will try
to shed some light on this issue.

Let us first recall what happens for geodesics in the usual time-independent case, where the Lax connection is $L_\pm = A_\pm^H + z^{\pm 1} \tfrac{1}{\sqrt{\l}} A_\pm^{G/H}$. In the trivial reduction we have $A_\pm = A_\pm (g(\t))$ so
the periodicity condition \rf{pe} is satisfied and the eigenvalues of the monodromy conserved even on a finite interval of length $a $. The path-ordered exponential trivializes to give $\M = \exp{ (a \, L_\s )}$ and hence, equivalently,
the eigenvalues of $L_\s = \ha (A_+^H - A_-^H + \tfrac{z}{\sqrt{\l}} A_+^{G/H} - \tfrac{z^{-1}}{\sqrt{\l}} A_-^{G/H})$ are conserved. For example,
let us consider the simplest $SU(2) \ov U(1)$ $\l$-model (with the  $U(1)$ subgroup generated by $\s_2$), parametrized after gauge fixing as (cf.\ eqs.\ \eqref{42},\eqref{43})
\begin{align}
&g = e^{i \a \s_3} e^{i \b \s_2} \ , \qquad\qquad \cos{\a} = \sqrt{p^2+q^2} \ , \qquad \ \tan{\b}=\frac{p}{q} \ , \\
&\L = \frac{k}{1-\l^2} \frac{(1-\l)^2 (\del p)^2 + (1+\l)^2 (\del q)^2}{1-p^2-q^2} \ . \la{2dl}
\end{align}
The trivial 1d reduction, or geodesic motion, is described by the Lagrangian ($\dot{} \equiv \del_\t$)
\be
L = \frac{k}{1-\l^2} \frac{(1-\l)^2 \dot{p}^2 + (1+\l)^2 \dot{q}^2}{1-p^2-q^2} \ . \la{t1d}
\ee
There is only one independent eigenvalue $l$ of $L_\s$ since it is a traceless $2\times 2$ matrix. Its expansion around, e.g.,
$z=1$ gives (at least) two independent conserved charges for the geodesic motion
\begin{align}
&\dot{Q}_1 = \dot{Q}_2 = 0 \ , \qquad \qquad l = \tfrac{1}{2(\l^2-1)}\sqrt{Q_1} + (z-1)^2 \tfrac{\l}{\l^2-1} \frac{Q_2}{\sqrt{Q_1}} + \ldots \ ,\\
&\quad Q_1 = \frac{- (\l-1)^4q^2\dot{p}^2 + 2(\l^2-1)^2 p q \dot{p}\dot{q}  + (1+\l)^2[4\l(q^2-1)-(\l-1)^2p^2]\dot{q}^2 }{(1-p^2-q^2)^2} \ , \la{q1} \\
&\quad Q_2 = \frac{ (1-\l)^2 \dot{p}^2+ (1+\l)^2\dot{q}^2 }{1-p^2-q^2} \ , \la{q2}
\end{align}
with $Q_2$ proportional to the Hamiltonian.

Returning to the time-dependent case, the monodromy is not defined on an infinite line, and on any finite interval the periodicity condition is not satisfied due to the explicit $\s$ dependence of the Lax connection for generic values of $w$. However, at certain special values of $w$ satisfying $\exp{(\tfrac{c}{k}w)}=\pm \infty$, the $\s$ dependence disappears to give flat connections
\begin{align}
&[\del_+ + A_+^H + \tfrac{1}{\l(\t)} A_+^{G/H} , \ \del_- + A_-^H + A_-^{G/H} ]=0 \ , \la{f1} \\
&[\del_+ + A_+^H + A_+^{G/H} , \ \del_- + A_-^H + \tfrac{1}{\l(\t)} A_-^{G/H} ] = 0 \ , \qquad \qquad \l(\t) = \exp{(-\tfrac{c}{k} \t)} \ , \la{f2}
\end{align}
generalizing the same expressions from the time-independent case ($\l(\t) \to \l$). At these values the periodicity condition \textit{is} satisfied on any finite interval. The monodromy trivializes as in the time-independent case to give $\M = \exp{(a\, L_\s)}$, so the eigenvalues of $L_\s$ are again conserved.

In fact, the two flat connections \rf{f1},\rf{f2} are related by a gauge transformation, so their conserved charges are the same. Hence, the maximum number of independent charges obtained from the flat connections \rf{f1},\rf{f2} is $r = {\rm rank}(G)$
for a symmetric space $G/H$. This number is generally less than the number of fields, $\dim{G/H} =\dim{G}-\dim{H}$ (e.g.\ for $SO(n+1)/SO(n)$ we get $r= n-1 < \dim{SO(n+1)/SO(n)} = n$), so this is not sufficient for integrability.

For example, in the ${SU(2) \ov U(1)} \sim {SO(3) \ov SO(2)}$ case \rf{2dl}, where there are 2 fields, we only obtain $r=1$ conserved charge,
\be
Q = \frac{ [\l(\t)-1]^4(p^2-1)\dot{p}^2 + 2[\l(\t)^2-1]^2 p q \dot{p}\dot{q}+ [\l(\t)+1]^4 (q^2-1) \dot{q}^2 }{[\l(\t)^2-1]^2(1-p^2-q^2)^2}  \ , \la{cnt}
\ee
for the trivial 1d reduction, which is  a time-dependent generalization of \rf{t1d},
\be
L = \frac{k}{1-\l^2} \frac{(1-\l(\t))^2 \dot{p}^2 + (1+\l(\t))^2 \dot{q}^2}{1-p^2-q^2} \ ,\quad  \qquad \l(\t) = \exp{(-\tfrac{c}{k} \t)}  \ .
\ee

In the time-independent limit $\l(\t) \to \l$ (obtained, e.g.,
by shifting $\t \to - \tfrac{k}{c}\log{\l} + b \t$ and taking $b\to 0$), this charge becomes a particular combination of the charges \rf{q1},\rf{q2} in the time-independent theory,
\be
Q \to Q_1 - (\l-1)^2 Q_2 \ .
\ee
The presence of the charge \rf{cnt} in the time-dependent 1d theory is a non-trivial property
However, at least upon restricting consideration to special values of $w$, we have not found
enough charges for the integrability of the geodesics.
Since the construction of charges is subtle, depending on the value of $w$,  boundary conditions and the choice of the spatial domain
(the periodicity condition \rf{pe} must be satisfied and, if the interval is infinite, the monodromy must converge at infinity),
it is possible that the monodromy constructed on some special domain would yield further conserved charges,
but this remains to be clarified.

%%%%%%%%%%%%%%%%%%%%%%%%%%%%%%%%%%%%%%%%
\chapter{Time-dependent 1d harmonic oscillator and conserved charge}\la{C}
%%%%%%%%%%%%%%%%%%%%%%%%%%%%%%%%%%%%%%%%

In Section \ref{2008s4} we came across a particular time-dependent, linear, 1d model \rf{lt}.
Starting with a general time-dependent linear 1d action\foot{We do not include the term $f(\t) \theta \dot{\theta}$
as it can be put in the form $k(\t) \theta^2$ by adding a total derivative.}
\be
S = \int d\t \, \big[ h(\t)\, \dot{\theta}^2 - k(\t)\, \theta^2 \big] \ , \qquad \dot{\theta} \equiv \del_\t \theta \ ,
\la{orig}
\ee
one may redefine $\t$ as $\t \to t(\t)$, $\dot{t}(\t) = h^{-1}(\t)$ to put all time dependence in the harmonic potential term
\be
S = \int dt \, \big[ {\theta'}^2 - m^2(t)\, \theta^2 ] \ , \qquad \qquad m^2(t) = k(\t(t)) h(\t(t)) \ , \ \ \ \theta ' \equiv \del_t \theta \ .
\ee
The corresponding equation of motion is
\be
\theta'' +m^2(t)\, \theta = 0 \ . \la{eoml}
\ee
It is easy to see that for a given function $\theta_0(t)$, the quantity
\be
Q = \theta_0 \, \theta' - {{\theta}'_0} \, \theta \ ,
\ee
is conserved on-shell if and only if $ \theta_0$ is a particular solution of the equation of motion \rf{eoml}.
Furthermore,
such a conserved charge provides a first integral for \rf{eoml},
\be
Q = {\theta_0} \, \theta' - \theta'_0 \, \theta = C_1 = \text{const} \ \
\quad \to \quad (\tfrac{\theta}{\theta_0})' = \tfrac{C_1}{\theta_0^2} \ . \la{exact}
\ee
Integrating this first-order equation
yields the general solution of \rf{eoml} ($C_2=\const$),
\be\la{d6}
\theta(t) = C_1\, \theta_0(t) \int \tfrac{dt }{\theta_0^2(t)} +\ C_2\, \theta_0(t) \ ,
\ee
with $\theta=\theta_0$ being, of course, a special case.
Thus \rf{eoml} is solvable if a particular solution $\theta_0$ can be constructed explicitly.

Changing back to the original parametrization \rf{orig} ($t \to \t$), the conserved charge \rf{exact}
takes the form
\be
Q = h(\t) \big[ {\theta_0} \, \dot{\theta} - \dot{\theta}_0 \, \theta \big] \ , \la{lic}
\ee
where $\theta_0=\theta_0(\t)$ is a particular solution of \rf{orig}, while \rf{d6} becomes
\be
\theta(\t) = C_1\, \theta_0(\t) \int \tfrac{d\t}{h(\t) \, \theta_0^2(\t)} + C_2\, \theta_0(\t) \ . \la{gena}
\ee
The linearized theory \rf{lt} corresponds to \rf{orig} with $h(\t) = k(\t) = \t$.
The conserved charge in \rf{ql},\rf{ql2} is indeed of the form \rf{lic}.
From the monodromy matrix one finds that
$Q = \t [ \g(\t) \, \theta + \bw(\t) \, \dot{\theta} ]$ where
\be
\hspace{-0.3cm}\g(\t) = \int_{-\infty}^{+\infty} d\s \,e^{-im \s} \, \tfrac{m (s_+ s_- -w-\s)}{2 \t s_+ s_- } ,\ \ \  \bw(\t) = \int_{-\infty}^{+\infty} d\s\, \, e^{-im \s}\, \tfrac{ i}{2 s_+ s_- } , \ \ \ s_\pm \equiv \sqrt{w+\s\pm \t} \ . \la{ints}
\ee
Let us show that $\g = - \dot \bw$, as required to match \rf{lic}. Indeed, the term proportional to $\dot{\theta}$ in the leading $\O(\e)$
expansion of the flatness equation $\del_\t L_\s - \del_\s L_\t + [L_\t,L_\s] = 0$ tells us that the integrands in \rf{ints}
satisfy $
\del_\t [ \tfrac{e^{-im \s} i}{2 s_+ s_- } ] + \tfrac{e^{-im \s} m (s_+ s_- -w-\s)}{2 \t s_+ s_- } = \del_\s\, [\tfrac{e^{-im \s}\, i (s_+ s_- -w-\s)}{2 \t s_+ s_- } ] $.
Integrating over $\s$ and noting that $ \tfrac{e^{-im \s}\, i (s_+ s_- -w-\s)}{2 \t s_+ s_- } $ vanishes at $\s=\pm \infty$,\foot{This follows from the periodicity condition \rf{pe} since it is the coefficient of $\dot{\theta}$ in a component of the leading $\O(\e)$ term in $L_\t$.}
we conclude that $\gamma + \dot\bw=0$.

Since the charge in \rf{ql} is conserved on-shell (from the monodromy matrix construction),
it follows that $\bw(\t) $ is a particular solution, and thus the general solution \rf{gena} is in this case given by
\be\la{genn}
\theta = C_1\, \bw(\t) \int \tfrac{d\t}{\t\, \bw^2(\t)} + C_2\, \bw (\t)\ .
\ee

%%%%%%%%%%%%%%%%%%%%%%%%%%%%%%%%%%%%%%%%
\chapter{Lax connection in time-dependent sine-Gordon model}\la{sing}
%%%%%%%%%%%%%%%%%%%%%%%%%%%%%%%%%%%%%%%%

As discussed in Section \ref{2008s6}, the sine-Gordon model,
\be
\L = \tfrac{1}{g^2} \big[ \ha \del_+ x \del_- x + m^2 \cos{x}\big] \ , \la{sgma}
\ee
displays the same pattern as the $\s$-models considered above: upon promoting the couplings $(g,m)$ to functions of 2d time $\t$, the Lax connection naturally generalizes to the resulting time-dependent model only if the time dependence
is given (up to rescalings of time) by the 1-loop RG flow of the original model,
\be m^2(\t) = e^{ \b(g)\, \t} m_0^2 \ , \ \ \ \qquad g(\t) = g \ , \ \ \ \ \qquad \b(g) = - 2 + g^2 \ . \la{52a} \ee
Below we shall justify this claim in more detail.

As was noted in section \ref{2008s6}, the time-dependent theory obtained from \rf{52a}
\be
\widehat{\L} = \tfrac{1}{g^2} \big[ \ha \del_+ x \del_- x + e^{\b(g)\, \t}\, m^2_0\, \cos{x}\big]\ , \la{53a}
\ee
is clearly integrable
since the explicit $\t$-dependence in \rf{53a} can
be removed by a 2d conformal transformation getting back to \rf{sgma}.
Indeed, starting with the Lax pair \cite{Lax:1968fm} for the original
sine-Gordon model \rf{sgma} ($\s_i$ are Pauli matrices)
\begin{align}
\hspace{-2cm} L_\pm = \pm \tfrac{i}{4} \del_\pm x \ \s_3 + \tfrac{i}{2} z^{\pm 1} \, m \cos{\tfrac{x}{2}} \ \s_1 \pm \tfrac{i}{2} z^{\pm 1} \, m \sin{\tfrac{x}{2}} \ \s_2 \ , \la{sgla}
\end{align}
and applying a conformal transformation $\xi^\pm \to f^\pm (\xi^\pm)$, one obtains a Lax connection
for the time-dependent theory \rf{53a},\foot{Note that, as for the $\s$-models discussed in Chapter \ref{loc} above, the dependence on the spectral parameter $z$ in \rf{larga} is again correlated with
a constant shift of $\s$.}
\be
\widehat{L}_\pm =\pm \tfrac{i}{4} \del_\pm x \ \s_3 + \tfrac{i}{2} z^{\pm 1} \, e^{\b(g)\, \xi^\pm} \, m_0 \cos{\tfrac{x}{2}} \ \s_1 \pm \tfrac{i}{2} z^{\pm 1} \,e^{\b(g) \, \xi^\pm} \, m_0 \sin{\tfrac{x}{2}} \ \s_2 \ . \la{larga}
\ee
Note that this Lax connection follows the same ansatz \rf{lac2} as in the \sm case:
it is obtained from the original Lax connection \rf{sgla} by the replacements
$z \to \widehat{z}= e^{\frac12 \b(g)\s}z$ and $m_0 \to m(\t) = e^{\frac12 \b(g)\t} m_0$.

Conversely, suppose we replace the couplings $(g,m)$ in \rf{sgma}
with general functions of $(\t,\s)$ and then demand that the resulting theory (cf.\ \rf{sth})
\be
\widehat{\L} = \tfrac{1}{g^2(\t,\s)} \big[ \ha \del_+ x \del_- x + m^2(\t,\s) \, \cos{x}\big] \la{54a}
\ee
admits a Lax representation. Motivated by \rf{sgla}, we shall assume the following ansatz for the Lax connection (cf.\ \rf{glt},\rf{slt})
\begin{align}
&\widehat{L}_\pm = f_{\pm}(\t,\s)\ \tfrac{i}{4} \del_\pm x \, \s_3 + \, v_\pm (\t,\s) \ \tfrac{i}{2}\cos{\tfrac{x}{2}} \, \s_1 + w_\pm (\t,\s) \ \tfrac{i}{2}\sin{\tfrac{x}{2}} \, \s_2\ . \la{sga}
\end{align}
Then
matching coefficients of various terms in the zero-curvature condition for \rf{sga} and in
the equation of motion corresponding to \rf{54a} leads to the following constraints on the coefficient functions in \rf{sga}
and the coupling functions,
\begin{align}
&f_\pm = \pm 1 \ , \qquad\qquad w_\pm = \pm v_\pm \ , \qquad\qquad \del_\mp v_\pm = 0 \ , \la{fe1} \\
&g(\t,\s) = g= \const \ , \qquad\qquad v_+ v_- = m^2(\t,\s) \ . \la{fe2}
\end{align}
It follows from \rf{fe1},\rf{fe2} that $v_\pm = \pm w_\pm = v_\pm(\xi^\pm)$, \ $m^2(\t,\s)=v_+(\xi^+)\, v_-(\xi^-)$.
Finally, applying a conformal transformation to \rf{sga},
we may set, e.g., $v_\pm(\xi^\pm) \to z^{\pm 1} \, e^{\b(g) \, \xi^\pm} \, m_0$, $m^2(\t,\s) \to e^{\b(g) \, \t} \, m_0^2$,
thus bringing the Lagrangian \rf{54a} and the Lax connection \rf{sga} to the form \rf{53a} and \rf{larga}
where the time dependence is given by the 1-loop RG flow \rf{52a} of the standard sine-Gordon model.

\renewcommand{\la}{\label}
\renewcommand{\rf}[1]{(\ref{#1})}

%%%%%%%%%%%%%%%%%%%%%%%%%%%%%%%
\end{appendices}
%%%%%%%%%%%%%%%%%%%%%%%%%%%%%%%

\newpage
\input{refs.tex}

\end{document}
%%%%%%%%%%%%%%%%%%%%%%%%%%%%%%%

%% file: 2103.tex
\numberwithin{equation}{chapter}

\renewcommand{\arraystretch}{1.3}

\makeatletter
\renewcommand\section{\@startsection {section}{1}{\z@}%
{-3.5ex \@plus -1ex \@minus -.2ex}%
{2.3ex \@plus.2ex}%
{\normalfont\large\bfseries}}
\renewcommand\subsection{\@startsection{subsection}{2}{\z@}%
{-3.25ex\@plus -1ex \@minus -.2ex}%
{1.5ex \@plus .2ex}%
{\normalfont\normalsize\bfseries}}
\makeatother

\expandafter\def\expandafter\bfseries\expandafter{\bfseries\ifmmode\else\boldmath\fi}
\expandafter\def\expandafter\mdseries\expandafter{\mdseries\ifmmode\else\unboldmath\fi}
\expandafter\def\expandafter\normalfont\expandafter{\normalfont\ifmmode\else\unboldmath\fi}

\providecommand{\href}[2]{#2}

\def\id{\protect{{1 \kern-.28em{\rm l}}}}
\def\be{\begin{eqnarray}}
\def\ee{\end{eqnarray}}
\def\bi{\bibitem}
\def\tr{{\rm tr}}
\def\ha{\tfrac{1}{2}}
\def\td{\tilde}
\def\ci{\cite}
\def\N{{\mathcal N}}
\def\ww{\Omega}
\def\S{{\mathcal S} }
\def\nn{\nu}
\def\z{\zeta}
\def\dg{\dagger}
\def\a{\alpha}
\def\b{\beta}
\def\ap{\alpha^\prime}
\def\aa{{\a'}}
\def\g{\gamma}
\def\ok{\frac{1}{k}}
\def\jL{{J}}
\def\cL{{\mathcal L}}
\def\cH{{\mathcal H}}
\def\E{{\mathcal E}}
\def\w{\omega}
\def\vep{\varepsilon}
\def\De{{\mathcal D}}
\def\k{\kappa}
\def\four{\tfrac14}
\def\third{\tfrac{1}{3}}
\def\det{\hbox{det}}
\def\tid{\tilde}
\def\vv{{\rm v}}
\def\ta{{\tilde \a}}
\def\fo{\frac{1}{4}}
\def\K{{\rm S}}
\def\el{\ell}
\def\Tr{{\rm Tr}}
\def\P{\Phi}
\def\l {\lambda}
\def\bl{{\tilde \l}}
\def\const{{\rm const}}
\def\vn{\vec n}
\def\Prod{\Pi}
\def\O{{\mathcal O}}
\def\m{\mu}
\def\vs{\vec \s}
\def\ie{i.e.}
\def\rS{{\rm S}}
\def\as{{\a}}

\def\foot{\footnote}
\def\ra{\rightarrow}
\def\F{{\cal F}}
\def\cc{\circ}
\def\eqv{\equiv}
\def\ni{\noindent}
\def\bw{{\rm w}}
\def\cT{{\cal T}}
\def\no{\nonumber}
\def\J{{\cal J}}
\def\om{\omega}
\def\l{\lambda}
\def\tl{{\tilde \l}}
\def\tk{{\tilde k}}
\def\sqtl{{\sqrt{\tilde \l}}}
\def\adss{$AdS_5 \times S^5$\ }
\def\D{\Delta}
\def\p{\phi}
\def\r{\rho}
\def\rN{{\rm N}}
\def\tw{{\tilde w}}
\def\vl{ \vec \ell}
\def\varpi{{\rm w}}
\def\bG{\bar \G}
\def\ve{\varepsilon}
\def\I{{\cal I}}
\def\La{{\Lambda}}
\def\R{{\rm R}}
\def\bt{\bar\theta}
\def\Z{{\cal Z}}
\def\pa{\partial}
\def\bea{\be}
\def\eea{\ee}
\def\DD{{\rm D}}
\def\chii{\varepsilon}
\def\th{\theta}
\def\t{\tau}
\def\beq{\be}
\def\eeq{\ee}
\def\beqa{\bea}
\def\eeqa{\eea}
\def\bs{\bigskip}
\def\c{{\rm a}}
\def\del{\partial}
\def\s{\sigma}
\def\eps{{\epsilon}}
\def\n{\nu}
\def\dag{\dagger}
\def\bd{\bar \del}
\def\ed{\end{document}}
\def\iffa{\iffalse}
\def\xp{x_+}
\def\xm{x_-}
\def\te{\textstyle}
\def\rx{{\rm x}}
\def\C {{\rm C}}
\def\hC {{\rm \hat C}}
\def\dd {{\rm d}}
\def\bb{{\rm b}}
\def\sql{{\sqrt{\l}}}
\def\ep{\epsilon}
\def\G{\Gamma}
\def\dx{\dot x}
\def\vp{\varphi}
\def\tx{{\tilde x}}
\def\ad{{\rm ad}}
\def\d{\delta}
\def\de {\del}
\def\L{\mathcal{L} }
\def\pcm {{\rm PCM}}
\def\wz{{\rm WZ}}
\def\pcmq {{\rm PCM}$_q$\ }
\def\ZMq {{\rm ZM}$_q$\ }
\def\ZM {{\rm ZM} }
\def\pp{{\rm p}}
\def\qq{{\rm q}}
\def\rk {{\rm u}}
\def\rrk{{\rm e}}
\def\ss{{\rm s}}
\def\tt{{\rm t}}
\def\uu{{\rm u}}
\def\sm{$\sigma$-model}
\def\sms{$\sigma$-models}
\def\ov{\over}

\def\Ad{\text{Ad}}
\def\lmp{\Lambda}
\def\bl{\bar \lambda}
\def\Lie{\operatorname{Lie}}
\def\hb{\hbar}
\def\k{\varkappa} \def \h {{\rm h}}
\def\dt{\tfrac{d}{d\t}}
\def\ddt{\frac{d}{d\t}}
\def\bb{{\rm b}}
\def\vv{{ v}}
\def\kk{{\rm k}}
\def\ka{{\kappa}}
\def\np{ }
\def\cM{\mathcal{M}}
\def\bh{\bar{h}}
\def\lam{$\lambda$-model}
\def\etm{$\eta$-model}
\def\bh{\bar h}
\def\hh{{\rm h}}
\def\PCM{\L_{_{\rm PCM}}}
\def\WZ{\L_{_{\rm WZ}}}
\def\M{{\cal M}}
\def\q{{\rm q}}
\def\RR{{J}}
\def\LL{{K}}
\def\kold{k_{\text{old}}}
\def\kaold{\ka_{\text{old}}}
\def\ed{\end{document}}
\def\F{H}
\def \cgg {\ccg}
\def\cg{c\,}
\def\cf{c_{_\F}}
\def\cff{c_{_\F}}
\def\HH{\F}
\def\XX{{X}}
\def\hhh{\widetilde {\hh}}
\def\e{\varepsilon}
\def\tih{{\hat h}}
\def\H{{\rm H}}
\def\GG{{\rm G}}
\def\BB{{\rm B}}
\def\g{\gamma}
\def\bl{\bar\lambda}
\def\ol{\underline}
\def\ul{\overline}
\def\MM{\mathcal M}
\def\hz{z}
\def\hhz{z} %\def\hhz{\widehat{z}} %v3
\def\wcL{\widehat{\L}}
\def\wL{\widehat{L}}
\def\XX{X}
\def\bw{\a}
\def\xx{{\rm x}}
\def\cR{{\mathcal R}}
\def\sms{$\s$-models\ }
\def\sm{$\s$-model\ }
\def\wLL{\widehat{\L}}
\def\br{\big[}
\def\bl{\big]}
\def \FF {h}
\def\GG{{\rm G}}
\def\wt{\widetilde}
\def \wh  {\wt \rh}

\def \foot{\footnote}\def \ci{cite}\def \l {\lambda}\def \iffa {\iffalse}
\def \RR {{R}} \def \ov {\over }\def \a  {\alpha} \def \ha {{1\ov 2}}
\def \ed {\end{document}}
\def \bi {\bibitem}
\def \adss {${\rm AdS}_5 \times S^5~$ }

\def \ov {\over}
\def \tr {{\rm tr}}
\def \ha {{1 \over 2}}
\def \td {\tilde}
\def \ci {\cite}
\def \bi {\bibitem}
\def \N  {{\cal N}}
\def \aa  {{\rm a}}
\def \te {\textstyle} 
\def \Z {{\cal Z}}
\def \RR {{L}}
\def \aa {{\rm a}} \def \eps {\epsilon}\def \rr {{\rm r}}

\def \rh {{\rm h}}
\def \cg {{\rm c}}

\def \ccg {{\cg_{_G}}} \def \ch {{\cg_{_H}}}
\def \cggg  {{\cg^2_{_G}}} 
 \def \ep {\epsilon}

\def \ww {{\rm w}}

  \def \bpsi {\bar \psi}
\def\cG{\ccg}

\def\ol{\overline}
\def \ha {\tfrac{1}{2}}

\def \B {{ B}}

\def \lh {{\rm h}}
    \def \ta  {U}   \def \tb {V} 

\def \hij  {p_{ij} }  \def\Hij  {\chi_{ij}}

\chapter{\texorpdfstring{$G\times G$}{GxG}  and \texorpdfstring{$G\times G/H$}{GxG/H}   sigma  models \label{GGS}}

It is  
 expected  that classically  integrable 2d \sms   should be   stable  under the  renormalization group  flow, 
 the intuition  being that hidden symmetries 
will constrain
 the  RG evolution. 
Constraints on coupling constants   required for integrability should thus be RG-invariant.

The aim of this chapter is to explore the  connection between integrability and the RG flow  on 
another class of examples ---   integrable $G\times G$ and $G\times G/H$ models that were derived from  the affine Gaudin construction \cite{DLMV, ABL}.  These models may be viewed as    generalizations  of the PCM$_k$,
\be
\L_{_{{\rm PCM}_k}} = {\rh} \, \L_{_{{\rm PCM}}}  + k \, \L_{_{\rm WZ}}(g)\ , \qquad  
\L_{_{{\rm PCM}}}  = -\ha \,  \Tr[ J_+ J_- ]\ ,   \qquad J \equiv g^{-1} dg \ , \ \ g\in G \ , \la{Pk}
\ee
i.e.\   the principal chiral model    (with inverse coupling $\rh$)   with  the WZ term  (with level $k$).
The conformal WZW model is obtained at the special points $\rh=\pm k$ (see Secrtions \ref{PkS}-\ref{WS}).

The PCM$_k$   admits  various  integrable deformations (see, e.g., \cite{Klimcik:2002zj,Klimcik:2008eq,Ki,Balog:1993es,Delduc:2013fga,DMV} and Section \ref{etas} above), which have been  interpreted 
  \cite{DLMV} as  particular  cases  of   integrable affine Gaudin models. 
The  affine Gaudin construction also produces    natural generalizations  of the  PCM$_k$  to integrable models on products of group  spaces $G^N = G\times \ldots \times G$ \cite{DLMV}. 

Here  we shall  consider  a  subclass of  such  models  defined  by 
\begin{align}
&\L = -\ha \, \rho_{ij} \, \Tr[ J^{(i)}_+ J^{(j)}_- ] + k_i  \, \L_{_{\rm WZ}}(g^{(i)}) \ ,  \qquad \qquad \la{cou}\\
&J^{(i)}\equiv {g^{(i)}}^{-1} d g^{(i)} \ , \qquad \qquad  g^{(i)} \in G^N  \ , \qquad i=1,\ldots,N \ . \no
\end{align}
We denote by  $J^{(i)}$   the Maurer-Cartan 1-form  corresponding to $i$-th copy of  $G$  and 
 $\rho_{ij}$ is a \textit{constant}  coupling   matrix (summation over repeated  $i,j$ is assumed). 

The PCM$_k$ \rf{Pk} corresponds to the special case $N=1$ (with $\rho_{11}=\rh$ and $k_1=k$), and is integrable for any values of its couplings. However, for $N>1$,   the model \rf{cou} is  classically integrable only for \textit{special}  couplings $(\rho_{ij}, k_i)$
 that correspond to  
 the  affine Gaudin models \cite{DLMV}. 
These are
selected  as the solutions 
 of certain polynomial equations.
We will focus  on the first  non-trivial case of $N=2$, i.e.\  on  $G\times G$  models.

 As we shall find in  Section \ref{21032},
  the 
classical integrability condition for 
 $G\times G$ theories \rf{cou} 
is 
 automatically stable 
 under the  \textit{2-loop} RG flow 
 in
 a particular  subtraction scheme (extending the 1-loop results of \cite{today}). 
 Here the  2-loop stability is  obtained without the need for any  finite counterterms.

\

The model  \rf{cou}  
is a special case   of the 2d 
 $\s$-model
 \eqref{13},
 whose 2-loop $\b$-functions are given in the ``\GB scheme'' in eq. \eqref{BMT} (see also Section \ref{SD} on subtraction schemes).
This special 2-loop   scheme  
 effectively  treats 
$\GG_{mn}$ and $\BB_{mn}$  as symmetrically as possible, with the respective 
$\beta$-functions   being the symmetric and antisymmetric parts of a single tensor expression \cite{Bos:1987mw,mt}  (see also \cite{Hull:1987pc,Ketov}). We note that in the context of this chapter, the diffeomorphism ($L_{\xi}(\GG+\BB)_{mn}$) and B-gauge transformation ($\del_{[m}\l_{n]}$) terms in \eqref{BMT} may be ignored
 since they  
automatically 
   vanish in the examples  considered below 
 due  to  manifest global $G_L \times G_L$  symmetry.

Applied  to the   case of the PCM$_k$  in  \rf{Pk},  the 2-loop $\b$-function in \GB scheme  \eqref{BMT}  gives (here we set $\a'=1$)\foot{In this chapter we denote $\t=\log\m$ (instead of $t$) since we will use a coupling constant called $t$.}
\be \ddt{\rm h} = \ccg\, (1-\tfrac{k^2}{\rh^2})\, \Big[ 1 +  \tfrac{\ccg}{2\rh} (1-  \tfrac{3\, k^2}{\rh^2}) \Big] \ ,  \qquad \qquad   \ddt{k}=0  
 \ , \la{Pkb}\ee
so that  the  position  of the WZW  fixed point $\h= \pm k$ remains unchanged at the 2-loop order. 
The  2-loop PCM$_k$ $\beta$-function \rf{Pkb} was 
found in  \cite{Bos:1987mw}  
using a scheme equivalent this one  at the 2-loop level. 
To recall, part of the scheme freedom
comes from the  prescription of how one treats  the  antisymmetric  2d tensor  $\e^{\m\n}$  appearing  in the $\BB$-term in \eqref{13} 
 in dimensional regularization. Ref.  \cite{Bos:1987mw}  used a  't Hooft-Veltman  prescription of treating $\e^{\m\n}$ as effectively 2-dimensional. 
 In  \cite{mt} it was  assumed  that, in $d=2+\ep$ dimensions,  $\e^{\m\n}\e_{\r\s} = 
  f(\ep) ( \delta^\m_\r \delta^\n_\s - \delta^\m_\s \delta^\n_\r ) $  where   $f= 1 + f_1 \ep +...$,    and then the  \GB scheme  \eqref{BMT}  corresponds to 
the choice  $f_1= -1$.  As noted in \cite{Bap}, the scheme used in \cite{Bos:1987mw}   
is equivalent (at least at the 2-loop level)  to  
$f(d) = \frac{1}{ d-1}= 1 - \ep + ...$, i.e.\  to the choice $f_1= -1$
 \ci{mt}
of the \GB scheme.

The  \GB   scheme    is naturally  ``adapted'' to the  vicinity of the WZW  conformal point: 
the derivative $\del_\rh \b_\rh\big|_{\rh=k}$ of the $\beta$-function for $\rh$ at the fixed point
correctly  reproduces \cite{Bos:1987mw}  the  anomalous  dimension of  the $\Tr (J_+J_-)$ operator (PCM  Lagrangian)
as computed \cite{KZ} using the  underlying  infinite dimensional 
Kac-Moody  symmetry  of the WZW   model.   Thus   this scheme is 
apparently 
 consistent  with the preservation of the KM symmetry  in 
  the vicinity of  
the conformal point. 

  It is then natural to expect that this scheme   should also  play  a special role   in  a more general  class  of 
 integrable  models \rf{cou}   containing   WZW  models as  special    limits,\foot{Similar logic was recently used 
    in \cite{shifman} in the  discussion of the 2-loop RG   evolution  of a   ``squashed''  $SU(2)$  variant  of  
   PCM$_k$.} 
   and should  facilitate  preservation   of the  hidden integrable structure of these models at the quantum level.  
We will indeed  see    evidence for this below:  the 
classical integrability  conditions for the $G\times G$ model  \rf{cou} 
will be automatically  preserved by the 2-loop   RG evolution 
 provided  one uses the $\beta$-functions in the \GB scheme  \eqref{BMT}. 

\

We shall also  study, 
in Section \ref{2103gu}, 
 a  gauged  analog  of the  models \rf{cou}  defined  on a coset space $G\times G/H$.
 This theory,  
  which
 was  recently derived from  affine Gaudin models  in  \cite{ABL},
may be   viewed as a
    generalization  of the standard  $G/H$ symmetric space $\s$-model,   
    also including 
 WZ terms.
    For these $G\times G/H$ theories to be gauge invariant, the  corresponding 
    couplings must satisfy certain linear relations. In addition, for a gauge invariant   model to be  classically integrable,
    the couplings  should  further satisfy certain polynomial relations \cite{ABL}. 
        
    We will 
    compute  the RG flow for these 
     integrable $G\times G/H$ theories, finding that  they  are stable under the 
      1-loop RG flow.  
      However, at the 2-loop level, 
RG stability 
does  not
 automatically   
 arise  and,  in general,   
          requires   certain 
      finite redefinitions of the couplings. These  are  
       equivalent to adding specific  finite counterterms,         
which  may  be  interpreted  as  required for preservation of integrability  
at the quantum level
 (analogous to what was observed  on other   examples in Chapter \ref{QCS}). 
       There are still a few special  cases, in particular the integrable $T^{1,1}$ model of  \cite{ABL},   that are  automatically stable 
       under the 2-loop RG flow (see Section \ref{2103T11}). 

\

\sloppy
In Section \ref{2103sq} we shall present a new  integrable \sm with 
target space  metric  
${T^{1,q} = SU(2) \times SU(2)/U(1)}$ \ci{romans} 
   and a particular $B$-field. The model admits as a special limit 
 the  conformal GMM model  with unequal levels \ci{GMM,PZT}.
Our central observation is that, in the case of the  subgroup $H$ in $G\times G/H$  being 
 abelian, the gauge invariance conditions of  \cite{ABL}  are too restrictive and  there is also a second 
 ``branch'' 
  of gauge invariant theories. This allows a natural generalization of the integrable $T^{1,1}$ model of \cite{ABL}  to $T^{1,q}$ 
  with a general  parameter 
$q$. We demonstrate that the resulting $T^{1,q}$ model is classically integrable, admitting a Lax representation. We  observe 
 that the $T^{1,q}$ model is self-dual  under T-duality in one  isometry direction, and argue that this  property 
forces it to be stable under the RG flow. 
We verify this fact  explicitly  by computing  the corresponding  2-loop RG flow of the two  coupling constants.

 \

A few concluding remarks  will be made in  Section \ref{2103CS}. 
Appendix \ref{2103A}  contains discussion of the integrability conditions for the $G^N$ model \rf{cou} relevant to this chapter. 
 The explicit formulae for the 2-loop $\beta$-functions of the $G\times G$ and $G\times G/H$ models used in this chapter, and their derivations, are presented in  Appendix \ref{2103B}.
  
 \
 
\noindent This chapter is based on the paper \cite{LT2103} and we closely follow the presentation there.

%%%%%%%%%%%%%%%%%%%%%%%%%%%%%%%%%%%%%%%%
\section{\texorpdfstring{$G\times G$}{GxG} models \la{2}}
%%%%%%%%%%%%%%%%%%%%%%%%%%%%%%%%%%%%%%%%

As 
 mentioned  above, the  $G^N$ model \rf{cou} is  classically integrable for \textit{special} values of its couplings $(\rho_{ij}, k_i)$ satisfying certain polynomial relations, which originate from the  affine Gaudin  construction
 \cite{DLMV}. For such values of the couplings the model  admits a  Lax connection of the form
\be
L_+ = \a_i \,  J^{(i)}_+ \ , \qquad\qquad  L_- = \b_i  \, J^{(i)}_- \ , \la{Lan}
\ee
whose flatness condition, $F_{+-}(L) \equiv \del_+ L_- - \del_- L_+ + [L_+, L_-] =0$, is equivalent to the equations of motion
following from \rf{cou}.  Moreover, the affine Gaudin construction 
guarantees 
the existence of  an infinite tower of conserved,  commuting higher-spin charges 
\cite{LMV}, as reviewed above in Section \ref{agm}.

Below  we shall consider  the simplest $N=2$ case of  the $G^N$ model  \rf{cou} for a simple Lie group $G$.  
We shall parametrize the $2\times 2$ matrix $\rho_{ij}$ in \rf{cou} in terms of the 4 components $s,t,u,b$ as follows 
\begin{align} 
\L = -\ha \,\begin{pmatrix}
s & t+b \\ t-b& u
\end{pmatrix}_{ij} \, \Tr[ J^{(i)}_+ J^{(j)}_- ] + k_i  \, \L_{_{\rm WZ}}(g^{(i)}) \ . \la{par}
%\qquad  
%\rho_{ij} \equiv  \rh_{(ij)} + b_{[ij]}
%= \begin{pmatrix}
%s & t \\ t& u
%\end{pmatrix}
%% \ , \quad b_{mn} = 
%+ \begin{pmatrix}
%0& b\\ -b& 0
%\end{pmatrix} \ .
\end{align}
Then the  affine Gaudin condition for   integrability is the vanishing of  a  cubic polynomial \cite{DLMV,today},
\be
f(s,t,u,b ,k_1,k_2) \equiv -t \, (s+t)\, (t+u)+b^2\, (s+t+u)+t \, k_1\,  k_2 + b \, (u\,  k_1 - s\,  k_2) = 0  \la{cond} \ .
\ee

Let us  note that the 
affine Gaudin conditions for integrability 
(e.g.\ \rf{cond} in the $N=2$ case) are certainly sufficient for integrability. However, it is not  a priori  clear  if they are  necessary,  
  since there could also be integrable theories of the form \rf{cou} that are unrelated to the affine Gaudin  construction 
of   \cite{DLMV}.
In Appendix \ref{2103A} we   presented a  check that the condition \rf{cond} is also \textit{necessary} for the integrability
of the $G\times G$ model \rf{par},  assuming the natural ansatz \rf{Lan} for the corresponding  Lax connection.

%%%%%%%%%%%%%%%%%%%%%%%%%%%%%
\subsection{RG flow in   \texorpdfstring{$G\times G$}{GxG} models}

The general $G^N$ model \rf{cou} has  global $(G_L)^N \times G_R$ symmetry acting as
\be
g^{(i)} \to u_L^{(i)} g^{(i)} u_R\ , \qquad\qquad   (u_L^{(i)}, u_R) \in (G_L)^N \times G_R \ . \la{GGs}
\ee
In fact,  \rf{cou} is  the most general 2-derivative  local Lagrangian  having  this symmetry. 
 This implies  that only $\rho_{ij}$ can  run  under the RG flow
(with the WZ parameters $k_i$  not renormalized as usual, since WZ terms do not have a local 2d representation that could be produced by the beta functions \eqref{BMT}).

Starting with the \sm couplings $(\GG_{mn},\BB_{mn})$   corresponding to the   $G \times G$ model 
\rf{par}   and  computing the 
corresponding 2-loop $\beta$-functions   in the \GB  scheme  \eqref{BMT}, 
we find  
\be\la{2c} \begin{aligned}
 & \qquad\qquad \ddt \rho_{ij}  = \a' \, \b_{ij}^{(1)} + \a'^2\,  \b_{ij}^{(2)} + \ldots \ , \\
 \b_{ij}^{(1)} = {}&{} \cgg \, F^{(4)}_{ij}(s,t,u,b ,k_1,k_2) \ ,  \qquad  
  \b_{ij}^{(2)} = {\rm c}_{_G}^2 \, (su-t^2)^{-5} \, F^{(9)}_{ij}(s,t,u,b ,k_1,k_2) \ , 
\end{aligned} \ee
where the matrices $F^{(4)}, F^{(9)}$ are  homogeneous polynomials of degrees 4 and 9 in their arguments 
and $\ccg$  is the dual Coxeter number of  the group $G$, as in  the PCM$_k$ case in \rf{Pkb}.
The   explicit  expressions 
for $F^{(4)}, F^{(9)}$
 are given in  Appendix \ref{2103GGapp}    and also in an example below.

Remarkably, despite the complicated   expressions  for the   $\beta$-functions,  one is able to verify   that 
  the integrability condition \rf{cond} is, in fact, preserved by the 2-loop RG flow:
\be { d f\ov d \tau} \Big|_{f=0} = \big( \a' \, \b_{ij}^{(1)} + \a'^2 \, \b_{ij}^{(2)} + ... \big) \, { \del f \ov \del {\rho_{ij}}  }\Big|_{f=0} = \a'\times 0+ \a'^2\times 0 + ... \ .  \la{GGstab}
\ee
The vanishing of the 1-loop $\O(\a')$ term in \rf{GGstab} was already established in \cite{today}, and the vanishing of the 2-loop 
 term is a  new  non-trivial  result. Let us stress   that 
 this property of 
 the integrability condition \rf{cond}
not being deformed  at the 2-loop level is specific to the 
  \GB   scheme \eqref{BMT}.

\iffa

\subsubsection{\texorpdfstring{$\rho_{21}=0$}{rho_21=0} and the \texorpdfstring{$G\times G$}{GxG} model related to \texorpdfstring{$\l$}{lambda}-model}
The most general integrable model with $\rho_{21}=0$  corresponds to the  following choice  of the parameters in \rf{par}
(this case was also considered in Appendix C of \cite{today})
\be
b=t \ , \qquad u = -k_2 \ .
\ee 
After the redefinition $  (g^{(1)},g^{(2)})\equiv(g,\wt{g}^{-1})$, the corresponding Lagrangian \rf{par}  
depending on $s,t, k_1, k_2$ 
may be written as 
  (cf. \rf{Pk}) 
\begin{align}
\L ={} &\big[ s \L_{_{\rm PCM}}(g) +k_1  \L_{_{\rm WZ}}(g)\big]  -  k_2 \big[  \L_{_{\rm PCM}}(\wt g) + \L_{_{\rm WZ}}(\wt g)
\big]+ t \, \Tr
\big[ J_+(g) K_-(\wt g)\big]  \ , \la{1}
\end{align}
where 
$J= g^{-1} d g,   \  K(\wt g) = d \wt{g}\,  \wt{g}^{-1}$. 
This is   just a  PCM$_{k_1}$    and  WZW model (with level $-k_2$) 
 coupled  via   the $J_+(g) K_-(\wt g)$ term. 
In this case the global $G_R$ symmetry in \rf{GGs} is enlarged to a chiral symmetry  $G_R(\xi^+)$,\foot{We denote by $G(\xi^+)$  right $G$  multiplications 
 depending on  light-cone coordinate $\xi^+= \ha (\xi^0 + \xi^1)$.\la{chir}}
\be\la{29}
(g,\, \wt{g}) \to (u\, g\, v, \ v^{-1} \wt{g}\,  w(\xi^+)) \ , \qquad \qquad (u,v,w(\xi^+))\in G_L \times G_L \times G_R(\xi^+)   \ .
\ee
These   symmetries protect the structure of \rf{1}  under
 renormalization so  that only the parameters $s$ and $t$ are expected to run  with the RG scale.  Indeed,  in this case the RG equations  \rf{2c}  take the following explicit  form\foot{As in \rf{Pkb},   here we set 
the loop counting parameter $\a'$ to  be 1. 
The 1-loop terms in \rf{sr},\rf{tr} match those in \cite{today} (after  reversing the sign  of the WZ terms $k_i \to - k_i$  to match the conventions).}
\begin{align}
\ddt s {}&{}= \frac{\cgg\, (s-k_1)  }{\left(k_2 s+t^2\right)^2} \Big[k_2^2 (k_1+s)+4 k_2 t^2-2 t^3\Big] \no \\
&\quad + \frac{\cggg\, (s-k_1) }{2 \left(k_2 s+t^2\right)^5} \Big[ 2 k_2 t^5 \left(38 t^2-11 k_1t +2 s^2+41 s t\right) \la{sr} \\
&\quad %quad \qquad \qquad
 +2 k_2^3 t^3 \left(-8 k_1 s-42 k_1 t+9 k_1^2-5 s^2+18 s t\right) % \no\\
%&\qquad\qquad \qquad
  -2 k_2^2 t^4 \left(-7 k_1 s-46 k_1 t+s^2+48 s t+28 t^2\right)\no\\
&\quad %\qquad \qquad \qquad 
+k_2^5 (k_1+s) \left(s^2-3 k_1^2\right)+2 k_2^4 t^2 (3 k_1+2 s) (3 s-5 k_1)-4 t^7 (5 s+6 t)\Big]\no  \ , \\
\ddt t{}&{} =\frac{ \cgg\, t (t-k_2)  }{\left(k_2 s+t^2\right)^2} \Big[ k_2 (k_1-s)+2 t (s+t)\Big] \no \\
&\quad + \frac{\cggg\, t (t-k_2)}{2 \left(k_2 s+t^2\right)^5}  \Big[4 t^5 \left(t (4 t-k_1)+5 s^2+10 s t\right)-k_2^4 (s-k_1) \left(s^2-3 k_1^2\right)\la{tr} \\
&\quad %\qquad \qquad \qquad
+2 k_2^2 t^2 \left(s^2 (28 t-3 k_1)+2 s t (13 t-16 k_1)+6 k_1 t (k_1-4 t)+3 s^3\right)\no\\
&\quad %\qquad \qquad \qquad
-2 k_2 t^3 \left(-k_1 t (13 s+19 t)+2 s^3+31 s^2 t+45 s t^2+10 t^3\right) %\no\\
%&\quad %\qquad \qquad \qquad
-2 k_2^3 t (s-k_1) (5 s t-3 k_1 (s+4 t))\Big] \ . \no % \la{tr}
\end{align}
At the obvious  fixed point   $s=k_1, \ t= k_2$, the model  \rf{1}   becomes  \cite{today}  
the sum of two decoupled WZW models,
$\L = (k_1+k_2) \L_{_{\rm WZW}} (g) - k_2 \L_{_{\rm WZW}}(g \wt g)$.
 As discussed in \cite{today}, the fixed points are all decoupled WZW models of this type.
  The RG trajectories either interpolate between such WZW-type fixed points or 
flow to them in the IR from the  asymptotically free  UV fixed  point $s, t \to \infty$.

An  interesting  special case of \rf{1}  is $s=k_1=- k_2\equiv - k' $, when it becomes 
\be
\L =- k'  \Big(  \L_{_{\rm WZW}}(g)  + \L_{_{\rm WZW}}(\wt {g})  - \l' \, \Tr\big[ J_+(g)  K_-(\wt g)\big]  \Big)  \ , \qquad \quad  \l' \equiv k'^{-1}t \ .\la{212}
\ee 
This particular $G\times G$ model  appears from the ``tripled''  version \cite{HLT2} 
of the $\l$-model \ci{Sfetsos:2013wia,Hollowood:2014rla}
after removing the  decoupled WZW part.  It is also 
 a special case of the ``doubly $\l$-deformed'' model of \cite{sfetsos,Georgiou:2017aei,Georgiou:2017oly,Georgiou:2017jfi}. 
 Here  the  $\beta$-functions   \rf{sr},\rf{tr} reduce to just $\l'$ running as 
\be\la{213}
\ddt \l' = \frac{2\cgg\, \l'^2}{k' (1+\l')^2} + \frac{4{\rm c}^2_{_{G}}\,  \l'^4(1-2\l')}{k'^2 (1-\l')(1+\l')^5} \ .
\ee
This is the 2-loop $\beta$-function \cite{HLT2}
for the $\l$-model 
based on  
the group $G$ 
with  parameters $(k, \lambda)$  related to $(k',\l')$ as 
$k' =k+2 \cgg$, $\l' = \tfrac{k}{k+2\cgg}\l^{-1}$.\foot{See also \cite{s19}. For the 1-loop $\b$-functions of the $\l$-models based on $G$ and $G/H$, see \cite{Sfetsos:2014jfa} and \cite{Appadu:2015nfa} respectively.} 

\fi
\subsection{Example with \texorpdfstring{$k_1=k_2=0$}{k_1=k_2=0}  \la{sc}}
Let us consider the special case of the  integrable $G\times G$ models \rf{par},\rf{cond}
 with vanishing WZ levels, $k_1=k_2=0$.
  In that case 
 the integrability condition \rf{cond}  implies that 
\be \la{214}
b = b(s,t,u) \equiv \Big[{ t(t+s)(t+u)\ov t + s+u}\Big]^{1/2} \ . \ee
 We thus obtain 
 from  \rf{par} 
 an integrable  $G \times G$ model  with  3  independent couplings $s,t,u$,
\be
\L = -\tfrac{1}{2} \begin{pmatrix}
s & t+ b(s,t,u) \\
t- b(s,t,u) & u
\end{pmatrix}_{ij} \Tr[ J^{(i)}_+ J^{(j)}_- ]   \ .
\ \la{matb}
\ee
Since $k_i$  do not run, this special case of the model \rf{par},\rf{cond}  should also  be stable under the RG flow, i.e.\ 
\rf{matb}   should be renormalizable  with only $s,t,u$  running. 
 Indeed, 
using for convenience the redefined 
couplings  $(s,t,u)\to (x,y,z)$ with
 \be\la{216}
x = s+t+u \ , \qquad y = \frac{s}{t} \ , \qquad z = \frac{u}{t} \ ,\la{rc}
\ee
 the 2-loop $\beta$-functions \rf{2c}  become 
 \begin{align}
&\ddt x = 2 \cgg - \frac{ \cggg }{2 x (y z-1)^3} \  \Big[ 16 +32 (y + z) + 16 ( y^2 + z^2) + 88 y z + 68 y z (y+z)    \la{br1} \\
 & \qquad \qquad  \qquad \qquad \qquad\qquad + 12 y z (y^2+z^2 + 5 y z )   + 8 y^2 z^2 (y+z)    - y^2 z^2 (y +  z)^2  \Big]\ ,  \no\\
%\no \\
%& \qquad \qquad \qquad\qquad \ \ \  + y^2 (z^4-8 z^3-60z^2-68z -16)-4 y (z+1) (z+4) (3 z+2)\Big]\ ,  \la{br1}\\
&\ddt y =F(x;y,z)\ , \qquad \qquad \ddt z = F(x;z,y) \ , \la{218} \\
&F(x;y,z)\equiv  \frac{  \, y  (y+1) (y+2) }{ \,  (zy-1)^2} \ \bigg(  \frac{ \cgg}{x} \big[ 1-zy- 3(z+1)^2 \big]   \la{br3} \\
& \qquad \ \   %\qquad\qquad \quad \qquad
 -\frac{ \cggg}{2x^2} \frac{1}{ (z y-1)^3 }\Big[  -z^6 y^2- y^6 (3z  y-38z-44) %\no\\
%& \qquad\qquad \qquad \quad \qquad \qquad 
 +2 z (y+1) (26 zy+101y+58) \no \\
&\qquad\qquad \qquad  \qquad  \qquad \ \ \ +20 (y+1)^2 -z^4 (y (3 y ((y-14) y-100)-296)-38) \no\\ 
&\qquad\qquad \qquad  \qquad  \qquad \ \ \ -z^3 (y (y (y ((y-4) y-178)-728)-708)-152) \no\\
& \qquad\qquad \qquad \qquad   \qquad \ \ \  +2 z^2 (y+1) (y (y (5 y+89)+262)+105)  \Big] \bigg)\ .\no 
\end{align}
The obvious symmetry  between $s$ and $u$ in \rf{matb}  is  translated into  the symmetry of the RG equations under 
$y\leftrightarrow z$.

The fact  that 
these 2-loop $\beta$-functions are much simpler than  
 the general (not necessarily integrable) case  of \rf{2c} 
 (see also  
 Appendix  \ref{2103GGapp}) 
 suggests that a   substantial simplification  happens 
 upon specifying the 
 couplings to be  at  the integrable locus $f=0$
 in 
  \rf{cond}
  (this   was already observed at the 1-loop order in \cite{today}).

\section{\texorpdfstring{$G\times G/H$}{GxG/H} models \la{gu}}

Let $H$ be a subgroup  of $G$ such  that $G/H$ is a symmetric space (we assume that both $G$ and $H$ are simple real Lie groups).  Then the 
 Lagrangian for the gauged $G\times G/H$ model  of \cite{ABL} takes the form\foot{Our conventions in 
\rf{couH} are related to the ones of  \cite{ABL}  by  $r_{ij} \to 2 \rho^{(0)}_{ij}$, $\rho_{ij} \to 2 \rho_{ij}^{(1)}$ and
 the  opposite sign  for 
the WZ terms, i.e. $k_i \to - k_i %^{\rm (ABL)}
$.}
\begin{align}
&\L = -\ha \, \rho_{ij} \, \Tr[ P^{(i)}_+ P^{(j)}_- ] -\ha \, r_{ij} \, \Tr[ \B^{(i)}_+ \B^{(j)}_- ] + k_i  \, \L_{_{\rm WZ}}(g^{(i)}) \ , \la{couH} \\
& P^{(i)} = P_{{G/H}}  J^{(i)} \ , \qquad   \B^{(i)} = P_H J^{(i)}\ ,  \ \ \qquad 
 J^{(i)} =  {g^{(i)}}^{-1} d g^{(i)}  \ . % \quad  (g^{(i)}) \in G^N  \ ,
  %\ \  i=1,2 \ ,
  \la{31}
\end{align}
Here  $P_{{G/H}}  $ and $P_H$ are projectors to the corresponding  parts of
 the algebra of $G$,   and 
$\r_{ij}$, $r_{ij}$ are  
 constant $2\times 2$ matrices.  
The global symmetry 
consists of 
 left multiplication $G_L\times G_L$, as well as the discrete  $\mathbb{Z}_2$  corresponding to
 the symmetric space structure of $G/H$.
The  action  for \rf{couH}
 is required  to be gauge invariant under the local  right action  by an element of $H$ (acting 
 the 
 same 
 on  both $g^{(i)}$) 
\be
g^{(i)} \to g^{(i)} \ww\ , \qquad \qquad  \ww(\xi^+, \xi^-)\in H \ .  \la{gI}
\ee
 For general choices of $G$ and $H$,  
   gauge invariance  imposes the 
   linear  
   constraints  \cite{ABL}\foot{The   special case of  abelian $H$   will be   
 discussed below in Section \ref{2103sq}.}
\be
 k_1 = - k_2\equiv k  \ , \qquad \qquad r_{ij} = \begin{pmatrix} r  & - r - k \\ -r +k & r  \end{pmatrix} \ . \la{bb1}
\ee
The remaining free parameters  of the gauge invariant model 
are then $r$, $k$ and  the $2\times 2$ matrix $\rho_{ij}$.

Requiring integrability imposes  further constraints  which,
as for the
$G^N$ models \rf{cou}, 
can be  obtained from the affine Gaudin  construction. 
The following parametrization of  the 6 constants $r,k, \rho_{ij}$ 
 in terms of   4 parameters $K,x,\z_+,\z_-$ was shown  in \cite{ABL} to be   sufficient for   integrability
\begin{align}
&r=r_{11} = r_{22} = K \frac{\z_-^2-\z_+^2}{(1-x^2)^2} \ , \ \ \qquad \qquad  r_{12} = 2K \frac{(1-\z_+^2)(x^2-\z_-^2)}{(1-x^2)^3}\ ,  \la{iH1}  \\
&r_{21} = -2K \frac{(1-\z_-^2)(x^2-\z_+^2)}{(1-x^2)^3} \ , \\
&\r_{11} = K\frac{1-2\z_+^2 +\z_-^2 \z_+^2}{(1-x^2)^2} \ , \ \ \qquad \qquad  \r_{12} = x \, r_{12}= 2 K \frac{x (1-\z_+^2)(x^2-\z_-^2)}{(1-x^2)^3} \ , \\
& \r_{21} = x^{-1} \, r_{21} = -2K \frac{(1-\z_-^2)(x^2-\z_+^2)}{x (1-x^2)^3}  \ , \qquad \qquad \r_{22} = K\frac{x^4-2\z_+^2 x^2 +\z_-^2 \z_+^2}{x^2(1-x^2)^2} \ , \\
&k=k_1 = -k_2 = - K \frac{2x^2 +2\z_-^2\z_+^2-(1+x^2)(\z_-^2+\z_+^2)}{(1-x^2)^3} \ . \la{iH4}
\end{align}
This parametrization  is simply equivalent to the gauge invariance conditions \rf{bb1} 
combined with the   two extra polynomial integrability conditions
\be \begin{aligned} 
f_1 \equiv {} &{} r^2 - k^2 - \r_{12} \r_{21}=0  \ , \\
 f_2 \equiv {}&{}  (r - k)^4  \r_{12} + (r - k)^2 (r -k -  2 \r_{11}) (r -k + 2 \r_{22})  \r_{21}    \\
&{} \qquad \qquad \ \ \ 
 - 2(r-k)   (\r_{11} + \r_{22})  \r_{12} \r_{21}^2 +  (\r_{12} +  \r_{21})\r_{12}  \r_{21}^3  =0   \ . 
\end{aligned}  \la{condH} \ee
Two  simple solutions of  these   conditions are found by setting 
 $r=k$ (i.e. $r_{21}=0$  in \rf{bb1})     and   
either 
    $\r_{21}=0$  or   $\r_{12}=0$.
    
We note that the integrable $G\times G/H$ models satisfying \rf{couH}\rf{bb1},\rf{couH}  were recently found  to be obtainable \cite{Fukushima:2021eni} from 4d Chern-Simons theory \cite{Costello:2019tri}.

\subsection{RG flow in  \texorpdfstring{$G\times G/H$}{GxG/H}  models}

The structure of the gauge invariant $G\times G/H$   action  \rf{couH},\rf{bb1} is protected 
by the  right $H$ gauge symmetry \rf{gI}  and the global $G_L\times G_L $ and  $\mathbb{Z}_2$ symmetry.
This rules out 
all
counterterms    
except those
  corresponding  to   renormalizations 
  of the 6 couplings $r,k,\rho_{ij}$ (of which $k$ is not renormalized as usual). 
Let us  parametrize $\rho_{ij}$ as in \rf{par},
\be\la{rrr}
\rho_{ij} = \begin{pmatrix}
s & t+b \\
t-b & u 
\end{pmatrix} \ .
\ee 
Computing the  $\beta$-functions \eqref{BMT} corresponding to the  \sm couplings $(\GG_{mn},\BB_{mn})$   for the model \rf{couH},\rf{bb1},\rf{rrr}, 
we find for  the 1-loop $\beta$-functions of the 5 running  couplings
\begin{align}
& \ddt \lh_p  = \a' \b_{\lh_p}^{(1)} \ , \qquad \qquad \lh_p\equiv (r,s,t,b,u) \ , \la{1H}  \\
&\b_r^{(1)} = \frac{\ccg- \ch}{(t^2 - s u)^2} \Big( r^2 s^2 - 2 b^2 t^2 - 2 r^2 t^2 + 2 t^4 -  2 b^2 s u - 2 s t^2 u + r^2 u^2 \\
   &\qquad \qquad \qquad \qquad + 4 b s t k + 4 b t u k - s^2 k^2 - 2 t^2 k^2 -  u^2 k^2 \Big) 
    + \ch \big(1-\tfrac{k^2}{r^2}\big) \ , \no \\
&\b_s^{(1)} = \frac{\ccg}{r (t^2 - s u)} \big[ b^2 s - s t^2 + r^2 u + 2 r (t^2 - s u) -  2 b t k + u k^2 \big]  \ , \\
  &\b_t^{(1)}= \frac{\ccg}{r (t^2 - s u)} \big[-b^2 t + b (s + u) k + t (r^2 - s u - k^2) \big] \ , \\
  & \b_b^{(1)} = \frac{\ccg}{r (t^2 - s u)} \big[-b (t^2 + s u) + t (s + u) k \big] \ , \\
  &\b_u^{(1)} =  \frac{\ccg}{r (t^2 - s u)} \big[ r^2 s + b^2 u - t^2 u + 2 r (t^2 - s u) - 2 b t k + s k^2 \big] \ . \la{fH}
\end{align}
 We observe   that the integrability conditions \rf{condH} are 
 stable under the 1-loop RG flow \rf{1H}--\rf{fH},
\be\la{317}
{d  f_a\ov d \tau} \Big|_{f_1=f_2=0} = \a'  \beta_{\lh_p} ^{(1)} {\del f_a\ov\del \lh_p} \Big|_{f_1=f_2=0} + \O(\a'^2)  = 0 + \O(\a'^2) \ , \qquad\qquad  a=1,2 \ .
\ee
However,  it turns out   that (as for   the  examples discussed in Chapter \ref{QCS})
 this property of RG stability does not automatically  extend to the 2-loop order. 
Computing the 2-loop $\beta$-functions  for the model \rf{couH},\rf{bb1} in the \GB scheme \eqref{BMT} (given explicitly in Appendix \ \ref{2103GGHapp}), 
we find that  
 the subleading correction to \rf{317} is non-zero at general values of the couplings, 
\be\la{318}
 \beta_{\lh_p} ^{(2)} {\del f_a\ov\del \lh_p} \Big|_{f_1=f_2=0} \not= 0\ ,  
 \qquad\qquad  a=1,2 \ .
\ee 
Moreover, we checked that 
 \rf{318} is also non-vanishing  in  
arbitrary  covariant  2-loop subtraction schemes.\foot{More precisely, we considered arbitrary subtraction schemes related to the \GB scheme \eqref{BMT} by covariant redefinitions of $\GG_{mn}$ and $\BB_{mn}$ (see Section \ref{SD}).}

As in other examples \cite{HLT1,HLT2}, one may expect to restore the property of  RG stability at the 2-loop order by adding certain finite quantum $\a'$-corrections to the target space geometry.
Because of the global and local symmetries, the only possible corrections would 
correspond to   redefinitions  $\lh_p \to \bar \lh_p$  of the couplings  $\lh_p= (r,s,t,b,u)$,
\be 
 \lh_p=\bar  \lh_p + \a' Q_p(\bar \lh) + \ldots \   \ . \la{319}
 \ee
 Such redefinitions may be interpreted as quantum corrections to the integrability conditions \rf{condH}:
if  the original couplings  $\lh_p$ satisfied
 $f_a(\lh)=0$, then the corrected ones $\bar \lh_p$ would satisfy a \textit{corrected} version of the integrability conditions,
\be
\bar{f}_a(\bar \lh)   = 0\ , \qquad \bar{f_a} \equiv  f_a + \a'  \, Q_p \,  \del_{\lh_p} f_a   + \ldots \ \ .
\ee

\subsection{Example: integrable deformation of GMM model on \texorpdfstring{$G\times G/H$}{GxG/H}   and  \texorpdfstring{$T^{1,1}$}{T11} model \la{T11}}
There are still special exceptional cases of the integrable $G \times G/H$ model \rf{couH},\rf{bb1},\rf{condH}  that are automatically stable under  the  2-loop  RG flow 
in the \GB  scheme. As we shall demonstrate below, one of them is the
 particular solution  of the integrability conditions \rf{condH} that was studied in \cite{ABL}, 
\be \la{3244}
r=k\ , \qquad \qquad \r_{12}=\r_{21}=0  \ , \ \ \ {\rm i.e.} \ \  t=b=0 \ . \ee 
The Lagrangian of the corresponding theory \rf{couH},\rf{bb1} is given by 
\be \begin{aligned}
\L ={}&-\ha \, \Tr \big[ \rh \, P_+ P_-  + \wt{\rh} \,  \widetilde P_+ \widetilde P_- \big] -\ha k \, \Tr \big[ \B_+ \B_-  + \widetilde{\B}_+ \widetilde{\B}_-  -2 \B_+ \widetilde \B_- \big]  \\
&{}\quad + k \, \big[ \L_{_{\rm WZ}} (g)  - \L_{_{\rm WZ}} (\widetilde g) \big]  \la{3c}   \ , 
\end{aligned} \ee
where  we  have  set\foot{Ref. \cite{ABL}  used the notation $(k, \, \rh, \, \wt \rh) \equiv (\l^2, \, \l_2^2,  \, \l_1^2 )$.}
\be\la{3255} \begin{aligned}
  & (g^{(1)}, g^{(2)})\equiv (g,\wt g)\ , \quad  \wt P_\pm =  P_\pm (\wt g)\ , \quad  \wt \B_\pm =  \B_\pm (\wt g)\ , \\
 &  \rh\equiv \rho_{11}=s\ , \quad  \wt \rh\equiv\r_{22}=u \ . \end{aligned} \ee 
This is an integrable deformation of the special point $\rh=\wt \rh = k$ that corresponds to the conformal GMM model \ci{GMM} on the homogeneous space $G\times G/H$ with equal levels.

 Specializing the 1-loop   and 2-loop $\beta$-functions in \rf{1H}--\rf{fH} and   Appendix \ref{2103GGHapp} to this case, 
 we find that the model \rf{3c} is automatically stable  under 2-loop renormalization 
  with only $\rh$ and $\wt \rh$ running,
\be \la{bg} \begin{aligned}
& \ddt \rh = 2 \cgg \big(1- \tfrac{k}{ \rh}\big) \Big( 1 + \tfrac{ 1}{ \rh} \Big[ 2(\cgg-\ch) -  (3\cgg-2\ch) \tfrac{k}{\rh}\Big] \Big)  \ , \qquad \\
&\ddt \wt \rh = 2 \cgg (1- \tfrac{k}{ \wt \rh}\big)  \Big( 1 + \tfrac{ 1}{ \wt \rh} \Big[ 2(\cgg-\ch) -  (3\cgg-2\ch) \tfrac{k}{\wt \rh}\Big]\Big)  \ , \quad \qquad
\ddt k =0 \ .
\end{aligned} \ee
Remarkably, the RG evolution of $\rh$ and $\wh$  is  decoupled.  Note that the  structure of their   $\beta$-functions  is  similar to the  one in the PCM$_k$ case   \rf{Pkb}. 
As expected, the GMM model $\rh=\wh=k$ is a fixed point. 

 Let us consider  the  simplest  example of this theory  \rf{3c}   with  $G= SU(2) $ and $H=U(1)$  and   
choose the parametrization 
\be g=e^{\p_1 T_1} e^{\theta_1 T_2} e^{\psi T_3} \ , \qquad \qquad   \wt g=e^{-\p_2 T_1} e^{-\theta_2 T_2} e^{-\wt \psi T_3} 
 \ , \la{coo} \ee
 where the $SU(2)$ generators are 
$T_A = \tfrac{i}{2} \s_A$   
 and  the generator of  $H=U(1)$ 
  is  $T_3$. We shall fix the $H$ gauge freedom   by setting $\wt \psi=0$. As a result,   we get an integrable 
   5-dimensional \sm  (cf. \eqref{13})
  \begin{align}
\L ={} & (G_{mn} + B_{mn})  \del_+ x^m \del_- x^n 
=\tfrac{1}{4} k \, \Big[ \del_+ \psi \del_- \psi + \cos^2{\th_1} \, \del_+\p_1 \del_- \p_1 + \cos^2{\th_2} \, \del_+\p_2 \del_- \p_2 \no \\
&\qquad +  2  \cos{\th_1}\,  \del_+ \p_1 \del_- \psi + 2   \cos{\th_2}\,  \del_+\psi \del_- \p_2 + 2 \cos{\th_1}\cos{\th_2} \,  \del_+ \p_1 \del_- \p_2 \Big]   \la{mod}  \\
&\qquad  + \tfrac{1}{4} \rh\,  \big[ \del_+ \th_1 \del_- \th_1 + \sin^2{\th_1}\,  \del_+ \p_1 \del_- \p_1 \big]+ \tfrac{1}{4} \wh \, \big[ \del_+ \th_2 \del_- \th_2 + \sin^2{\th_2} \, \del_+ \p_2 \del_- \p_2  \big] \ . \no
  \end{align}
  The resulting  target space  geometry 
corresponds to 
 the  $T^{1,1}$  metric  
     and a  particular $B$-field \cite{ABL}\foot{Due to 
   differing conventions, the $B$-field here  is opposite in sign to that  in  \cite{ABL}.
   This difference is not significant, and can  be removed by a parity transformation.}
 \begin{align}
ds^2_{T^{1,1}} ={} & G_{mn} dx^m dx^n  = \tfrac{1}{4} k \,  (d\psi + \cos{\th_1} \, d\phi_1 + \cos{\th_2} \, d\phi_2)^2  \la{t11}  \\
 &\qquad \qquad \qquad\ \ + \tfrac{1}{4} \rh \, (d \th_1^2 + \sin^2{\th_1} \, d \p_1^2)+ \tfrac{1}{4} \wh \,  ( d  \th_2^2 + \sin^2{\th_2} \, d \p_2^2 )  \ ,  \no\\
 B ={}& \ha B_{mn} dx^m  \wedge dx^n 
  = \tfrac{1}{4} k \, (d\psi + \cos{\th_1} \, d\phi_1) \wedge (d\psi + \cos{\th_2} \, d\phi_2) \ . \la{b11}
\end{align}
The 3 parameters $\rh,\wt \rh, k$ of \rf{3c} are thus mapped to the 3 parameters of the $T^{1,1}$ metric  in    \cite{romans}. To recall, the $T^{1,1}$   metric 
\rf{t11} 
is
 an Einstein space 
 if $\rh=\wt \rh= {3\ov 2} k $. 
 It then serves  as a  base of a  Ricci flat 6d conifold  with metric $dr^2 + r^2 ds^2_{T^{1,1}}$ if we formally 
set 
$k=\tfrac{1}{9}$
 so that  $R_{ij} = 4 g_{ij}$. 
 In general, the non-zero components  of the  Ricci tensor of  the
 cone geometry 
 $ds^2= G_{mn}(X) dX^m dX^n= dr^2 + r^2  g_{ij}(x) dx^i dx^j$ (with $i=1, ..., d$) 
 are  ${R_{ij} (G)  = R_{ij} (g) -  (d-1) g_{ij}}$. Thus  it vanishes if $g_{ij}$ is an Einstein metric with a particular value of the scalar curvature 
 $R(g) = d (d-1)$ (this condition  is satisfied, e.g., for a unit-radius  sphere $S^d$  when  $G_{mn}$ is flat).

For the integrable $T^{1,1}$ model \rf{mod}, the 2-loop  RG equations  \rf{bg} become (putting $ \cgg=2, \ \ch=0$)
\be \la{bet} \begin{aligned}
 \ddt \rh = 4 \big(1- \tfrac{k}{ \rh}\big) \big[ 1 + \tfrac{ 4}{ \rh} \big( 1 -  \tfrac{3\, k}{2\,  \rh}\big) \big]  \ , \qquad \qquad   \ 
&\ddt \wt h = 4 \big(1- \tfrac{k}{ \wt \rh}\big) \big[ 1 +  \tfrac{4 }{ \wt \rh} \big( 1 -  \tfrac{3\, k}{ 2\, \wt \rh}\big) \big]  \ .
\end{aligned} \ee
As we shall discuss in Section \ref{2103T}, the 2-loop RG stability of this  $T^{1,1}$ model  may be  understood 
 as  consequence 
of  the fact 
 that  the \sm \rf{mod} is self-dual  under T-duality  in the $\psi$-direction.

%%%%%%%%%%%%%%%%%%%

\section{Integrable \texorpdfstring{$T^{1,q}$}{T1q}  model \la{sq}}

Let us  now  introduce a new  integrable \sm  with  
target space metric 
 $T^{1,q}$
  and a particular $B$-field,  which is a
  one-parameter generalization of  the $T^{1,1}$ model \rf{mod} of \cite{ABL}. 
  Its   special  conformal  case will be the $SU(2) \times SU(2)/U(1)$  GMM  model, 
  now 
  with \textit{unequal}  
   WZ levels \ci{GMM,PZT} (with their ratio related to
  the parameter  $q$). 

Our central observation is that,  starting with the $G \times G/H$ model \rf{couH}  and considering 
the case   when  the subgroup $H$ is {\it abelian},  the gauge invariance condition \rf{bb1} of \cite{ABL}    is too restrictive.
 At the particular point $\rho_{12}=\rho_{21}=0$, there is also a second 
 ``branch'' 
 of gauge invariant  models,
\be
\begin{aligned}
&\L = -\ha \, \rho_{ij} \, \Tr[ P^{(i)}_+ P^{(j)}_- ] -\ha \, r_{ij} \, \Tr[ \B^{(i)}_+ \B^{(j)}_- ] + k_i  \, \L_{_{\rm WZ}}(g^{(i)}) \ ,
 \\
 &\rho_{12}=\rho_{21}=0 \ , \qquad r_{ij} = \begin{pmatrix} r  & q( - r - k_1) \\ q(-r +k_1) & q^2 r  \end{pmatrix} \  , \qquad q^2\equiv -k_2/k_1  \ , 
\end{aligned}
\la{bb2}
\ee
where $k_1$, $k_2$   are assumed to  be  of opposite sign. 
 The theory \rf{bb2} 
 is  invariant under a modified  gauge transformation
\be\la{42} 
(g^{(1)}, \, g^{(2)}) \to (g^{(1)} \ww^q , \,  g^{(2)} \ww) \ , \qquad \qquad  \ww=\ww(\xi^+, \xi^-)\in H \ .
\ee
Here $\ww^q$ is the  $q$-th power of the abelian  group element $\ww$.  
 In the case  when   the  abelian $H$  is compact
 then, 
to make   $\ww^q$   single-valued,  one  should  
 assume   that 
$q= \sqrt{ |{k_2\ov k_1}|}$ is an {integer}.\foot{More generally,  one could consider   a ``twisted'' action of the abelian subgroup, $(g^{(1)}, \, g^{(2)}) \to (g^{(1)} \ww^q , \,  g^{(2)} \ww^p)$  
characterized by integers $p,q$ satisfying $q^2/p^2=-k_2/k_1$.
In the $SU(2) \times SU(2)/U(1)$   example discussed below,   that would  lead to the  $T^{p,q}$  model.}
The reason for the 
  restriction of  $H$ to be abelian if $q\neq 1$ is that the  variation of the Lagrangian \rf{couH}  under \rf{42}  with 
 $\ww\in H$ will  be proportional to $(q-1)\L_{_{\rm WZ}}(\ww)$, which vanishes for abelian $H$   for any $q$. We also need to assume $\rho_{12} = \rho_{21}$ to prevent mixing between $P^{(1)}$ and $P^{(2)}$ terms, which are rotated differently  by $\ww^q$ and $\ww$ respectively.
At the value $q=1$  (i.e. $k_1=-k_2$),  this model 
 intersects with 
 the first ``branch'' of gauge-invariant $G\times G/H$ theories \rf{couH},\rf{bb1} considered above.
 
 Note that, in generic cases, WZ terms present a topological obstruction to gauging \cite{HS}. There are, however, special ``anomaly-free'' subgroups of the  WZ term's global symmetry $G_L\times G_R$ that can be gauged \cite{witten}, satisfying $\Tr_L[T_A T_B] - \Tr_R[T_A T_B] = 0$. This condition is satisfied here by the gauge transformations \rf{gI} and  \rf{42} on both ``branches'' of theories, due to cancellation between the two copies of $G$ in $G\times G/H$.
 
We 
claim 
that  model \rf{bb2} is integrable (admitting a Lax representation) 
 if\,\foot{The case $r=-k$ is also integrable since it is related to \rf{rk} by parity.}
 \be r=k  \ . \la{rk}
   \ee
 In this case it becomes a 
generalization of \rf{3c} to  
 the case of  
 unequal  levels $k,\wt k$,
\begin{align}
\hspace{-0.2cm}\L ={}&{}-\ha \, \Tr \big[ \rh \, P_+ P_-  + \wh \,  \widetilde P_+ \widetilde P_- \big]
-\ha  \, \Tr \big[ k\,  \B_+ \B_-  +  \wt k \, \widetilde{\B}_+ \widetilde{\B}_-  -2  \sqrt{k \wt k} \, \B_+ \widetilde \B_- \big]  \la{4c}  \\
&\quad  + k \L_{_{\rm WZ}} (g)   - \wt k  \L_{_{\rm WZ}} (\widetilde g)  \  ,\no \end{align}
where we have set (cf.  \rf{3255})
  \be \la{455}
    \begin{aligned}  &
    (g^{(1)}, g^{(2)})\equiv (g,\wt g)\ , \qquad  \wt P_\pm =  P_\pm (\wt g), \qquad   \wt \B_\pm =  \B_\pm 
  (\wt g),\ \\
& \rh\equiv \r_{11}, \qquad \wh\equiv \r_{22} \ , \qquad  k\equiv k_1, \quad \wt k \equiv - k_2 \ , \qquad 
q= \sqrt{ \tfrac{\wt k}{k}} \ .
 \end{aligned}\ee
 
 The fact that the $\wt k =k$ limit  \rf{3c}  is an integrable theory provides a first check of the integrability of \rf{4c}:
its Lax connection was obtained \cite{ABL} by taking the limit $r=k$, $\r_{12}=\r_{21}=0$ of the affine Gaudin Lax connection for the general  integrable $G\times G/H$ model \rf{couH},\rf{bb1},\rf{condH}.
 It was found   that certain components of the $\Lie(G)$-valued Lax connection degenerate to zero and thus the flatness condition  of the resulting 
 connection $L_\pm$ does not imply  all of the equations of motion.  
 However, one can  consider a generalized limiting procedure 
   by  infinitely rescaling  the spectral parameter while taking this limit, thus 
   obtaining a second  connection $\wt L_\pm$ that ``misses''  a different  subset of equations of motion. 
   The  flatness conditions of the two Lax connections   together  encode the full set of the equations
  of motion \cite{ABL},
  \begin{align}
&L_+(z) = \B_+ + z^{-1} P_+ \ ,\qquad  \ \ \  \ L_- (z) = \frac{1}{k z^2 - \rh} \Big[ (k-\rh)(\B_- + z P_-) + k (z^2-1) \wt \B_- \Big] \ ,\la{44}\\
&\wt L_-(z) = \wt \B_- + z \wt P_- \ , \quad  \ \ \ \,
\wt L_+(z) = \frac{1}{k z^{-2} - \wh} \Big[ (k-\wh)(\wt \B_+ + z^{-1} \wt P_+) + k (z^{-2}-1) \B_+ \Big]  \ .\la{45}
\end{align}
  The fact of having two separate Lax connections  may  seem   unusual but should be sufficient for the 
   integrability applications: for example, each Lax connection  will lead to its own family of conserved charges.
  They may also be regarded as a single   block-diagonal connection valued in $\Lie(G) \oplus \Lie(G)$. Indeed for $k=0$  the  two Lax connections \rf{44},\rf{45}  become the  familiar ones \eqref{sll}
    of the two decoupled  $G/H$ 
$\sigma$-models.\foot{I thank B. Hoare for this comment.}

We observe that the Lax connections \rf{44},\rf{45} naturally  generalize to the new model  \rf{4c} with unequal levels by replacing certain  factors of $k 
\to \wt k$,
\begin{align}
&L_+(z) = \B_+ + z^{-1} P_+ \ , \la{Lf}
\qquad   L_- (z) = \frac{1}{k z^2 - \rh} \Big[ (k-\rh)(\B_- + z P_-) +  \sqrt{k\wt k}(z^2-1) \wt \B_- \Big] \ ,\\
&\wt L_-(z) = \wt \B_- + z \wt P_- \ , \la{Ll} \ \ \,   
\wt L_+(z) = \frac{1}{\wt k z^{-2} - \wh} \Big[ (\wt k-\wh)(\wt \B_+ + z^{-1} \wt P_+) + \sqrt{k\wt k} (z^{-2}-1) \B_+ \Big]  \ .
\end{align}

 Assuming the  simplest case  $G=SU(2)$, $H=U(1)$, 
 using the same  coordinate parametrization of this $SU(2) \times SU(2)/U(1)$ model as in \rf{coo}, 
and  fixing again the  $H=U(1)$  gauge  as  $\wt \psi=0$,  we find the following generalization of \rf{mod}
 \begin{align}
\L ={} & (G_{mn} + B_{mn})  \del_+ x^m \del_- x^n 
=\tfrac{1}{4} k \, \Big[ \del_+ \psi \del_- \psi + \cos^2{\th_1} \, \del_+\p_1 \del_- \p_1 + q^2 \cos^2{\th_2} \, \del_+\p_2 \del_- \p_2 \no \\
&\qquad +  2  \cos{\th_1}\,  \del_+ \p_1 \del_- \psi + 2q   \cos{\th_2}\,  \del_+\psi \del_- \p_2 + 2q \cos{\th_1}\cos{\th_2} \,  \del_+ \p_1 \del_- \p_2 \Big]  \la{modq} \\
&+ \tfrac{1}{4} \rh\,  \big[ \del_+ \th_1 \del_- \th_1 + \sin^2{\th_1}\,  \del_+ \p_1 \del_- \p_1 \big]+ \tfrac{1}{4} \wh \, \big[ \del_+ \th_2 \del_- \th_2 + \sin^2{\th_2} \, \del_+ \p_2 \del_- \p_2  \big] \ , \qquad  q=\sqrt{\tfrac{\wt k}{k}} \  . \no 
  \end{align}
The resulting  target space metric is that of the $T^{1,q}$  space \cite{romans}
 and the $B$-field is a natural generalization of the one in \rf{b11},
\begin{align}
ds^2_{T^{1,q}} ={} & G_{mn} dx^m dx^n  = \tfrac{1}{4} k \,  (d\psi + \cos{\th_1} \, d\phi_1 +q \cos{\th_2} \, d\phi_2)^2  \la{gq}
\\
 &\qquad \qquad \qquad
 + \tfrac{1}{4} \rh \, (d \th_1^2 + \sin^2{\th_1} \, d \p_1^2)+ \tfrac{1}{4} \wh \,  ( d \th_2^2 + \sin^2{\th_2} \, d \p_2^2 )  \ , \no \\
 B ={}& \ha B_{mn} dx^m  \wedge dx^n 
  = \tfrac{1}{4} k \, (d\psi + \cos{\th_1} \, d\phi_1) \wedge (d\psi + q\cos{\th_2} \, d\phi_2) \ . \la{bq}
\end{align}
 Like the $q=1$   case \cite{ABL}  in \rf{mod}, 
       the   presence of  the $B$-field  is  crucial here for   integrability
(the $T^{1,q}$ \sm  without $B$-field  is not  integrable \cite{bpz,Basu:2011fw}).
The   coordinate form of the Lax connections \rf{Lf},\rf{Ll} is  ($T_A$ 
 are the $SU(2)$ generators  in \rf{coo}) 
\begin{align}
&{L'}_+(z) = \cos{\theta_1} \, \del_+ \phi_1 \, T_3 + z^{-1} \big( \del_+ \th_1  \, T_2 + \sin \th_1 \, \del_+ \p_1  \, T_1 \big) \ , \\
&{L'}_- (z) = \tfrac{1}{k z^2 - \rh} \Big[ (k-\rh)\big( \cos \th_1  \, \del_- \p_1  \, T_3 + z  ( \del_- \th_1  \, T_2 + \sin \th_1 \, \del_- \p_1  \, T_1 )   \big) \no  \\
&\qquad \qquad \qquad \qquad \qquad 
 - (z^2-1) \big( \sqrt{k \wt k}   \, \cos \th_2 \, \del_- \p_2 + k \,  \del_- \psi \big)  \Big]    \ ,  \no\\
 \ \no \\ 
&\wt L_+(z) = \tfrac{1}{\wt k z^{-2} - \wt h} \Big[ (\wt k-\wh)\big( - \cos \th_2  \, \del_- \p_2  \, T_3 + z^{-1}  ( - \del_+ \th_2  \, T_2 + \sin \th_3 \, \del_+ \p_3  \, T_1 )   \big)  \\
&\qquad \qquad \qquad \qquad \qquad \  + \sqrt{k\wt k} (z^{-2}-1)\big( \cos \th_1 \, \del_+ \p_1 + \del_+ \psi   \big)\Big]  \ , \no\\
&\wt L_-(z) =  - \cos{\theta_2} \, \del_- \phi_2 \, T_3 + z \big( - \del_- \th_2  \, T_2 + \sin \th_2 \, \del_- \p_2  \, T_1 \big)  \ .\no
\end{align}
To simplify the expressions   we followed   \cite{ABL}  here in   replacing $L_\pm $ by its gauge transformed   version 
 ${L'}_\pm = \ww^{-1} L_\pm \ww + \ww^{-1} \del_\pm \ww$,   with $\ww=\exp (- \psi T_3)$. 

At the special point 
 $\rh=k, \ \wh=\wt k= q^2 k $,  the model  \rf{modq} 
 becomes the $SU(2)\times SU(2)/U(1)$  case of the  conformal GMM   model  with   levels $k_1=k, \ k_2 =-\wt k$. 
 It was  pointed out in \cite{PZT} that the $SU(2)\times SU(2)/U(1)$  GMM model corresponds to the 
 $T^{1,q}$ 
  metric  and 
  a particular $B$-field,   and its 2-loop conformality was explicitly  checked (see also \ci{belo}). 
  The  general GMM model has a  current algebra  symmetry \cite{GMM} and is  also  integrable in the    Lax connection   sense  \cite{Bardakci:1996gs}.
 What we have shown above  is that 
  it   admits an  integrable extension \rf{modq} 
    away from the conformal point $\rh=k, \ \wh=\wt k$.

\subsection{Stability under the 2-loop RG flow}

Let us   now show   that the integrable  
$T^{1,q}$ model \rf{modq}   is  stable under the 2-loop RG flow. 

The general gauge invariant model \rf{bb2} (with the  $r=k$ 
condition   \rf{rk} relaxed) must be 
stable  under the RG   with  only $(r,\rh,\wt\rh)$   running,  due to  its $H$ gauge invariance and  global $G_L \times G_L$ symmetry. Even at the point $q=1$ or $\wt k = k$ where the two 
 ``branches'' 
 of gauge invariant theories \rf{bb2},\rf{bb1} intersect, the RG flow cannot `turn on' the extra couplings  
 $\r_{12}$, $\r_{21}$ since this is forbidden by an extra global ``center'' symmetry  \cite{romans, ABL},
$ (g,\, \widetilde g) \to (g\, z ,\, \widetilde g) , \   z\in 
{\rm Z}(H) 
 \subset H$  when $H$ is abelian.\foot{Note, however, that  the center symmetry alone is not sufficient to explain the stability of the $T^{1,q}$ model since it does not prevent $r$ from running.}

 We shall see that the integrable $T^{1,q}$ model, obtained by fixing $r=k$, is a ``fixed line'' of the RG flow.
Relaxing  $r=k$  has the effect of replacing $k\to r$ in the metric,\foot{It was recently argued by observing chaotic behaviour \cite{Ishii:2021asw} that relaxing the $r= k$ condition indeed destroys classical integrability, as  least in the equal level case $q=1$ (i.e.\ $\wt k =k$). This implies that, with $q=1$, the fixed line $r=k$ and its parity conjugate $r=-k$ are the only integrable models among the  $\s$-models on the class of backgrounds \rf{bQ2}.}
 with $k$ still appearing in the $B$-field (cf. \rf{gq},\rf{bq}),\foot{Rescaling $r\to  r' k  $  and $\psi\to { 1\ov \sqrt k   } \psi'$   the  background \rf{bQ2}  can be put into the  form 
    symmetric under 
   $k\leftrightarrow \wt k, \ \rh \leftrightarrow  \wt \rh$:
   
   $\qquad  \ ds^2  = \tfrac{1}{4} r' \,  (d\psi' + \sqrt k \cos{\th_1} \, d\phi_1 + \sqrt {\wt k}  \cos{\th_2} \, d\phi_2)^2 + %\la{gQ2}  % \\
%&\quad +
 \tfrac{1}{4} \rh \, (d \th_1^2 + \sin^2{\th_1} \, d \p_1^2)+ \tfrac{1}{4} \wh\,  (  d\th_2^2 + \sin^2{\th_2} \, d \p_2^2 )  \ , $
 \ \ 
 
 $
\qquad \ \ \    B = \tfrac{1}{4}  \, (d\psi' + \sqrt k \cos{\th_1} \, d\phi_1) \wedge (d\psi' + \sqrt {\wt k}  \cos{\th_2} \, d\phi_2) \ . $
   }
\begin{align}
ds^2  ={}& \tfrac{1}{4} r \,  (d\psi + \cos{\th_1} \, d\phi_1 + q \cos{\th_2} \, d\phi_2)^2 + %\la{gQ2}  % \\
%&\quad +
 \tfrac{1}{4} \rh \, (d \th_1^2 + \sin^2{\th_1} \, d \p_1^2)+ \tfrac{1}{4} \wh\,  (  d\th_2^2 + \sin^2{\th_2} \, d \p_2^2 )  \ \no ,\\
  \la{bQ2} B ={}& \tfrac{1}{4} k \, (d\psi + \cos{\th_1} \, d\phi_1) \wedge (d\psi + q \cos{\th_2} \, d\phi_2)  \ , \qquad \ \  q= \sqrt {\tfrac{\wt k}{ k} }\ .  
\end{align}
   The corresponding 
      2-loop  $\beta$-functions in the \GB scheme   following from \eqref{BMT}  are 
     ($k, \wt k$ do not run)
  \begin{align}
%&\ddt k =0 \ , \\
&\ddt r = 2(r^2-k^2) \Big[  \tfrac{1}{\rh^2 }  + 
\tfrac{\wt k}{k} \tfrac{1}{{\wt \rh}^2 }  + %\frac{1}{  \rh^2 {\wt \rh}^2 k r} 
  \tfrac{r^2-3 k^2}{ r } \big(\,  \tfrac{1}{\rh^4}  +   \tfrac{{\wt k}^2}{k^2} \tfrac{ 1}{\wt \rh^4}
  \big)\Big] \ ,\la{b3}\\
&\ddt \rh = 4\big[ 1 - \tfrac{r}{2\rh}(1+\tfrac{k^2}{r^2}) \big] %\no \\
%&\qquad \qquad 
+ \tfrac{2}{ \rh^3   r^2} \Big[   8 \rh^2 r^2     - 8 \rh k^2r  
  - 12  \rh r^3  + 4 r^2  k^2 +    3 k^4  + 5 r^4  \no  \\
 &\qquad \qquad\qquad \qquad \qquad \qquad \qquad \qquad  \qquad +   \tfrac{\wt k}{k}
  \tfrac{{\rh}^2}{\wt \rh^2} (r^2-k^2) (3r^2- k^2)   \Big] \ ,  \la{b4}  \\
&\ddt {\wt\rh} =  4\big[ 1- \tfrac{ \wt k}{k} \tfrac{ r }{{2\wt \rh  }}(1+\tfrac{k^2}{r^2})\big]  %\no  \\
%&\qquad \qquad
 +\tfrac{2 }{  {\wt \rh}^3  r^2}\Big[8  {\wt \rh}^2 r^2 -8 {\wt \rh}  {\wt k}k r -12 \tfrac{\wt k}{k}  {\wt \rh} r^3 
  + 4 r^2 {\wt k}^2  
 +3 {\wt k}^2 k^2+5 \tfrac{{\wt k}^2}{k^2} r^4   \no\\
&\qquad \qquad\qquad \qquad \qquad \qquad \qquad \qquad  \qquad +
  \tfrac{\wt k}{k} \tfrac{{\wt \rh}^2}{\rh^2}  (r^2-k^2) (3r^2- k^2) \Big] \ . \la{b5}
\end{align}
Thus   
$r=\pm k$  are fixed lines 
 of 
 \rf{b3}, 
both at 1-loop and 2-loop order. 
 The  couplings $(r,\rh,{\wt \rh})$ grow   linearly with $\tau\to \infty $ in the UV (reflecting asymptotic  freedom), 
 while they decrease to the GMM fixed point $(r,\rh,{\wt \rh}) = (k,k,\wt k)$ in the IR. In fact, as  argued in \cite{GMM}, the GMM model on $G\times G'/H$ is an exact CFT, 
assuming at least one of  the  cosets  $G/H$ or $G'/H$ is a symmetric space (as is indeed the case for 
 $SU(2) \times SU(2)/U(1)$).

Specialising to the fixed line $r=k$  of \rf{b3},  
 the expressions \rf{b4},\rf{b5}  simplify, giving  the 
 2-loop  $\beta$-functions  of  the  integrable $T^{1,q}$ model \rf{4c},\rf{modq},
\begin{align}
& 
\ddt \rh = 4 \big(1- \tfrac{k}{ \rh}\big) \Big[ 1 + \tfrac{ 4}{ \rh} \big( 1 -  \tfrac{3\,k}{ 2\,\rh}\big) \Big]  \ ,  %\no \\
 \qquad \qquad 
 \ddt \wt h = 4 \big(1- \tfrac{\wt k}{ \wt \rh}\big) \Big[ 1 +  \tfrac{4 }{ \wt \rh} \big( 1 - \tfrac{3\, \wt k}{ 2\, \wt \rh}\big) \Big]  
%\ddt k = \ddt \wt k  =0
\ .  \la{betq} 
\end{align}
These   are   a  natural generalization of the  $\beta$-functions  for the $T^{1,1}$ model     in \rf{bet} to the case of $\wt k \not= k$.
 Like  in \rf{bet}, the  RG evolution of $\rh$ and $\wh$  happens to be  decoupled (while
 this is  not the case for $r\not=k$
 in \rf{b4},\rf{b5}).

We note that,  in addition to  the $r=k$  case of the $T^{1,q}$ model, 
  the $\s$-model   corresponding to    \rf{bQ2} admits  \textit{another} integrable limit,
 $\wt k = 0$. 
 In this case it  factorizes  into 
 a squashed $S^3$   with  WZ  term    and a round $S^2$.
   The $\b$-functions
   \rf{b3} and \rf{b4} 
   become those of this
   squashed $S^3$  WZ-model  
 in
 \ci{shifman} (for 1-loop $\beta$-functions see   \ci{yosh,Demulder:2017zhz})\foot{The relation to the 
   notation used in  \ci{shifman} is 
   $\eta= k \rh^{-1}, \ \lambda^2= {2\pi}  \rh ^{-1} $, \ $ \kappa= 1 - r \rh^{-1} $. } and the 
    $\beta$-function \rf{b5} for  the  coefficient $\wt \rh$ becomes that of  the $S^2={SU(2) \ov U(1)}$
 $\s$-model (eq.\ \eqref{sb} with $\cgg=2, \ \ch=0$).
 Taking  further limits, the $\beta$-functions   \rf{b3},\rf{b4},\rf{b5} agree with  
 other  previously known expressions:
\begin{enumerate}[(i)]
\item   Setting  $\wt k=0$   
     and $r=\rh$, we get from \rf{bQ2}    
       the direct  sum of the  PCM$_k$  (round $S^3$ with a WZ term) 
 and the $S^2$  $\s$-model.  In this  case \rf{b3},\rf{b4} are  indeed equivalent to the $\b$-function 
 of PCM$_k$, i.e.  \rf{Pkb}  with $\cgg=2$. 
 
\item Setting  $\wt k = 0$ (i.e.\  $q=0$) 
   and  then   $k=0$, we 
   instead 
   get the  direct sum of a squashed $S^3$ (with no WZ term) and a round $S^2$.
    The  $\beta$-functions  for $r$ and $\rh$ agree with those of the   ``squashed'' PCM in  \cite{HLT2}  
  (with $G=SU(2)$ 
   and the  ``squashing'' parameter $\varepsilon= {r\ov \rh}$):
    
 \noindent
 $
 \qquad \ddt r = {2r^2 }{\rh^{-2} }    +  {r^3}{\rh^{-4}}  \ , \qquad
 \ddt \rh = 4\big( 1 - \tfrac{1}{2} r \rh^{-1} \big) + {2}{ \rh^{-3 } } \big(   8 \rh^2      
  - 12  \rh r   + 5 r^2 \big)  \ .
%\ddt {\wt\rh} =  4 + 16 \wt h^{-1}
$ 

\end{enumerate}
        
\subsection{Covariance under T-duality \la{T}}
One can argue that  the RG stability of the  integrable  $T^{1,q}$ model \rf{modq}, i.e.\ the presence of the fixed line $r=k$ of \rf{b3},   is related  to
its property of being  self-dual under T-duality in the isometric $\psi$-direction.
To see this, let us write the  Lagrangian \rf{modq}  
in the following form\foot{Note that  $ \widehat{\L}$
   becomes   simply quadratic in the fields 
 at the  GMM point $\rh=k,\  \wh= \wt k= q^2  k$. }
\begin{align}
&\L = \tfrac{1}{4} k \, \Big[ (\del_+ \psi +\ta_+) ( \del_- \psi + \tb_- ) - \ha \ta_+ \tb_-\Big]  + \widehat{\L}  \ , \la{413} \\
&\qquad \qquad  \ta_\pm \equiv   2\cos{\th_1}\, \del_\pm \phi_1\ , \qquad  \qquad   \tb_\pm\equiv   2 q \cos{\th_2}\, \del_\pm \phi_2\ ,   \la{notU}\\
&\widehat{\L} 
\equiv   
 \tfrac{1}{4}\rh  \Big[   \del_+ \th_1 \del_- \th_1 + (  \sin^2{\th_1}  + \tfrac{k }{\rh}  \cos^2{\th_1} )   \del_+ \p_1 \del_- \p_1 \Big]  
% +    \tfrac{1}{4}k \, \cos^2{\th_1} \, \del_+\p_1 \del_- \p_1 
 \la{mmm} \\ 
 &\qquad 
+  \tfrac{1}{4}  \wh  \Big[  \del_+ \th_2 \del_- \th_2 +  (  \sin^2{\th_2}   
+ \tfrac{k q^2}{\wh}  \cos^2{\th_2} )\, \del_+\p_2 \del_- \p_2 \Big]  \no\ . 
% \la{mmm}
  \end{align}
\sloppy
Starting from the interpolating Lagrangian (obtained by $\del_\pm \psi \to A_\pm$   and adding the condition ${\del_+ A_-  - \del_- A_+=0}$  with a  Lagrange multiplier $\bpsi$), 
\be
\L_{\rm int} =  \tfrac{1}{4} k \Big[ (A_+ + \ta_+) (A_- + \tb_- ) - \ha \ta_+ \tb_- - {\bpsi}(\del_+ A_-  - \del_- A_+) \Big] +  \widehat{\L}   \ , \la{int}
\ee
 and integrating out $A_\pm$, we obtain the  following  T-dual  Lagrangian 
\be\la{417}
\bar{\L} = \tfrac{1}{4}  k\Big[ (\del_+ \bpsi + \ta_+) ( \del_- \bpsi - \tb_- ) + \ha \ta_+ \tb_-\Big] + \widehat{\L} \ .
\ee
This  is the same  as
the original theory 
 \rf{413}, 
with $\psi \to \bpsi$ and a coordinate redefinition $\phi_2 \to -\phi_2$ (under which $\tb_- \to -\tb_-$).

To appreciate  the special  structure of 
\rf{413},
   let us  relax the condition $r=k$ and go back to  the 
 general non-integrable  model \rf{bb2} 
  corresponding to the background \rf{bQ2}.
 Using again the notation \rf{notU}, 
  we find the following generalization of \rf{413}
 \begin{align} \la{m1}
\L = &\tfrac{r+k}{8} \Big[ (\del_+ \psi +\ta_+) ( \del_- \psi + \tb_- ) - \ha \ta_+ \tb_-\Big]   \\
&\qquad \qquad + \tfrac{r-k}{8} \Big[ (\del_- \psi +\ta_-) ( \del_+ \psi + \tb_+ ) - \ha \ta_- \tb_+\Big]   + \widehat \L \ . \no
\end{align}
Applying   the T-duality  $\psi\to \bpsi$  to \rf{m1} we get, instead of \rf{417},
\begin{align} \la{m2}
\bar \L ={} & \tfrac{1}{4} r \Big[ \Big (\del_+ \bpsi +\tfrac{r+k}{2r}\ta_+ + \tfrac{r-k}{2r}\tb_+ \Big) \Big(\del_- \bpsi - \tfrac{r-k}{2r}\ta_-  - \tfrac{r+k}{2r}\tb_- \Big) \\
 &\qquad \qquad \qquad \qquad \qquad \qquad \qquad \qquad + \tfrac{r+k}{4r}\ta_+ \tb_- + \tfrac{r-k}{4r} \ta_- \tb_+ \Big] +  \widehat \L    \ .  \no
%\bar \L ={} & \tfrac{1}{4} r \Big[ \Big (\del_+ \bpsi + \ha (1+\tfrac{k}{r})a_+  + \ha (1-\tfrac{k}{r})b_+ \Big) \Big(\del_- \bpsi - \ha (1 -\tfrac{k}{r})a_-  - \ha (1+\tfrac{k}{r})b_- \Big)  \\
%& \qquad + \tfrac{1}{4}(1+\tfrac{k}{r})a_+ b_- + \tfrac{1}{4}(1-\tfrac{k}{r}) a_- b_+ \Big] +  \widehat \L  \no 
%
%\tfrac{r}{4}  \del_+ \bpsi \del_- \bpsi  + \tfrac{r+k}{8} \big( a_+  \del_- \bpsi  - b_-  \del_+ \bpsi )  -
%\tfrac{r-k}{8} \big( a_-  \del_+ \bpsi  
% + b_+  \del_- \bpsi )  \no \\
%& - \tfrac{r^2-k^2}{16r}\big( a_+ a_- -  b_+ b_-) - \tfrac{r+k}{16}\tfrac{k}{r}a_+ b_- -  \tfrac{r-k}{16}(2+\tfrac{k}{r})a_- b_+ + \widehat \L \  .
\end{align}
For  general values of $r$  and  $k$,  
\rf{m2} is different from  \rf{m1}; the only self-dual theory where  \rf{m1} and \rf{m2} coincide is the $T^{1,q}$ model  \rf{413} corresponding to $r=k$ (or its parity-conjugate $r=-k$). 

\sloppy
By the standard  path integral argument,  the  T-dual models \rf{m1},\rf{m2}   should be  quantum-equivalent. In general,  the 
  T-duality  transformation rules  are
   subject to quantum  $\a'$ corrections  \ci{Tseytlin:1991wr,Haagensen:1997er,Kaloper:1997ux}
    that may be attributed to extra  finite counterterms  resulting  from 
     integration over  the auxiliary gauge field $A_\pm$ (see eq.\ \eqref{7} above).  If
      the kinetic term of the  isometric coordinate  is non-trivial, i.e.\ 
      the
      term 
      quadratic in $A_\pm $ 
      is  $A_+ M(x) A_-$, then the leading  quantum correction to the effective Lagrangian 
      is represented by   the   term  $\Delta \L \sim \a' \, \del_+ \log M \, \del_- \log M$ (as well as a shift of the dilaton \ci{Buscher:1987qj,Schwarz:1992te}). 
     In the case of \rf{m1} we have $M=1$ and thus  this  correction  is absent.
     
Since the model  \rf{m1}  is stable
under the RG 
 due to its symmetries,  
 with  the 3  running couplings $r,\rh,\wt\rh$,  
its T-dual \rf{m2} must also be stable.
 Given that   the self-dual  points ${r=\pm k}$ are part of both RG-stable families 
 \rf{m1} and \rf{m2}  then they must also 
 remain in both families after the renormalization. Since they are the \textit{only} such self-dual  models, then $r=\pm k$ must be   fixed lines  of the RG flow.
     This   was  indeed  confirmed above 
     by the explicit computation of the $\beta$-functions  leading to 
    \rf{b3}.

\section{Discussion \la{CS}}
%%%%%%%%%%%%%%%%%%%%%%

In this   chapter we discussed  further
 instances 
 of  a close connection    between 
the conditions of  integrability and  a  consistent restriction of the RG flow to a subspace of couplings. 

We have found the  2-loop  $\b$-functions of  the 6-parameter  $G \times G$ model \rf{par} 
and  have shown that  its integrability condition \rf{cond} is automatically preserved by the RG flow. 
 In \cite{today}, the 1-loop $\beta$-functions for this integrable 
   model were written in a universal form in terms of the twist function, revealing a hidden simplicity. 
   It would be  interesting to see if the complicated  expressions we have found 
   for the 2-loop $\b$-functions (see  Appendix \ref{2103GGapp}) 
simplify on the 
``integrable surface''
 once expressed in terms of the 
    twist function. Indeed, one may try to extend  the method of \cite{today} from 1-loop to 2-loop order, i.e.\  computing the Riemann tensor and thus the 2-loop  $\beta$-function in terms of the twist function. 
    It   would  also be interesting   to investigate  the connection  
    to  the ``doubled''  approach of   \cite{hassler}  which   studied the model \rf{par},\rf{cond}  with 
 additional  integrable $\eta$- or $\l$-deformation parameters   turned on.
  
 We   also studied  the  6-parameter gauged $G\times G/H$  model \rf{couH},\rf{bb1},
 which is integrable   under the conditions \rf{condH}. 
 The latter  were  found
  to be stable under the 1-loop  RG flow,  but generally require   a certain  deformation 
 (i.e.\ the addition of  finite counterterms)  at the 2-loop level
to preserve  renormalizability. 

 There are  still  some special cases   in which integrable $G\times G/H$  models
are automatically stable under the 2-loop RG flow. One simple example is  the $T^{1,1}$ model of \ci{ABL}. 
We also   constructed   a    new   class of integrable $G\times G/H$ models  \rf{4c} in the case  when the subgroup  $H$ 
is abelian (see \rf{bb2},\rf{rk},\rf{4c}). 
For $G=SU(2)$ and $ H=U(1)$,  this  led to 
 an integrable $T^{1,q}$ 
 model  generalizing  the  $T^{1,1}$ model, 
which   
we also found to be stable under the 2-loop  RG flow  
for general values of the parameter $q$. 
 This  model 
 may be interpreted as
an integrable  deformation of the 
conformal 
GMM model 
 with unequal levels  \ci{GMM}. 
An open question is whether the integrable $T^{1,q}$ model  admits a description in terms of affine Gaudin models (like the $T^{1,1}$ case) or if it is outside of that formalism. 
 
Since the GMM model  admits a  $G\times G'/H$   generalization (with $G\neq G'$),
this raises   the question of whether  there is a larger class of integrable  $G\times G'/H$ 
models that flow to 
such 
conformal 
theories. One obvious possibility is to   consider   some analytic continuations, e.g., 
 take  $G'$  to be a different  real form of the complexification of $G$.
  For example,   the  counterpart  of  the $SU(2)\times SU(2)/U(1)$ model 
 would be   $SL(2,\mathbb{R})\times SU(2)/U(1)$.\la{an}

Another interesting open question is why the integrable $G\times G$ model was automatically stable under the 2-loop RG flow, while the gauged $G\times G/H$ analog requires a non-trivial quantum deformation of the ``integrable surface'' in the space of couplings. This is analogous to WZW models, which are conformal without any $\a'$-corrections, and gauged WZW models, which require a non-trivial tower of $\a'$ corrections (e.g.\ in eq.\ \eqref{6}) to remain conformal beyond 1-loop.
One possible resolution, which we shall consider below in Section \ref{1910TO}, is to reformulate the $G\times G/H$ model as a theory with a gauge field and redefine the gauge field in terms of $H$-valued scalars; as for the ``tripled'' formulation of the  $\l$-model in Chapter \ref{CT}, such a reformulation could  
 restore higher-loop renormalizability without needing any quantum corrections to the geometry.

%% file: 1910v2.tex
\def\id{\protect{{1 \kern-.28em{\rm l}}}}
\def\be{\begin{eqnarray}}
\def\ee{\end{eqnarray}}
\def\bi{\bibitem}
\def\tr{{\rm tr}}
\def\ha{\tfrac{1}{2}}
\def\td{\tilde}
\def\ci{\cite}
\def\N{{\mathcal N}}
\def\ww{\Omega}
\def\S{{\mathcal S} }
\def\nn{\nu}
\def\z{\zeta}
\def\dg{\dagger}
\def\a{\alpha}
\def\b{\beta}
\def\ap{\alpha^\prime}
\def\aa{{\a'}}
\def\g{\gamma}
\def\ok{\frac{1}{k}}
\def\jL{{J}}
\def\cL{{\mathcal L}}
\def\cH{{\mathcal H}}
\def\E{{\mathcal E}}
\def\w{\omega}
\def\vep{\varepsilon}
\def\De{{\mathcal D}}
\def\k{\kappa}
\def\four{\tfrac14}
\def\third{\tfrac{1}{3}}
\def\det{\hbox{det}}
\def\tid{\tilde}
\def\vv{{\rm v}}
\def\XX{{\rm X}}
\def\ta{{\tilde \a}}
\def\fo{\frac{1}{4}}
\def\K{{\rm S}}
\def\el{\ell}
\def\Tr{{\rm Tr}}
\def\P{\Phi}
\def\l {\lambda}
\def\bl{{\tilde \l}}
\def\const{{\rm const}}
\def\vn{\vec n}
\def\Prod{\Pi}
\def\O{{\mathcal O}}
\def\m{\mu}
\def\vs{\vec \s}
\def\ie{i.e.}
\def\rS{{\rm S}}
\def\as{{\a}}
\def\e{\epsilon}
\def\foot{\footnote}
\def\ra{\rightarrow}
\def\F{{\cal F}}
\def\cc{\circ}
\def\eqv{\equiv}
\def\ni{\noindent}
\def\bw{{\rm w}}
\def\cT{{\cal T}}
\def\no{\nonumber}
\def\J{{\cal J}}
\def\om{\omega}
\def\l{\lambda}
\def\tk{{\tilde k}}
\def\tl{{\tilde \l}}
\def\sqtl{{\sqrt{\tilde \l}}}
\def\adss{$AdS_5 \times S^5$\ }
\def\D{\Delta}
\def\p{\phi}
\def\r{\rho}
\def\rN{{\rm N}}
\def\tw{{\tilde w}}
\def\vl{ \vec \ell}
\def\varpi{{\rm w}}
\def\bG{\bar \G}
\def\ve{\varepsilon}
\def\I{{\cal I}}
\def\La{{\Lambda}}
\def\R{{\rm R}}
\def\bt{\bar\theta}
\def\Z{{\cal Z}}
\def\pa{\partial}
\def\bea{\be}
\def\eea{\ee}
\def\DD{{\rm D}}
\def\chii{\varepsilon}
\def\th{\theta}
\def\t{\tau}
\def\beq{\be}
\def\eeq{\ee}
\def\beqa{\bea}
\def\eeqa{\eea}
\def\bs{\bigskip}
\def\c{{\rm a}}
\def\del{\partial}
\def\s{\sigma}
\def\eps{{\epsilon}}
\def\n{\nu}
\def\dag{\dagger}
\def\bd{\bar \del}
\def\ed{\end{document}}
\def\iffa{\iffalse}
\def\xp{x_+}
\def\xm{x_-}
\def\te{\textstyle}
\def\rx{{\rm x}}
\def\C {{\rm C}}
\def\hC {{\rm \hat C}}
\def\dd {{\rm d}}
\def\bb{{\rm b}}
\def\sql{{\sqrt{\l}}}
\def\ep{\epsilon}
\def\G{\Gamma}
\def\dx{\dot x}
\def\vp{\varphi}
\def\tx{{\tilde x}}
\def\ad{{\rm ad}}
\def\d{\delta}
\def\de {\del}
\def\L{\mathcal{L} }
\def\pcm {{\rm PCM}}
\def\wz{{\rm WZ}}
\def\pcmq {{\rm PCM}$_q$\ }
\def\ZMq {{\rm ZM}$_q$\ }
\def\ZM {{\rm ZM} }
\def\pp{{\rm p}}
\def\qq{{\rm q}}
\def\rk {{\rm u}}
\def\rrk{{\rm e}}
\def\ss{{\rm s}}
\def\tt{{\rm t}}
\def\uu{{\rm u}}
\def\sm{$\sigma$-model }
\def\sms{$\sigma$-models }
\def \ov {\over}
\def\Ad{\text{Ad}}
\def\lmp{\Lambda}
\def\bl{\bar \lambda}
\def \hb {\hbar}
\def \k {\varkappa} \def \h {{\rm h}}
\def \dt {\dt}
\def\ddt{\frac{d}{dt}}
\def \bb {{\rm b}}
\def \vv {{ v}}
\def \kk {{\rm k}}
\def \ka {{\kappa}}
\def \np { }
\def \cM {\mathcal{M}}
\def\H{{\rm H}}
\def \BB {{\rm B}}
\def \HH {{\rm H}}
\def \GG {{\rm G}}
\def \PCM {{\rm PCM}}
\def \WZ {{\rm WZ}}
\def \Lie {{\rm Lie}}
\def\lam{$\lambda$-model}
\def\etm{$\eta$-model}
\def\bh{\bar h}
\def\hh{{\rm h}}
\def \OO {{\cal O}}
\def \etm {$\eta$-model\ }
\def \lam {$\l$-model\ }
\def \dt {\frac{d}{dt}}
\def \rrr {x} \def \qp {{\rm q}}
\def \qp {{\mathrm h} }
\def \bq {{q'}} \def \barp {x} \def \barq {y}
\def \rk {{\rm k}}
\def\F{H}
\def\cg{c_{_G}}
\def\cf{c_{_\F}}
\def\cff{c_{_\F}}
\def\HH{\F}
\def\XX{{X}}
\def\hhh{\widetilde {\hh}}
\def\q{{\rm q}}
\def\RR{{J}}
\def\LL{{K}}
\def\M{{\cal M}}
\def\e{\varepsilon}
\def\tih{{\hat h}}
\def\bl{\bar\lambda}

\chapter{``Tripled'' formulation of the \texorpdfstring{$\l$}{lambda}-model \label{CT}}

We have been studying a close link between 
renormalizablility, 
or stability under the RG flow,
and
 integrability in 2d $\s$-models.
 Having previously been observed
\cite{Fateev:1992tk,fateev96,lukyanov,Fateev:2019xuq,Fateev:2018yos,Litvinov:2018bou} only at the 1-loop (Ricci flow) level, it
is important to study this relation also at higher loop orders.

This question of higher loops was addressed above in Chapters \ref{QCS} and \ref{GGS},
 where we found that, starting from 2 loops,
renormalizability typically requires a specific deformation of the classical target space
geometry, which may be interpreted as the result of adding
finite
local counterterms. In Chapter \ref{QCS},  we considered the simplest examples of 
integrable $\eta$- and $\l$-deformed bosonic \sms with 2- and 3-dimensional target spaces.
In Chapter \ref{GGS}, we looked at another class of  integrable \sms with target spaces $G\times G$ and $G\times G/H$.

 Here we shall concentrate on the $\l$-deformation \cite{Sfetsos:2013wia,Hollowood:2014rla} and address the \textit{general case} where the model is based on an arbitrary 
group
$G$ or  symmetric space $G/\HH$. Its Lagrangian is (repeating eq.\ \eqref{lam} for clarity)
\begin{align} 
&\L  = k \, \Big( \L_G(g) \la{1}  + \Tr\big[ J_+ A_- - A_+ K_- + g^{-1} A_+ g A_- - A_+ A_- - ( \lambda^{-1}-1) A_+ P A_-\big] \Big) , \\
& %{\PCM} = - \ha \Tr[J_+ J_-] \ , \quad dB_{\rm WZ}=\tfrac{1}{6} \Tr [J\wedge J \wedge J] \ , \quad
J=g^{-1} dg \ , \quad K = dg \, g^{-1} \ ,\no\\
& P = \begin{cases}
1 \ , \quad \ \ \ \ \ \text{group space }G\\
P_{G/\HH} \ , \quad \text{symmetric space }G/\HH \ ,
\end{cases} \la{1111}
\end{align}
where $\L_G=\L_\PCM+ \L_\WZ$ is the WZW Lagrangian, $g \in G$, $A_\pm \in \Lie(G)$ and $P_{G/\HH}$ is the projector onto the orthogonal complement of $\Lie(\HH)$ in $\Lie(G)$.
Instead of $\l$ it is often convenient to use
a redefined parameter,
\begin{equation}\la{133}
%\g = \lambda^{-1}-1
%\ , \qquad \qquad 
\ka = \frac{1-\lambda}{1+\lambda} \ .
\end{equation}
The limit $\lambda \to 0$ (or $\ka \to 1$)
yields the conformal  WZW model  on $G$ (or  gauged WZW model on $G/\HH$)
 with level $k$,
which should thus be a fixed point of the RG flow.
The $\l$-model's $\mathbb{Z}_2$ transformation \eqref{z2sym} then implies that the opposite limit $\l \to\infty$ (or $\ka \to -1$)  should also be a fixed point.
Indeed, in the group case the $\l \to \infty$ limit of \rf{1} is conformal:
it is the $G/G$ chiral gauged WZW model \cite{Chung:1992mj}, which, on
integrating out $A_\pm$, gives the $G$ WZW model at level $-k$.
Similarly, in the coset case we find in this limit the $G/H$ gauged WZW model at level $-k$.

Integrating out the 2d gauge field $A_\pm$ in \rf{1}, i.e. reducing the model to the standard (or ``physical'') configuration space,
one finds a \sm with parameters $k$ and $\l$ (see eq. \eqref{lamc}).  
As we found in Chapter \ref{QCS}, to preserve renormalizability of the $\l$-model 
at the 2-loop level,
 one must make a non-trivial modification of
the classical target space geometry.

In this chapter our central observation will be that, \textit{before} integrating out $ A_\pm$,
the \lam \rf{1} is renormalizable without any deformation.
As was done in the context of gWZW models \cite{Karabali,Tseytlin:1992ri,Sfetsos:1993bh,Sfetsos:19932}, we shall change variables from $A_\pm$ to scalar fields $h, \bar h \in G$ as
\be A_+ = h \,  \del_+ h^{-1}, \ A_- = \bar{h} \, \del_+ \bar h^{-1}  \ , \la{cov} \ee
resulting in  a \sm
on the extended or ``tripled'' ($G \times G \times G$) configuration space $(g, h, \bh)$.
It may be interpreted as the sum of two decoupled parts: (i) a  WZW model on $G$ and (ii) a deformation of a 
WZW model on $G \times G$ by a particular left-right current interaction.
This special decoupling is expected to be a manifestation of integrability and, as a result, 
the \textit{manifest} global symmetries become sufficient to protect  
 the model
 under the RG flow
without the need for any additional finite counterterms.

\

The rest of the chapter is organized as follows. In Section \ref{revg} we shall review the previous application of the change of variables \rf{cov} to gauged  and chiral gauged WZW models.

In Sections \ref{1910extended} and \ref{trips}, we consider the $\lambda$-models, respectively based on groups and symmetric spaces, on the configuration space $(g,A_+,A_-)$, i.e.\ with the gauge field unintegrated.
By using the change of variables \rf{cov} to write the $\l$-models as  certain ``tripled'' $\s$-models
with extra manifest symmetries, we shall argue that they are automatically renormalizable to all orders and explicitly compute their 2-loop $\b$-functions.
In the
 non-abelian dual (NAD)
  limit \eqref{limr}, these  will correctly reproduce the (scheme-invariant) 2-loop $\b$-functions of the PCM and symmetric space $\s$-model, suggesting that the interpolating models for NAD are also automatically renormalizable without needing corrections.

\sloppy
In Section \ref{lowd} we consider some simple examples of the models obtained in the ``tripled'' formulation of the $\l$-models based on $SU(2) \ov U(1)$ and $SU(2)$. In Section \ref{1910TO} we construct  analogous ``tripled'' or ``extended'' space formulations of some other integrable $\s$-models (a variant of the $\l$-model built on $G^N$ \cite{Bassi:2019aaf} and the \mbox{$G\times G/H$} models of \cite{ABL}).
Some concluding remarks are made in Section \ref{1910conclusions}.

\

\noindent This chapter is based on the paper \cite{HLT2} and we closely follow the presentation there.

\section{Review of gWZW model \label{revg}}
First let us review the  original use of the change of variables \rf{cov} in the context of the gauged WZW models \cite{Karabali,Tseytlin:1992ri,Sfetsos:1993bh,Sfetsos:19932}. The gauged WZW model is closely related to the $\l$-model \rf{1} but with the gauge field being valued in some subalgebra $\Lie(F)\subset \Lie(G)$ (see Section \ref{gWZW}),
\be \begin{aligned}
&\L  = k \, \Big( \L_G(g) + \Tr\big[ J_+ A_- - A_+ K_- + g^{-1} A_+ g A_- - A_+ A_- ] \Big) \ , \\
&g\in G \ , \ \ A_\pm \in \Lie(F) \ . \la{gg}
\end{aligned} \ee
Changing variables $(A_+, A_-) \to (h,\bar h)$ as in \rf{cov}, now with with $h,\bar h \in F$, and using the Polyakov-Wiegmann identity\foot{The Polyakov-Wiegmann identity is a multiplicative  identity for the WZW Lagrangian \cite{Polyakov:1983tt},

$\qquad \L_G(u v) = \L_G(u) + \L_G(v) - \Tr[ u^{-1} \del_+ u \, \del_- v v^{-1} ]  \ . $  \la{PW}} 
 \cite{Polyakov:1983tt},
 the Lagrangian  \rf{gg} can be written as a combination of WZW models, 
\begin{equation}\begin{split}\la{22g}
\L &= k \Big( \L_G (h^{-1} g \bar{h}) - \L_F ( h^{-1} \bar{h} )\Big) \ .
\end{split}\end{equation}
The change of variables
in \rf{cov} results in a Jacobian contributing to the action as
\cite{Polyakov:1983tt,Polyakov:1988qz}
\begin{align}
\Delta \L & = - 2 c_{_F} \Big( \L_F(h^{-1} \bar{h} ) + \q \Tr [A_+ A_-]\Big)\la{jac1} \\
&= - 2 c_{_F} \Big( \L_F(h^{-1} \bar{h} )+  \q \Tr [ \del_+ h \, h^{-1} \, \del_- \bar{h} \, \bar{h}^{-1} ]\Big) \ , \la{jacg}
\end{align}
where $c_{_F}$ is the dual Coxeter number\foot{More precisely, $c_{_F}$ denotes here a rescaled version of the dual Coxeter number of $F$ by an index factor relating to the embedding of $F$ into $G$ (see discussion in the final paragraph of App.\ \ref{conv} for  subgroups $H \subset G$). If $G$ and $F$ are both classical groups with the same index for their fundamental representation (e.g. both $A$-type) then this index factor is trivial and $c_{_F}$ is the dual Coxeter number of $F$.} of $F$ and
an arbitrary coefficient $\q$ parametrizes the ambiguity of adding a finite local  counterterm $\Tr [A_+ A_-]$.
Combining \rf{jacg} and the classical Lagrangian \rf{22g} gives
\be
\L = k \L_G (h^{-1} g \bar{h}) - (k+2 c_{_F}) \L_F ( h^{-1} \bar{h} ) -  2 \q c_{_F} \big[ \L_F (h^{-1} \bar{h}) - \L_F ( h^{-1} )- \L_F (\bar{h} ) \big] \ . \ \la{5termg}
\ee
Now one can choose the coefficient $q=0$ to obtain the sum of two decoupled WZW models (e.g. after defining $\tilde{g}=h^{-1} g h$, $\tilde h=h^{-1} \bar h$)
\be
L = k \L_G (h^{-1} g \bar{h}) - (k+2 c_{_F}) \L_F ( h^{-1} \bar{h} ) \la{ggr}
\ee
Expressing the gWZW model as this sum of decoupled WZW models makes manifest its properties of (i) exact conformal invariance and (ii) gauge invariance under $g\to u g u^{-1} $, $h\to u h$, $\bar h \to u \bar h$.

The exact form of the  gWZW metric and $\BB$-field can be derived \cite{Tseytlin:1992ri} from \rf{ggr} (agreeing with \cite{Dijkgraaf:1991ba,Bars:1992sr}) by considering the quantum effective action. The WZW model's effective action is just the classical one with a shifted level (respectively $k \to k+ c_{_G}$  or $k\to k+c_{_F}$), leading to
\be \begin{aligned}
\L_{\rm eff} {}&{}= (k+c_{_G}) \L_G (h^{-1} g \bar{h}) + [ - (k+2 c_{_F}) + c_{_F} ]\L_F ( h^{-1} \bar{h} ) \\
 &{}=  (k+c_{_G}) \L_G (h^{-1} g \bar{h}) - (k+ c_{_F}) \L_F ( h^{-1} \bar{h} ) \ .
\end{aligned} \ee
Transforming back to $A_\pm$, the resulting theory is not local but the ``local part'' gives a prescription for a corrected version of \rf{gg} in which $A_\pm$ can be integrated out \textit{classically} to get an exactly conformal $\s$-model (of which the $SL(2,R)/U(1)$ example is given in \eqref{6}).

\bigskip

Now let us briefly consider the related  $G/F$ \textit{chiral gauged WZW model}  (cWZW) \cite{Chung:1992mj}, for which a similar argument was carried out in \cite{Sfetsos:1993bh},
\be \begin{aligned}
&\L  = k \, \Big( \L_G(g) + \Tr\big[ J_+ A_- - A_+ K_- + g^{-1} A_+ g A_-  ] \Big) \ , \\
&g\in G \ , \ \ A_\pm \in \Lie(F) \ . \la{cg}
\end{aligned} \ee
 This theory differs from the gWZW model \rf{gg} by the addition of $k \, \Tr[A_+A_-]$. Again changing variables as in \rf{cov}, the extra classical term in \rf{cg} is proportional to the ambiguous local term with coefficient $\q$ in \rf{jac1}. Thus the result for the cWZW model is obtained by replacing  $-2\q c_{_F} \to k-2\q c_{_F}$ in \rf{5termg},
\be
\L = k \L_G (h^{-1} g \bar{h}) - (k+2 c_{_F}) \L_F ( h^{-1} \bar{h} ) +  (k-2 \q c_{_F}) \big[ \L_F (h^{-1} \bar{h}) - \L_F ( h^{-1} )- \L_F (\bar{h} ) \big] \ . \ \la{5termc}
\ee
Again in this case, there is a choice $\q=-1$ that leads to cancellations and gives a sum of decoupled WZW models \cite{Sfetsos:1993bh},
\be
\L = k \L_G (h^{-1} g \bar{h}) +  (k+ 2 c_{_F}) \big[ \L_F ( h^{-1} )+ \L_F (\bar{h} ) \big] \ , \la{cgg}
\ee
illustrating that the cWZW model is also exactly conformal.

%%%%%%%%%%%%%%%%%%%%%%%%%%%%%%%%%%%%%%%%
\section{``Tripled'' \texorpdfstring{$\l$}{lambda}-model based on a group}\la{extended}
%%%%%%%%%%%%%%%%%%%%%%%%%%%%%%%%%%%%%%%%

Now let us turn to the $\l$-model based on a group $G$ (putting $P=1$ in eq.\ \rf{1}),
\begin{align} 
&\L = k \, \Big( \L_G(g) + \Tr\big[ J_+ A_- - A_+ K_- + g^{-1} A_+ g A_- - \lambda^{-1} A_+  A_-\big] \Big) , \la{1G}  \\
&g\in G \ , \quad A_\pm \in \Lie(G) \ . \no
\end{align}
This model interpolates between the gWZW model \rf{gg} ($\l=1$) and the cWZW model \rf{cg} ($\l=\infty$) in the particular case $F=G$. Hence, upon making the same change of variables $(A_+, A_-)\to (h,\bar h)$ as in \rf{cov}, the resulting theory is just the weighted average of the gWZW and cWZW results \rf{5termg},\rf{5termc} above,
\be \begin{aligned}
\L = &k \, \L_G (h^{-1} g \bar{h}) - (k+2 c_{_G}) \L_G ( h^{-1} \bar{h} )\\
 &\qquad \qquad -  (k(\l^{-1}-1)+2 \q c_{_G}) \big[ \L_G (h^{-1} \bar{h}) - \L_G ( h^{-1} )- \L_G (\bar{h} ) \big] \ , \ \la{5term}
\end{aligned} \ee
with the new scalar fields $h,\bh$ now valued in the full group $G$. This expression includes the Jacobian contribution \rf{jacg}, with the coefficient $\q$ parametrizing the local term ambiguity. In this case $\q$ just specifies a particular redefinition of the coupling $\l$, so the choice of $\q$ is not important and we shall pick $\q=-1$ (as in the cWZW case above).

Again using the Polyakov-Wiegmann identity (see footnote \ref{1910PW}), we may rewrite \rf{5term} as
\begin{align}
&\L= k \L_G (\tilde{g}) + \L'(\tih, \bh) \ ,\qquad \qquad \qquad \qquad \tilde{g} \equiv h^{-1} g \bar{h}\in G \ ,  \qquad \tih\equiv h^{-1} \in G \ , \la{277} \\
&\L'= - (k+2 \cg) \big[ \L_G ( \tih) + \L_G (\bar{h} ) \big] + k\l^{-1} \Tr [ \tih^{-1} \del_+ \tih \, \del_- \bh \bh^{-1}] \ .
\la{27}
\end{align}
This is a \sm with fields valued in  the extended configuration space $(\tilde{g},\tih,\bar{h}) \in G \times G\times G $. 
As for the gauged WZW models considered  above,
the first WZW term  $k \L_G (\tilde{g})$ still decouples from the other fields in \rf{277}. This decoupling is a special property expected to be associated with integrability: indeed if one deformed the $\l$-model \rf{1G} in an arbitrary way preserving its global symmetry but breaking integrability, the decoupling would be lost.\foot{I thank N. Gromov for a discussion of this point.}  Since the decoupled WZW term is conformal on its own, then its coefficient $k$ will not run, and the RG flow of $\l$ is determined by the other ``truncated'' part $\L'$.

Redefining the couplings as
\be
\la{redef}
\tk \equiv k + 2 \cg \ ,\qquad \qquad \tl \equiv
 \frac{k}{\tk}\l^{-1}  \ ,
\ee
then the truncated part,
\begin{align} \la{6dmodel}
& \L'(\tih, \bar h)
= - \tk \Big( \L_G ( \tih ) + \L_G (\bar{h} ) - \tl \Tr [ \tih^{-1}\del_+ \tih \, \del_- \bh \bh^{-1} ] \Big) \ , \qquad \quad \tih \equiv h^{-1}, \ \bh \in G\ , 
\end{align}
 may be interpreted as the sum of
  two WZW models for the group $G $
with the same level $-\tk$
perturbed by a current-current interaction.
Classical integrability of the model \rf{6dmodel} (implying integrability of the group space $\l$-model) 
was first shown in \cite{Bardakci:1996gs} (see also \cite{Hull:1995gj}).
In fact \rf{6dmodel} is a particular instance of the integrable $G\times G$ models \cite{DLMV} in \eqref{2103par},\eqref{2103cond} above, with the couplings and fields  related by 
\be
(k_1,k_2,s,u,t,b) =(-\tk, +\tk, -\tk, -\tk, \tk \tl^{-1},\tk \tl^{-1}) \ , \qquad (g^{(1)},g^{(2)})=(\tih, \bh^{-1}) \ . \la{crel}
\ee
At the value $\wt \l=1$ of the coupling, the model \rf{6dmodel} becomes a special case ($G\times G/G$ with equal levels) of the conformal GMM model \cite{Guadagnini:1987ty}.
A generalization of the model \rf{6dmodel} was also considered in
\cite{sfetsos,s19},
where it was interpreted as a special case of a ``doubly $\l$-deformed'' $\s$-model.
Our path-integral relation
between the $\l$-model \rf{1} and the truncated model \rf{6dmodel}
should be equivalent (at least at the classical and 1-loop level)
to the canonical equivalence between the doubly $\l$-deformed model
and two copies of the $\l$-model found in \cite{Georgiou:2017oly}
(upon setting one of the two $\l$-parameters to zero).
The leading order in $1/k$ (1-loop) renormalization of similar models
was also studied earlier in \cite{Gerganov:2000mt,LeClair:2001yp}.

Our central observation is that the ``truncated'' model \rf{6dmodel}
is automatically renormalizable to all orders with only $\tl$ (or equivalently $\l$) running. 
This is because, thanks to the decoupling of the first WZW term in \rf{277}, the truncated model is fully protected under the RG flow by its manifest symmetries:
the global $G$ symmetry (present in the usual formulation \rf{1G} of the $\l$-model),
\be \tih \to \tih \,  g_0 \ , \qquad \bh \to g_0^{-1} \, \bh \ , \qquad\qquad g_0 \in G \ , \ee
and chiral gauge symmetries (cf. the symmetry \eqref{cgs} of the WZW model),
\be \la{210}
\tih \to u(\s^-)\, \tih \ , \qquad \bh \to \bh\, w(\s^+) \ , \qquad \qquad u, w \in G\ , \ee
arising ubiquitously from the change of variables \rf{cov}. These symmetries uniquely  protect the structure of  \rf{6dmodel}, with $\tl$ and $\tk$ being the only free parameters.
 In particular, the chiral symmetries \rf{210} prohibit the appearance of other current-current interaction terms under the RG flow.

We shall now directly compute the 2-loop $\b$-function for the coupling $\tl$ of the truncated \sm \rf{6dmodel} in large $\tk$ perturbation theory, demonstrating its renormalizability with only  $\tl$ (or $\l$) running. Let us emphasize that  --- modulo a finite renormalization, or redefinition, of $\l$ encoded in the choice of the coefficient $\q$ in \rf{jac1} --- the RG evolution of  the tripled \sm \rf{277} 
 is exactly equivalent to that of the $\l$-model \rf{1G},  since they were related at the level of the path integral including the necessary Jacobian \rf{jac1}.\foot{The
  $\b$-function of the truncated $G\times G$ model \rf{6dmodel} can actually be extracted from that of the more general $G\times G$ model studied in Chapter \ref{GGS} (given explicitly in App.\ \ref{2103GGapp}). For completeness, we explicitly run through the computation again here. The results are in agreement those in App.\ \ref{2103GGapp} for the special values of the parameters in \rf{crel}.}

Let us introduce the basis $\{T_A\}$ for $\Lie(G)$ (see Appendix \ref{conv} for conventions).
We shall use the indices $A$ and $\bar{A}$ respectively for the two
$G$-valued fields $\tih$ and $\bar{h}$ in \rf{6dmodel} (with the tangent space index for $G \times G$ denoted as
$M = \{A,\bar A\}$). As in
\cite{Georgiou:2017jfi} we introduce the vielbein
\begin{align}
&E^M = (e^A, e^{\bar{A}}) = (\sqrt{1-\tl^2}\, \RR^A,\, \LL^{\bar{A}} + \tl \RR^A) \ , \la{vielbein}\\
&\RR  \equiv T_A \RR^A  = \tih^{-1} d \tih\ , \qquad  \LL^{\bar{a}}  \equiv  T_A \LL^A =  d \bar{h} \bar{h}^{-1} \ ,
\end{align}
where $\RR$ and $\LL$ are the
currents that appear in the deformation term in \rf{6dmodel}.
Up to permutations, the non-zero components of the metric and $\H$-tensor of the $G\times G$ model \rf{6dmodel} are (cf. \eqref{13})
\begin{gather} \GG_{AB} = \GG_{\bar{A}\bar{B}} = \tfrac{\tk}{2} \, \Tr[T_A T_B]\ , \la{vielmet}
\\
\H_{ABC} = -i \tfrac{\tk}{2} \tfrac{\sqrt{1-\tl^2}(1+2\tl)}{(1+\tl)^2} f_{ABC} \ , \qquad
\H_{\bar{A}\bar{B}\bar{C}} = -i \tfrac{\tk}{2} f_{ABC} \ , \qquad
\H_{\bar{A}BC} = -i \tfrac{\tk}{2} \tfrac{\tl}{1+\tl} f_{ABC} \ . \la{Hflux}
\end{gather}
Our aim is to compute the corresponding 2-loop $\b$-function in \eqref{BMT}.
Let us formally define the torsion as
$T^M = \ha {\H^M}_{NP} E^N \wedge E^P $ where the tangent space index is raised with the inverse of the metric \rf{vielmet}.
Then, from the Cartan structure equation
$d E^M + {\widehat{\w}^M}{}_{N} \wedge E^N = T^M$, we obtain the torsionful spin connection
${{\widehat \w}^M}{}_{N} = {\widehat{\w}^M}{}_{NP} E^P$ with non-zero components
\be
(\tl^{-1}+1){{\widehat \w}^A}_{ \ B} =
\widehat \w^{\bar{A}}{}_{\bar{B}} = i f^{A}{}_{BC} \big(- \tfrac{\tl}{\sqrt{1-\tl^2}} e^C + e^{\bar{C}} \big) \ .
\ee
The non-zero components of the curvature,
${\widehat{R}^M}{}_{N} = d{\widehat{\w}^M}{}_{N} + {\widehat{\w}^M}{}_{P} \wedge {\widehat{\w}^P}{}_{N}=\ha {\widehat{R}^M}{}_{NPQ} E^P \wedge E^Q$, are then found to be
\begin{equation}\begin{split}\la{torsionriem}
(\tl^{-2}-1){\widehat{R}^A}{}_{BDE} =
-\sqrt{\tl^{-2}-1}{\widehat{R}^A}{}_{BD \bar{E} }
= {\widehat{R}^A}{}_{B \bar{D} \bar{E} } = - \frac{\tl}{(1+\tl)^2} f^A{}_{BC} f^C{}_{DE} \ .
\end{split}\end{equation}
Substituting \rf{torsionriem} into the 2-loop RG equation \eqref{BMT} in the \GB scheme, one obtains
\begin{equation}\la{220}
%(\beta^{(1)}_{\m\n} + \beta^{(2)}_{\m\n} )\,
\big[ \ddt(G_{mn} + B_{mn})\big]  dx^m \otimes
dx^n = \sum^{{\rm dim}\, G}_{A=\bar A=1}\Big[ \frac{2 \cg \tl^2}{(1+\tl)^2}
+ \frac{4 \cg^2}{\tk} \frac{\tl^4(1-2\tl)}{(1+\tl)^5(1-\tl)} \Big] \, \Tr[ \RR \,  \otimes \, \LL ] \ .
\end{equation}
Here $\otimes$ indicates that the product is not symmetrized.
We conclude that only the $\tl$-dependent term in \rf{6dmodel} gets renormalized, i.e.
the 2-loop RG equation \eqref{BMT} is solved with $\xi_m=\l_m=0$ and
\be\la{01}
\ddt \tk =0 \ ,\qquad \qquad \ddt \tl = \frac{2 \cg}{\tk} \Big[ \frac{\tl^2}{(1+\tl)^2} + \frac{2 \cg}{\tk} \frac{\tl^4(1-2\tl)}{(1+\tl)^5(1-\tl)} \Big] \ .
\ee
The 1-loop term here agrees with previous results, both for the standard $\l$-model (eq.\ \eqref{la1} with $\ka = \frac{\tl-1}{\tl+1}$)  \cite{Sfetsos:2014jfa,Appadu:2015nfa} and  the ``truncated'' $G\times G$ model \rf{6dmodel}
\cite{Georgiou:2017aei,Georgiou:2017jfi}).

The level $k=\tk-2\cg$ is thus RG-invariant as it should be. 
The fixed points of the RG flow are as expected:
 $\tl=\infty$ (i.e. $\l=0$), corresponding to the $G$ WZW model with level $k$, and 
$\tl=0$ (i.e. $\l=\infty$) corresponding to the $G/G$ cWZW model, or the $G$ WZW model with level $-k$ after integrating out the gauge field.
\unskip\foot{Taking
$\tl=1$ (i.e. $\l=1$) with finite $k$ (in contrast to the NAD limit
\eqref{limr}) gives the trivial $G/G$ gWZW model
in which one can fix a gauge so that the remaining degrees of
freedom correspond to the Cartan torus.}

Expressing $\tk$ and $\tl$ in \rf{redef} in terms of $k$ and the coupling
$\ka$ using \rf{133}, we find
\be\ddt k =0 \ ,\qquad \qquad
\ddt \ka = \frac{\cg}{4k}(1-\ka^2)^2 \Big[ 1 + \frac{\cg(1-\ka)^2 (1 -10\ka - 3 \ka^2)}{8k \ka} \Big] \ . \la{24}
\ee
Note that the 2-loop term in \rf{24} is scheme-dependent: it can be changed by
redefining $\ka$ by a $1/k$ term (or shifting $k$ by a finite term): for example, it will depend on the choice of the  parameter $\q$ in
\rf{jac1}.
Even though $k$ is not running, we effectively have a 2-coupling theory
(with $1/k$ playing the role of a loop-counting parameter)
so only the 1-loop term in the $\b$-function \rf{24} is scheme-independent.

In a general scheme, the symmetry $k \to - k, \ \l \to \l^{-1}$ \eqref{z2sym}
of the 1-loop RG equation in \rf{01},\rf{24} is not manifest 
at the 2-loop level.
To preserve it requires a particular formulation of the quantum theory, i.e.\
a specific definition of the couplings, or choice of scheme.
For example, if we redefine the parameters as
\begin{align} \la{2777}
(\tk, \tl) \to (\kk, \bl) \ , \qquad \qquad \tk =\kk + \cg\ , \qquad \quad \tl=\bl\Big[ 1 + \cg \kk^{-1} \frac{1-\bl}{1 +\bl} \Big]
\ , \end{align}
then the RG equation for $\bl$ resulting from \rf{01} is
\be \la{2789}
\frac{d}{dt}\bl = \frac{2 \cg}{\kk}\frac{\bl^2}{(1+\bl)^2}\Big[1 - \frac{2\cg}{\kk} \frac{\bl^2}{(1+\bl)^2} \frac{\bl+\bl^{-1}-1 }{1-\bl^2}
\Big] \ ,
\ee
which is invariant under
the following quantum version of \eqref{z2sym}
\be\la{227}
\kk \to -\kk \ , \qquad \qquad \bl \to \bl^{-1} \ .\ee
Note that here $\kk=k+ \cg$ is the same as the shifted level of the WZW theory.
It is also worth observing that \rf{2777} is not the only redefinition that
restores the symmetry \eqref{z2sym}. 
Indeed, even restricting to those that preserve the
existence of the fixed points and the NAD limit discussed below, there are many.

In the non-abelian dual (NAD) limit \eqref{limr} (i.e. $\ka= \tfrac14 \hh k^{-1}, \ k\to \infty$) when the $\l$-model \rf{1G} with unintegrated gauge field becomes the interpolating model \eqref{NADi} for non-abelian duality of the PCM,
the $\b$-function \rf{24} becomes
\be
\ddt \hh = {\cg} + \ha {\cg^2}\hh^{-1} \ . \la{25}
\ee
This result matches the standard expression  \eqref{Prg} for the 2-loop $\b$-function of the PCM on a simple group $G$ (see, e.g., \cite{McKane:1979cm,Hikami:1980hi,friedan}).  The 2-loop term in \rf{25} is scheme-independent as the NAD limit produces a single-coupling theory. Hence the scheme-dependent 2-loop term in the $\l$-model $\b$-function \rf{24} does still contain some invariant information, which passes a non-trivial test in this limit.

%%%%%%%%%%%%%%%%%%%%%%%%%%%%%%%%%%%%%%%%
\section{``Tripled'' \texorpdfstring{$\l$}{lambda}-model based on a symmetric space \label{trips}}
%%%%%%%%%%%%%%%%%%%%%%%%%%%%%%%%%%%%%%%%

Next let us consider the $\lambda$-model based on a symmetric space
 $G/\F$ (setting $P=P_{G/\F}$ in \rf{1}),
\be \begin{aligned}
%&\L  = k \, \Big( \L_G(g)   + \Tr\big[ J_+ A_- - A_+ K_- + g^{-1} A_+ g A_- - A_+\big( 1 - (\l^{-1}-1)P_{G/H}\big) A_- ] \Big) \ , \\  %v2
&\L  = k \, \Big( \L_G(g)   + \Tr\big[ J_+ A_- - A_+ K_- + g^{-1} A_+ g A_- - A_+\big( 1 + (\l^{-1}-1)P_{G/H}\big) A_- ] \Big) \ , \\
&g\in G \ , \ \ A_\pm \in \Lie(G) \ . \la{ggu}
\end{aligned} \ee
This model has a special subgroup $H\subset G$ corresponding to its gauge symmetry, however we note that the 2d gauge field $A_\pm$ is still  valued in the algebra of the full group $G$. 

We shall repeat the same steps as in the group case above,   obtaining a  similar tripled formulation:
redefining $(A_+, A_-) \to (h,\bh)$ according to  \rf{cov}, using the Polyakov-Wiegmann identity (see footnote \ref{1910PW}), 
including the contribution from the Jacobian \rf{jac1}, and
setting $\tilde{g} = h^{-1} g \bar h$ and $ \tih=h^{-1}$, we now get
\be\la{effectivecoset1} \begin{aligned}
\L = {}&{}k \L_G (\tilde{g}) - (k+2\cg) \L_G ( \tih\bar{h} )+ k (\l^{-1}-1) \Tr [ \tih^{-1}\del_+ \tih \, P_{G/H} \, \del_- \bh \bh^{-1} ] \\
{}&{}
+ 2 \q \cg \Tr [ \tih^{-1}\del_+ \tih \, \del_- \bh \bh^{-1} ]\ . 
\end{aligned} \ee
Classically (i.e. for large $k$)
this model has the
expected $\F$ gauge symmetry
\begin{equation}\la{gsym}
\tih \to \tih \,  f \ , \qquad \bar h \to f^{-1}  \, \bar h \ , \qquad\qquad f=f(\s^+,\s^-) \in \F \ .
\end{equation}
Since the term proportional to $q$ in \rf{effectivecoset1}
 does not respect this gauge symmetry, then we
 should make the
 choice  $\q=0$ in the ambiguous Jacobian term in \rf{jac1}, 
corresponding to a natural scheme
preserving the gauge symmetry (this is the same choice as in the gWZW case \rf{ggr} above). Again applying the Polyakov-Wiegmann identity, we then obtain
\begin{equation}\begin{split}\la{effectivecoset2}
\L 
%& = {}&{} k \L_G (\tilde{g}) - (k+2\cg) \L_G ( \tih\bar{h} ) + k (\l^{-1}-1) \Tr [ \tih^{-1}\del_+ \tih \, P_{G/H}\, \del_- \bh \bh^{-1} ] \ ,
%\\
 &= k \L_G (\tilde{g}) - (k+2\cg) \big[\L_G ( \tih)
+ \L_G(\bar{h} )\big]\\
{}&{} \qquad \qquad \qquad  + \Tr [ \tih^{-1}\del_+ \tih\, \big(k+2\cg + k (\l^{-1}-1)  P_{G/H}\big)\, \del_- \bh \bh^{-1}] \ .
\end{split}\end{equation}
As in the group case, the first WZW term $k \L_G(\tilde{g})$ decouples,
leaving the RG flow  to be determined by the ``truncated'' $\s$-model,
\begin{align}
&\L'(\tih, \bh)
=- \tk \Big(\L_G ( \tih) + \L_G(\bar{h} ) - \Tr \Big[\tih^{-1}\del_+\tih\, (P_H + \tl \, P_{G/H})\, \del_-\bh\bh^{-1} \Big] \Big) \la{214} \ ,
% -\tl \Tr [ \tih^{-1}\del_+ \tih \, P_{G/H} \,\del_- \bh \bh^{-1}] \Big), 
\\
& \la{215}
\qquad \tk \equiv k + 2 \cg \ , \qquad \tl \equiv %1+ \frac{k {\g}}{k+2\cg}
\frac{k}{\tk}\l^{-1} +\frac{2 \cg}{\tk} \ .
%= \l^{-1} + \O (k^{-1} ) \ .
\end{align}
This  is a WZW model on  $G \times G$, deformed a left-right current interaction term, now containing projectors $P_\F$,$P_{G/\F}$ onto the subalgebra $\Lie(H)$ its orthogonal complement.
The classical integrability of the model \rf{214} (implying also the integrability of the $G/H$ $\l$-model)
was argued in \cite{Bardakci:1996gs} (example 4 there). It is also a special case of the integrable $G\times G/H$ models \cite{ABL} considered above in eqs.\ \eqref{2103couH},\eqref{2103bb1},\eqref{2103condH}, with the couplings and fields related according to
\be
\la{grel}
(k,r,s,u,t,b)=(-\tk, -\tk,-\tk,-\tk, \tk\tl,\tk\tl) \ , \qquad (g^{(1)},g^{(2)})=(\tih, \bh^{-1}) \ .
\ee

The  truncated model \rf{214}  is again automatically renormalizable due to manifest symmetries.
As well as the gauge symmetry \rf{gsym} and discrete symmetries,
it is also invariant under the same chiral gauge transformations \rf{210} as the group case. These symmetries protect the structure of \rf{214}, with $\tl$ and $\tk$ being the only free parameters.

We shall explicitly demonstrate the renormalizability of the symmetric space \lam \rf{ggu} by computing the 2-loop $\beta$-function for the truncated model \rf{214}.\foot{Like the group case above, the RG flow for the truncated model \rf{215} is a special case of the that of the $G\times G/H$ models in Chapter \ref{GGS} (given explicitly in App.\ \ref{2103GGHapp}). The results here are in agreement with those in App.\ \ref{2103GGHapp} for the special values of the parameters in \rf{grel}.}
The computation runs analogously to the group case above, with the only difference being that one must correctly handle the gauge invariance.
We decompose the generators $\{T_A\}$ of $G$
into the generators $\{T_\a\}$ of $H$ and the remaining generators $\{T_a\}$ chosen to be orthogonal to $\Lie{}(H)$ under the trace, i.e.\ splitting the algebra index of $G$ as $A=\{\a,a\}$ (see Appendix \ref{conv} for conventions). Since $G/H$ is a symmetric space then the $\mathbb{Z}_2$ grading of $\Lie(G)$ implies $f_{\a \b c} = f_{abc} =0$ (i.e. the only non-zero structure constants are $f_{
\a\b\g}$ and $f_{\a b c}$).
The tangent space index for $G \times G$ now splits as $M = \{ A; \bar{A} \} =\{\a, a; \bar{\a}, \bar{a} \}$, with the unbarred and barred indices corresponding respectively to the two $G$-valued fields $\hat h$ and $\bar{h}$.

Here we use the vielbein,
\be
E^M = (e^a, e^{\bar{a}}, e^\a, e^{\bar\a}) = (\sqrt{1-\tl^2} \RR^a, \, \LL^{\bar{a}} + \tl \RR^a, \,
\RR^\a + \LL^{\bar\a}, \, \RR^\a - \LL^{\bar \a}) \ . \la{vielbein2}
\ee
Up to permutations, the non-zero components of the metric and $\H$-tensor are
\begin{gather}
\GG_{ab} = \GG_{\bar{a}\bar{b}} = \tfrac{\tilde k}{2} \, \Tr[T_a T_b] \ , \qquad \GG_{\a\b} = \tfrac{\tilde k}{2}  \, \Tr[T_\a T_\b] \ ,
\\
\H_{\a\b\g} = -i \tfrac{\tilde k}{2} f_{\a\b\g} \ , \qquad \H_{\a bc} = - i \tfrac{\tilde k}{2} f_{\a bc} \ , \qquad \H_{\a \bar{b} \bar{c}} = - i \tfrac{\tilde k}{2} f_{\a bc} \ . \la{Hcoset}
\end{gather}
The $\F$ gauge symmetry of the model \rf{214} leads to the 
vanishing of all $\bar{\a}$-components of the metric and the $\H$-tensor.

As discussed in App.\ \ref{2103GGHapp},
there are various possible approaches  to treat the $H$ gauge symmetry.
Here we shall use the trick of lifting the degeneracy of the metric with a small parameter $\e$ acting as a ``regulator'',
\be \GG_{\bar \a \bar \b} = \e \, \tfrac{\tilde k}{2} \, \Tr[ T_\a T_\b] \ , \ee
computing the (torsionful) Riemann tensor, projecting out the $\bar{\a}$ directions and setting the regulator $\e$ to zero. This is a short-cut for a proper gauge-fixing procedure explained in footnote \ref{2103fgs}.

The resulting torsionful curvature has
the non-zero components 
\begin{equation}\begin{split}\la{torsionriem2}
(\tl^{-2}-1){\widehat{R}^a}{}_{bde} =
-\sqrt{\tl^{-2}-1}{\widehat{R}^a}{}_{b d \bar{e} }
= {\widehat R}^a{}_{b \bar{d} \bar{e} } = - f^a{}_{b\g} f^{\g}{}_{de} \ .
\end{split}\end{equation}
Substituting this
into the RG equation \eqref{BMT}, one obtains
\begin{equation}\la{2202}
{\big[ \ddt(\GG_{mn} + \BB_{mn})\big]} \, dx^m \otimes dx^n = \sum^{{\rm dim}\, G - {\rm dim}\, \F}_{i=\bar \imath=1}\Big[ \cg \tl + \frac{\cg}{\tk} \frac{ \tl(\cf - (2\cg - \cf)\tl^2)}{1-\tl^2} \Big] \, \Tr[\RR \,  \otimes \, P_{G/H}  \LL ] \ .
\end{equation}

We conclude that only the $\tl$-dependent term in \rf{214} gets renormalized, i.e.\
the 2-loop RG equation \eqref{BMT} is solved with $\xi_m=\l_m=0$ and
\be \la{02}
\ddt \tk =0 \ , \qquad \qquad \ddt \tl = \frac{\cg \tl}{\tk} \Big[ 1 + \frac{1}{\tk} \frac{ \cf - (2\cg - \cf)\tl^2}{1-\tl^2}\Big] \ .
\ee
The 1-loop term here agrees with the standard results for the symmetric space $\l$-model (eq.\ \eqref{la2} with $\ka = \frac{\tl-1}{\tl+1}$) \cite{Sfetsos:2014jfa,Appadu:2015nfa}.
The level $k$ is RG-invariant, and, as expected, $\tl =\infty$ and $\tl=0$ are fixed points.
At the fixed point $\tl = \infty$ ($\l=0$) the \lam reduces to the $G/\F$ gWZW model.
The other fixed point $\tl=0$ ($\l=-\tfrac{k}{2\cg}$) is a quantum-corrected version of the 1-loop fixed point $\l=\infty$; this model is like a version of the $G/G$ cWZW model with an additional $H$ gauge symmetry, and after integrating out $A_\pm$
it corresponds to   the $G/ H$ gWZW model with level $-k$.

Expressing $\tk$ and $\tl$ in \rf{215} in terms of $k$ and the coupling
$\ka$ using \rf{133}, we find
\begin{align}
&\ddt k =0 \ , \la{6d2loopbetacoset}
\qquad\qquad
\ddt \ka = \frac{\cg}{2k} \Big[ (1-\ka^2) + \frac{\cg(1-\ka)^2(1+4\ka-\ka^2) - \cf(1-\ka^4)}{2k\ka} \Big] \ .
\end{align}

As in the group space case \rf{24}, the 2-loop term in the
$\b$-function \rf{02} is, in general, scheme-dependent.
Again, we find that the symmetry under $k \to - k$, $\l \to \l^{-1}$ in \eqref{z2sym},
present at the 1-loop order is absent in the 2-loop term of \rf{02}.
However, after introducing the shifted level $\kk = \tk- \cg = k + \cg$ as in \rf{2777}, the 2-loop RG equation for $\tl$ \rf{02} becomes
\be
\ddt \tl = \frac{\cg \tl}{\kk} \Big[ 1 - \frac{1}{\kk} (\cg-\cf)
\frac{1+\tl^2}{1-\tl^2} \Big] \ ,
\ee
which is manifestly invariant under $\kk \to -\kk$, $\tl \to \tl^{-1}$, i.e.
a quantum version of the symmetry \eqref{z2sym} of the original couplings (cf. \rf{227}).

In the NAD limit \eqref{limr}, when the extended-space $\l$-model becomes the interpolating model \eqref{NADi} for NAD of the symmetric space, we get from \rf{6d2loopbetacoset}
\be
\la{555}
\ddt \hh = 2 {\cg} + 4{\cg (\cg - \cf)}{\hh}^{-1} \ .
\ee
As verified in Appendix B of \cite{HLT2}, equation \rf{555} is the correct 2-loop $\beta$-function for the $G/\F$ symmetric space $\s$-model (matching eq.\ \eqref{sb} above and, in particular cases, the results in \cite{Brezin:1975sq,Hikami:1980hi}). Like the group case above, the 2-loop coefficient here is also scheme-independent in this single-coupling limit, providing a non-trivial, scheme-invariant test of the $\l$-model's 2-loop $\b$-function.

%%%%%%%%%%%%%%%%%%%%%%%%%%%%%%%%%%%%%%%%
\section{Low-dimensional examples: \texorpdfstring{$\l$}{lambda}-models for \texorpdfstring{$SU(2)$}{SU(2)} and
\texorpdfstring{$SU(2)/U(1)$}{SU(2)/U(1)}\label{lowd}}
%%%%%%%%%%%%%%%%%%%%%%%%%%%%%%%%%%%%%%%%
Let us study the simplest case of the truncated model \rf{6dmodel} in the group case with $G = SU(2)$.
Parametrizing
\begin{equation}\la{hhbpar}
\tih = \exp(i\phi \sigma_3) \exp(i \theta \sigma_1) \exp(i \psi \sigma_3) \ , \qquad\quad
\bh = \exp(i\bar \phi \sigma_3) \exp(i \bar \theta \sigma_1) \exp(i \bar \psi \sigma_3) \ ,
\end{equation}
where $\sigma_A$ are the standard Pauli matrices, we find the following 6-dimensional metric and $\BB$-field (with $\BB = \ha \BB_{mn} dx^m \wedge dx^n$)
\begin{align}
&\qquad \qquad ds^2  = (ds^2)_{G\times G} + (ds^2)_{\rm int} \ , \qquad \quad  \BB = \BB_{G\times G} +  \BB_{\rm int} \ , \la{su2geom}
\\ \no
&(ds^2)_{G\times G}   = - \tk (d\theta^2 + d\phi^2 + d\psi^2 + 2\cos 2\theta d\phi d\psi +
d\bar\theta^2 + d\bar\phi^2 + d\bar\psi^2 + 2\cos 2\bar\theta d\bar\phi d\bar\psi) \ ,
\\ \no
&\BB_{G\times G}   = - \tk(\cos2\theta d\phi\wedge d\psi + \cos 2\bar \theta d\bar \phi \wedge d\bar \psi) \ ,
\\ \no
&(ds^2)_{\rm int}  = - 2\tk\tl \big[\cos 2(\psi+\bar\phi) (d\theta d\bar \theta - \sin 2\theta \sin 2\bar \theta d \phi d\bar \psi) + ( d\psi +\cos2\theta d\phi)(d\bar \phi + \cos 2\bar \theta d \bar \psi)
\\ \no & \qquad  \qquad \qquad \qquad + \sin 2(\psi+\bar \phi)(\sin 2 \theta d\phi d \bar \theta + \sin 2 \bar \theta d \theta d \bar \psi)\big] \ ,
\\ \no
&\BB_{\rm int}  = -\tk\tl \big[\cos 2(\psi+\bar\phi) (d\theta \wedge d\bar \theta - \sin 2\theta \sin 2\bar \theta d \phi \wedge d\bar \psi) + ( d\psi +\cos2\theta d\phi)\wedge(d\bar \phi + \cos 2\bar \theta d \bar \psi)
\\ & \qquad \qquad \qquad \qquad + \sin 2(\psi+\bar \phi)(\sin 2 \theta d\phi \wedge d \bar \theta + \sin 2 \bar \theta d \theta \wedge d \bar \psi)\big] \ . \no
\end{align}
Here the ``$G\times G$'' terms come from the WZW models and the ``int'' terms come from the current-current interactions between them.
One can check explicitly that this metric and $\BB$-field solve the 2-loop RG equation \eqref{BMT} with vanishing $\xi^m$ and $\l_m$, and the couplings running as in \rf{01} (with the dual Coxeter number $c_{_{SU(2)}} = 2$).

Next, let us consider the truncated model \rf{214} obtained from the $\l$-model  based on the symmetric space $G/\F = SU(2)/ U(1)$.
Using the parametrization for $\tih$ and $\bar{h}$ given in eq.\ \rf{hhbpar},
we find that the corresponding
target space background depends on $\psi$ and $\bar\phi$ only
through the combination $\chi = \psi + \bar \phi$, which is a manifestation of the $U(1)$ gauge symmetry.
The resulting 5-dimensional metric and $\BB$-field are
\begin{align}
&\qquad \qquad ds^2  = (ds^2)_{G\times G/H} +(ds^2)_{\rm int} \ , \qquad \qquad \BB = \BB_{G\times G/H} + \BB_{\rm int} \ ,
\\ \no
&(ds^2)_{G\times G/H}  = - \tk (d\theta^2 + d\phi^2 + d\bar\theta^2 + d\bar\psi^2 ) \ , \qquad \BB_{G\times G/H} = 0 \ ,
\\ \no
&(ds^2)_{\rm int}  = - 2\tk\tl \big[\cos 2\chi (d\theta d\bar \theta - \sin 2\theta \sin 2\bar \theta d \phi d\bar \psi)
+ \sin 2\chi(\sin 2 \theta d\phi d \bar \theta + \sin 2 \bar \theta d \theta d \bar \psi)\big]
\\ \no & \qquad  \qquad \qquad \qquad + \tk d\chi^2 -2\tk( d\chi +\cos2\theta d\phi)(d \chi + \cos 2\bar \theta d \bar \psi)
\ ,
\\ \no
&\BB_{\rm int} = -\tk\tl \big[\cos 2\chi (d\theta \wedge d\bar \theta - \sin 2\theta \sin 2\bar \theta d \phi \wedge d\bar \psi)
+ \sin 2\chi(\sin 2 \theta d\phi \wedge d \bar \theta + \sin 2 \bar \theta d \theta \wedge d \bar \psi)\big]
\\ & \qquad  \qquad \qquad \qquad -\tk( d\chi +\cos2\theta d\phi)\wedge(d\chi + \cos 2\bar \theta d \bar \psi)\ .\no
\end{align}
Again one can check explicitly that this metric and $\BB$-field solve the 2-loop RG equation
\eqref{BMT} with vanishing $\xi^m$ and $\l_m$, and the couplings running as in \rf{02}
(with the dual Coxeter numbers  $c_{_{SU(2)}}=2$ and $c_{_{U(1)}}=0$).

\section{``Tripled'' formulations of other models \la{TO}}
In principle the trick of rewriting 2d gauge fields in terms of group-valued scalars can be applied to any theory with 2d gauge fields. In this section, containing unpublished work, we shall construct analogous ``tripled'' or ``extended'' formulations of two other classes of $\s$-models.

\subsection{\texorpdfstring{$G^N$}{G^N} \texorpdfstring{$\l$}{lambda}-models}
First let us consider a  generalization of the $\l$-model to $N$ copies of the group $G$, which was recently obtained from affine Gaudin models \cite{Bassi:2019aaf},
\be
\L = k_i  \, \L_{G/G}(g^{(i)}, A^{(i)}) - \ha \,  \g_{ij} \, \Tr[ A^{(i)}_+ A^{(j)}_- ] \ ,  \qquad \qquad i=1,\ldots n \ .\la{ln}
\ee
Here $k_i$ are the WZ levels and  $\g_{ij}$ is $2\times 2$ matrix of coupling constants. Integrating out the quadratic gauge fields $A_\pm^{(i)}$ yields a $\s$-model with scalar fields $g^{(i)} \in G$.

This may be viewed as an example of  Tseytlin's ``universal class'' model \eqref{uc} adapted to $G^N$, with the constant operator $Q$ chosen to preserve the  global  $G$  symmetry $(g^{(i)}, A^{(i)}) \to (u g^{(i)} u^{-1},u A^{(i)} u^{-1})$, $u\in G$. Like the $G^N$ models discussed in Chapter \ref{GGS}, the theory \rf{ln} is classically integrable only for special values of the coupling constants $(k_i,\g_{ij})$, parametrized algebraically in terms of the twist function in \cite{Bassi:2019aaf}. Although we shall not work out the details here, this parametrization may be recast as a set of polynomial equations in $(k_i,\g_{ij})$ (cf. the equation \eqref{2103cond} for undeformed $G\times G$ models).

Like standard $\l$-models, the $G^N$ $\l$-model \rf{ln} is a deformation of a WZW model (when $\g_{ij}$ is taken to infinity while being non-degenerate)
and becomes in a particular limit, $k_i \to \infty$, $g^{(i)} = e^{-\tfrac{v^{(i)}}{2k_i}}$, the non-abelian dual of a particular $G^N$ model \eqref{2103cou} in its $(G_L)^N$ symmetry,
\be
\L = - \ha  \, \Tr[ \g_{ij}  A^{(i)}_+ A^{(j)}_-   -  v^{(i)} F_{+-}(A^{(i)})] \ , \qquad F_{+-}(A) \equiv  \del_+ A_- - \del_- A_+ + [A_+, A_-] \ .
\ee

Our observation is that the tripled formulation of the $\l$-model generalizes very easily to this $G^N$  case. Let us perform the same change of variables $(A^{(i)}_+,A^{(i)}_-) \to (h^{(i)}, \bh^{(i)})$ on each copy of the group, where  $A^{(i)}_+= h^{(i)} \del_+ \big(h^{(i)}\big)^{-1} $, $A^{(i)}_- = \bh^{(i)} \del_- \big(\bh^{(i)}\big)^{-1}$ (cf.\ eq.\ \rf{cov}). Including the Jacobian \rf{jac1} for this transformation (and picking again $\q=-1$), the resulting theory is a ``tripled'' $\s$-model with fields $(g^{(i)},h^{(i)},\bh^{(i)}) \in G^{3N}$,
\be \begin{aligned} \la{ne}
&\L = k_i  \,\L_G(\tilde{g}^{(i)}) + \L'(\hat{h}^{(i)}, \bh^{(i)}) \ , \qquad \qquad \tilde{g}^{(i)} \equiv\big( h^{(i)}\big)^{-1} g^{(i)} \bh^{(i)} \ , \quad \hat{h}^{(i)}\equiv \big(h^{(i)}\big) \ , \\
&\L' = -(k_i+2\cg)\big[ \L_G(\hat{h}^{(i)}) + \L_G(\bh^{(i)}) \big] + \Tr\big[ (k_i  \d_{ij} -\ha \g_{ij}) \, J_+^{(i)} \, \bar{K}_-^{(j)} \big] \ , \\
&\qquad J_+^{(i)} \equiv \big(\hat{h}^{(i)} \big)^{-1} \del_+ \hat{h}^{(i)} \ , \qquad \bar{K}_-^{(i)} \equiv \del_+ \bh^{(i)} \big( \bh^{(i)} \big)^{-1} \ . \la{trin}
\end{aligned} \ee
This is a particular case of the undeformed $G^M$ model  \eqref{2103cou} with $M=3N$ (after a redefinition $\bh^{(i)} \to \big( \bh^{(i)}\big)^{-1} $).

As in the $\l$-model considered above, the first WZW terms $k_i \, \L_G(\tilde{g}^{(i)})$ in \rf{ne} decouple, leaving the RG flow to be determined by the ``truncated'' Lagrangian $\L'$, which is a $\s$-model with fields $(h^{(i)},\bh^{(i)}) \in G^{2N}$. This truncated theory is also a particular case of  the undeformed  $G^M$ model  \eqref{2103cou} with $M=2N$.

Hence we have produced a quantum equivalence between (i) the $\l$-deformation \rf{ln} based on $G^N$ models  and (ii) certain $G^M$ models with $M=2N$ (plus $N$ decoupled WZW models). In effect, moving from the $G^N$ model to the related $\l$-deformation is equivalent to replacing $N\to 2N$, restricting to particular class of the $G^{2N}$ model's couplings, and adding $N$ decoupled WZW models.

\bigskip 
Let us make some observations. As mentioned above, the $G^N$ $\l$-model is only integrable for special values of the couplings solving polynomial equations. This is indeed consistent with the equivalence found here, since the same is true for the undeformed $G^M$ models of which the tripled formulation \rf{trin} is a special case. Hence one should be able to understand the integrability conditions for $G^N$ $\l$-models in terms of those of undeformed $G^M$ models; in this respect the $\l$-deformation does not introduce any new elements. It would be interesting to understand the relation between these models at the level of affine Gaudin models and the twist functions.

Renormalizability of the $G^N$ $\l$-model  \rf{ln} does not seem obvious based alone on its manifest $G$ symmetry, which  is not  sufficient to protect the form of the action. However, in the tripled formulation \rf{trin}, there is again a decoupling of some fields $\tilde{g}^i$ and the remaining  truncated theory $\L'$ is protected by $G(\xi^-)^N \times G(\xi^+)^N \times G$ chiral gauge symmetry,
\be
\begin{aligned}
&(\hat{h}^{(i)}, \ \bh^{(i)}) \to (u^{(i)}(\xi^-) \,  \hat{h}^{(i)}  \, w,  \   w^{-1} \, \bh^{(i)}  \, v^{(i)}(\xi^+))  \ ,\\
&u^{(i)}(\xi^-) \in G(\xi^-) \ , \quad v^{(i)}(\xi^+) \in G(\xi^+) \ , \quad w\in G \ ,
\end{aligned}
\ee
which uniquely fixes the theory up to the parameters $(k_i, \g_{ij})$.  Hence it will be renormalizable to all orders with just $\g_{ij}$ running (while the WZ levels do not run for general reasons).  Since the original and tripled formulations are equivalent at the quantum level then we have proved  the renormalizability of the $G^N$ $\l$-model \rf{ln} with  unintegrated gauge field  to all orders.

So far we have established the renormalizability of the  $G^N$ $\l$-model \rf{ln} before imposing the integrability condition. We expect the ``integrable surface'' in the space of couplings $(k_i,\g_{ij})$ to be a stable  subspace under the RG flow (cf. the $G\times G$ and $G\times G/H$ models studied in Chapter \ref{GGS}). It would be interesting to check this fact and to see whether the integrability conditions require quantum corrections to remain stable beyond 1-loop. The tripled formulation \rf{trin} is the natural arena to study these questions since it is already RG-stable with a finite set of couplings  before imposing the integrability condition.

\subsection{\texorpdfstring{$G\times G/H$}{GxG/H}  models}
Now let us consider the $H$-gauge-invariant  $G\times G/H$ models considered in Section \ref{2103gu}, with Lagrangian,
\be \begin{aligned}
&\L = -\ha \, \rho_{ij} \, \Tr[ P^{(i)}_+ P^{(j)}_- ] -\ha \, r_{ij} \, \Tr[ \B^{(i)}_+ \B^{(j)}_- ] + k_i  \, \L_{_{\rm WZ}}(g^{(i)}) \ ,  \\
&
 k_1 = - k_2\equiv k  \ , \qquad  r_{ij} = \begin{pmatrix} r  & - r - k \\ -r +k & r  \end{pmatrix} \ , \\
& P^{(i)} = P_{{G/H}}  J^{(i)} \ , \qquad   \B^{(i)} = P_H J^{(i)}\ ,  \ \ \qquad 
 J^{(i)} =  {g^{(i)}}^{-1} d g^{(i)}  \ . % \quad  (g^{(i)}) \in G^N  \ ,
  %\ \  i=1,2 \ , 
  \la{ghr}
\end{aligned} \ee
First let us re-write the theory in terms of a 2d gauge field by starting with an ungauged ($G\times G$)-type model (here $R_{ij}$ are arbitrary constants and $k_1=-k_2=-k$),
\be
\L = -\ha \, \rho_{ij} \, \Tr[ P^{(i)}_+ P^{(j)}_- ] -\ha \, R_{ij} \, \Tr[ \B^{(i)}_+ \B^{(j)}_- ] + k  \, \big( \L_{_{\rm WZ}}(g^{(1)})- \L_{_{\rm WZ}}(g^{(2)}) \big) \ ,
\ee
and gauging the $H$  global symmetry $(g^{(1)}, g^{(2)}) \to (g^{(1)} h, g^{(2)} h)$, $h\in H$ by the same gauging procedure discussed for coset spaces in Section \ref{sssm}. Promoting $h \to h(\xi)$ to a local parameter, the  Lagrangian transforms as
\be\begin{aligned}
&\L \to -\ha \, \rho_{ij} \, \Tr[ P^{(i)}_+ P^{(j)}_- ] + k  \, \big( \L_{_{\rm WZ}}(g^{(1)})- \L_{_{\rm WZ}}(g^{(2)}) \big) \\
&\qquad   -\ha \, \Tr[ R_{ij} \, \B^{(i)}_+ \B^{(j)}_-  + \del_+ h h^{-1} (\a \,  B_-^{(1)} +\b \, B_-^{(2)}) 
\\
&\qquad  \qquad\qquad  \qquad\qquad  \qquad \quad 
+ \del_- h h^{-1} (\g  \, B_+^{(1)} +\d  \, B_+^{(2)} ) + \e \, \del_+ h h^{-1} \, \del_- h h^{-1} ] , \\
&\a = R_{11} + R_{21} - k \ , \quad \b = R_{12}+R_{22}+k \ , \\
&\g = R_{11} + R_{12} + k \ , \quad \d = R_{21}+R_{22}-k \ , \quad \e = R_{11}+R_{12}+R_{21}+R_{22} \ . \la{prev}
\end{aligned}\ee
Now promoting $\del_+ h h^{-1} \to - A_\pm$ in the transformed Lagrangian \rf{prev}, the resulting theory is gauge invariant (under $g \to h^{-1} g h$, $A_\pm \to h^{-1} A_\pm h + h^{-1} \del_\pm h$,\ \, $h(\xi)\in H$),
\begin{align}
\L(g,A) = {}&{}-\ha \, \rho_{ij} \, \Tr[ P^{(i)}_+ P^{(j)}_- ] + k  \, \big( \L_{_{\rm WZ}}(g^{(1)})- \L_{_{\rm WZ}}(g^{(2)}) \big)   \la{Am} \\
&   -\ha \, \Tr[ R_{ij} \, \B^{(i)}_+ \B^{(j)}_-  - A_+ (\a \,  B_-^{(1)} +\b \, B_-^{(2)}) 
%\\
%&  \qquad\quad  
-  A_- (\g  \, B_+^{(1)} +\d  \, B_+^{(2)} ) + \e \, A_+ A_- ] \ . \no
\end{align}
Upon integrating out the 2d gauge field $A_\pm$, one indeed obtains the desired  $G\times G/H$ model  \rf{ghr} with the following identification of parameters
\be
r = \frac{k^2 + k (R_{12}-R_{21}) + R_{11} R_{22} - R_{12} R_{21}}{R_{11}+R_{12}+R_{21}+R_{22}} \ , \qquad k=k \ . \la{ip}
\ee

Hence  the model \rf{Am} with the gauge field is equivalent to the $G\times G/H$ model \rf{ghr} and, moreover, only one combination of the couplings $R_{ij}$ appears after integrating out $A_\pm$. This means one is free to fix any 3 components (or combinations of components) of $R_{ij}$ to arbitrary values, as long as  $R_{11}+R_{12}+R_{21}+R_{22} \neq 0$ so that \rf{ip} does not diverge. 

For general $R_{ij}$ we may again change variables $(A_+, A_-) \to (h,\bh)$ according to \rf{cov} (this is like replacing $\del_- h h^{-1} \to \del_- \bh \bh^{-1}$ in \rf{prev}). The resulting theory is a $\s$-model with an enlarged configuration space $(g^{(1)}, g^{(2)}, h, \bh) \in G\times G \times H \times H$,
\be \begin{aligned}
\L(g^{(1)},g^{(2)},h,\bh) = {}&{}-\ha \, \rho_{ij} \, \Tr[ P^{(i)}_+ P^{(j)}_- ] + k  \, \big( \L_{_{\rm WZ}}(g^{(1)})- \L_{_{\rm WZ}}(g^{(2)}) \big) \\
&   -\ha \, \Tr[ R_{ij} \, \B^{(i)}_+ \B^{(j)}_-  + \del_+ h h^{-1} (\a \,  B_-^{(1)} +\b \, B_-^{(2)}) \\
&  \qquad\qquad\qquad\quad  + \del_- \bh \bh^{-1}  (\g  \, B_+^{(1)} +\d  \, B_+^{(2)} ) + \e \, \del_+ h h^{-1}  \del_- \bh \bh^{-1} ] \ .  \la{AM}
\end{aligned}\ee
One may fix the gauge symmetry as, e.g., $\bh=1$ (or $A_-=0$) to obtain
\be \begin{aligned}
&\L(g^{(1)},g^{(2)},h) = -\ha \, \rho_{ij} \, \Tr[ P^{(i)}_+ P^{(j)}_- ] + k  \, \big( \L_{_{\rm WZ}}(g^{(1)})- \L_{_{\rm WZ}}(g^{(2)}) \big) \\
& \qquad \qquad \qquad \qquad \qquad \quad   -\ha \, \Tr[ R_{ij} \, \B^{(i)}_+ \B^{(j)}_-  + \del_+ h h^{-1} (\a \,  B_-^{(1)} +\b \, B_-^{(2)}) ] \ , \la{gff} \\
&\a = R_{11} + R_{21} - k \ , \qquad \b = R_{12}+R_{22}+k \ . 
\end{aligned}\ee
The resulting gauge-fixed model \rf{gff} is a $\s$-model with configuration space $(g^{(1)}, g^{(2)}, h) \in G \times G \times H$.  It may be described as a particular ``squashed'' $G\times G\times H$ model, since the $G\times G$ sector includes dependence on the projectors $P_H$ and $P_{G/H}$.

One simple example of \rf{gff} obtained from the choice
\be
R_{ij} = \begin{pmatrix}
r & 0 \\
-r+k & 0 
\end{pmatrix} \ , \la{Rc}
\ee
is given by
\be\begin{aligned}
\L(g^{(1)},g^{(2)},h) ={}&{}-\ha \, \rho_{ij} \, \Tr[ P^{(i)}_+ P^{(j)}_- ] + k  \, \big( \L_{_{\rm WZ}}(g^{(1)})- \L_{_{\rm WZ}}(g^{(2)}) \big) \\
&   -\ha \, \Tr[ (r \B^{(1)}_+ + (k-r) \B^{(2)}_+ )\B^{(1)}_-  + k \, \del_+ h h^{-1} B_-^{(2)}  ] \ . \la{gfm}
\end{aligned}\ee

Another natural choice consistent with \rf{ip} is 
\be
R_{ij} = \eps \,  \rho_{ij} \  , \qquad r = \frac{k^2 \e^{-1} + k  (R_{12}-R_{21}) + \e (R_{11} R_{22} - R_{12} R_{21})}{R_{11}+R_{12}+R_{21}+R_{22}}  \ .
\ee
Then \rf{gff} becomes 
\be\begin{aligned}
&\L(g^{(1)},g^{(2)},h) = -\ha \, \rho_{ij} \, \Tr[ \e B^{(i)}_+ B^{(j)} + P^{(i)}_+ P^{(j)}_- ] + k  \, \big( \L_{_{\rm WZ}}(g^{(1)})- \L_{_{\rm WZ}}(g^{(2)}) \big) \\
& \qquad \qquad \qquad \quad   -\ha  \, \Tr[  \del_+ h h^{-1}  \, \big(\a \,  B_-^{(1)} +\b \, B_-^{(2)}\big) ] \ ,  \la{gfm2} \\
&\a = \e( \r_{11} + \r_{21}) - k \ , \qquad \b = \e(\r_{12}+\r_{22}) +k \ . 
\end{aligned}\ee

These extended-space reformulations of the $G\times G/H$ models will inherit the same classical integrability and RG flow properties as the original theory (modulo the inclusion of a Jacobian of the same form as \rf{jac1}). Hence we learn that the extended $G\times G \times H$ models \rf{gff} (and the particular cases \rf{gfm} and \rf{gfm2}) should also be integrable when the couplings $r,k,\rho_{ij}$ satisfy the integrability conditions \eqref{2103condH} of the $G\times G/H$ model.
Speculatively, one might hope that, after imposing the integrability conditions, there could be a decoupling of some fields (as for the $\l$-models studied above), which may leave the remaining theory  manifestly renormalizable due to its global symmetries. Such a phenomenon would be interesting as (i) it could provide  another origin for the integrability conditions \eqref{2103condH} of the $G\times G /H$ model and (ii) in such a reformulation, the $G\times G/H$ model's integrability condition would become automatically stable at higher-loop orders (cf. eq. \eqref{2103318} in the standard formulation).

%%%%%%%%%%%%%%%%%%%%%%%%%%%%%%%%%%%%%%%%
\section{Discussion}\la{conclusions}
%%%%%%%%%%%%%%%%%%%%%%%%%%%%%%%%%%%%%%%%

Formulating the \lam on an extended
($G\times G\times G$)
configuration space
``linearizes'' the RG flow, i.e.\ makes its renormalizability manifest without the need
for extra local counterterms apart from the running of the coupling $\l$. 
This implies that the $\l$-model with {unintegrated} gauge field is renormalizable to all orders.

 As well as eliminating  the need for quantum corrections to remain renormalizable at higher-loop orders,  the tripled formulation also explains  the $\l$-model's renormalizability already at the 1-loop order. Whereas the standard formulation \rf{1} is apparently protected under the RG flow by \textit{hidden} symmetries,  the tripled formulation is protected by \textit{manifest} symmetries.
 
The decoupling of one of the WZW models $k \L_G(\tilde g)$ in \rf{277} and \rf{effectivecoset2} appears to be the special property that renders the manifest global symmetries sufficient for renormalizability of the remaining ``truncated'' models. This property should be associated with integrability, since it would be lost under generic deformations of the $\l$-model \rf{1} preserving its manifest symmetries but breaking integrability. It would be interesting in future to understand  the precise relationship between integrability and this special decoupling in the tripled formulation.

\

In the non-abelian dual limit \eqref{limr}, the resulting scheme-invariant 2-loop $\b$-functions  matched the known $\b$-functions for the PCM and symmetric space. One may regard this tripled computation as a particular scheme for  the renormalization of  the interpolating models \eqref{NADi} for non-abelian duality (obtained from the tripled $\l$-model in this limit). One might worry that the NAD limit could be too singular to commute with the RG flow (e.g., the decoupling of the $\L_G(\tilde g)$ term in \rf{277} breaks down in that limit),
but  this is not likely since one can independently argue the renormalizability of the interpolating models  \eqref{NADi}
 through their relation to the original PCM/symmetric space models
by 
 an integral over the linear Lagrange multiplier $v$.
 This integral is not summing any loops as $v$ only appears linearly, so the classical result $\delta(F_{+-}(A))$ is expected to be consistent with the RG flow, without needing any extra counterterms.\foot{I thank A.A. Tseytlin for a discussion of this point.} It would be interesting in future to check this fact directly, e.g.\  computing the RG flow for a ``tripled'' formulation of the interpolating models for NAD.
 
 If one regards the  interpolating models, or the corresponding tripled $\s$-models, as  particular formulations of the corresponding NAD theories at the quantum level,
 then 
it is natural to conclude that non-abelian duality  commutes
with the RG flow beyond the 1-loop order on arbitrary  groups $G$  and symmetric spaces $G/H$,
with the original and NAD models having the same 2-loop $\beta$-function. This is in line with  previous 1-loop conclusions \cite{Fridling:1983ha,Fradkin:1984ai}  and resolves related confusions at the 2-loop level \cite{n1,n2,n3,Bonneau:2001za}.

\

Hence, for  (i) the $\l$-models with unintegrated gauge field and (ii) the  interpolating models for NAD,
  the non-trivial corrections that were required in the standard configuration space in Chapter \ref{QCS} should be
understood as finite counterterms resulting from the integral over the 2d gauge field appearing quadratically in the action.
It remains to understand the general structure of those counterterms, or equivalently the 
correct prescription for the 2d gauge field integral consistent with integrability
(generalizing \eqref{7}, which was sufficient only in the simple $SU(2) \ov U(1)$ example).

 The 2-loop $\b$-functions \rf{24} and \rf{6d2loopbetacoset} obtained through the tripled formulation are consistent with those found for
particular examples of the $\l$-model in the standard formulation in eqs.\ \eqref{49} and \eqref{19100}, modulo a redefinition of the couplings $k,\l$, i.e.\ scheme change (see \cite{HLT2} for the precise redefinitions that relate them). In particular this means that the
scheme-invariant 1-loop terms match exactly; moreover the 2-loop terms also match in the single-coupling  NAD limit when they become scheme-invariant.

\

\sloppy
A similar ``tripled'' approach may also be useful
for clarifying the higher-loop deformation in abelian T-duality.
For example, consider the metric \mbox{$ds^2 = dy^2 + a(y) dx^2$} and its classical dual
\mbox{$\wt{ds^2} = dy^2 + a(y)^{-1} d\td x^2$,}
with the interpolating model given by
\be
-\L= (\del_\m y)^2 + a(y) (p_\m)^2 + \td x \, \e^{\m\n}  \del_\m p_\n \ ,
\ee
such that $y$ is a spectator coordinate (cf.\ eq.\ \eqref{int}).
If we integrate out $\td x$ to give $p_\m = \del_\m x$, we recover the
original model for $x$.
If we integrate out $p_\m$, we find the T-dual model for $\td x$.
If instead we set $p_1=\del_1 x $ and integrate out $p_0$ we get the
``doubled'' model of \cite{Tseytlin:1990va} for $x$ and $\td x$ (equivalent
to the ``axial'' gauge choice in the Appendix of \cite{Rocek:1997hi}).
The  ``tripled'' model for ($x, \td x, \bar x$), analogous to the ones discussed above, is obtained by setting \mbox{$p_\m= \del_\m x + \e_{\m\n}\del^\n \bar x$:}
\be
-\L =  (\del_\m y)^2 + a(y) [ (\del_\m x)^2 - (\del_\m \bar x)^2 + 2 \e^{\m\n} \del_\m x \del_\n \bar x ] + \del_\m \td x \del^\m \bar x  \ .
\ee
This may be interpreted as a \sm on 4-dimensional target space with pp-wave metric and $\BB$-field.

\

We also considered similar ``tripled'' or ``extended'' formulations for other models.  
We found that the tripled formulation of the $\l$-model generalizes easily to $\l$-models based on $G^N$. In this tripled formulation, it is clear that the $G^N$ $\l$-models are renormalizable to all orders due to manifest symmetries (before imposing the integrability condition). 
The resulting tripled $\s$-models are particular class of $G^{3N}$ models, or  $G^{2N}$ models after removing decoupled WZW models. 
 Hence in this case we find a correspondence between the integrability and RG flow properties of $G^N$ $\l$-models and certain undeformed $G^{2N}$ models.

 We reformulated the $G\times G/H$ models studied in Chapter \ref{GGS} in terms of a gauge field and  found a 3-parameter family of extended formulations (3 of the components of $R_{ij}$, with the 4th being fixed by the relation \rf{ip}). The resulting $\s$-model has configuration space $G\times G\times H \times H$, or $G\times G\times H$ after fixing the $H$ gauge symmetry. Hence we find a correspondence between the integrability and RG properties of $G\times G/H$ models and certain  ``squashed'' $G\times G \times H$ models.  It would be interesting to study the RG flow for both   $G^N$ $\l$-models and $G\times G/H$ models in these new, extended formulations.

A useful extension of the tripled formulation  would be to the $\eta$-model. 
Up to analytic continuation, the $\eta$-model and \lam are related by limits and T-duality \cite{Hoare:2015gda,Hoare:2017ukq}
or by Poisson-Lie duality \cite{Hoare:2015gda,Klimcik:2015gba,Vicedo:2015pna,Hoare:2017ukq},
so it seems plausible that the tripled formulation of the $\l$-model could lead to a similar approach for the $\eta$-model.
This could allow a generalizaton of our results for the $SU(2) \ov U(1)$ $\eta$-model in Section \ref{Seta} to general groups and symmetric spaces.

Another obvious generalization would be  to $\s$-models on supergroups and
supercosets (see, e.g.,
\cite{Kagan:2005wt,Babichenko:2006uc,Zarembo:2010sg,Appadu:2015nfa}), and to integrable-deformed superstring $\s$-models.
See Section \ref{stric} for further discussion of this possibility.

%% file: 2008.tex
%\documentclass[12pt,a4paper]{article}
%\pdfoutput=1
%\usepackage[english]{babel}
%\usepackage[latin1]{inputenc}
%\usepackage[T1]{fontenc}
%\usepackage{amsfonts,amsbsy,bm,euscript,mathrsfs}
%\usepackage{amssymb,faktor,slashed}
%\usepackage{color}
%\usepackage[tbtags]{amsmath}
%\usepackage[bookmarks=true,colorlinks=true,linkcolor=black,citecolor=black,urlcolor=black,bookmarksnumbered,linktocpage=true]{hyperref}
%\usepackage{graphicx}
%\usepackage{chngcntr}
%\usepackage{makecell}
%\usepackage{mathtools}

%\usepackage[left]{showlabels}
%\usepackage{showkeys}

%\usepackage[a4paper,text={170mm,257mm},centering]{geometry}
%
%\numberwithin{equation}{section}
%
%\renewcommand{\arraystretch}{1.3}

%\makeatletter
%\renewcommand\section{\@startsection {section}{1}{\z@}%
%{-3.5ex \@plus -1ex \@minus -.2ex}%
%{2.3ex \@plus.2ex}%
%{\normalfont\large\bfseries}}
%\renewcommand\subsection{\@startsection{subsection}{2}{\z@}%
%{-3.25ex\@plus -1ex \@minus -.2ex}%
%{1.5ex \@plus .2ex}%
%{\normalfont\normalsize\bfseries}}
%\makeatother

%\expandafter\def\expandafter\bfseries\expandafter{\bfseries\ifmmode\else\boldmath\fi}
%\expandafter\def\expandafter\mdseries\expandafter{\mdseries\ifmmode\else\unboldmath\fi}
%\expandafter\def\expandafter\normalfont\expandafter{\normalfont\ifmmode\else\unboldmath\fi}

%\providecommand{\href}[2]{#2}

%\newcommand{\arxivlink}[1]{\href{http://arxiv.org/abs/#1}{[arXiv:#1]}}
%\newcommand{\doilink}[2]{\href{http://doi.org/#2}{#1}}
%
%\newcommand{\mathsym}[1]{{}}
\def\id{\protect{{1 \kern-.28em{\rm l}}}}
\def\be{\begin{eqnarray}}
\def\ee{\end{eqnarray}}
\def\bi{\bibitem}
\def\tr{{\rm tr}}
\def\ha{\tfrac{1}{2}}
\def\td{\tilde}
\def\ci{\cite}
\def\N{{\mathcal N}}
\def\ww{\Omega}
\def\S{{\mathcal S} }
\def\nn{\nu}
\def\z{\zeta}
\def\dg{\dagger}
\def\a{\alpha}
\def\b{\beta}
\def\ap{\alpha^\prime}
\def\aa{{\a'}}
\def\g{\gamma}
\def\ok{\frac{1}{k}}
\def\jL{{J}}
\def\cL{{\mathcal L}}
\def\cH{{\mathcal H}}
\def\E{{\mathcal E}}
\def\w{\omega}
\def\vep{\varepsilon}
\def\De{{\mathcal D}}
\def\k{\kappa}
\def\four{\tfrac14}
\def\third{\tfrac{1}{3}}
\def\det{\hbox{det}}
\def\tid{\tilde}
\def\vv{{\rm v}}
\def\ta{{\tilde \a}}
\def\fo{\frac{1}{4}}
\def\K{{\rm S}}
\def\el{\ell}
\def\Tr{{\rm Tr}}
\def\P{\Phi}
\def\l {\lambda}
\def\bl{{\tilde \l}}
\def\const{{\rm const}}
\def\vn{\vec n}
\def\Prod{\Pi}
\def\O{{\mathcal O}}
\def\m{\mu}
\def\vs{\vec \s}
\def\ie{i.e.}
\def\rS{{\rm S}}
\def\as{{\a}}
\def\foot{\footnote}
\def\ra{\rightarrow}
\def\F{{\cal F}}
\def\cc{\circ}
\def\eqv{\equiv}
\def\ni{\noindent}
\def\bw{{\rm w}}
\def\cT{{\cal T}}
\def\no{\nonumber}
\def\J{{\cal J}}
\def\om{\omega}
\def\l{\lambda}
\def\tl{{\tilde \l}}
\def\tk{{\tilde k}}
\def\sqtl{{\sqrt{\tilde \l}}}
\def\adss{$AdS_5 \times S^5$\ }
\def\D{\Delta}
\def\p{\phi}
\def\r{\rho}
\def\rN{{\rm N}}
\def\tw{{\tilde w}}
\def\vl{ \vec \ell}
\def\varpi{{\rm w}}
\def\bG{\bar \G}
\def\ve{\varepsilon}
\def\I{{\cal I}}
\def\La{{\Lambda}}
\def\R{{\rm R}}
\def\bt{\bar\theta}
\def\Z{{\cal Z}}
\def\pa{\partial}
\def\bea{\be}
\def\eea{\ee}
\def\DD{{\rm D}}
\def\chii{\varepsilon}
\def\th{\theta}
\def\t{\tau}
\def\beq{\be}
\def\eeq{\ee}
\def\beqa{\bea}
\def\eeqa{\eea}
\def\bs{\bigskip}
\def\c{{\rm a}}
\def\del{\partial}
\def\s{\sigma}
\def\eps{{\epsilon}}
\def\n{\nu}
\def\dag{\dagger}
\def\bd{\bar \del}
\def\ed{\end{document}}
\def\iffa{\iffalse}
\def\xp{x_+}
\def\xm{x_-}
\def\te{\textstyle}
\def\rx{{\rm x}}
\def\C {{\rm C}}
\def\hC {{\rm \hat C}}
\def\dd {{\rm d}}
\def\bb{{\rm b}}
\def\sql{{\sqrt{\l}}}
\def\ep{\epsilon}
\def\G{\Gamma}
\def\dx{\dot x}
\def\vp{\varphi}
\def\tx{{\tilde x}}
\def\ad{{\rm ad}}
\def\d{\delta}
\def\de {\del}
\def\L{\mathcal{L} }
\def\pcm {{\rm PCM}}
\def\wz{{\rm WZ}}
\def\pcmq {{\rm PCM}$_q$\ }
\def\ZMq {{\rm ZM}$_q$\ }
\def\ZM {{\rm ZM} }
\def\pp{{\rm p}}
\def\qq{{\rm q}}
\def\rk {{\rm u}}
\def\rrk{{\rm e}}
\def\ss{{\rm s}}
\def\tt{{\rm t}}
\def\uu{{\rm u}}
\def\sm{$\sigma$-model}
\def\sms{$\sigma$-models}
\def\ov{\over}
\def\Ad{\text{Ad}}
\def\lmp{\Lambda}
\def\bl{\bar \lambda}
\def\Lie{\operatorname{Lie}}
\def\hb{\hbar}
\def\k{\varkappa} \def \h {{\rm h}}
\def\dt{\tfrac{d}{dt}}
\def\ddt{\frac{d}{dt}}
\def\bb{{\rm b}}
\def\vv{{ v}}
\def\kk{{\rm k}}
\def\ka{{\kappa}}
\def\np{ }
\def\cM{\mathcal{M}}
\def\bh{\bar{h}}
\def\lam{$\lambda$-model}
\def\etm{$\eta$-model}
\def\bh{\bar h}
\def\hh{{\rm h}}
\def\PCM{\L_{\rm PCM}}
\def\WZ{\L_{\rm WZ}}
\def\M{{\cal M}}
\def\q{{\rm q}}
\def\RR{{J}}
\def\LL{{K}}
\def\kold{k_{\text{old}}}
\def\kaold{\ka_{\text{old}}}
\def\ed{\end{document}}
\def\F{H}
\def\cgg{c_{_G}}
\def\cg{c\,}
\def\cf{c_{_\F}}
\def\cff{c_{_\F}}
\def\HH{\F}
\def\XX{{X}}
\def\hhh{\widetilde {\hh}}
\def\e{\varepsilon}
\def\tih{{\hat h}}
\def\H{{\rm H}}
\def\GG{{\rm G}}
\def\BB{{\rm B}}
\def\g{\gamma}
\def\bl{\bar\lambda}
\def\ol{\underline}
\def\ul{\overline}
\def\MM{\mathcal M}
\def\hz{z}
\def\hhz{z} %\def\hhz{\widehat{z}} %v3
\def\wcL{\widehat{\L}}
\def\wL{\widehat{L}}
\def\XX{X}
\def\bw{\a}
\def\xx{{\rm x}}
\def\cR{{\mathcal R}}
\def\sms{$\s$-models\ }
\def\sm{$\s$-model\ }
\def\wLL{\widehat{\L}}
\def\br{\big[}
\def\bl{\big]}
\newcommand\ei[1]{\mathbf{\underline{#1}}}
\def \FF {h}
\def\GG{{\rm G}}
%%%%%%%%%%%%%%%%%%%%%%%%%%%%%%%%%%%%%%%%%

\chapter{Sigma models with local couplings \label{loc}}

There is a remarkable link between the 
integrability of 2d $\s$-models and their stability under RG flow, i.e.\ their
renormalizability with finitely many running couplings. 
 However, it
remains rather mysterious how the integrability translates into simple
 RG behaviour.

In this chapter we shall present a new and different link between
classical integrability and the RG flow (based on the paper \cite{HLT2008} and closely following the presentation there).
Consider a familiar example of an integrable 2d $\s$-model, the 
principal chiral model (PCM). Suppose we
formally replace its overall coupling $h$
by a function of 2d time $\t$,
obtaining
the
Lagrangian
\be\la{1}
\wcL = -\ha h(\t)  \, \Tr[J_+ J_-] \ .
\ee
The resulting time-dependent theory is not Lorentz-invariant but one may wonder if
some notion of classical integrability still applies even for a non-constant $h(\t)$.
As we shall see below,
this is indeed the case {\it provided} the function $h(\t)$ is special:
the same as the solution of the 1-loop RG equation of the standard PCM,
i.e.\ $h(\t) \sim \t$.

More generally, starting with a Lorentz-invariant
2d theory
that is classically integrable (i.e.\ admitting a Lax connection),
one may formally promote
its coupling constants $h_\a$ to functions $h_\a (\t)$ depending on 2d time.
We address the question of which functions $h_\a (\t)$
allow the integrability to survive in the resulting time-dependent theory, in the sense that the resulting equations of motion still
admit a Lax representation.
We shall suggest a certain natural ansatz for the corresponding Lax connection
generalizing that of the time-independent theory and
will find that the unique functions $h_\a (\t)$ preserving the integrability under this ansatz are the ones solving the 1-loop RG equations of the original theory,
\be
\del_\t h_\a = \b_\a(h) \ , \la{1lr}
\ee
with
2d time $\t$ playing the role of
RG time $t=\log \mu$.
We observe this new `integrability--RG flow' connection
in a number of non-trivial examples but will not give a general proof of it.

One may wonder how such
time-dependent models (where, e.g.,
energy is no longer conserved)
can be integrable. In fact, integrable examples of time-dependent 1d
mechanical models
are known (see, e.g., \cite{t-dep-1d} and refs. there).
Our present interest in such 2d models is motivated by the desire to find new solvable examples
of
conformal $\s$-models representing consistent string backgrounds with Minkowski signature.
One class of examples would be conformal \sms with $(D+2)$-dimensional
target space metric
$ds^2 =\GG_{mn} (x) dx^m dx^n = -2 dudv + K(u,x) du^2 + dx^i dx^i$ \ ($x^m=(u,v,x^i)$, \ $i=1, ..., D$),
admitting a covariantly constant null Killing vector.
In the light-cone (l.c.)\ gauge $u=\tau$, one then gets
a  
 model with a time-dependent potential $K(\t,x)$, which is formally solvable in some cases if
$K$ is quadratic in $x^i$ (see, e.g., \cite{Papadopoulos:2002bg} and refs. there).\foot{Another solvable example obtained as a certain Yang-Baxter deformation of flat 4d Minkowski space was found in \cite{yoshida}.
Examples with $K$ not depending on $\t$ corresponding to
integrable l.c.-gauge theories (sine-Gordon, Liouville,
etc.) were discussed, e.g., in
\cite{Maldacena:2002fy,Russo:2002qj,Bakas:2002kt}. }

The most general metric with a covariantly constant
null Killing vector has the Brinkmann-Walker form,
$ds^2 = -2 dudv + K(u,x) du^2 + A_i(u,x) du dx^i + \GG_{ij} (u,x) dx^i dx^j$.  Once one relaxes the assumption that $\GG_{ij}$ is flat, the $K$ and $A_i$
terms can be absorbed into  $\GG_{ij}(u,x)$ by a coordinate transformation
to give 
$ds^2 = - 2 dud v + \GG_{ij}(u,x) d x^i d x^j$.
We shall focus on the remarkable class of string \sms on such backgrounds \cite{Tseytlin:1992pq,Tseytlin:1992ee},
\begin{align}
\L =- 2 \del_+ u \del_- v + \GG_{ij}(u,x) \del_+ x^i \del_- x^j \ . \la{csm}
\end{align}
Moving to a curved 2d background $h_{\m\n}$ and incuding also a dilaton term $\L_\p= - \a' \sqrt{-h} \,  R^{(2)} \phi(v,u,x)$,
this $\s$-model is Weyl-invariant, i.e.\ $\bar \b_{mn} = R_{mn} + 2 D_m D_n\p + ...=0$,
provided $\phi$ is linear in the null coordinate $v$
and $\GG_{ij}(u,x)$ depends on $u$ according to the RG equation of the  `transverse' $\s$-model
$\L=\GG_{ij}(x) \del_+ x^i \del_- x^j $ \cite{Tseytlin:1992pq,Tseytlin:1992ee},
\begin{align}
& \del_u \GG_{ij} = \bar{\beta}_{ij} (\GG)= \beta_{ij} + 2 D_i D_j \bar{\phi}
\ , \qquad \quad \beta_{ij} = R_{ij} + \ldots
\ , \la{bd} \\
&\phi = v + \bar{\phi}(u,x) \ . \la{dilv}
\end{align}
The `first-order' equation \rf{bd} comes from the contribution of the connection in the $D_m D_n\p$ term
in the Weyl-invariance condition due to the null Killing form of the metric and the $v$-term in the dilaton.\foot{The exact
RG evolution equation \rf{bd} in the l.c.\ direction $u$
may be compared to an approximate RG equation
(see, e.g., \cite{Schmidhuber:1994bv}) appearing in the
case of a `cosmological' metric
and a time-dependent dilaton which provides a `friction' term in the string
generalization of the Einstein equations.
}
We have ignored an additional
diffeomorphism term $D_{(i}W_{j)}$  in \rf{bd} that appears only at higher loop orders.
The model \rf{csm}  also admits a natural generalization to the case of non-zero `transverse'
$\BB$-field coupling $\BB_{ij}(u,x)$: the resulting $(D+2)$-dimensional $\s$-model is Weyl invariant
provided $\del_u(\GG_{ij} + \BB_{ij}) = \bar{\beta}_{ij}(\GG,\BB)$
where $ \bar{\beta}_{ij}$ is the corresponding Weyl anomaly coefficient 
\cite{Tseytlin:1992pq,Tseytlin:1992ee}. 

One example of such a conformal background with a sphere $S^D$ as the transverse space is (to 1-loop order)\,
$ds^2 = - 2 du dv + u \, d\Omega^2_D, \ \ \p= v  + \tfrac{D-24}{6} u+ \frac{1}{8} D \log u $ .

This construction is also naturally generalized to supersymmetric $\s$-models \cite{Tseytlin:1992ee}. On certain \textit{special} K\"{a}hler symmetric spaces, the supersymmetric $\s$-models (which then have $\mathcal{N}=(2,2)$ supersymmetry \cite{Zumino:1979et}) 
 are known to have a 1-loop exact $\b$-function \cite{Novikov}.\foot{In general, $\mathcal{N}=(2,2)$ supersymmetric $\s$-models on K\"{a}hler manifolds have a non-trivial 4-loop $\b$-function coefficient \cite{Grisaru:1986px,Grisaru2} (which may be absorbed into a non-local redefinition of the metric in the case of Ricci-flat K\"{a}hler (Calabi-Yau) manifolds \cite{Nemeschansky:1986yx}).}
 Hence, taking one of those as the transverse theory, the  background in \rf{csm},\rf{bd},\rf{dilv} resulting from the 1-loop RG flow will be \textit{exact} string backgrounds \cite{Tseytlin:1992ee}.

Starting with the classical string $\s$-model \rf{csm} in conformal gauge and fixing the residual conformal symmetry by the l.c.\ gauge condition $u=\t$, we get a time-dependent
theory for the transverse coordinates\foot{Here
we ignore
the `diffeomorphism vector'
term that will also contribute to the time dependence of the metric; however,
this term can be eliminated or modified as desired by an appropriate field redefinition
in \rf{lct}
depending on time, $x^i \to y^i(x,\t)$ (see discussion in Section \ref{2008sec2} below).}
\be
\L_{l.c.} = \GG_{ij}(\t ,x) \del_+ x^i \del_- x^j \ , \qquad\qquad
\del_\t \GG_{ij} = \beta_{ij}(\GG) \ . \la{lct}
\ee
If we now assume that the transverse $\s$-model
$\L=
\GG_{ij}(x) \del_+ x^i \del_- x^j$ is classically
integrable, then the fact that $\GG_{ij}(\t ,x)$ is subject to the RG equation in \rf{lct}
suggests, in view of the above `integrability--RG flow' connection,
that the l.c.\ gauge theory \rf{lct} is classically integrable
(admits a flat Lax connection)
and thus may potentially be solvable.

\

Let us note that a special class of the $\s$-models \rf{csm} with $\GG_{ij}(u,x) = u\, g_{ij}(x) $ 
are obtained upon fixing the conformal gauge $h_{\m\n} d\xi^\m d \xi^\n = e^{-2v } \eta_{\m\n}$ (so that $\sqrt {-h} \, R^{(2)} = - 2 \del_+ \del_- v$) in 
a theory of 
 2d dilaton gravity (with dilaton $u$) coupled to a $\s$-model \ci{Tseytlin:1992pq} with $g_{ij}(x)$ as the target space metric,
\be \la{dig}
\L = - \sqrt {-h} \, u \big[R^{(2)} + g_{ij}(x) \, h^{\m\n} \, \del_\m x^i \del_\n x^j \big] \ .
\ee
Particular 2d models of the type \rf{dig}, with $g_{ij}(x)$
corresponding to specific symmetric spaces,
appear upon dimensional reduction (in 2 Killing vector directions) of 4d Einstein gravity
\ci{Belinsky:1971nt,Breitenlohner:1986um,Nicolai:1991tt}.
After solving $\del_+ \del_- u=0$ (the equation from varying $v$) to obtain $u(\t,\s) = f^+ (\xi^+) + f^- (\xi^-)$,
the resulting `local coupling' $\s$-model,
\be \la{ff}
\L = u(\t, \s)\, g_{ij}(x) \del_+ x^i \del_- x^j \ , \qquad \quad u(\t,\s) = f^+ (\xi^+) + f^- (\xi^-) \ , \ee
was shown, in some special cases
\ci{Belinsky:1971nt} and for general symmetric spaces
\ci{Breitenlohner:1986um} (see also \ci{Nicolai:1991tt}), to be integrable in the sense of admitting a Lax
pair.
The spectral parameter is now replaced by a function of $\t,\s$ expressed in terms of $f^+ (\xi^+) $, $
f^- (\xi^-)$ and an integration constant $w$ (which may be viewed as a new spectral parameter ---
see below).

Here will go beyond this related earlier work in two important ways:
(i) showing that there are other integrable  $\s$-models with `local couplings' (e.g., related to integrable-deformed models)
for which the dependence of $\GG_{ij} (u,x)$ on $ u(\t,\s)$ is not simply through an overall factor as in \rf{ff};
(ii) establishing the link between the dependence on $u$ required for
integrability and the RG evolution of couplings in the original $\s$-model defined by the transverse metric $\GG_{ij} (u,x)\big|_{u=\const}$.

\ 

The rest of this chapter is organized as follows.
In Section \ref{2008sec2} we will show that, starting from a certain class of classically integrable $\s$-models 
admitting a flat Lax connection, it is possible to find a Lax connection for the corresponding
time-dependent theory obtained from the original one by replacing the couplings
by functions of $\t$ satisfying the 1-loop RG equations. We will explicitly discuss the case of the PCM, and a similar statement is
also true for several classes of integrable $\s$-models (PCM with WZ term, symmetric spaces,
$\eta$-model, $\l$-models) presented in Table \ref{2008tab1}.

In Section \ref{2008s3} we will demonstrate that the converse is also true: demanding the existence of a Lax pair for the time-dependent model implies that the couplings must solve the 1-loop RG equations
of the original theory. Details of the derivation of the RG flow will be presented in Appendix \ref{2008A}.

In Section \ref{2008s4} we will discuss how the monodromy matrix associated with
the Lax connection can be used to construct, at least in principle, non-local
conserved charges for time-dependent models.\foot{See also Section 5 of \cite{HLT2008} for an attempt to construct higher-spin, local charges for  the PCM with a local ($\s$-dependent) coupling.} 
We will concentrate on the PCM example
with comments on other models relegated to Appendix \ref{2008om}.
To shed light on the notion of integrability in the time-dependent case we
will also consider consistent 1d
reductions of some simple 2d $\s$-models, investigating their solvability.
In Appendix \ref{2008C} we will further discuss a particular linear 1d model which is a special case
of a time-dependent harmonic oscillator.

Section \ref{2008s6} contains a discussion of the results and possible extensions.
In particular, we will mention that the relation between integrability of the time-dependent
theory and the RG flow should extend to models with potentials, illustrating this
on the example of the sine-Gordon model (see also Appendix \ref{2008sing}).

%%%%%%%%%%%%%%%%%%%%%%%%%%%%%%%%%%%%%%%%
\section{Lax connection for time-dependent 
sigma models}\la{sec2}
%%%%%%%%%%%%%%%%%%%%%%%%%%%%%%%%%%%%%%%%

Below we shall discuss how,
starting from a classically integrable model (with Lagrangian $\L$) admitting a Lax pair $L_\pm$,
it is possible to find also a Lax pair $ \widehat{L}_\pm $ for the corresponding
time-dependent theory $\widehat{\L}$ obtained from the original one
by replacing the couplings $h_\a$ by functions $ h_\a (\t)$ depending on $\t$
according to the 1-loop RG equation \rf{1lr},\foot{Here and below for simplicity
we are only indicating the dependence of the Lagrangian on the couplings, suppressing dependence on fields.}
\begin{align}
&h_\a \to h_\a (\t) \ , \qquad \qquad \tfrac{d}{d\t} h_\a (\t) = \b_\a(h(\t)) \la{tins} \ ,\\
&\L(h_\a)\ \to\ \widehat{\L} = \L(h_\a (\t)) \ . \la{lcg}
\end{align}
The Lax connection must be modified to account for `extra' terms in the equations of motion following from \rf{lcg} that are proportional to the derivatives $\del_\t h_\a$ of the couplings.
We propose a very simple ansatz for this modification: the extra terms can be effectively `absorbed'
into the spectral parameter $z$
with the structure
of the Lax connection remaining the same, i.e.
\begin{align}
&L_\pm(h_\a, z) \to \widehat{L}_{\pm} = L_\pm (h_\a(\t), \hhz(w; \t,\s) )\ , \la{lac2}\\
&z \to \hhz = \hhz(w;\t,\s) \ , \ \ \ \qquad h_\a \to h_\a (\t) \ .
\la{lac1}
\end{align}
While $z$ was regarded as constant
in the original Lax connection,
it is now promoted to a particular function $\hhz(w;\t,\s)$ of the 2d coordinates.
Here $w$ is an arbitrary (complex) constant playing the role of a new spectral parameter.
As we shall see below, the explicit form of the function $\hhz(w;\t,\s)$ will be model-dependent.

Let us note that if the original theory is a \sm on flat 2d Minkowski space,
which
is classically invariant under the 2d conformal transformations $\xi^\pm \to f^{\pm}(\xi^\pm)$,  
then the Lagrangian of the corresponding time-dependent theory \rf{lcg} changes
under such transformations as
\be
\widehat{\L}= \L(h_\a (\t))\ \to\ \L( h_\a(\t')) \ , \qquad \ \ \ \ \ \t = \xi^+ + \xi^-\ \to\ \t'=
f^+ (\xi^+)+ f^- (\xi^-) \ . \la{nlc}
\ee
In the case of the
string
\sm \rf{csm},\rf{lct},
this is equivalent to replacing the l.c.\ gauge $u=\t$ with a
more general one given by the solution $u = f^+ (\xi^+)+ f^- (\xi^-)$ of the
equation of motion $\del_+ \del_- u = 0$.
One special choice is $u=\xi^-$ for which
we get
$
\wcL_{\rm ch} = \L(h_\a (\xi^-))
$.
One may formally consider that theory,
which only fixes one `half' of the conformal reparametrizations, as an extreme limit $(f^+(\xi^+),f^-(\xi^-)) =(0, \xi^-)$ of \rf{nlc}.

To provide evidence for the above proposal, we have considered
six examples of integrable $\s$-models (each introduced in Section \ref{rev}) with the results summarized in Table \ref{2008tab1} below.
There, for each model, we give the original Lagrangian, original Lax pair, 1-loop RG equation and its solution, and the expression for the function $\hhz(w;\t,\s)$ that generalizes the original spectral parameter in such a way
that \rf{lac2} gives a Lax pair for the time-dependent theory \rf{lcg}.

The classically integrable $\s$-models in Table \ref{2008tab1} are each 1-loop renormalizable
 assuming a particular choice of
`renormalizable'
diffeomorphism vector. This means that the insertion of time dependence according to the RG flow in \rf{tins},\rf{lcg}
will only apply to a finite set of couplings. 
As discussed
at the beginning of this chapter,
a time-dependent $\s$-model
arises in the light-cone gauge from a conformal string $\s$-model.
There, the time dependence is governed by the RG flow with
a specific diffeomorphism vector (generally different from the renormalizable one)
following from the $(D+2)$-dimensional dilaton solving the Weyl-invariance condition. 
However,
these two patterns of time dependence
(according to the RG flow with different diffeomorphism vectors)
are simply related by a time-dependent field redefinition in
the corresponding time-dependent models, and  such a redefinition preserves the existence of a Lax representation.

All of our examples are built on either a group $G$ or a symmetric space $G/H$. Our notation is the same as that of Section \ref{rev}, except that here $h$ denotes a coupling constant, or $h_\a$ to denote many of them (whereas before they were called  $\h$ to distinguish from group elements).
 We again denote $J_\pm = g^{-1} \del_\pm g$, where $g\in G$ is scalar field.
PCM$_k$ stands for the principal chiral model plus a Wess-Zumino ($\BB$-field)
term with coefficient $k$
(and the special case $h=\pm k$ is the conformal WZW model).
The group space $\eta$-model is an integrable deformation of the PCM
while the group space and symmetric space $\l$-models are integrable
deformations of the WZW and gauged WZW models.

\begin{table}
{
\renewcommand{\arraystretch}{1.5}
\begin{tabular}{l || l | l | l }
& PCM& PCM$_k$ \\ \hline\hline
Lagrangian& $\L =-\ha h \Tr[J_+ J_-]$ & $\L =-\ha h \Tr[J_+ J_-]+ k\, \L_{\rm WZ}$\\ \hline

Lax connection & $L_\pm =\ha (1+z^{\pm 1})J_\pm$ & $L_\pm =\ha (1 + z^{\pm 1}) (1\pm \frac{k}{h})J_\pm$ \\ \hline

\makecell[l]{1-loop RG equation \\
and its solution}& $\begin{matrix*}[l]\ddt h = \cg \\ h(t) =\cg t \end{matrix*}$ & $\begin{matrix*}[l]\ddt h = \cg(1-\frac{k^2}{h^2}), \ \ \ \ddt k = 0 \\ h(t) -k \tanh^{-1}{\big(\frac{h(t)}{k}\big)} = \cg t \end{matrix*}$\\ \hline

\makecell[l]{Spectral parameter\\function }& $\hz=
\sqrt{\frac{w+\s-\t}{w+\s+\t}}$ & $\begin{matrix*}[l] \cg (w+\s-\t) + 2zh(\t)\, \frac{k+\hz(k+h(\t))}{k(1+\hz)^2+h(\t)(\hz^2-1)} \\ \quad \quad \qquad + k \log{\big(\frac{(1+\hz)(k-h(\t))}{(k-h(\t))+\hz(k+h(\t))}\big)} = 0 \end{matrix*}$
\vspace{0.5cm}
\end{tabular}

\begin{tabular}{l || l | l | l}
& Group space $\eta$-model & Group space $\l$-model \\ \hline\hline
Lagrangian& $\L =-\ha h \Tr [ J_+ \frac{1}{1-\k \cR} J_- ] $ & $\begin{matrix*}[l]\L =k\big[ \L_{G/G}(g,A)\\
\qquad \qquad - (\l^{-1}-1) \Tr( A_+ A_-)\big] \end{matrix*}$\\ \hline

Lax connection & $\begin{matrix*}[l] L_\pm = \ha (1+z^{\pm 1}) C_\pm \\ C_\pm = - (1+\k^2)\Ad_g\frac{1}{1\pm \k \cR} J_\pm \end{matrix*}$ & $ \begin{matrix*}[l] L_\pm = \ha (1+z^{\pm 1}) \frac{2}{1+\l} A_\pm \\
A_\pm \text{ takes on-shell value}\end{matrix*}$ \\ \hline

\makecell[l]{1-loop RG equation \\
and its solution}
& $\begin{matrix*}[l] \ddt h = \cg (1+\k^2)^2 ,\ \ \ddt (h \k^{-1}) = 0\\
\nu \equiv  \tfrac{h(t)}{4\k(t)} = \text{const} \\
{\tan}^{-1}\, {\k(t)} + \frac{\k(t)}{1+\k(t)^2} = \frac{\cg}{2\nu} t
\end{matrix*}$
& $\begin{matrix*}[l]
\ddt \l = - \frac{2\cg}{k}\frac{\l^2}{(1+\l)^2}, \ \ddt k =0, \\
- 2\log{\l(t)} - \l(t) + \l(t)^{-1} = \frac{2\cg}{k} t
\end{matrix*}$
\\ \hline

\makecell[l]{Spectral parameter \\
function}
& $\begin{matrix*}[l]
\frac{\cg}{2 \nu} (w+\s) - \frac{1-\hz}{1+\hz} \frac{\k(\t)}{1+\k(\t)^2} \\
\qquad \qquad + \tan^{-1}{ \big(\frac{1}{\k(\t)} \frac{1-\hz}{1+\hz} \big)} = 0
\end{matrix*}$ &
$\begin{matrix*}[l]
\frac{2\cg}{k} (w+\s) + \frac{1-\hz}{1+\hz} [\l(\t)-\l(\t)^{-1}] \\
\qquad \qquad - 2\log{\big( \frac{\hz\l(\t)-1}{\hz-\l(\t)} \big)} = 0\end{matrix*}$
\vspace{0.5cm}
\end{tabular}

\begin{tabular}{l || l | l | l}
& Symmetric space $\s$-model & Symmetric space $\l$-model \\ \hline\hline

Lagrangian& $\L = -\ha h \Tr [ J_+P_{G/H} J_- ] $ & $\begin{matrix*}[l]\L =k\big[ \L_{G/G}(g,A)\\
\qquad \qquad - \Tr \big[ A_+ (\l^{-1} P_{G/H} -1) A_-\big] \end{matrix*}$\\ \hline

Lax connection &
$\begin{matrix*}[l] L_\pm = J^{H}_\pm + z^{\pm 1} J^{G/H}_\pm \\ J^{H}_\pm = P_H J_\pm , \ J^{G/H}_\pm = P_{G/H} J_\pm \end{matrix*}$
& $\begin{matrix*}[l] L_\pm = A^{H}_\pm + z^{\pm 1} \frac{1}{\sqrt{\l}} A^{G/H}_\pm \\ A^{H}_\pm = P_H A_\pm , \ A^{G/H}_\pm = P_{G/H} A_\pm \\
A_\pm \text{ takes on-shell value} \end{matrix*}$ \\ \hline

\makecell[l]{1-loop RG equation \\
and its solution} & $\begin{matrix*}[l] \ddt h = 2\cg \\
h(t) = 2\cg t
\end{matrix*}$
& $\begin{matrix*}[l]
\ddt\l = -\frac{\cg}{k} \l \\
\l(t) = \exp{(- \frac{\cg}{k} t )}
\end{matrix*}$
\\ \hline

\makecell[l]{Spectral parameter \\
function}
& $\hz=
\sqrt{\frac{w+\s-\t}{w+\s+\t}}$ &
$ \hz =
\exp{( - \frac{\cg}{2k} \t)} \sqrt{\frac{1+\exp{(\frac{\cg}{k} [w+\s-\t] )}}{1+\exp{( \frac{\cg}{k} [w+\s+\t] )}}}
$
\vspace{0.5cm}
\end{tabular}

\caption{\small Examples of integrable \sms and their time-dependent generalizations.
$t$ is the RG time which is replaced by the 2d time $\t$ in the couplings entering the
Lagrangian $\wcL$ in \rf{lcg}.
\la{tab1}}
}
\end{table}

In this section we shall explicitly consider only the basic PCM example,
postponing a more general discussion to Section \ref{2008s3},
where we shall also explain that the existence of the spectral function $\hhz(w;\t,\s)$ is a highly non-trivial feature, depending on
specifically choosing the coupling functions $h_\a(\t)$ to solve the 1-loop RG equation.

For the PCM corresponding to a simple Lie group $G$
\be
\L =-\ha h \Tr[J_+ J_-] \ , \qquad J=g^{-1} dg \ , \ \ g\in G \ , \la{pl}
\ee
the global $G\times G$ symmetry implies
that only the overall coupling $h$ runs under the 1-loop
RG,\foot{The constant $c\equiv c_{_G}$ is the dual Coxeter number of the group $G$ (see Appendix \ref{conv} for the group theory conventions).
\la{fo8}
}
\be
\ddt h = \cg \ ,\qquad \qquad h(t) =\cg t \ , \qquad \qquad \cg \equiv \cgg.
\ee
The corresponding time-dependent theory
\rf{lcg} is then
given
by
\be
\widehat{\L} = - \tfrac{1}{2} \cg \t \, \Tr[J_+ J_-] \ . \la{lcPCM}
\ee
As explained above, this theory arises naturally in the l.c.\ gauge $u=\t$ from the conformal string $\s$-model (cf.\ \rf{csm},\rf{lct})
\be
\L =-2 \del_+ u \del_- v - \tfrac{1}{2} \cg u \Tr[J_+ J_-] \ , \la{nPCM}
\ee
where the constant $\cg$ can be eliminated by a rescaling $(u,v) \to ( \cg^{-1} u,\, \cg v)$.

The equations of motion for the model \rf{lcPCM} written in first-order form are
\begin{align}
&\del_+ (\t J_- ) + \del_- (\t J_+) = 0 \ , \la{pe1} \\
&F_{+-}(J) \equiv \del_+ J_- - \del_- J_+ + [J_+, J_-] = 0 \ . \la{pe2}
\end{align}
Starting from the Lax connection of the original PCM \rf{pl}
\be
L_\pm = \ha(1+z^{\pm 1}) J_\pm \ , \la{pla}
\ee
and making the replacement
$z \to \sqrt{\frac{{w+\s-\t}}{{w+\s+\t}}} $ where $w$ is a constant spectral parameter
(see Table \ref{2008tab1}), we obtain the following expression for the
Lax connection \rf{lac2} of the time-dependent theory
\be
\widehat{L}_\pm = \ha \big( 1+ {{z}\, }^{\pm 1}
\big) J_\pm \ , \qquad \qquad {z}=\sqrt{\frac{{w+\s-\t}}{{w+\s+\t}}} \ . \la{lpl}
\ee
Indeed,
its curvature is
\begin{align}
&F_{+-}(\widehat{L}) \equiv \del_+ \wL_- - \del_- \wL_+ + [\wL_+, \wL_-] \la{curv} \\
&\qquad \ \ \ \, \, = \frac{{z}}{2(w+\s-\t)} \Big[ \del_+ (\t J_- ) + \del_- (\t J_+) \Big]
+ \tfrac{1}{4} (1+{z})(1+{{z}\, }^{-1} ) \ F_{+-}(J)
\ , \no
\end{align}
so that, at
any $(\t, \s)$, its vanishing for all $w$ implies
the equations of motion \rf{pe1},\rf{pe2} of the time-dependent theory.

A surprising feature
(shared also by the other examples in Table \ref{2008tab1})
is that, despite the Lagrangian \rf{lcPCM} explicitly involving only $\t$,
the Lax connection \rf{lpl}
also depends on the 2d spatial coordinate $\s$.
The spatial coordinate $\s$ and the spectral parameter $w$ appear only through
the combination $w+\s$. Since $w$ is a complex constant, it cannot be
eliminated by a shift of real $\s$. Moreover, what is important is the
existence of a Lax connection depending on $w$, which can in principle be used
to construct conserved charges, etc. Indeed, we can instead interpret the
formal possibility to shift $\s$ as the freedom to introduce a spectral
parameter in the first place.

Another unusual property (shared by all the examples) is that the Lax connection has branch cuts in the spectral $w$-plane
(e.g., for the PCM the branch cuts end at $w=-\s\pm \t$).
Normally, one could simply remove square roots by redefining the spectral parameter (or more formally moving to an appropriate Riemann surface on which the Lax connection is meromorphic).
However, this is not possible since the positions of the branch cuts depend on $(\t,\s)$ and one cannot redefine the spectral parameter in a way depending on $(\t,\s)$ without changing the equations of motion encoded in the zero curvature condition.

In each case in Table \ref{2008tab1}, one can freely choose any branch of the function $z(w;\t,\s)$ (in the $w$-plane) while still encoding the correct equations of motion in the zero-curvature equation. For example, in the PCM case, one may choose either sign of the square root ${z} =\pm \sqrt{\tfrac{{w+\s-\t}}{{w+\s+\t}}}$ in the Lax connection \rf{lpl}.
All similar square roots in the spectral functions below
have the same branch ambiguity, corresponding to the option to reverse their sign.

\medskip

Applying a general conformal transformation to the action corresponding to \rf{lcPCM},\rf{lpl} we get (cf.\ \rf{nlc})
\begin{align}
& \widehat{\L} =- \tfrac{1}{2}\cg \, \big(f^+( \xi^+) + f^-(\xi^-)\big) \, \Tr[J_+ J_-] \ , \qquad \la{lac} \\
&\widehat{L}_\pm= \ha \big( \, 1+ z^{\pm 1} \big) J_\pm \ , \qquad \quad
z= \sqrt{ \frac{{w- 2f^-(\xi^-)}}{{w+ 2 f^+ (\xi^+)}}} \ . \la{laca}
\end{align}
This theory may equivalently be obtained from the string $\s$-model \rf{nPCM}
by picking up a generalized l.c.\ gauge $u =\t'= f^+(\xi^+) + f^-(\xi^-)$.
A (degenerate) limit $u=a+b \t$, \ $a\to 1$, $b\to 0 $ eliminates the explicit $\tau$-dependence and gives back the original PCM \rf{pl}, with \rf{lac} becoming
the original Lax connection in \rf{pla} with a redefined spectral parameter $z\to \sqrt{\frac{{w-2 a_-}}{{w+2 a_+}}}$, where $a_\pm$ are constants satisfying $a_+ + a_- = 1$.
We note that the theory \rf{lac} is a special case of \rf{ff} where $\GG_{ij}$ is the group-space metric. Indeed, an equivalent expression for the Lax connection
of this `local' PCM (or symmetric space $\s$-model, see Table \ref{2008tab1})
was originally found in \cite{Breitenlohner:1986um}
where the dependence of the analog of the
spectral parameter on the functions $f^+(\xi^+) $ and $f^-(\xi^-)$ was
discovered. The spectral function $z$ in \rf{laca} depends separately on $u$ and its dual field $\td u =f^+( \xi^+) - f^-(\xi^-) $, \
$d u= *d\td u $,
and thus separately on $f^+$ and $f^-$ \cite{Breitenlohner:1986um,Nicolai:1991tt}.

%%%%%%%%%%%%%%%%%%%%%%%%%%%%%%%%%%%%%%%%
\section{RG flow from condition of integrability of time-dependent theory}\la{s3}
%%%%%%%%%%%%%%%%%%%%%%%%%%%%%%%%%%%%%%%%

As we have found above on several examples,
if the couplings of an integrable $\s$-model are promoted to functions of time that
solve the 1-loop RG equations, $h_\a \to h_\a(\t),\ \del_\t h_\a = \b_\a(h)$,
then the Lax connection of the original model $\L(h_\a)$ admits a natural generalization
to a classical Lax connection for the time-dependent model $\L(h_\a(\t))$.
Here we shall argue that the converse is also true:
demanding the existence of a Lax pair for the time-dependent theory implies that $h_\a (\t)$ must solve
the 1-loop RG equations.

It is useful to start with a more general theory with local couplings\foot{We generally expect integrable models to have a finite number of couplings $h_\a$.
A natural way to identify these couplings is
as the parameters that run under the RG flow.}
$h_\a(\t,\s)$ depending on both $\t$ and $\s$, i.e.
\begin{align}
&\widehat{\L} = \L(h_\a(\t,\s)) \ , \la{sth}
\end{align}
and demand the existence of a Lax representation for this
theory.
For the examples in Table \ref{2008tab1}, the original Lax connection
takes a particular form
(which is generic to many integrable $\s$-models built on groups $G$ or symmetric spaces $G/H$):
\begin{alignat}{2}
&\hspace{-2.5cm}G \ : \qquad\qquad & &L_\pm = \ha (1+z^{\pm 1}) \mc{A}_\pm \la{gl} \ ,\\
&\hspace{-2.5cm} G/H \ :
\qquad\qquad & &L_\pm = \mc{B}_\pm + z^{\pm 1} \mc{P}_\pm \la{sl} \ .
\end{alignat}
Here the connection components $\mc{A}_\pm \in \Lie(G)$,
$\mc{B}_\pm \in \Lie(H)$, $\mc{P}_\pm \in \Lie(G)/\Lie(H)$ depend implicitly on the fields and their derivatives
and the couplings $h_\a$ (e.g.\ for PCM, $\mc{A}_\pm = J_\pm $, cf.\ \rf{pla}), while the dependence on the
spectral parameter $z$ is indicated explicitly.
Let us now take the following natural ansatz for the Lax connection $\widehat{L}_\pm$
corresponding to the $(\t,\s)$-dependent theory \rf{sth}
\begin{alignat}{2}
&G \ : \qquad \qquad & &\widehat{L}_\pm = p_\pm(w;\t,\s)\, \mc{A}_\pm\ , \la{glt}\\
& G/H \ : \qquad \qquad & &\widehat{L}_\pm = q_\pm(w;\t,\s)\, \mc{B}_\pm + r_\pm(w;\t,\s)\, \mc{P}_\pm \la{slt} \ ,
\end{alignat}
where $p_\pm, q_\pm, r_\pm$ are some functions, and $\mc{A}_\pm$, $\mc{B}_\pm$, $\mc{P}_\pm$ are assumed to take the same form as in \rf{gl},\rf{sl}, but now depending on the coupling \textit{functions} $h_\a(\t,\s)$ in place of the couplings, i.e.
$\mc{A}_\pm = \mc{A}_\pm (x,\del x; h_\a(\t,\s))$, etc.

Demanding that the flatness of the connection \rf{glt},\rf{slt}
is equivalent to the equations of motion of the `local coupling' theory \rf{sth},
we shall first prove that it is sufficient to use just
a single function
$\hz(w;\t,\s)$, so that the Lax connections \rf{glt},\rf{slt} become the same as the original ones \rf{gl},\rf{sl}
with the replacement $z\to \hz(w;\t,\s)$
as in \rf{lac2},
\be
\widehat{L}_\pm = L_\pm\big( h_\a(\t,\s), \hz(w;\t,\s)\big) \ . \la{laxs}
\ee
Indeed, the equations of motion for the theory \rf{sth} contain 
two types of terms:
the ``original'' terms (not depending on $\del_\pm h_\a$), and ``new'' terms (depending on $\del_\pm h_\a$).
To match the ``original'' terms with the corresponding terms in the zero-curvature condition for $\widehat{L}_\pm$, one needs
to impose the following conditions on $p_\pm$, $q_\pm$, $r_\pm$
(see Appendix \ref{2008det} for details)
\begin{alignat}{2}
& G \ : \qquad \qquad & &p_+^{-1} + p_-^{-1} = 2 \ , \la{alg}\\
& G/H \ : \qquad \qquad & &q_\pm = 1 \ , \qquad r_+ r_- = 1 \ . \la{als}
\end{alignat}
Substituting \rf{alg},\rf{als} into the Lax ansatz \rf{glt},\rf{slt} then gives \rf{gl},\rf{sl} but with $z\to \hz(w;\t,\s)$, where we set
 $\hz(w;\t,\s) = 2 p_+ - 1$ (for models on $G$) and
$\hz(w;\t,\s) = 2 r_+ $ (for models on $G/H$).

All of our examples are single-coupling theories\foot{We do not promote the WZ level $k$ in PCM$_k$ or $\l$-models to
a function of $(\t,\s)$ since the resulting model
would not be well-defined: starting with the
3d representation of the WZ term and replacing $k \H \to k(\t,\s) \H$
would no longer give a local 2d action
while $k \BB \to k(\t,\s) \BB $ in the 2d action would not be  globally well-defined, or locally consistent with global symmetry
(invariance of the 2d action under global symmetry transformations requires dropping total derivatives).}
except for the group $\eta$-model. Here we shall
specialize to the simplest case with one coupling $h(\tau,\sigma)$
but the same result will also be true for the group $\eta$-model (see Appendix \ref{2008mr}).
To reproduce the ``new''  terms (proportional to  $\del_\pm h$) in the
equations of motion for \rf{sth} from the flatness condition for the generalized Lax connection \rf{laxs}, one needs to additionally impose certain constraints on both $h(\t,\s)$ and $\hz(w;\t,\s)$.
For the models in Table \ref{2008tab1}
this leads to the following two first-order differential equations for the function $\hz(w;\t,\s)$
(here $h=h(\t,\s)$)
\begin{align}
\del_\t z = V_\t (z,h) \ , \qquad\qquad \del_\s z = V_\s (z,h) \ , \la{pd1}
\end{align}
where $V_{\t,\s} (z,h)$ are model-dependent functions. For all five single-coupling examples in Table \ref{2008tab1}, the consistency condition ($\del_\t V_\s - \del_\s V_\t =0$) for the system \rf{pd1} takes the
remarkable form,
\be
\del_+ \del_- h - \frac{\b'(h)}{\b(h)}\, \del_+ h \del_- h = 0 \la{2w} \ ,
\ee
where $\b(h)$ is precisely the 1-loop $\b$-function for the coupling $h$ in the original model.
In addition
to the appearance (in this purely classical context)
of the 1-loop $\beta$-functions, a remarkable feature of \rf{2w}
that it does not depend on $z$ (which completely factors out of the consistency condition).

The condition \rf{2w} may be written simply as $ \del_- ( \frac{\del_+ h}{\b(h)}) =0$ or
$ \del_+ ( \frac{\del_- h}{\b(h)}) =0$ and thus leads to two first-order equations,
\be
\frac{\del_+ h}{\b(h)} = s_+(\xi^+) \ , \qquad \qquad \frac{\del_- h}{\b(h)} = s_-(\xi^-) 	\ , \la{cfs}
\ee
where $s_\pm (\xi^\pm)$ are arbitrary functions of the 2d light-cone coordinates $\xi^\pm =\ha ( \t \pm \s)$.
By applying a conformal transformation (i.e.\ redefining $\t$ and $\s$), one can absorb $s_\pm$ into $\del_\pm$, i.e.\ replace
$s_\pm \to 1$,
so that in terms of the redefined coordinates the first-order equations \rf{cfs}
take the form of the 1-loop RG equation in $\tau$
\be
\del_\t h = \b(h) \ , \qquad \qquad \del_\s h = 0 \ . \la{RGs2}
\ee
Thus, modulo a conformal transformation,
the 1-loop RG solution is the only choice
of local coupling
$h(\t,\s)$
for which the Lax connection can be uplifted to the $(\t,\s)$-dependent theory according to \rf{laxs}.
This argument (in eqs.\ \rf{pd1}-\rf{RGs2}) is demonstrated explicitly for the example of the PCM in Appendix \ref{2008fdr}.

The same conclusion is reached of course if one starts directly with the theory where
the couplings $h_\a$ depend only on time (giving 
 $\del_\pm \to \del_\t$
and $s_+=s_-=\const$ in eqs.\ \rf{2w},\rf{cfs}):
using the same ansatz for the Lax connection
one finds that the only functions $h_\a(\t)$ consistent with preserving integrability
are the solutions of the 1-loop RG flow ---  modulo
a rescaling of time, i.e.\ the part of the conformal group that does not introduce spatial dependence.
By performing a classical conformal transformation, one could instead assign a preferred role to space $\s$ rather than time $\t$, and in that case the RG equation \rf{RGs2} will hold with $\t \to \s$.

To finish the construction of the generalized Lax pair let us now explain how to obtain the explicit form of the spectral parameter function
$z(w;\t,\s)$ in \rf{laxs}.
Starting with $h=h (\t)$ that satisfies the 1-loop RG equation in \rf{RGs2}
(so that the consistency conditions \rf{cfs} of \rf{pd1}
are satisfied) one can solve the second equation in \rf{pd1} as
\be
\del_\s z = V_\s (z, h(\t)) \qquad \rightarrow \qquad \int \frac{dz}{V_\s(z,h_\a(\t))} =\s + \ell (\t) \ . \la{sold}
\ee
The function $\ell(\t)$ is then fixed by substituting this solution into the first equation in \rf{pd1}.
The solution for $\ell(\t)$ leaves one free integration constant, which we call
$w$. Finally, eq.~\rf{sold} becomes an algebraic equation for $z(w;\t,\s)$
that can be explicitly solved in some simple cases (see Table \ref{2008tab1}).
Note that since the parameter $w$ appears as an integration constant in the function $\ell(\t)$,
it always appears in the combination $w+\s$ with the spatial coordinate,
implying that constant shifts of $\s$ can be compensated by shifts of $w$.

Although we do not have a proof that the 1-loop RG flow always follows from requiring
integrability of the time-dependent generalizations of integrable models, the examples
in Table \ref{2008tab1} reveal a highly non-trivial pattern.
In particular, 
some of the 1-loop RG equations obtained in this way are quite non-trivial, not admitting simple closed form solutions.

%%%%%%%%%%%%%%%%%%%%%%%%%%%%%%%%%%%%%%%%
\section{Non-local conserved charges in time-dependent integrable models}\la{s4}
%%%%%%%%%%%%%%%%%%%%%%%%%%%%%%%%%%%%%%%%

We have shown in the previous sections that some well-known examples of classically integrable $\s$-models admit generalizations of their Lax connections to models with time-dependent couplings under the condition that the latter solve
the 1-loop RG equation in $\tau$.
However, one may question the usefulness of the resulting Lax connections since they have several unusual properties: they depend on $\t$ and $\s$ explicitly; the spectral parameter only enters as a `constant piece' of the spatial coordinate $\s$;  and they
have branch cuts in the spectral plane (whose positions depend on $\t$ and $\s$).

In this section we shall argue that, nevertheless, the Lax connections can still be used to construct the `non-local' charges typical of integrable models. However, the space-time dependence of the Lax connection renders these charges difficult to compute explicitly. We shall focus here on the case of the PCM, while the other examples from Table \ref{2008tab1} are discussed in Appendix \ref{2008om}.

\subsection{Non-local charges}\la{s41}

The derivation of non-local conserved charges from the  monodromy matrix $\M = \mathcal P \,  \exp \, \int^b_a d\s \, L_\s$  discussed in Section \ref{ci} still applies for Lax connections with explicit space-time dependence. Indeed, the zero-curvature condition $\del_\s L_\t - \del_\t L_\s + [L_\s,L_\t]=0$ implies that
\be
\del_\t \M = \M \, L_\t(b) - L_\t(a) \, \M \ . \la{ce}
\ee
Hence, if we assume a periodicity condition on the spatial domain $a<\s<b$,
\be
L_\t(a) = L_\t(b)\ , \la{pe}
\ee
then we obtain
\begin{align}
\del_\t \M = [\M, L_\t(a)] \ , \la{lp}
\end{align}
and so the eigenvalues of $\M$ are conserved ($\del_\t \Tr[\M^n]=0$). If also the value at the endpoints vanishes, $L_\t(b) = L_\t(a)=0$, then all of the components of $\M$ are conserved ($\del_\t \M=0$).

We shall focus on the time-dependent PCM on an infinite spatial interval $(a,b)=( -\infty,\infty)$.\foot{It seems much harder to satisfy the periodicity condition \rf{pe} on a spatial circle due to the explicit non-periodic $\s$ dependence in the Lax connections for the time-dependent models.}
Picking the `negative' branch\foot{We note that it is consistent to `pick a branch' here: e.g., by assuming
that $w$ has an imaginary part, it then follows that the sign of the square root does not change from $\s=-\infty$ to $\s=\infty$. The choice of a branch is arbitrary but this negative choice is more useful for satisfying the periodicity condition and constructing conserved charges. Also, for the group $\eta$-model and PCM$_k$, which are deformations of PCM, the function
${z}$ is single-valued at $\s=\pm \infty$, with $z_{\infty}\equiv z\big|_{\s=\infty}=-1$. It is then natural to obtain $z_{\infty}=-1$
for the PCM as a limit of these models.} of the square root,
then the spectral function ${z}(w;\t,\s)$ in \rf{lpl} has the finite limit ${z} =\sqrt{\frac{w+\s+\t}{w+\s-\t}} \to z_{\infty} = -1$ at $\s \to \pm \infty$.
Hence, assuming $J_\pm$ is bounded for large $|\s|$, the Lax connection $\widehat{L}_\pm = \ha (1+{z}^{\pm 1} ) J_\pm$
vanishes at spatial infinity. The periodicity condition \rf{pe} is satisfied with $\widehat{L}_\t$  vanishing at spatial infinity,
so it follows  that all the components of $\M$ are conserved.

The boundary condition (that $J_\pm$ is bounded), needed above for the conservation of $\M$, is quite weak. A further, more stringent boundary condition comes from the requirement that $\M$ converges
as $(a,b)\to (-\infty, +\infty)$
to produce well-defined charges.

Although the monodromy is conserved, it is only defined implicitly by the first-order ordinary differential equation \eqref{px}, i.e.
\begin{align}
&\M^{-1} \del_\s \M = \widehat{L}_\s = \tfrac{1}{4}\big(1+\big[ \tfrac{w+\s-\t}{w+\s+\t} \big]^{1/2} \big)J_+ - \tfrac{1}{4}\big(1+\big[ \tfrac{w+\s-\t}{w+\s+\t} \big]^{-1/2}\big)J_- \ , \qquad \M\big|_{\s=-\infty} = I \ . \la{mpe}
\end{align}
Such an equation generally admits a solution but it is hard to find its explicit form. One possible approach
is to develop an expansion in the spectral parameter around a point where $\wL_\s$ vanishes (and thus the monodromy is trivial, $\M=I$).

In the usual time-independent PCM case (obtained by replacing $\sqrt{\tfrac{w+\s-\t}{w+\s+\t}} $ by a constant $z$ in \rf{mpe}), such an expansion around $z=-1$ yields the conserved `multi-local' Yangian charges in eq.\ \eqref{me}.
Each term in the expansion  around $z=-1$ is individually conserved because each term in the corresponding expansion of $L_\t$ (and hence the right hand side of \rf{ce}) vanishes at spatial infinity.

For the time-dependent theory, the only zero around which to expand $\widehat{L}_\s$ in \rf{mpe} is $w=\infty$ (again taking the negative branch of the square root). But, due to its explicit spatial dependence, the corresponding expansion of the Lax connection
is a sum of terms $w^{-n} P_{n-1}$, where $P_{n-1}$ is a polynomial of degree $n-1$ in $\s$. Then for any polynomial decaying boundary conditions on the fields at spatial infinity, the periodicity condition \rf{pe} on $\widehat{L}_\t$ will be broken at sufficiently high orders in this expansion.

Hence, while we have shown the formal existence of conserved non-local charges defined implicitly by \rf{mpe}, it appears to be hard to compute the monodromy matrix $\M$ explicitly: due to the explicit $\s$ dependence in the Lax connection, the usual expansion trick \eqref{me} does not easily work.
It is not clear at the moment how
to verify
if the resulting conserved charges are infinite in number (and independent)
as they should be for an integrable 2d theory.

\subsection{1d reductions}\la{s43}

One useful test of 2d integrability is to check the integrability of various 1d mechanical theories obtained as
consistent reductions of the 2d equations of motion.

Assuming an ansatz for a classical solution of an integrable 2d model
yielding a 1d system of equations (say with $\t$ as the remaining independent variable), one may expect
the resulting 1d system also to be integrable in the sense of admitting a Lax pair, $\tfrac{d}{d\t} A = [B, A]$. One should further demand that the conserved charges are in involution.

In general, if the solution satisfies the periodicity condition \rf{pe}, then  the flatness of the 2d Lax connection leads to a 1d Lax pair
given by $(A,B)=(\M, -L_\t(\infty))$ in \rf{lp}, evaluated on the reduction ansatz. For example, in \cite{Arutyunov:2003za}, such a reduction was performed for a certain `spinning string' ansatz in $AdS_5 \times S^5$, resulting in an integrable 1d Neumann-Rosochatius model. There (reversing the roles of $\t$ and $\s$ from that paper to match the above discussion) the $\s$-dependence of the ansatz was simple and it was possible to completely remove the $\s$-dependence from the Lax connection using a gauge transformation --- which is essentially the same problem as computing the monodromy matrix. In our case, with explicit $\s$-dependence already in the Lax connection, the problem of directly removing the $\s$-dependence by gauge transformations would appear to be much more difficult.

\paragraph{Trivial reduction.}
In the simplest `trivial' 1d reduction one assumes that the
fields do not depend on $\s$. This is indeed a consistent truncation of the above time-dependent models, obtained by replacing $\del_\pm \to \t$ in their Lagrangians.
For example, in the PCM case, setting $g=g(\tau)$ in \rf{lcPCM} gives the 1d action for
geodesic motion on a group space with a time-dependent radius.
The solvability of geodesics in this model is
obvious since the explicit time dependence may be eliminated by a redefinition of $\t$
\be
\la{48} S_1 \sim \int d\t \ \t \ \Tr\big[ (g^{-1} \del_\t g )^2 \big] = \int d\t' \ \Tr\big[ (g^{-1} \del_{\t'} g )^2 \big] \ , \qquad \ \ \t'=\log \t \ .
\ee
The resulting equation of motion $\del_{\t'} (g^{-1} \del_{\t'} g) = 0$ is solved by
\be
g = g_0\, e^{\t' u_0} = g_0\, \t^{u_0 } \ , \qquad \quad g_0 = \const \in G \ , \quad \ u_0=\const \in \Lie(G) \ .
\ee
The same argument equally applies for any case where the time dependence only appears as an overall factor rescaling the Lagrangian -- which will follow if there are enough global symmetries that only the overall `radius' can run under the RG flow (e.g.\ also in the symmetric space $\s$-model case).

In general, the global symmetry charges may be found from the monodromy $\M$. Formally, the monodromy on an infinite line does not converge upon the trivial reduction since $J_\t = g^{-1} \del_\t g$ does not decay at spatial infinity (as it is now independent of $\s$).
On a finite space interval, the periodicity condition \rf{pe} is not satisfied so charges will not be conserved. However, making an expansion around the zero of the Lax connection at $w=\infty$ (cf.\ \eqref{me}), the global symmetry is reflected in the $\s$ independence of the leading $\O(w^{-1})$ term in $\wL_\t$,
\begin{align}
\wL_\t &= \tfrac{1}{4}(1+\big[ \tfrac{w+\s-\t}{w+\s+\t} \big]^{1/2})J_+ + \tfrac{1}{4}(1+\big[ \tfrac{w+\s-\t}{w+\s+\t} \big]^{-1/2})J_-
= \tfrac{1}{2} \t J_\s \, w^{-1} + \O(w^{-2})\ .
\end{align}
This puts the leading term on the right hand side of \rf{ce} in the commutator form \rf{lp}. Since the monodromy matrix is the identity at the leading order,
the right hand side of \rf{ce} is of $\O(w^{-2})$. It then follows that the $\O(w^{-1})$ term in the monodromy matrix, which is the global $G_R$ Noether charge, is
still conserved\foot{The Noether charge for $G_L$ symmetry is obtained similarly after a gauge transformation of the Lax connection and choosing instead the positive branch of the square root.}
\begin{align}
&\M = 1 + \ha w^{-1} Q_R + \O(w^{-2}) \ , \qquad \del_\t Q_R = 0 \ , \qquad Q_R = \int d\s \ \t J_\t \ . \la{noet}
\end{align}

\paragraph{Non-trivial reduction.}
Now let us consider a non-trivial 1d reduction when the 2d fields \textit{do} depend on $\s$ in a particular prescribed way. Such a reduction will lead to a
non-trivial 1d model, giving a more stringent test of the integrability of the time-dependent 2d model.

Starting from the time-dependent $SU(2)$ PCM \rf{lcPCM} parametrized as
\begin{align}
&\wLL = -\tfrac{c}{2} \, \t \, \Tr [ J_+ J_- ] = c\, \t \, \big[ \del_+ \theta \del_- \theta + \sin^2{\theta} (\del_+ \phi \del_- \phi + \sin^2{\phi}\, \del_+ \psi \del_- \psi)\big] \ , \la{412} \\
&\qquad g = n^a \s_a \ , \qquad \s_a = (I, \s_i ) \ , \ \ i=1,2,3 \ , \no \\
&\qquad n^0 n^0 - n^i n^i = 1 \ , \qquad n^a = ( \cos{\theta} , \ i \sin{\theta} \cos{\phi} , \ i \sin{\theta} \sin{\phi} \cos{\psi} , \ i \sin{\theta} \sin{\phi} \sin{\psi} ) \ , \no
\end{align}
let us consider, e.g., the following ansatz for a classical solution
\be
\theta = \theta(\t) \ , \qquad \phi = m \s \ , \qquad \psi = 0 \ . \la{wan}
\ee
This leads to a consistent reduction of the 2d theory: the $\phi$ and $\psi$ equations are both satisfied,
while the equation for $\theta(\t) $ follows from the 1d Lagrangian ($\dot{\theta} \equiv \del_\t \theta$)
\be
\L_1 = {c} \, \t\, \big( \dot{\theta}^2 - m^2 \sin^2 {\theta}\big) \ . \la{s1}
\ee
Thus the effective 1d model is a time-dependent analog of the `sine-Gordon' mechanics.

Let us assume the $\s$-direction is an infinite line, still using the ansatz \rf{wan} (e.g.\ by taking $m$ to be a continuous parameter and formally decompactifying $\phi$).
Then
one can check that $J_\pm$ is oscillating but bounded at spatial infinity. Hence, as discussed above, the right hand side of \rf{ce} vanishes and the entries of the monodromy matrix are conserved.
While again it is not easy to compute the monodromy explicitly in terms of $\theta$ and $\dot{\theta}$,
it can be done in the small field expansion $(\theta,\dot{\theta}) \to \e (\theta,\dot{\theta})$, $\e \ll 1$ since the Lax component $\wL_\s$ vanishes as $(\theta, \dot{\theta}) \to 0$. In this expansion the Lagrangian \rf{s1} becomes (after rescaling it by a factor of $c^{-1} \e^{-2}$)
\begin{align}
&\L_1' = \t\big[ \dot{\theta}^2 - m^2 \e^{-2} \sin^2 {(\e \theta)}\big] = \L^{lin} + \tfrac{1}{3} \e^2\, m^2\, \theta^4 + \O(\e^4) \ , \la{ne}\\
&\L^{lin} = \t \, \big[ \dot{\theta}^2 - m^2 \theta^2 \big] \ . \la{lt}
\end{align}
The leading (`linearized') Lagrangian \rf{lt} describes (after the same redefinition $\tau = e^{\tau'}$ as in \rf{48})
a harmonic oscillator with time-dependent frequency $m^2 e^{2 \tau'}$.
The corresponding $\O(\e)$ terms in the monodromy matrix lead to the conserved charge
\begin{align}
&\dot{Q}^{lin} = 0 \ , \qquad Q^{lin} = \t \big[ \bw(\t; w) \, \dot{\theta} - \dot \bw(\t; w) \, \theta \big] \ , \la{ql} \\
& \bw(\t;w) = \int_{-\infty}^{+\infty} d\s\, e^{-im \s} \tfrac{ i}{2 s_+ s_- }
\ , \qquad \qquad s_\pm \equiv \sqrt{w+\s \pm \t}\ . \la{ql2}
\end{align}
In Appendix \ref{2008C} we use this conserved charge to construct the general solution of the linearized equation of motion
(with $\theta=\bw$ being a particular solution).

Computing the perturbative corrections to the monodromy, one then finds for the
conserved charge of the non-linear theory \rf{ne}
\begin{align}
&\dot{Q} = 0 \ , \ \ \ Q = Q^{lin} + \e^2 \Big[\int d\s \, q^{(1)} + \int_{\s_1<\s_2} d\s_1 d\s_2 \, q^{(2)} + \int_{\s_1<\s_2<\s_3} d\s_1 d\s_2 d\s_3 \, q^{(3)} \Big] + \O(\e^4) \ , \no \\
&\ \ \ q^{(1)} = - \tfrac{m \, e^{-im\s} }{3s_+ s_-} (s_+ s_- -w-\s) \, \theta^3 \ , \la{cq}\\
&\ \ \ q^{(2)} = \tfrac{i m }{4s^1_+ s^1_-s^2_+ s^2_-} \Big[ e^{-im \s_2} ( s_+^1 s_-^1 - w - \s_1)\big[m(s_+^2 s_-^2 -w- \s_2)\theta +i \t \dot{\theta}\big] - (\s_1 \leftrightarrow \s_2) \Big] \, \theta^2\ , \no \\
&\ \ \ q^{(3)} = \tfrac{e^{-im(\s_1 - \s_2 + \s_3)}}{8 s^1_+ s^1_-s^2_+ s^2_-s^3_+ s^3_-} \big[m(s_+^1 s_-^1 -w- \s_1)+i \t \dot{\theta}\big] \no \\
&\qquad \qquad \times \big[ - m(s_+^2 s_-^2 -w- \s_2)+i \t \dot{\theta}\big]\big[m(s_+^3 s_-^3 -w- \s_3)+i \t \dot{\theta}\big] \ , \ \ \quad s_\pm^n \equiv \sqrt{w+\s_n \pm \t} \ . \no
\end{align}
This perturbative procedure of constructing the conserved charge can be extended to higher orders,
computing more and more $\s$ integrals at each step.

Let us also mention that the time-dependent $SL(2,\mathbb{R})$ PCM (cf.\ \rf{412}),
\be\la{420}
\wLL = - c \, \t \Big[ \del_+ \r \del_- \r + \ha e^{-2\r} \, (\del_+ x^+\del_- x^- + \del_- x^+\del_+ x^-) \Big] \ ,
\ee
admits a `non-trivial' reduction that is manifestly integrable.
Indeed, if we set $x^+ = a \s, \ x^- = b\s, \ \r=\r(\t)$ then the $x^\pm$ equations are solved
and the equation for $\r(\t)$ follows from the 1d action (cf.\ \rf{s1})
\be\la{421}
S_1 = - c \,\int d \t \, \t \big[ (\del_\t{\r})^2 - ab \, e^{-2\r} \big] \ .
\ee
Here the explicit time dependence can be eliminated by redefining $\t=e^{\t'}$ and $\r = \r' + \t'$
so that we end up with the standard 1d Liouville mechanics with the energy being a conserved charge.

%%%%%%%%%%%%%%%%%%%%%%%%%%%%%%%%%%%%%%%%
\section{Discussion}\la{s6}
%%%%%%%%%%%%%%%%%%%%%%%%%%%%%%%%%%%%%%%%

We have
observed
a surprising new connection between
classical integrability and the RG flow in 2d theories:
starting with an integrable theory and promoting its couplings to time-dependent functions, its
Lax connection generalizes naturally to the resulting
time-dependent theory only if the coupling functions
solve the 1-loop RG equations
of the original theory.
We demonstrated this on six classes of integrable $\s$-models.

One interesting implication is
that the 1-loop $\beta$-functions, which are normally associated with 1-loop divergences in quantum theory,
can thus be obtained in these models through the classical procedure of requiring the existence of a flat Lax connection in
the corresponding time-dependent theory.

As we discussed at the start of the chapter,
such time-dependent models can be naturally embedded into string theory by
starting with a Weyl-invariant $\s$-model \rf{csm} with two extra null directions $(u,v)$
and fixing a l.c.\ gauge $u=\t$.
If the l.c.\ theory \rf{lct} is classically solvable, it is natural to expect that the corresponding
 string $\s$-model \rf{csm}  should be solvable,
since one may move
from $u=\t$ to
any other
solution of the $u$-equation $\del_+\del_- u = 0 $
by a conformal transformation. The $v$-equations
$ \del_\pm v =\ha \GG_{ij}(\t,x) \, \del_\pm x^i \del_\pm x^j$
(following from the conformal constraints)
are
linear and so are also readily solvable. It is not immediately clear, however,
if integrability in the sense of admitting a Lax representation should similarly lift up to the $(D+2)$-dimensional string $\s$-model directly.

The Lax connections for the time-dependent theories have an
unusual form (depending explicitly on $\t$ and $\s$, as in the special
examples discussed in \cite{Belinsky:1971nt,Breitenlohner:1986um}). However, given certain boundary conditions, the entries or
the eigenvalues
of the monodromy matrix on an infinite spatial line are conserved.
Due to the explicit $\s$ dependence in the Lax connection, it is hard to
evaluate the monodromy and find the conserved charges explicitly.
We considered some consistent reductions of the 2d equations of motion to time-dependent 1d systems.
The `trivial' ($\s$-independent)
reduction of the PCM is clearly integrable due to the remaining global symmetry.

For the `non-trivial' 1d reduction in Section \ref{2008s43} we evaluated the monodromy, and hence constructed a conserved charge \rf{cq} in perturbation theory in a small field expansion. The perturbative existence of this conserved charge (following from the existence of a Lax connection)
is a non-trivial property, suggesting
that the 1d theory \rf{s1} may be interpreted as integrable.  In contrast, with generic (time-dependent) potentials added to the linearized theory \rf{lt}, we would not expect the linearized conserved charge \rf{ql} to admit even a perturbative extension to the non-linear theory.
The time-dependent 1d theory \rf{s1} is certainly not integrable in the Liouville sense as it would fail the Kovacic algorithm test used in \cite{Basu:2011fw,Stepanchuk:2012xi}.
Here we are suggesting a broader notion of integrability relating  to the explicit solvability of the equations of motion (allowing non-Liouvillian solutions like Bessel functions, etc.).

One way to fully establish the classical integrability of the $\s$-models with local couplings
would be to construct an infinite tower of local conserved charges in involution, in the spirit of \cite{Goldschmidt:1980wq,Evans:1999mj,Evans:2000qx,LMV}.
In Section 5 of \cite{HLT2008} we  attempted this
for the PCM with `local couplings'
in the Hamiltonian formulation.
A naive application of the method used for constant couplings does not yield such charges, except in the case of the `chiral' theories only depending on $\xi^+$ or on $\xi^-$,
where it works with a slight modification (see Appendix D of \cite{HLT2008}).
Nevertheless, the form of the Lax matrix and $R$-matrix is suggestive of an underlying algebraic structure.
Understanding it may provide additional insights into the question of integrability.

While we considered time-dependent models at the classical level only, one may wonder if they themselves
are renormalizable, i.e.\ stable under RG flow. Renormalization of generalized
$\s$-models with target-space metric depending on the 2d coordinates was considered in \cite{Osborn:1987au}. In the case
we discussed (cf.\ \rf{lct}) when $\GG_{ij}$ depends only on $\t$, i.e. $\L=\GG_{ij}(\t, x) \del_+ x^i \del_- x^j$, the 1-loop logarithmic counterterm
is proportional to $ K_1= \GG_{ml} \GG^{jk} \del_\t x^m\, \del_\t \Gamma^l_{jk} + \frac{1}{4} \del_\t \GG^{ij} \del_\t \GG_{ij} $, where
$\Gamma^l_{jk}$ is the Christoffel connection of $\GG_{ij}$. Thus in general one needs to add also counterterms with one and no derivatives, i.e. $V_m(x) \del_\t x^m + T(x)$. However, in the `factorized' case when $\GG_{ij} (\t, x) = f(\t) \GG_{ij}(x),$
like in the time-dependent PCM \rf{lcPCM}, the above counterterm becomes an $x$-independent function,
$K_1\sim ( \del_\t f)^2 $. Furthermore, if $f$ is linear in $\tau$ as in \rf{lcPCM} then this $K_1$ is just a constant
and thus such a model is renormalizable.
This suggests that at least some time-dependent integrable models
discussed in this chapter may have
well-defined quantum generalizations.

\

Among other possible directions, it would be important to generalize the present investigation to
other integrable $\sigma$-models, e.g.\ the class of models based on complex homogeneous spaces
whose  Ricci flow was studied in \ci{byk,byk2} and  the $\mathbb{Z}_T$ coset models of \cite{Young:2005jv,yo}.
Another interesting class of examples would be the integrable $G\times G$ and $G\times G/H$ models studied in Chapter \ref{GGS} above.
Furthermore, while in this chapter we have only studied 2-derivative $\s$-models,
we expect the connection between
the requirement of integrability of the time-dependent theory and the RG flow to be more general and to apply also to
models with potentials.
Indeed,
one also finds this remarkable connection in the case of the sine-Gordon model,
\be
S= \tfrac{1}{4\pi} \int d\t d\s \, \tfrac{1}{g^2}\big[ \ha \del_+ x \del_- x + m^2 \cos{x}\big]\ . \la{sgm}
\ee
Replacing the couplings $(g,m)$ by functions\foot{For a discussion of the sine-Gordon model with a particular `non-integrable'
local coupling $m=m(\s)$ see \cite{Levkov:2020lfa} and references there.}
of $\t$
one can show
that, for the resulting theory to be integrable
(assuming a natural ansatz for a generalization of the Lax connection to the time-dependent theory
---
see Appendix \ref{2008sing}),
the functions $(g(\t),m(\t))$ should be solutions of the 1-loop RG flow equation for \rf{sgm}, i.e.
should be given by
\be m^2(\t) = e^{ \b(g)\, \t} \, m_0^2 \ , \ \ \ \qquad g(\t) = g \ , \ \ \ \ \qquad \b(g) = - 2 + g^2 \ . \la{52}\ee
The time-dependent theory,
\be
\widehat{S}= \tfrac{1}{4\pi} \int d\t d\s \, \tfrac{1}{g^2} \big[ \ha \del_+ x \del_- x + e^{\b(g)\, \t}\, m^2_0\, \cos{x}\big]\ ,\la{53}
\ee
is indeed classically integrable since the explicit $\t$-dependence in \rf{53} can
be removed by a 2d conformal transformation getting back to \rf{sgm}:
under $\xi^\pm \to f^\pm (\xi^\pm)$ the action
\rf{sgm} retains its form with
$m^2 \to {f^+}'(\xi^+)\, {f^-}'(\xi^-)\, m^2 $.
Note that
the non-trivial 1d reduction of the $SU(2)$ PCM in eq.\ \rf{s1} is also obtained as the `trivial' reduction $x=x(\t)$
of \rf{53}
after a redefinition $\t \to \log \t $, supporting the expectation that \rf{s1} is an integrable 1d theory.
It would be interesting to explore
whether
this time-dependent integrability--RG flow connection applies also to
other examples of integrable massive theories, such as complex sine-Gordon and Toda models.

\

One of the features of our construction of the Lax connection for time-dependent theories is that
the constant spectral parameter of the original theory
is replaced by
a function
$z\to z(w;\t,\s)$ of a new spectral parameter $w$ and the 2d coordinates (cf.
\rf{pla},\rf{lpl}).
This puts the spectral parameter and 2d space-time coordinates on a more equal footing,
suggesting a possible interpretation from the point of view of the
construction \cite{Costello:2019tri}
of many
integrable 2d theories from a 4d Chern-Simons theory,
with the two extra directions related to the complex spectral parameter (see also \cite{Delduc:2019whp} and refs.\ there).
In that context the redefinition of $z$ is like changing
the differential structure of the 4d space, i.e.\ replacing $\del_{\t,\s} z = 0$
with $\del_{\t,\s} z = V_{\t,\s}(z;\t,\s)$ (cf.\ \rf{pd1}).
It would be interesting to see if the time-dependent models considered above
can  be reproduced from the 4d Chern-Simons construction. 
Indeed, the $\s$-dependent version of the PCM discussed above
has already been obtained \cite{Penna:2020uky} from a closely related Chern-Simons theory on twistor space.

%% file: refs.tex
%%%%%%%%%%%%%%%%%%%%%%%%%%%%%%%

%%%%%%%%%%%%%%%%%%%%%%%%%%%%%%%

%% file: main.bbl
\begin{thebibliography}{999}
\renewcommand{\bibname}{References}
\addcontentsline{toc}{chapter}{\bibname}
%
%%%refs
%
%\cite{Hoare:2018jim}
\bibitem{HLTS}{
B.~Hoare, N.~Levine and A.~A.~Tseytlin,
``On the massless tree-level S-matrix in 2d sigma models,''
\doilink{J. Phys. A \textbf{52}, no.14, 144005 (2019)}{
doi:10.1088/1751-8121/ab0b79}
\arxivlink{1812.02549}.
%16 citations counted in INSPIRE as of 11 Jul 2021
%
%
}\bibitem{HLT1}{
B.~Hoare, N.~Levine and A.~A.~Tseytlin,
``Integrable 2d sigma models: quantum corrections to geometry from RG flow,''
\doilink{Nucl. Phys. B \textbf{949}, 114798 (2019)}{10.1016/j.nuclphysb.2019.114798}
\arxivlink{1907.04737}.
%
}\bibitem{HLT2}{
B.~Hoare, N.~Levine and A.~A.~Tseytlin,
``Integrable sigma models and 2-loop RG flow,''
\doilink{JHEP {\bf 1912}, 146 (2019)}{doi:10.1007/JHEP12(2019)146}
\arxivlink{1910.00397}.
%
%\cite{Hoare:2020fye}
}\bibitem{HLT2008}{
B.~Hoare, N.~Levine and A.~A.~Tseytlin,
``Sigma models with local couplings: a new integrability -- RG flow connection,''
\doilink{JHEP \textbf{11}, 020 (2020)}{10.1007/JHEP11(2020)020}
\arxivlink{2008.01112}.
%8 citations counted in INSPIRE as of 01 Jul 2021
%
%
%\cite{Levine:2021fof}
}\bibitem{LT2103}{
N.~Levine and A.~A.~Tseytlin,
``Integrability vs. RG flow in $G \times G$ and $G \times G /H$ sigma models,''
\doilink{JHEP \textbf{05}, 076 (2021)}{
doi:10.1007/JHEP05(2021)076}
\arxivlink{2103.10513}.
%3 citations counted in INSPIRE as of 11 Jul 2021
%
%
%
%
%
%
%
%\cite{Maldacena:1997re}
}
\bibitem{Maldacena:1997re}{
J.~M.~Maldacena,
``The Large N limit of superconformal field theories and supergravity,''
\doilink{Adv. Theor. Math. Phys. \textbf{2}, 231-252 (1998)}{10.1023/A:1026654312961}
\arxivlink{hep-th/9711200}.
%16616 citations counted in INSPIRE as of 13 May 2021
%
%\cite{Witten:1998qj}
}\bibitem{Witten:1998qj}{
E.~Witten,
``Anti-de Sitter space and holography,''
\doilink{Adv. Theor. Math. Phys. \textbf{2}, 253-291 (1998)}{10.4310/ATMP.1998.v2.n2.a2}
\arxivlink{hep-th/9802150}.
%10776 citations counted in INSPIRE as of 05 Jul 2021
%
%\cite{Gubser:1998bc}
}\bibitem{Gubser:1998bc}{
S.~S.~Gubser, I.~R.~Klebanov and A.~M.~Polyakov,
``Gauge theory correlators from noncritical string theory,''
\doilink{Phys. Lett. B \textbf{428}, 105-114 (1998)}{10.1016/S0370-2693(98)00377-3}
\arxivlink{hep-th/9802109}.
%9133 citations counted in INSPIRE as of 05 Jul 2021
%
%
%%\cite{Metsaev:1998it}
}\bibitem{Metsaev:1998it}{
R.~R.~Metsaev and A.~A.~Tseytlin,
``Type IIB superstring action in AdS(5) x S**5 background,''
\doilink{Nucl. Phys. B \textbf{533}, 109-126 (1998)}{
doi:10.1016/S0550-3213(98)00570-7}
\arxivlink{hep-th/9805028}.
%%763 citations counted in INSPIRE as of 13 May 2021
%%
%%\cite{Bena:2003wd}
}
\bibitem{Bena:2003wd}{
I.~Bena, J.~Polchinski and R.~Roiban,
``Hidden symmetries of the AdS(5) x S**5 superstring,''
\doilink{Phys. Rev. D \textbf{69}, 046002 (2004)}{
doi:10.1103/PhysRevD.69.046002}
\arxivlink{hep-th/0305116}.
%925 citations counted in INSPIRE as of 13 May 2021
%
%\cite{Das:2004hy}
}\bibitem{Das:2004hy}{
A.~K.~Das, J.~Maharana, A.~Melikyan and M.~Sato,
``The algebra of transition matrices for the AdS(5) x S**5 superstring,''
\doilink{JHEP \textbf{12}, 055 (2004)}{
doi:10.1088/1126-6708/2004/12/055}
\arxivlink{hep-th/0411200}.
%72 citations counted in INSPIRE as of 13 May 2021
%
%\cite{Young:2005jv}
}\bibitem{Tseytlin:1988rr}{
A.~A.~Tseytlin,
``Sigma model approach to string theory,''
\doilink{Int. J. Mod. Phys. A \textbf{4}, 1257 (1989)}{
doi:10.1142/S0217751X8900056X}.
}\bibitem{Young:2005jv}{
C.~A.~S.~Young,
``Non-local charges, Z(m) gradings and coset space actions,''
\doilink{Phys. Lett. B \textbf{632}, 559-565 (2006)}{doi:10.1016/j.physletb.2005.10.090}
\arxivlink{hep-th/0503008}.
%
%\cite{Magro:2008dv}
}\bibitem{Magro:2008dv}{
M.~Magro,
``The classical exchange algebra of AdS(5) x S**5,''
\doilink{JHEP \textbf{01}, 021 (2009)}{
doi:10.1088/1126-6708/2009/01/021}
\arxivlink{0810.4136}.
%41 citations counted in INSPIRE as of 01 Jun 2021
%
%\cite{Vicedo:2009sn}
}\bibitem{Vicedo:2009sn}{
B.~Vicedo,
``Hamiltonian dynamics and the hidden symmetries of the AdS(5) x S**5 superstring,''
\doilink{JHEP \textbf{01}, 102 (2010)}{
doi:10.1007/JHEP01(2010)102}
\arxivlink{0910.0221}.
%29 citations counted in INSPIRE as of 01 Jun 2021
%
%
%
%\cite{Zamolodchikov:1989cf}
}\bibitem{ZTBA}{
A.~B.~Zamolodchikov,
``Thermodynamic Bethe Ansatz in relativistic models. Scaling three state Potts and Lee-Yang models,''
\doilink{Nucl. Phys. B \textbf{342}, 695-720 (1990)}
{doi:10.1016/0550-3213(90)90333-9}.
%642 citations counted in INSPIRE as of 02 Jul 2021
%
%\cite{Yang:1968rm}
}\bibitem{YTBA}{
C.~N.~Yang and C.~P.~Yang,
``Thermodynamics of one-dimensional system of bosons with repulsive delta function interaction,''
\doilink{J. Math. Phys. \textbf{10}, 1115-1122 (1969)}{
doi:10.1063/1.1664947}.
%585 citations counted in INSPIRE as of 02 Jul 2021
%
%\cite{Gromov:2013pga}
}\bibitem{Gromov:2013pga}{
N.~Gromov, V.~Kazakov, S.~Leurent and D.~Volin,
``Quantum spectral curve for Planar $\mathcal{N} = 4$ super-Yang-Mills theory,''
\doilink{Phys. Rev. Lett. \textbf{112}, no.1, 011602 (2014)}{
doi:10.1103/PhysRevLett.112.011602}
\arxivlink{1305.1939}.
%238 citations counted in INSPIRE as of 05 Jul 2021
%
%\cite{Levkovich-Maslyuk:2019awk}
}\bibitem{Levkovich-Maslyuk:2019awk}{
F.~Levkovich-Maslyuk,
``A review of the AdS/CFT Quantum Spectral Curve,''
\doilink{J. Phys. A \textbf{53}, no.28, 283004 (2020)}{
doi:10.1088/1751-8121/ab7137}
\arxivlink{1911.13065}.
%6 citations counted in INSPIRE as of 05 Jul 2021
%
%\cite{Minahan:2002ve}
}\bibitem{Minahan:2002ve}{
J.~A.~Minahan and K.~Zarembo,
``The Bethe ansatz for N=4 superYang-Mills,''
\doilink{JHEP \textbf{03}, 013 (2003)}{
doi:10.1088/1126-6708/2003/03/013}
\arxivlink{hep-th/0212208}.\\
%1317 citations counted in INSPIRE as of 01 Jun 2021
%\cite{Beisert:2003yb}
%}\bibitem{Beisert:2003yb}
%
N.~Beisert and M.~Staudacher,
``The N=4 SYM integrable super spin chain,''
\doilink{Nucl. Phys. B \textbf{670}, 439-463 (2003)}{
doi:10.1016/j.nuclphysb.2003.08.015}
\arxivlink{hep-th/0307042}.
%684 citations counted in INSPIRE as of 02 Jul 2021
%
%\cite{Beisert:2003tq}
}\bibitem{BeisertHL}{
N.~Beisert, C.~Kristjansen and M.~Staudacher,
``The Dilatation operator of conformal N=4 superYang-Mills theory,''
\doilink{Nucl. Phys. B \textbf{664}, 131-184 (2003)}{
doi:10.1016/S0550-3213(03)00406-1}
\arxivlink{hep-th/0303060}.
%714 citations counted in INSPIRE as of 02 Jul 2021
%\cite{Beisert:2005fw}
}\bibitem{BeisertBA}{
N.~Beisert and M.~Staudacher,
``Long-range psu(2,2|4) Bethe Ansatze for gauge theory and strings,''
\doilink{Nucl. Phys. B \textbf{727}, 1-62 (2005)}{
doi:10.1016/j.nuclphysb.2005.06.038}
\arxivlink{hep-th/0504190}.
%620 citations counted in INSPIRE as of 02 Jul 2021
%
%\cite{Beisert:2005tm}
}\bibitem{BeisertS}{
N.~Beisert,
``The SU(2|2) dynamic S-matrix,''
\doilink{Adv. Theor. Math. Phys. \textbf{12}, 945-979 (2008)}{
doi:10.4310/ATMP.2008.v12.n5.a1}
\arxivlink{hep-th/0511082}.
%641 citations counted in INSPIRE as of 02 Jul 2021
%
%
%\cite{Eden:2012fe}
}\bibitem{Eden:2012fe}{
B.~Eden, P.~Heslop, G.~P.~Korchemsky, V.~A.~Smirnov and E.~Sokatchev,
``Five-loop Konishi in N=4 SYM,''
\doilink{Nucl. Phys. B \textbf{862}, 123-166 (2012)}{
doi:10.1016/j.nuclphysb.2012.04.015}
\arxivlink{1202.5733}.\\
%54 citations counted in INSPIRE as of 05 Jul 2021
%%\cite{Gromov:2015dfa}
%}\bibitem{Gromov:2015dfa}
%
N.~Gromov and F.~Levkovich-Maslyuk,
``Quantum Spectral Curve for a cusped Wilson line in $ \mathcal{N}=4 $ SYM,''
\doilink{JHEP \textbf{04}, 134 (2016)}{
doi:10.1007/JHEP04(2016)134}
\arxivlink{1510.02098}.
%72 citations counted in INSPIRE as of 05 Jul 2021
%\cite{Lovelace:1983yv}
}\bibitem{Lovelace:1983yv}{
C.~Lovelace,
``Strings in curved space,''
\doilink{Phys. Lett. B \textbf{135}, 75-77 (1984)}{
doi:10.1016/0370-2693(84)90456-8}.\\
%400 citations counted in INSPIRE as of 05 Jul 2021
%\cite{Candelas:1985en}
%}\bibitem{Candelas:1985en}
P.~Candelas, G.~T.~Horowitz, A.~Strominger and E.~Witten,
``Vacuum configurations for superstrings,''
\doilink{Nucl. Phys. B \textbf{258}, 46-74 (1985)}{
doi:10.1016/0550-3213(85)90602-9}.\\
%2870 citations counted in INSPIRE as of 05 Jul 2021
%
%\cite{Sen:1985eb}
%}\bibitem{Sen:1985eb}
A.~Sen,
``The heterotic string in arbitrary background field,''
\doilink{Phys. Rev. D \textbf{32}, 2102 (1985)}{
doi:10.1103/PhysRevD.32.2102}.\\
%427 citations counted in INSPIRE as of 05 Jul 2021
%\cite{Callan:1985ia}
%}\bibitem{Callan:1985ia}
C.~G.~Callan, Jr., E.~J.~Martinec, M.~J.~Perry and D.~Friedan,
``Strings in background fields,''
\doilink{Nucl. Phys. B \textbf{262}, 593-609 (1985)}{
doi:10.1016/0550-3213(85)90506-1}.
%1779 citations counted in INSPIRE as of 05 Jul 2021
%
}\bibitem{kdv}{
C.~S.~Gardner, J.~M.~Greene, M.~D.~Kruskal and R.~M.~Miura,
``Method for solving the Korteweg-deVries equation,''
\doilink{Phys.\ Rev.\ Lett.\ 19(19), 1095 (1967)}{10.1103/PhysRevLett.19.1095}.
%
}\bibitem{lax}{
P.~D.~Lax,
``Integrals of nonlinear equations of evolution and solitary waves,''
\doilink{Commun. Pure Appl. Math. \textbf{21}, 467-490 (1968)}{10.1002/cpa.3160210503}.
%365 citations counted in INSPIRE as of 29 Jul 2021
%
%
}\bibitem{zs}{
V.~E.~E.~Zakharov and A.~B.~Shabat,
 ``Exact theory of two-dimensional self-focusing and one-dimensional self-modulation of waves in nonlinear media,''
  Soviet physics JETP, 34(1), 62 (1972);
``A scheme for integrating the nonlinear equations of mathematical physics by the method of the inverse scattering problem. I,''
\doilink{Functional Analysis and Its Applications 8(3), 226-235 (1974)}{10.1007/BF01075696};
``Integration of nonlinear equations of mathematical physics by the method of inverse scattering. II,''
\doilink{Functional Analysis and Its Applications, 13(3), 166-174 (1979)}{10.1007/BF01077483}.
%
%\cite{Zakharov:1973pp}
}\bibitem{zm}{
V.~E.~Zakharov and A.~V.~Mikhailov,
``Relativistically invariant two-dimensional models in field theory integrable by the inverse problem technique. (In Russian),''
Sov. Phys. JETP \textbf{47}, 1017-1027 (1978).
%398 citations counted in INSPIRE as of 05 Jul 2021
%
%\cite{Bernard:1990jw}
}\bibitem{Bernard:1990jw}{
D.~Bernard,
``Hidden Yangians in 2-D massive current algebras,''
\doilink{Commun. Math. Phys. \textbf{137}, 191-208 (1991)}{
doi:10.1007/BF02099123}.
%115 citations counted in INSPIRE as of 06 Jul 2021
%
%\cite{Eichenherr:1979ci}
}\bibitem{Eichenherr:1979ci}{
H.~Eichenherr and M.~Forger,
``On the dual symmetry of the nonlinear sigma models,''
\doilink{Nucl. Phys. B \textbf{155}, 381-393 (1979)}{
doi:10.1016/0550-3213(79)90276-1}.
%237 citations counted in INSPIRE as of 06 Jul 2021
%
%\cite{Luscher:1977rq}
}\bibitem{lp}{
M.~Luscher and K.~Pohlmeyer,
``Scattering of massless lumps and nonlocal charges in the two-dimensional classical nonlinear sigma Model,''
\doilink{Nucl. Phys. B \textbf{137}, 46-54 (1978)}{
doi:10.1016/0550-3213(78)90049-4}.\\
%318 citations counted in INSPIRE as of 06 Jul 2021
%\cite{Pohlmeyer:1975nb}
%}\bibitem{Pohlmeyer:1975nb}
%
K.~Pohlmeyer,
``Integrable Hamiltonian systems and interactions through quadratic constraints,''
\doilink{Commun. Math. Phys. \textbf{46}, 207-221 (1976)}{
doi:10.1007/BF01609119}.
%628 citations counted in INSPIRE as of 06 Jul 2021
%
%\cite{Alvarez:1994dn}
}\bibitem{Alvarez:1994dn}{
E.~Alvarez, L.~Alvarez-Gaume and Y.~Lozano,
``An introduction to T duality in string theory,''
\doilink{Nucl. Phys. B Proc. Suppl. \textbf{41}, 1-20 (1995)}{
doi:10.1016/0920-5632(95)00429-D}
\arxivlink{hep-th/9410237}.
%260 citations counted in INSPIRE as of 06 Jul 2021
%
%\cite{Buscher:1987sk}
%}\bibitem{Buscher:1987sk}{
%T.~H.~Buscher:
%``A symmetry of the string background field equations,''
%\doilink{Phys. Lett. B \textbf{194}, 59-62 (1987)}{
%doi:10.1016/0370-2693(87)90769-6};
%%898 citations counted in INSPIRE as of 06 Jul 2021
%%\cite{Buscher:1987qj}
%%}\bibitem{Buscher:1987qj}
%%T.~H.~Buscher,
%``Path Integral Derivation of Quantum Duality in Nonlinear Sigma Models,''
%\doilink{Phys. Lett. B \textbf{201}, 466-472 (1988)}{
%doi:10.1016/0370-2693(88)90602-8}.
%%918 citations counted in INSPIRE as of 06 Jul 2021
%%
%%
%%\cite{delaOssa:1992vci}
}\bibitem{delaOssa:1992vci}{
X.~C.~de la Ossa and F.~Quevedo,
``Duality symmetries from nonAbelian isometries in string theory,''
\doilink{Nucl. Phys. B \textbf{403}, 377-394 (1993)}{
doi:10.1016/0550-3213(93)90041-M}
\arxivlink{hep-th/9210021}.
%311 citations counted in INSPIRE as of 06 Jul 2021
%
%\cite{Wess:1971yu}
}\bibitem{Wess:1971yu}{
J.~Wess and B.~Zumino,
``Consequences of anomalous Ward identities,''
\doilink{Phys. Lett. B \textbf{37}, 95-97 (1971)}{
doi:10.1016/0370-2693(71)90582-X}.
%2832 citations counted in INSPIRE as of 06 Jul 2021
%
}\bibitem{novikov}{
 S. P. Novikov, 
 ``Multivalued Functions and Functionals. An Analogue of the Morse Theory,''
Dokl. Akad. Nauk SSSR Sect. Matem. 260, 31 (1981).
%
%}\bibitem{Witten:1983ar}
%E.~Witten,
%``Nonabelian Bosonization in Two-Dimensions,''
%Commun. Math. Phys. \textbf{92}, 455-472 (1984)
%doi:10.1007/BF01215276
%%2136 citations counted in INSPIRE as of 06 Jul 2021
%
%
%\cite{Piette:1987bt}
}\bibitem{cherednik}{
I.~V.~Cherednik,
``Integrability of the equation of a two-dimesional asymmetric $O(3)$ field and of its quantum analog,''
Yad. Fiz. \textbf{33}, 278 (1981) [Sov. J. Nucl. Phys. \textbf{33}, 144 (1981)].
}\bibitem{Piette:1987bt}{
B.~Piette, R.~A.~Zait and W.~J.~Zakrzewski,
``Solutions of the U(n) sigma models with the Wess-Zumino term,''
Z. Phys. C \textbf{39}, 359 (1988),
DTP-87/33.
%2 citations counted in INSPIRE as of 06 Jul 2021
%
%
%\cite{Knizhnik:1984nr}
}\bibitem{Knizhnik:1984nr}{
V.~G.~Knizhnik and A.~B.~Zamolodchikov,
``Current algebra and Wess-Zumino model in two-dimensions,''
\doilink{Nucl. Phys. B \textbf{247}, 83-103 (1984)}{
doi:10.1016/0550-3213(84)90374-2}.
%1533 citations counted in INSPIRE as of 06 Jul 2021
%
%
}\bibitem{Polyakov:1983tt}{
A.~M.~Polyakov and P.~B.~Wiegmann,
``Theory of nonabelian Goldstone bosons,''
\doilink{Phys.\ Lett.\ B {\bf 131}, 121 (1983)}{10.1016/0370-2693(83)91104-8}.
%%CITATION = doi:10.1016/0370-2693(83)91104-8;%%
%
%\cite{Polyakov:1984et}
}\bibitem{Polyakov:1984et}{
A.~M.~Polyakov and P.~B.~Wiegmann,
``Goldstone fields in two-dimensions with multivalued actions,''
\doilink{Phys. Lett. B \textbf{141}, 223-228 (1984)}{
doi:10.1016/0370-2693(84)90206-5}.
%429 citations counted in INSPIRE as of 06 Jul 2021
%
%\cite{Goddard:1984vk}
}\bibitem{Goddard:1984vk}{
P.~Goddard, A.~Kent and D.~I.~Olive,
``Virasoro algebras and coset space models,''
\doilink{Phys. Lett. B \textbf{152}, 88-92 (1985)}{
doi:10.1016/0370-2693(85)91145-1}.
%752 citations counted in INSPIRE as of 06 Jul 2021
%
%\cite{Frishman:1992mr}
}\bibitem{Frishman:1992mr}{
Y.~Frishman and J.~Sonnenschein,
``Bosonization and QCD in two-dimensions,''
\doilink{Phys. Rept. \textbf{223}, 309-348 (1993)}{
doi:10.1016/0370-1573(93)90145-4}
\arxivlink{hep-th/9207017}.\\
%78 citations counted in INSPIRE as of 06 Jul 2021
%
P.~Bowcock,
``Canonical quantization of the gauged {Wess-Zumino} model,''
\doilink{Nucl. Phys. B \textbf{316}, 80-100 (1989)}{
doi:10.1016/0550-3213(89)90387-8}.\\
%84 citations counted in INSPIRE as of 06 Jul 2021
%
%\cite{Gawedzki:1988nj}
%}\bibitem{Gawedzki:1988nj}
K.~Gawedzki and A.~Kupiainen,
``Coset construction from functional integrals,''
\doilink{Nucl. Phys. B \textbf{320}, 625-668 (1989)}{
doi:10.1016/0550-3213(89)90015-1}.
%\\
%321 citations counted in INSPIRE as of 06 Jul 2021
%
%\cite{Karabali:1988au}
%}\bibitem{Karabali:1988au}
%D.~Karabali, Q.~H.~Park, H.~J.~Schnitzer and Z.~Yang,
%``A GKO construction based on a path integral formulation of gauged Wess-Zumino-Witten actions,''
%\doilink{Phys. Lett. B \textbf{216}, 307-312 (1989)}{
%doi:10.1016/0370-2693(89)91120-9}.
%241 citations counted in INSPIRE as of 06 Jul 2021
%
%
%
}\bibitem{Delduc:2017fib}{
F.~Delduc, B.~Hoare, T.~Kameyama and M.~Magro,
``Combining the bi-Yang-Baxter deformation, the Wess-Zumino term and TsT transformations in one integrable $\sigma$-model,''
\doilink{JHEP \textbf{10}, 212 (2017)}{
doi:10.1007/JHEP10(2017)212}
\arxivlink{1707.08371}.
%34 citations counted in INSPIRE as of 06 Jul 2021
%
%
%
%%\cite{Demulder:2017zhz}
%}\bibitem{Demulder:2017zhz}
%S.~Demulder, S.~Driezen, A.~Sevrin and D.~C.~Thompson,
%``Classical and Quantum Aspects of Yang-Baxter Wess-Zumino Models,''
%JHEP \textbf{03}, 041 (2018)
%doi:10.1007/JHEP03(2018)041
%[arXiv:1711.00084 [hep-th]].
%%31 citations counted in INSPIRE as of 07 Jul 2021
%
%\cite{Delduc:2016ihq}
}\bibitem{Delduc:2016ihq}{
F.~Delduc, S.~Lacroix, M.~Magro and B.~Vicedo,
``On q-deformed symmetries as Poisson\textendash{}Lie symmetries and application to Yang\textendash{}Baxter type models,''
\doilink{J. Phys. A \textbf{49}, no.41, 415402 (2016)}{
doi:10.1088/1751-8113/49/41/415402}
\arxivlink{1606.01712}.
%23 citations counted in INSPIRE as of 07 Jul 2021
%
%\cite{Delduc:2014kha}
}\bibitem{Delduc:2014kha}{
F.~Delduc, M.~Magro and B.~Vicedo,
``Derivation of the action and symmetries of the $q$-deformed $AdS_{5} \times S^{5}$ superstring,''
\doilink{JHEP \textbf{10}, 132 (2014)}{
doi:10.1007/JHEP10(2014)132}
\arxivlink{1406.6286}.
%143 citations counted in INSPIRE as of 08 Jul 2021
%
%
%\cite{
}\bibitem{Arutyunov:2015qva}{
G.~Arutyunov, R.~Borsato and S.~Frolov,
``Puzzles of $\eta$-deformed AdS$_5 \times$ S$^5$,''
\doilink{JHEP \textbf{12}, 049 (2015)}{
doi:10.1007/JHEP12(2015)049}
\arxivlink{1507.04239}.
%97 citations counted in INSPIRE as of 08 Jul 2021
%
%\cite{Arutyunov:2015mqj}
}\bibitem{Arutyunov:2015mqj}{
G.~Arutyunov, S.~Frolov, B.~Hoare, R.~Roiban and A.~A.~Tseytlin,
``Scale invariance of the $\eta$-deformed $AdS_5\times S^5$ superstring, T-duality and modified type II equations,''
\doilink{Nucl. Phys. B \textbf{903}, 262-303 (2016)}{
doi:10.1016/j.nuclphysb.2015.12.012}
\arxivlink{1511.05795}.
%138 citations counted in INSPIRE as of 08 Jul 2021
%
%
}\bibitem{etaspec}{
G.~Arutyunov, M.~de Leeuw and S.~J.~van Tongeren,
``The exact spectrum and mirror duality of the $(\text{AdS}_5{\times}S^5)_\eta$ superstring,''
\doilink{Theor. Math. Phys. \textbf{182}, no.1, 23-51 (2015)}{10.1007/s11232-015-0243-9} \arxivlink{1403.6104}.\\
%89 citations counted in INSPIRE as of 17 Oct 2020
%
R.~Klabbers and S.~J.~van Tongeren,
``Quantum Spectral Curve for the eta-deformed AdS$_5$xS$^5$ superstring,''
\doilink{Nucl. Phys. B \textbf{925}, 252-318 (2017)}{10.1016/j.nuclphysb.2017.10.005}
\arxivlink{1708.02894}.
%
%
%
}\bibitem{Parke:1980ki}{
S.~J.~Parke,
``Absence of particle production and factorization of the $S$ matrix in (1+1)-dimensional models,''
\doilink{Nucl. Phys. B \textbf{174}, 166-182 (1980)}{
doi:10.1016/0550-3213(80)90196-0}.
%62 citations counted in INSPIRE as of 08 Jul 2021
%
%\cite{}
}\bibitem{Zamolodchikov:1978xm}{
A.~B.~Zamolodchikov and A.~B.~Zamolodchikov,
``Factorized S matrices in two-dimensions as the exact solutions of certain relativistic quantum field models,''
\doilink{Annals Phys. \textbf{120}, 253-291 (1979)}{
doi:10.1016/0003-4916(79)90391-9}.
%1464 citations counted in INSPIRE as of 08 Jul 2021
%
%
%\cite{Polyakov:1977vm}
}\bibitem{Polyakov:1977vm}{
A.~M.~Polyakov,
``Hidden symmetry of the two-dimensional chiral fields,''
\doilink{Phys. Lett. B \textbf{72}, 224-226 (1977)}{
doi:10.1016/0370-2693(77)90707-9}.
%124 citations counted in INSPIRE as of 16 Jul 2021
%
%\cite{Luscher:1977uq}
}\bibitem{Luscher:1977uq}{
M.~Luscher,
``Quantum nonlocal charges and absence of particle production in the two-dimensional nonlinear sigma model,''
\doilink{Nucl. Phys. B \textbf{135}, 1-19 (1978)}{
doi:10.1016/0550-3213(78)90211-0}.
%283 citations counted in INSPIRE as of 16 Jul 2021
%
%\cite{Dorey:1996gd}
}\bibitem{Dorey:1996gd}{
P.~Dorey,
``Exact S matrices,''
\arxivlink{hep-th/9810026}.
%126 citations counted in INSPIRE as of 08 Jul 2021
%
%\cite{Wiegmann:1984pw} %all pcm
}\bibitem{Wiegmann:1984pw}{
P.~B.~Wiegmann,
``On the theory of nonabelian Goldstone bosons in two-dimensions: exact solution of the SU(N)xSU(n) nonlinear $\sigma$ model,''
\doilink{Phys. Lett. B \textbf{141}, 217 (1984)}{
doi:10.1016/0370-2693(84)90205-3}.\\
%
%
%112 citations counted in INSPIRE as of 08 Jul 2021
%\cite{Wiegmann:1984ec}
%}\bibitem{Wiegmann:1984ec}
P.~Wiegmann,
``Exact factorizaed S-matrix of the chiral field in two dimensions,''
\doilink{Phys. Lett. B \textbf{142}, 173-176 (1984)}{
doi:10.1016/0370-2693(84)91256-5}.
%88 citations counted in INSPIRE as of 08 Jul 2021
%
%
%\cite{Fateev:1994ai}
}\bibitem{Fateev:1994ai}{
V.~A.~Fateev, V.~A.~Kazakov and P.~B.~Wiegmann,
``Principal chiral field at large N,''
\doilink{Nucl. Phys. B \textbf{424}, 505-520 (1994)}{
doi:10.1016/0550-3213(94)90405-7}
\arxivlink{hep-th/9403099}.
%61 citations counted in INSPIRE as of 08 Jul 2021
%
%
%
%
%%\cite{Beisert:2010kk}
}\bibitem{qds}{
%\cite{Beisert:2008tw}
%}\bibitem{Beisert:2008tw}
N.~Beisert and P.~Koroteev,
``Quantum deformations of the one-dimensional Hubbard model,''
\doilink{J. Phys. A \textbf{41}, 255204 (2008)}{
doi:10.1088/1751-8113/41/25/255204}
\arxivlink{0802.0777}.\\
%118 citations counted in INSPIRE as of 08 Jul 2021
%
N.~Beisert,
``The classical trigonometric r-matrix for the quantum-deformed Hubbard chain,''
\doilink{J. Phys. A \textbf{44}, 265202 (2011)}{
doi:10.1088/1751-8113/44/26/265202}
\arxivlink{1002.1097}.\\
%37 citations counted in INSPIRE as of 08 Jul 2021
%
%\cite{Hoare:2011wr}
%}\bibitem{Hoare:2011wr}
B.~Hoare, T.~J.~Hollowood and J.~L.~Miramontes,
``q-deformation of the $AdS_5 x S^5$ superstring S-matrix and its relativistic limit,''
\doilink{JHEP \textbf{03}, 015 (2012)}{
doi:10.1007/JHEP03(2012)015}
\arxivlink{1112.4485}.
%71 citations counted in INSPIRE as of 08 Jul 2021
%
%
%\cite{Seibold:2020ywq}
}\bibitem{ses}{
F.~K.~Seibold, S.~J.~Van Tongeren and Y.~Zimmermann,
``The twisted story of worldsheet scattering in $\eta$-deformed $AdS_5 \times S^5$,''
\doilink{JHEP \textbf{12}, 043 (2020)}{
doi:10.1007/JHEP12(2020)043}
\arxivlink{2007.09136}.\\
%4 citations counted in INSPIRE as of 08 Jul 2021
%
%\cite{Bocconcello:2020qkt}
%}\bibitem{Bocconcello:2020qkt}
M.~Bocconcello, I.~Masuda, F.~K.~Seibold and A.~Sfondrini,
``S matrix for a three-parameter integrable deformation of AdS$_{3} \times$  S$^{3}$ strings,''
\doilink{JHEP \textbf{11}, 022 (2020)}{
doi:10.1007/JHEP11(2020)022}
\arxivlink{2008.07603}.
%3 citations counted in INSPIRE as of 08 Jul 2021
% 
%
}\bibitem{Vicedo:2017cge}{
B.~Vicedo,
``On integrable field theories as dihedral affine Gaudin models,''
\doilink{Int. Math. Res. Not. \textbf{2020}, no.15, 4513-4601 (2020)}{
doi:10.1093/imrn/rny128}
\arxivlink{1701.04856}.
%34 citations counted in INSPIRE as of 09 Jul 2021}
%14 citations counted in INSPIRE as of 09 Jul 2021
%
%\cite{Lacroix:2018njs}
}\bibitem{Lacroix:2018njs}{
S.~Lacroix,
``Integrable models with twist function and affine Gaudin models,''
\arxivlink{1809.06811}.
%9 citations counted in INSPIRE as of 09 Jul 2021
%
%
%\cite{Lacroix:2020flf}
}\bibitem{Lacroix:2020flf}{
S.~Lacroix and B.~Vicedo,
``Integrable $\mathcal{E}$-models, 4d Chern-Simons theory and affine Gaudin models. I.~Lagrangian aspects,''
\doilink{SIGMA \textbf{17}, 058 (2021)}{
doi:10.3842/SIGMA.2021.058}
\arxivlink{2011.13809}.
%8 citations counted in INSPIRE as of 09 Jul 2021
%
%\cite{Vicedo:2019dej}
}\bibitem{Vicedo:2019dej}{
B.~Vicedo,
``Holomorphic Chern-Simons theory and affine Gaudin models,''
\arxivlink{1908.07511}.
%22 citations counted in INSPIRE as of 09 Jul 2021
%
%
%%\cite{Delduc:2016ihq}
%}\bibitem{Delduc:2016ihq}
%F.~Delduc, S.~Lacroix, M.~Magro and B.~Vicedo,
%``On q-deformed symmetries as Poisson\textendash{}Lie symmetries and application to Yang\textendash{}Baxter type models,''
%J. Phys. A \textbf{49}, no.41, 415402 (2016)
%doi:10.1088/1751-8113/49/41/415402
%[arXiv:1606.01712 [hep-th]].
%%23 citations counted in INSPIRE as of 09 Jul 2021
%
%
%\cite{Sfetsos:1996xj}
}\bibitem{Sfetsos:1996xj}{
K.~Sfetsos,
``Poisson-Lie T duality and supersymmetry,''
\doilink{Nucl. Phys. B Proc. Suppl. \textbf{56}, 302-309 (1997)}{
doi:10.1016/S0920-5632(97)00339-3}
\arxivlink{hep-th/9611199}.
%45 citations counted in INSPIRE as of 12 Jul 2021
%
%
%
%%\cite{Wulff:2016tju}
}\bibitem{Wulff:2016tju}{
L.~Wulff and A.~A.~Tseytlin,
``Kappa-symmetry of superstring sigma model and generalized 10d supergravity equations,''
\doilink{JHEP \textbf{06}, 174 (2016)}{
doi:10.1007/JHEP06(2016)174}
\arxivlink{1605.04884}.
%114 citations counted in INSPIRE as of 13 Jul 2021
%
%\cite{Bassi:2019aaf}
}\bibitem{Bassi:2019aaf}{
C.~Bassi and S.~Lacroix,
``Integrable deformations of coupled $\sigma$-models,''
\doilink{JHEP \textbf{05}, 059 (2020)}{
doi:10.1007/JHEP05(2020)059}
\arxivlink{1912.06157}.
%15 citations counted in INSPIRE as of 14 Jul 2021
%
%
%%\cite{Hoare:2010fb}
%}\bibitem{Hoare:2010fb}
%B.~Hoare and A.~A.~Tseytlin,
%``On the perturbative S-matrix of generalized sine-Gordon models,''
%JHEP \textbf{11}, 111 (2010)
%doi:10.1007/JHEP11(2010)111
%[arXiv:1008.4914 [hep-th]].
%%27 citations counted in INSPIRE as of 15 Jul 2021
%
%
%\cite{Coleman:1973ci}
}\bibitem{Coleman:1973ci}{
S.~R.~Coleman,
``There are no Goldstone bosons in two-dimensions,''
\doilink{Commun. Math. Phys. \textbf{31}, 259-264 (1973)}{
doi:10.1007/BF01646487}.
%1025 citations counted in INSPIRE as of 15 Jul 2021
%
}\bibitem{Zamolodchikov:1992zr}{
A.~B.~Zamolodchikov and A.~B.~Zamolodchikov,
``Massless factorized scattering and \sms with topological terms,''
\doilink{Nucl.\ Phys.\ B {\bf 379}, 602 (1992)}{10.1016/0550-3213(92)90136-Y}.\\
%
%doi:10.1016/0550-3213(92)90136-Y
%%CITATION = doi:10.1016/0550-3213(92)90136-Y;%%
P.~Fendley and H.~Saleur,
``Massless integrable quantum field theories and massless scattering in (1+1)-dimensions,''
\arxivlink{hep-th/9310058}.
%%CITATION = HEP-TH/9310058;%%
%
}\bibitem{Nappi:1979ig}{C.~R.~Nappi,
``Some properties of an analog of the nonlinear $\sigma$ model,''
\doilink{Phys.\ Rev.\ D {\bf 21}, 418 (1980)}{doi:10.1103/PhysRevD.21.418}.\\
%%CITATION = doi:10.1103/PhysRevD.21.418;%%
F.~E.~Figueirido,
``Particle creation in a conformally invariant supersymmetric model,''
\doilink{Phys.\ Lett.\ B {\bf 227}, 392 (1989)}{
doi:10.1016/0370-2693(89)90949-0}.
%%CITATION = doi:10.1016/0370-2693(89)90949-0;%%
}\bibitem{Wulff}{
L.~Wulff,
``Condition on Ramond-Ramond fluxes for factorization of worldsheet scattering in anti\textendash{}de Sitter space,''
\doilink{Phys. Rev. D \textbf{96}, no.10, 101901 (2017)}{
doi:10.1103/PhysRevD.96.101901}
\arxivlink{1708.09673};
%14 citations counted in INSPIRE as of 15 Jul 2021
``Classifying integrable symmetric space strings via factorized scattering,''
\doilink{JHEP \textbf{02}, 106 (2018)}{
doi:10.1007/JHEP02(2018)106}
\arxivlink{1711.00296};
%10 citations counted in INSPIRE as of 15 Jul 2021
%L.~Wulff,
``Constraining integrable AdS/CFT with factorized scattering,''
\doilink{JHEP \textbf{04}, 133 (2019)}{
doi:10.1007/JHEP04(2019)133}
\arxivlink{1903.08660}.
%11 citations counted in INSPIRE as of 15 Jul 2021
%
%\cite{Berenstein:2002jq}
}\bibitem{BMN}{
D.~E.~Berenstein, J.~M.~Maldacena and H.~S.~Nastase,
``Strings in flat space and pp waves from N=4 superYang-Mills,''
\doilink{JHEP \textbf{04}, 013 (2002)}{
doi:10.1088/1126-6708/2002/04/013}
\arxivlink{hep-th/0202021}.
%1849 citations counted in INSPIRE as of 15 Jul 2021
%
%\cite{Gubser:2002tv}
}\bibitem{Gubser:2002tv}{
S.~S.~Gubser, I.~R.~Klebanov and A.~M.~Polyakov,
``A semiclassical limit of the gauge / string correspondence,''
\doilink{Nucl. Phys. B \textbf{636}, 99-114 (2002)}{
doi:10.1016/S0550-3213(02)00373-5}
\arxivlink{hep-th/0204051}.
%1013 citations counted in INSPIRE as of 15 Jul 2021
%
%\cite{Kruczenski:2002fb}
}\bibitem{Kruczenski}{
M.~Kruczenski,
``A Note on twist two operators in N=4 SYM and Wilson loops in Minkowski signature,''
\doilink{JHEP \textbf{12}, 024 (2002)}{
doi:10.1088/1126-6708/2002/12/024}
\arxivlink{hep-th/0210115}.
%200 citations counted in INSPIRE as of 15 Jul 2021
%
M.~Kruczenski, R.~Roiban, A.~Tirziu and A.~A.~Tseytlin,
``Strong-coupling expansion of cusp anomaly and gluon amplitudes from quantum open strings in AdS(5) x S**5,''
\doilink{Nucl. Phys. B \textbf{791}, 93-124 (2008)}{
doi:10.1016/j.nuclphysb.2007.09.005}
\arxivlink{0707.4254}.
%107 citations counted in INSPIRE as of 15 Jul 2021
%
%
}\bibitem{Dubovsky}{
S.~Dubovsky, R.~Flauger and V.~Gorbenko,
``Effective string theory revisited,''
\doilink{JHEP {\bf 1209}, 044 (2012)}{
doi:10.1007/JHEP09(2012)044}
\arxivlink{1203.1054};
%%CITATION = doi:10.1007/JHEP09(2012)044;%%
%S.~Dubovsky, R.~Flauger and V.~Gorbenko,
``Solving the simplest theory of quantum gravity,''
\doilink{JHEP {\bf 1209}, 133 (2012)}{
doi:10.1007/JHEP09(2012)133}
\arxivlink{1205.6805}.\\
%%CITATION = doi:10.1007/JHEP09(2012)133;%%
%
%}\bibitem{Cooper:2014noa}
 P.~Cooper, S.~Dubovsky, V.~Gorbenko, A.~Mohsen and S.~Storace,
``Looking for integrability on the worldsheet of confining strings,''
\doilink{JHEP {\bf 1504}, 127 (2015)}{
doi:10.1007/JHEP04(2015)127}
\arxivlink{1411.0703}.
%%CITATION = doi:10.1007/JHEP04(2015)127;%%
%
}\bibitem{Fradkin:1985ys}{
E.~S.~Fradkin and A.~A.~Tseytlin,
``Quantum string theory effective action,''
\doilink{Nucl. Phys. B \textbf{261}, 1-27 (1985)}{doi:10.1016/0550-3213(85)90559-0} 
Erratum: [Nucl. Phys. B \textbf{269}, 745-745 (1986)].
%doi:10.1016/0550-3213(85)90559-0
%850 citations counted in INSPIRE as of 16 Jul 2021
%
%%\cite{Borsato:2016ose}
%}\bibitem{Borsato:2016ose}
%R.~Borsato and L.~Wulff,
%``Target space supergeometry of $\eta$ and $\lambda$-deformed strings,''
%JHEP \textbf{10}, 045 (2016)
%doi:10.1007/JHEP10(2016)045
%[arXiv:1608.03570 [hep-th]].
%%98 citations counted in INSPIRE as of 16 Jul 2021
%
%\cite{Gross:1986iv}
}\bibitem{Gross:1986iv}{
D.~J.~Gross and E.~Witten,
``Superstring modifications of Einstein's equations,''
\doilink{Nucl. Phys. B \textbf{277}, 1 (1986)}{
doi:10.1016/0550-3213(86)90429-3}.
%862 citations counted in INSPIRE as of 16 Jul 2021
%
%\cite{Roiban:2007jf}
}\bibitem{Roiban:2007jf}{
R.~Roiban, A.~Tirziu and A.~A.~Tseytlin,
``Two-loop world-sheet corrections in AdS(5) x S**5 superstring,''
\doilink{JHEP \textbf{07}, 056 (2007)}{
doi:10.1088/1126-6708/2007/07/056}
\arxivlink{0704.3638}.
%119 citations counted in INSPIRE as of 16 Jul 2021
%
%\cite{Valent:1984rj}
}\bibitem{Valent:1984rj}{
G.~Valent,
``Perturbative renormalization of the CP(N) model,''
\doilink{Nucl. Phys. B \textbf{238}, 142-166 (1984)}{
doi:10.1016/0550-3213(84)90470-X}.\\
%14 citations counted in INSPIRE as of 16 Jul 2021
%
%}\bibitem{Hikami:1979ih}
S.~Hikami,
``Renormalization group functions of {CP}$^{N-1}$ nonlinear sigma model and $N$ component scalar {QED} model,''
\doilink{Prog. Theor. Phys. \textbf{62}, 226 (1979)}{
doi:10.1143/PTP.62.226}.
%50 citations counted in INSPIRE as of 16 Jul 2021
%
%
%\cite{Rosenhaus:2019utc}
}\bibitem{Rosenhaus:2019utc}{
V.~Rosenhaus and M.~Smolkin,
``Integrability and renormalization under $T \bar T$,''
\doilink{Phys. Rev. D \textbf{102}, no.6, 065009 (2020)}{
doi:10.1103/PhysRevD.102.065009}
\arxivlink{1909.02640}.
%23 citations counted in INSPIRE as of 24 Jul 2021
%
%\cite{Zamolodchikov:2004ce}
}\bibitem{TTbar}{
A.~B.~Zamolodchikov,
``Expectation value of composite field T anti-T in two-dimensional quantum field theory,''
\arxivlink{hep-th/0401146}.\\
%187 citations counted in INSPIRE as of 24 Jul 202
%\cite{Cavaglia:2016oda}
%}\bibitem{Cavaglia:2016oda}
%
A.~Cavagli\`a, S.~Negro, I.~M.~Sz\'ecs\'enyi and R.~Tateo,
``$T \bar{T}$-deformed 2D quantum field theories,''
\doilink{JHEP \textbf{10}, 112 (2016)}{
doi:10.1007/JHEP10(2016)112}
\arxivlink{1608.05534}.\\
%220 citations counted in INSPIRE as of 24 Jul 2021
%\cite{Smirnov:2016lqw}
%}\bibitem{Smirnov:2016lqw}
%
F.~A.~Smirnov and A.~B.~Zamolodchikov,
``On space of integrable quantum field theories,''
\doilink{Nucl. Phys. B \textbf{915}, 363-383 (2017)}{
doi:10.1016/j.nuclphysb.2016.12.014}
\arxivlink{1608.05499}.
%243 citations counted in INSPIRE as of 24 Jul 2021
%
%
%%\cite{Chung:1992mj}
%}\bibitem{Chung:1992mj}
%S.~w.~Chung and S.~H.~H.~Tye,
%``Chiral gauged WZW theories and coset models in conformal field theory,''
%Phys. Rev. D \textbf{47}, 4546-4566 (1993)
%doi:10.1103/PhysRevD.47.4546
%[arXiv:hep-th/9202002 [hep-th]].
%%17 citations counted in INSPIRE as of 26 Jul 2021
%
%
%\cite{Sfetsos:1993sr}
%}\bibitem{Sfetsos:1993sr}
%6 citations counted in INSPIRE as of 26 Jul 2021
%
%}\bibitem{Sfetsos:1993bh}
%K.~Sfetsos and A.~A.~Tseytlin,
%``Chiral gauged WZNW models and heterotic string backgrounds,''
%Nucl. Phys. B \textbf{415}, 116-154 (1994)
%doi:10.1016/0550-3213(94)90069-8
%[arXiv:hep-th/9308018 [hep-th]].
%%20 citations counted in INSPIRE as of 26 Jul 2021
%
%\cite{Arutyunov:2009ga}
}\bibitem{Arutyunov:2009ga}{
G.~Arutyunov and S.~Frolov,
``Foundations of the AdS$_{5} \times S^{5}$ superstring. Part I,''
\doilink{J. Phys. A \textbf{42}, 254003 (2009)}{
doi:10.1088/1751-8113/42/25/254003}
\arxivlink{0901.4937}.
%345 citations counted in INSPIRE as of 26 Jul 2021
%
%\cite{vanTongeren:2013gva}
}\bibitem{vanTongeren:2013gva}{
S.~J.~van Tongeren,
``Integrability of the ${\rm Ad}{{{\rm S}}_{5}}\times {{{\rm S}}^{5}}$ superstring and its deformations,''
\doilink{J. Phys. A \textbf{47}, 433001 (2014)}{
doi:10.1088/1751-8113/47/43/433001}
\arxivlink{1310.4854}.
%53 citations counted in INSPIRE as of 26 Jul 2021
%
%\cite{Bombardelli:2016rwb}
}\bibitem{Bombardelli:2016rwb}{
D.~Bombardelli, A.~Cagnazzo, R.~Frassek, F.~Levkovich-Maslyuk, F.~Loebbert, S.~Negro, I.~M.~Sz\'ecs\'enyi, A.~Sfondrini, S.~J.~van Tongeren and A.~Torrielli,
``An integrability primer for the gauge-gravity correspondence: An introduction,''
\doilink{J. Phys. A \textbf{49}, no.32, 320301 (2016)}{
doi:10.1088/1751-8113/49/32/320301}
\arxivlink{1606.02945}.
%46 citations counted in INSPIRE as of 26 Jul 2021
%
%
%\cite{Tseytlin:2004xa}
}\bibitem{Tseytlin:2004xa}{
A.~A.~Tseytlin,
``Semiclassical strings and AdS/CFT,''
\arxivlink{hep-th/0409296}.
%151 citations counted in INSPIRE as of 26 Jul 2021
%
%\cite{Nemeschansky:1986yx}
}\bibitem{Nemeschansky:1986yx}{
D.~Nemeschansky and A.~Sen,
``Conformal invariance of supersymmetric $\sigma$ models on Calabi-yau manifolds,''
\doilink{Phys. Lett. B \textbf{178}, 365-369 (1986)}{
doi:10.1016/0370-2693(86)91394-8}.
%127 citations counted in INSPIRE as of 27 Jul 2021
%
%\cite{Novikov}
}\bibitem{Novikov}{
V.~A.~Novikov, M.~A.~Shifman, A.~I.~Vainshtein and V.~I.~Zakharov,
``Exact Gell-Mann-Low function of supersymmetric Yang-Mills theories from instanton calculus,''
\doilink{Nucl. Phys. B \textbf{229}, 381-393 (1983)}{
doi:10.1016/0550-3213(83)90338-3}.\\
%766 citations counted in INSPIRE as of 27 Jul 2021
%
V.~A.~Novikov, M.~A.~Shifman, A.~I.~Vainshtein and V.~I.~Zakharov,
``Two-dimensional sigma models: modeling nonperturbative effects of quantum chromodynamics,''
\doilink{Phys. Rept. \textbf{116}, 103 (1984)}{
doi:10.1016/0370-1573(84)90021-8}.
%199 citations counted in INSPIRE as of 27 Jul 2021
%
%\cite{Grisaru:1986px}
}\bibitem{Grisaru2}{
M.~T.~Grisaru, A.~E.~M.~van de Ven and D.~Zanon,
%``Four Loop beta Function for the N=1 and N=2 Supersymmetric Nonlinear Sigma Model in Two-Dimensions,''
%Phys. Lett. B \textbf{173}, 423-428 (1986)
%doi:10.1016/0370-2693(86)90408-9;
``Two-dimensional supersymmetric sigma models on Ricci flat Kahler manifolds are not finite,''
\doilink{Nucl. Phys. B \textbf{277}, 388-408 (1986)}{
doi:10.1016/0550-3213(86)90448-7}; 
``Four loop divergences for the N=1 supersymmetric nonlinear sigma model in two-dimensions,''
\doilink{Nucl. Phys. B \textbf{277}, 409-428 (1986)}{
doi:10.1016/0550-3213(86)90449-9}.
%236 citations counted in INSPIRE as of 27 Jul 2021
%233 citations counted in INSPIRE as of 27 Jul 2021
%
%\cite{Zumino:1979et}
%\cite{Borsato:2019oip}
}\bibitem{Borsato:2019oip}{
R.~Borsato and L.~Wulff,
``Two-loop conformal invariance for Yang-Baxter deformed strings,''
\doilink{JHEP \textbf{03}, 126 (2020)}{
doi:10.1007/JHEP03(2020)126}
\arxivlink{1910.02011}.
}\bibitem{Babelon}{
O.~Babelon, D.~Bernard and M.~Talon,
``Introduction to classical integrable systems,''
\doilink{Cambridge University Press (2009)}{10.1017/CBO9780511535024}.
}\bibitem{Zumino:1979et}{
B.~Zumino,
``Supersymmetry and Kahler manifolds,''
\doilink{Phys. Lett. B \textbf{87}, 203 (1979)}{
doi:10.1016/0370-2693(79)90964-X}.
%676 citations counted in INSPIRE as of 27 Jul 2021
%
%\cite{Abdalla:1985nm}
}\bibitem{Abdalla:1985nm}{
E.~Abdalla and M.~Forger,
``Integrable nonlinear $\sigma$ models with fermions,''
\doilink{Commun. Math. Phys. \textbf{104}, 123 (1986)}{
doi:10.1007/BF01210796}.
%21 citations counted in INSPIRE as of 28 Jul 2021
%
%\cite{Borsato:2016zcf}
}\bibitem{Borsato:2016zcf}{
R.~Borsato, A.~A.~Tseytlin and L.~Wulff,
``Supergravity background of $\lambda$-deformed model for AdS$_2 \times$  S$^2$ supercoset,''
\doilink{Nucl. Phys. B \textbf{905}, 264-292 (2016)}{
doi:10.1016/j.nuclphysb.2016.02.018}
\arxivlink{1601.08192}.
%61 citations counted in INSPIRE as of 28 Jul 2021
%
%\cite{Chervonyi:2016ajp}
}\bibitem{Lunin}{
O.~Lunin, R.~Roiban and A.~A.~Tseytlin,
``Supergravity backgrounds for deformations of AdS$_{n} \times S^n$ supercoset string models,''
\doilink{Nucl. Phys. B \textbf{891}, 106-127 (2015)}{
doi:10.1016/j.nuclphysb.2014.12.006}
\arxivlink{1411.1066}.\\
%67 citations counted in INSPIRE as of 28 Jul 2021
%
%
Y.~Chervonyi and O.~Lunin,
``Supergravity background of the $\lambda$-deformed AdS$_3 \times$ S$^3$ supercoset,''
\doilink{Nucl. Phys. B \textbf{910}, 685-711 (2016)}{
doi:10.1016/j.nuclphysb.2016.07.023}
\arxivlink{1606.00394}.
%48 citations counted in INSPIRE as of 28 Jul 2021
%
%\cite{Seibold:2019dvf}
}\bibitem{Seibold:2019dvf}{
F.~K.~Seibold,
``Two-parameter integrable deformations of the $AdS_3 \times S^3 \times T^4$ superstring,''
\doilink{JHEP \textbf{10}, 049 (2019)}{
doi:10.1007/JHEP10(2019)049}
\arxivlink{1907.05430}.
%11 citations counted in INSPIRE as of 28 Jul 2021
%
%
%\cite{Osten:2016dvf}
}\bibitem{Osten:2016dvf}{
D.~Osten and S.~J.~van Tongeren,
``Abelian Yang\textendash{}Baxter deformations and TsT transformations,''
\doilink{Nucl. Phys. B \textbf{915}, 184-205 (2017)}{
doi:10.1016/j.nuclphysb.2016.12.007}
\arxivlink{1608.08504}.
%78 citations counted in INSPIRE as of 28 Jul 2021
%
}\bibitem{homog}{
I.~Kawaguchi, T.~Matsumoto and K.~Yoshida,
``Jordanian deformations of the $AdS_5 x S^5$ superstring,''
\doilink{JHEP \textbf{04}, 153 (2014)}{
doi:10.1007/JHEP04(2014)153}
\arxivlink{1401.4855}.\\
%176 citations counted in INSPIRE as of 28 Jul 2021
%
T.~Matsumoto and K.~Yoshida,
``Yang\textendash{}Baxter sigma models based on the CYBE,''
\doilink{Nucl. Phys. B \textbf{893}, 287-304 (2015)}{
doi:10.1016/j.nuclphysb.2015.02.009}
\arxivlink{1501.03665}.\\
%79 citations counted in INSPIRE as of 28 Jul 2021
%
S.~J.~van Tongeren,
``Yang\textendash{}Baxter deformations, AdS/CFT, and twist-noncommutative gauge theory,''
\doilink{Nucl. Phys. B \textbf{904}, 148-175 (2016)}{
doi:10.1016/j.nuclphysb.2016.01.012}
\arxivlink{1506.01023}.
%93 citations counted in INSPIRE as of 28 Jul 2021
%
}\bibitem{homog2}{
B.~Hoare and A.~A.~Tseytlin,
``Homogeneous Yang-Baxter deformations as non-abelian duals of the $\mathrm{AdS}_5 \sigma$-model,''
J. Phys. A \textbf{49}, no.49, 494001 (2016)
doi:10.1088/1751-8113/49/49/494001
[arXiv:1609.02550 [hep-th]].
%63 citations counted in INSPIRE as of 28 Jul 2021
%
}\bibitem{homog3}{
R.~Borsato and L.~Wulff,
``Integrable feformations of $T$-dual $\sigma$ models,''
\doilink{Phys. Rev. Lett. \textbf{117}, no.25, 251602 (2016)}{
doi:10.1103/PhysRevLett.117.251602}
\arxivlink{1609.09834}; 
%72 citations counted in INSPIRE as of 31 Jul 2021
``On non-abelian T-duality and deformations of supercoset string sigma-models,''
\doilink{JHEP \textbf{10}, 024 (2017)}{
doi:10.1007/JHEP10(2017)024}
\arxivlink{1706.10169}.
%45 citations counted in INSPIRE as of 28 Jul 2021
%
%%%%%%%%%%%%%%%%%%here
%
%
%}\bibitem{HLT1} B.~Hoare, N.~Levine and A.~A.~Tseytlin,
%``Integrable 2d sigma models: quantum corrections to geometry from RG flow,''
%\arxivlink{1907.04737}.
%%%CITATION = ARXIV:1907.04737;%%
%
}\bibitem{Fukushima:2021eni}{
O.~Fukushima, J.~i.~Sakamoto and K.~Yoshida,
``Integrable deformed $T^{1,1}$ sigma models from 4D Chern-Simons theory,''
\arxivlink{2105.14920}.
}\bibitem{ecker}{G.~Ecker and J.~Honerkamp,
``Application of invariant renormalization to the nonlinear chiral invariant pion lagrangian in the one-loop approximation,''
\doilink{Nucl.\ Phys.\ B {\bf 35}, 481 (1971)}{10.1016/0550-3213(71)90468-8}.\\
%%CITATION = %doi:10.1016/0550-3213(71)90468-8;%%
%
%
J.~Honerkamp,
``Chiral multiloops,''
\doilink{Nucl.\ Phys.\ B {\bf 36}, 130 (1972)}{10.1016/0550-3213(72)90299-4}.
%%CITATION = doi:10.1016/0550-3213(72)90299-4;%%
%
}\bibitem{friedan}{D.~H.~Friedan,
``Nonlinear models in two + epsilon dimensions,''
\doilink{Annals Phys.\ {\bf 163}, 318 (1985)}{doi:10.1016/0003-4916(85)90384-7};
%%CITATION = %doi:10.1016/0003-4916(85)90384-7;%%
``Nonlinear models in two + epsilon dimensions,''
\doilink{Phys.\ Rev.\ Lett.\ {\bf 45}, 1057 (1980)}{10.1103/PhysRevLett.45.1057}.
%%CITATION = doi:10.1103/PhysRevLett.45.1057;%%
%
%
%
%
%}\bibitem{ricci}{R.~Ricci, A.~A.~Tseytlin and M.~Wolf,
%``On T-duality and integrability for strings on AdS backgrounds,''
%\doilink{JHEP {\bf 0712}, 082 (2007)}{doi:10.1088/1126-6708/2007/12/082}
%\arxivlink{0711.0707}.
%%CITATION = %doi:10.1088/1126-6708/2007/12/082;%%
%
}\bibitem{Fateev:1992tk}{V.~A.~Fateev, E.~Onofri and A.~B.~Zamolodchikov,
``Integrable deformations of the $O(3)$ sigma model. The sausage model,''
\doilink{Nucl.\ Phys.\ B {\bf 406}, 521 (1993)}{10.1016/0550-3213(93)90001-6}.
%%CITATION = doi:10.1016/0550-3213(93)90001-6;%%
%
}\bibitem{fateev96}{V.~A.~Fateev,
``The sigma model (dual) representation for a two-parameter family of integrable quantum field theories,''
\doilink{Nucl.\ Phys.\ B {\bf 473}, 509 (1996)}{10.1016/0550-3213(96)00256-8}.
%%CITATION = %doi:10.1016/0550-3213(96)00256-8;%%
%
}\bibitem{lukyanov}{S.~L.~Lukyanov,
``The integrable harmonic map problem versus Ricci flow,''
\doilink{Nucl.\ Phys.\ B {\bf 865}, 308 (2012)}{10.1016/j.nuclphysb.2012.08.002}
\arxivlink{1205.3201}.
%%CITATION = doi:10.1016/j.nuclphysb.2012.08.002;%%
%
}\bibitem{Fateev:2019xuq}{V.~Fateev,
``Classical and quantum integrable sigma models. Ricci flow, ``nice duality" and perturbed rational conformal field theories,''
\doilink{J. Exp. Theor. Phys. \textbf{129}, no.4, 566-590 (2019)}{
doi:10.1134/S1063776119100042}
\arxivlink{1902.02811}.
%%CITATION = ARXIV:1902.02811;%%
%
}\bibitem{Fateev:2018yos}{V.~A.~Fateev and A.~V.~Litvinov,
``Integrability, duality and sigma models,''
\doilink{JHEP {\bf 1811}, 204 (2018)}{10.1007/JHEP11(2018)204}
\arxivlink{1804.03399}.
%%CITATION = doi:10.1007/JHEP11(2018)204;%%
%
}\bibitem{Demulder:2017zhz}{S.~Demulder, S.~Driezen, A.~Sevrin and D.~C.~Thompson,
``Classical and quantum aspects of Yang-Baxter Wess-Zumino models,''
\doilink{JHEP {\bf 1803}, 041 (2018)}{10.1007/JHEP03(2018)041}
\arxivlink{1711.00084}.
%%CITATION = doi:10.1007/JHEP03(2018)041;%%
%
}\bibitem{Klimcik:2019kkf}{C.~Klimcik,
``Dressing cosets and multi-parametric integrable deformations,''
\doilink{JHEP \textbf{07}, 176 (2019)}{
doi:10.1007/JHEP07(2019)176}
\arxivlink{1903.00439}.
%%CITATION = ARXIV:1903.00439;%%
%
}\bibitem{Litvinov:2018bou}{A.~V.~Litvinov and L.~A.~Spodyneiko,
``On dual description of the deformed $O(N)$ sigma model,''
\doilink{JHEP {\bf 1811}, 139 (2018)}{10.1007/JHEP11(2018)139}
\arxivlink{1804.07084}.
%%CITATION = doi:10.1007/JHEP11(2018)139;%%
%
}\bibitem{Tseytlin:1991wr}{A.~A.~Tseytlin,
``Duality and dilaton,''
\doilink{Mod.\ Phys.\ Lett.\ A {\bf 6}, 1721 (1991)}{10.1142/S021773239100186X}, \\
{\small \url{https://www.academia.edu/40796716/Tseytlin1991_duality_dilaton_MPLA} }\ .
%%CITATION = %doi:10.1142/S021773239100186X;%%
%
}\bibitem{Panvel:1992he}{J.~Panvel,
``Higher order conformal invariance of string backgrounds obtained by $O(d,d)$ transformations,''
\doilink{Phys.\ Lett.\ B {\bf 284}, 50 (1992)}{10.1016/0370-2693(92)91923-W}
\arxivlink{hep-th/9204024}.
%%CITATION = doi:10.1016/0370-2693(92)91923-W;%%
%
}\bibitem{Haagensen:1997er}{P.~E.~Haagensen and K.~Olsen,
``T duality and two loop renormalization flows,''
\doilink{Nucl.\ Phys.\ B {\bf 504}, 326 (1997)}{10.1016/S0550-3213(97)00496-3}
\arxivlink{hep-th/9704157}.
%%CITATION = doi:10.1016/S0550-3213(97)00496-3;%%
%
}\bibitem{Haagensen:1997bh}{P.~E.~Haagensen,
``Duality and the renormalization group,''
\arxivlink{hep-th/9708110}.
%%CITATION = HEP-TH/9708110;%%
%
}\bibitem{Dijkgraaf:1991ba}{R.~Dijkgraaf, H.~L.~Verlinde and E.~P.~Verlinde,
``String propagation in a black hole geometry,''
\doilink{Nucl.\ Phys.\ B {\bf 371}, 269 (1992)}{10.1016/0550-3213(92)90237-6}.
%%CITATION = %doi:10.1016/0550-3213(92)90237-6;%%
%
}\bibitem{Bars:1992sr}{I.~Bars and K.~Sfetsos,
``Conformally exact metric and dilaton in string theory on curved space-time,''
\doilink{Phys.\ Rev.\ D {\bf 46}, 4510 (1992)}{10.1103/PhysRevD.46.4510}
\arxivlink{hep-th/9206006}.
%%CITATION = doi:10.1103/PhysRevD.46.4510;%%
%
}\bibitem{Tseytlin:1992ri}{A.~A.~Tseytlin,
``Effective action of gauged WZW model and exact string solutions,''
\doilink{Nucl.\ Phys.\ B {\bf 399}, 601 (1993)}{10.1016/0550-3213(93)90511-M}
\arxivlink{hep-th/9301015}.
%%CITATION = %doi:10.1016/0550-3213(93)90511-M;%%
%
%\cite{Karabali:1988au}
}\bibitem{Karabali}{
D.~Karabali, Q.~H.~Park, H.~J.~Schnitzer and Z.~Yang,
``A GKO construction based on a path integral formulation of gauged Wess-Zumino-Witten actions,''
\doilink{Phys. Lett. B \textbf{216}, 307-312 (1989)}{
doi:10.1016/0370-2693(89)91120-9}.\\
%241 citations counted in INSPIRE as of 29 Jun 2021
%\cite{Karabali:1989dk}
%
D.~Karabali and H.~J.~Schnitzer,
``BRST quantization of the gauged WZW action and coset conformal field theories,''
\doilink{Nucl. Phys. B \textbf{329}, 649-666 (1990)}{
doi:10.1016/0550-3213(90)90075-O}.
%259 citations counted in INSPIRE as of 29 Jun 2021
%
}\bibitem{bs}{I.~Bars and K.~Sfetsos,
``Exact effective action and space-time geometry in gauged WZW models,''
\doilink{Phys.\ Rev.\ D {\bf 48}, 844 (1993)}{10.1103/PhysRevD.48.844}
\arxivlink{hep-th/9301047}.
%%CITATION = %doi:10.1103/PhysRevD.48.844;%%
%
}\bibitem{Tseytlin:1993my}{A.~A.~Tseytlin,
``Conformal sigma models corresponding to gauged Wess-Zumino-Witten theories,''
\doilink{Nucl.\ Phys.\ B {\bf 411}, 509 (1994)}{10.1016/0550-3213(94)90461-8}
\arxivlink{hep-th/9302083}.
%%CITATION = %doi:10.1016/0550-3213(94)90461-8;%%
%
}\bibitem{Tseytlin:1993df}{A.~A.~Tseytlin,
``On field redefinitions and exact solutions in string theory,''
\doilink{Phys.\ Lett.\ B {\bf 317}, 559 (1993)}{10.1016/0370-2693(93)91372-T}
\arxivlink{hep-th/9308042}.
%%CITATION = doi:10.1016/0370-2693(93)91372-T;%%
%
}\bibitem{Horowitz:1994ei}{G.~T.~Horowitz and A.~A.~Tseytlin,
``On exact solutions and singularities in string theory,''
\doilink{Phys.\ Rev.\ D {\bf 50}, 5204 (1994)}{10.1103/PhysRevD.50.5204}
\arxivlink{hep-th/9406067}.
%%CITATION = doi:10.1103/PhysRevD.50.5204;%%
%
}\bibitem{Tseytlin:1995fh}{A.~A.~Tseytlin,
``Exact solutions of closed string theory,''
\doilink{Class.\ Quant.\ Grav.\ {\bf 12}, 2365 (1995)}{10.1088/0264-9381/12/10/003},
\arxivlink{hep-th/9505052}.
%%CITATION = doi:10.1088/0264-9381/12/10/003;%%
%
}\bibitem{Bardacki:1990wj}{K.~Bardakci, M.~J.~Crescimanno and E.~Rabinovici,
``Parafermions From coset models,''
\doilink{Nucl.\ Phys.\ B {\bf 344}, 344 (1990)}{10.1016/0550-3213(90)90365-K}.
%%CITATION = doi:10.1016/0550-3213(90)90365-K;%%
%
}\bibitem{Witten:1991yr}{E.~Witten,
``On string theory and black holes,''
\doilink{Phys.\ Rev.\ D {\bf 44}, 314 (1991)}{10.1103/PhysRevD.44.314}.
%%CITATION = doi:10.1103/PhysRevD.44.314;%%
%
}\bibitem{Tseytlin:1991ht}{A.~A.~Tseytlin,
``On the form of the black hole solution in D = 2 theory,''
\doilink{Phys.\ Lett.\ B {\bf 268}, 175 (1991)}{10.1016/0370-2693(91)90800-6}.
%%CITATION = doi:10.1016/0370-2693(91)90800-6;%%
%
}\bibitem{Jack:1992mk}{I.~Jack, D.~R.~T.~Jones and J.~Panvel,
``Exact bosonic and supersymmetric string black hole solutions,''
\doilink{Nucl.\ Phys.\ B {\bf 393}, 95 (1993)}{10.1016/0550-3213(93)90239-L}
\arxivlink{hep-th/9201039}.
%%CITATION = doi:10.1016/0550-3213(93)90239-L;%%
%
%}\bibitem{FernandezPousa:1996hi}{C.~R.~Fernandez-Pousa, M.~V.~Gallas, T.~J.~Hollowood and J.~L.~Miramontes,
%``The symmetric space and homogeneous sine-Gordon theories,''
%\doilink{Nucl.\ Phys.\ B {\bf 484}, 609 (1997)}{10.1016/S0550-3213(96)00603-7}
%\arxivlink{hep-th/9606032}.
%%%CITATION = doi:10.1016/S0550-3213(96)00603-7;%%
%
}\bibitem{Hoare:2010fb}{B.~Hoare and A.~A.~Tseytlin,
``On the perturbative S-matrix of generalized sine-Gordon models,''
\doilink{JHEP {\bf 1011}, 111 (2010)}{10.1007/JHEP11(2010)111}
\arxivlink{1008.4914}.
%%CITATION = doi:10.1007/JHEP11(2010)111;%%
%
%}\bibitem{deVega:1982sh}{H.~J.~de Vega and J.~M.~Maillet,
%``Semiclassical quantization of the complex {sine-Gordon} field theory,''
%\doilink{Phys.\ Rev.\ D {\bf 28}, 1441 (1983)}{10.1103/PhysRevD.28.1441}.
%%%CITATION = doi:10.1103/PhysRevD.28.1441;%%
%
%}\bibitem{Bonneau:1984pj}{G.~Bonneau and F.~Delduc,
%``S matrix properties versus renormalizability in two-dimensional $O(N)$ symmetric models,''
%\doilink{Nucl.\ Phys.\ B {\bf 250}, 561 (1985)}{10.1016/0550-3213(85)90495-X}.
%%%CITATION = doi:10.1016/0550-3213(85)90495-X;%%
%%
}\bibitem{Schwarz:1992te}{A.~S.~Schwarz and A.~A.~Tseytlin,
``Dilaton shift under duality and torsion of elliptic complex,''
\doilink{Nucl.\ Phys.\ B {\bf 399}, 691 (1993)}{10.1016/0550-3213(93)90514-P}
\arxivlink{hep-th/9210015}.
%%CITATION = %doi:10.1016/0550-3213(93)90514-P;%%
%
}\bibitem{Gerasimov:1990fi}{A.~Gerasimov, A.~Morozov, M.~Olshanetsky, A.~Marshakov and S.~L.~Shatashvili,
``Wess-Zumino-Witten model as a theory of free fields,''
\doilink{Int.\ J.\ Mod.\ Phys.\ A {\bf 5}, 2495 (1990)}{10.1142/S0217751X9000115X}.
%%CITATION = doi:10.1142/S0217751X9000115X;%%
%
%}\bibitem{Buscher:1987qj} T.~H.~Buscher,
%``Path integral Derivation of Quantum Duality in Nonlinear Sigma Models,''
%\doilink{Phys.\ Lett.\ B {\bf 201}, 466 (1988)}{10.1016/0370-2693(88)90602-8}.
%%%CITATION = doi:10.1016/0370-2693(88)90602-8;%%
%
%
}\bibitem{Buscher:1987qj}{
T.~H.~Buscher,
``A symmetry of the string background field equations,''
\doilink{Phys. Lett. B \textbf{194}, 59-62 (1987)}{
doi:10.1016/0370-2693(87)90769-6};
%898 citations counted in INSPIRE as of 06 Jul 2021
%\cite{Buscher:1987qj}
%}\bibitem{Buscher:1987qj}
%T.~H.~Buscher,
``Path integral derivation of quantum duality in nonlinear sigma models,''
\doilink{Phys. Lett. B \textbf{201}, 466-472 (1988)}{
doi:10.1016/0370-2693(88)90602-8}.
%918 citations counted in INSPIRE as of 06 Jul 2021
%
}\bibitem{DeJaegher:1998pp}{J.~De Jaegher, J.~Raeymaekers, A.~Sevrin and W.~Troost,
``Dilaton transformation under Abelian and nonAbelian T duality in the path integral approach,''
\doilink{Nucl.\ Phys.\ B {\bf 548}, 563 (1999)}{10.1016/S0550-3213(99)00157-1}
\arxivlink{hep-th/9812207}.
%%CITATION = doi:10.1016/S0550-3213(99)00157-1;%%
%
}\bibitem{Klimcik:2002zj}{C.~Klimcik,
``Yang-Baxter sigma models and dS/AdS T duality,''
\doilink{JHEP {\bf 0212}, 051 (2002)}{10.1088/1126-6708/2002/12/051}
\arxivlink{hep-th/0210095}.
%%CITATION = doi:10.1088/1126-6708/2002/12/051;%%
%
}\bibitem{Penna:2020uky}{
R.~F.~Penna,
``A Twistor Action for Integrable Systems,''
\arxivlink{2011.05831}.
%5 citations counted in INSPIRE as of 04 Aug 2021
}\bibitem{Klimcik:2008eq}{C.~Klimcik,
``On integrability of the Yang-Baxter sigma-model,''
\doilink{J.\ Math.\ Phys.\ {\bf 50}, 043508 (2009)}{10.1063/1.3116242}
\arxivlink{0802.3518}.
%%%CITATION = doi:10.1063/1.3116242;%
%
}\bibitem{Ki}{C.~Klimcik,
``Integrability of the bi-Yang-Baxter sigma-model,''
\doilink{Lett.\ Math.\ Phys.\ {\bf 104}, 1095 (2014)}{10.1007/s11005-014-0709-y}
\arxivlink{1402.2105}.
%%CITATION = doi:10.1007/s11005-014-0709-y;%%
%\cite{Alfimov:2020jpy}
}\bibitem{Alfimov:2020jpy}{
M.~Alfimov, B.~Feigin, B.~Hoare and A.~Litvinov,
``Dual description of $\eta$-deformed OSP sigma models,''
\doilink{JHEP \textbf{12}, 040 (2020)}{
doi:10.1007/JHEP12(2020)040}
\arxivlink{2010.11927}.
%0 citations counted in INSPIRE as of 31 Jul 2021
}\bibitem{Bazhanov:1984gu}{
V.~V.~Bazhanov,
``Trigonometric solution of triangle equations and classical Lie algebras,''
\doilink{Phys. Lett. B \textbf{159}, 321-324 (1985)}{
doi:10.1016/0370-2693(85)90259-X}.\\
M.~Jimbo,
``Quantum r matrix for the generalized Toda system,''
\doilink{Commun. Math. Phys. \textbf{102}, 537-547 (1986)}{
doi:10.1007/BF01221646}.
%529 citations counted in INSPIRE as of 31 Jul 2021
%
%118 citations counted in INSPIRE as of 31 Jul 2021
}\bibitem{Delduc:2013fga}{F.~Delduc, M.~Magro and B.~Vicedo,
``On classical $q$-deformations of integrable sigma-models,''
\doilink{JHEP {\bf 1311}, 192 (2013)}{10.1007/JHEP11(2013)192}
\arxivlink{1308.3581}.
%%CITATION = doi:10.1007/JHEP11(2013)192;%%
%
}\bibitem{Sfetsos:2013wia}{K.~Sfetsos,
``Integrable interpolations: From exact CFTs to non-Abelian T-duals,''
\doilink{Nucl.\ Phys.\ B {\bf 880}, 225 (2014)}{10.1016/j.nuclphysb.2014.01.004}
\arxivlink{1312.4560}.
%%CITATION = doi:10.1016/j.nuclphysb.2014.01.004;%%
%
}\bibitem{Hollowood:2014rla}{T.~J.~Hollowood, J.~L.~Miramontes and D.~M.~Schmidtt,
``Integrable deformations of strings on symmetric spaces,''
\doilink{JHEP {\bf 1411}, 009 (2014)}{10.1007/JHEP11(2014)009}
\arxivlink{1407.2840}.
%%CITATION = doi:10.1007/JHEP11(2014)009;%%
%
}\bibitem{n1}{A.~Subbotin and I.~V.~Tyutin,
``On the equivalence of dual theories,''
\doilink{Int.\ J.\ Mod.\ Phys.\ A {\bf 11}, 1315 (1996)}{10.1142/S0217751X96000596},
Erratum: [Int.\ J.\ Mod.\ Phys.\ A {\bf 11}, 2231 (1996)]
\arxivlink{hep-th/9506132}.
%%CITATION = %doi:10.1142/S0217751X96000596;%%
%
}\bibitem{n2}{J.~Balog, P.~Forgacs, Z.~Horvath and L.~Palla,
``Perturbative quantum (in)equivalence of dual sigma models in two-dimensions,''
\doilink{Nucl.\ Phys.\ Proc.\ Suppl.\ {\bf 49}, 16 (1996)}{10.1016/0920-5632(96)00311-8}
\arxivlink{hep-th/9601091}.
%%CITATION = %doi:10.1016/0920-5632(96)00311-8;%%
%
}\bibitem{n3}{L.~K.~Balazs, J.~Balog, P.~Forgacs, N.~Mohammedi, L.~Palla and J.~Schnittger,
``Quantum equivalence of sigma models related by nonAbelian duality transformations,''
\doilink{Phys.\ Rev.\ D {\bf 57}, 3585 (1998)}{10.1103/PhysRevD.57.3585}
\arxivlink{hep-th/9704137}.
%%CITATION = %doi:10.1103/PhysRevD.57.3585;%%
%
}\bibitem{Bonneau:2001za}{G.~Bonneau and P.~Y.~Casteill,
``Dualized sigma models at the two loop order,''
\doilink{Nucl.\ Phys.\ B {\bf 607}, 293 (2001)}{10.1016/S0550-3213(01)00216-4}
\arxivlink{hep-th/0103260}.
%%CITATION = doi:10.1016/S0550-3213(01)00216-4;%%
%
}\bibitem{Zamolodchikov:1980ku}{A.~B.~Zamolodchikov and V.~A.~Fateev,
``Model factorized S matrix and an integrable Heisenberg chain with spin 1. (in Russian),''
Sov.\ J.\ Nucl.\ Phys.\ {\bf 32}, 298 (1980),
[Yad.\ Fiz.\ {\bf 32}, 581 (1980)].
%%CITATION = SJNCA,32,298;%%
%
}\bibitem{Hoare:2014pna}{B.~Hoare, R.~Roiban and A.~A.~Tseytlin,
``On deformations of $AdS_n$ x $S^n$ supercosets,''
\doilink{JHEP {\bf 1406}, 002 (2014)}{10.1007/JHEP06(2014)002}
\arxivlink{1403.5517}.
%%CITATION = doi:10.1007/JHEP06(2014)002;%%
%
}\bibitem{Graham:1987ep}{S.~J.~Graham,
``Three loop beta function for the bosonic nonlinear $\sigma$ model,''
\doilink{Phys.\ Lett.\ B {\bf 197}, 543 (1987)}{10.1016/0370-2693(87)91052-5}.
%%CITATION = doi:10.1016/0370-2693(87)91052-5;%%
%
}\bibitem{Foakes:1987ij}{A.~P.~Foakes and N.~Mohammedi,
``Three loop calculation of the beta function for the purely metric nonlinear $\sigma$ model,''
\doilink{Phys.\ Lett.\ B {\bf 198}, 359 (1987)}{10.1016/0370-2693(87)90679-4}.
%%CITATION = %doi:10.1016/0370-2693(87)90679-4;%%
%
}\bibitem{Foakes:1987gg}{A.~P.~Foakes and N.~Mohammedi,
``An explicit three loop calculation for the purely metric two-dimensional nonlinear $\sigma$ model,''
\doilink{Nucl.\ Phys.\ B {\bf 306}, 343 (1988)}{10.1016/0550-3213(88)90696-7}.
%%CITATION = %doi:10.1016/0550-3213(88)90696-7;%%
%
%}\bibitem{Zamolodchikov:1978xm} A.~B.~Zamolodchikov and A.~B.~Zamolodchikov,
%``Factorized S-Matrices in Two-Dimensions as the Exact Solutions of Certain Relativistic Quantum Field Models,''
%\doilink{Annals Phys.\ {\bf 120}, 253 (1979)}{10.1016/0003-4916(79)90391-9}.
%%%CITATION = doi:10.1016/0003-4916(79)90391-9;%%
%
}\bibitem{Grisaru:1986px}{M.~T.~Grisaru, A.~E.~M.~van de Ven and D.~Zanon,
``Four loop beta function for the N=1 and N=2 supersymmetric nonlinear sigma model in two-dimensions,''
\doilink{Phys.\ Lett.\ B {\bf 173}, 423 (1986)}{10.1016/0370-2693(86)90408-9}.
%%CITATION = doi:10.1016/0370-2693(86)90408-9;%%
%
}\bibitem{BDCGR}{A.~A.~Belavin and V.~G.~Drinfel'd,
``Triangle equations and simple Lie algebras''
\doilink{Sov. Sci. Rev. {\bf C4}, 93 (1984)}{10.1007/BF01041913}.\\
%
M.~Cahen, S.~Gutt and J.~Rawnsley,
``Some remarks on the classification of Poisson Lie groups,''
\doilink{Contemp. Math. {\bf 179} (1994) 1}{10.1090/conm/179/01932}.
%
}\bibitem{Hoare:2015gda}{B.~Hoare and A.~A.~Tseytlin,
``On integrable deformations of superstring sigma models related to $AdS_n \times S^n$ supercosets,''
\doilink{Nucl.\ Phys.\ B {\bf 897}, 448 (2015)}{10.1016/j.nuclphysb.2015.06.001}
\arxivlink{1504.07213}.
%%CITATION = doi:10.1016/j.nuclphysb.2015.06.001;%%
%
}\bibitem{Itsios:2014lca}{G.~Itsios, K.~Sfetsos and K.~Siampos,
``The all-loop non-Abelian Thirring model and its RG flow``,
\doilink{Phys.\ Lett.\ B {\bf 733}, 265 (2014)}{10.1016/j.physletb.2014.04.061}
\arxivlink{1404.3748}.
%%CITATION = ARXIV:1404.3748;%%
%
}\bibitem{Appadu:2015nfa}{C. Appadu and T.J. Hollowood,
``Beta function of k deformed ${\text AdS}_{5} \times S^5$ string theory``,
\doilink{JHEP {\bf 1511}, 095 (2015)}{10.1007/JHEP11(2015)095}
\arxivlink{1507.05420}.
%%CITATION = doi:10.1007/JHEP11(2015)095;%%
%
}\bibitem{Fridling:1983ha}{B.~E.~Fridling and A.~Jevicki,
``Dual representations and ultraviolet divergences in nonlinear $\sigma$ models,''
\doilink{Phys.\ Lett.\ {\bf 134B}, 70 (1984)}{10.1016/0370-2693(84)90987-0}.
%%CITATION = doi:10.1016/0370-2693(84)90987-0;%%
%
}\bibitem{Fradkin:1984ai}{E.~S.~Fradkin and A.~A.~Tseytlin,
``Quantum equivalence of dual field theories,''
\doilink{Annals Phys.\ {\bf 162}, 31 (1985)}{10.1016/0003-4916(85)90225-8}.
%%CITATION = doi:10.1016/0003-4916(85)90225-8;%%
%
}\bibitem{Vicedo:2015pna}{B.~Vicedo,
``Deformed integrable $\sigma$-models, classical $R$-matrices and classical exchange algebra on Drinfel'd doubles,''
\doilink{J.\ Phys.\ A {\bf 48}, no. 35, 355203 (2015)}{10.1088/1751-8113/48/35/355203}
\arxivlink{1504.06303}.
%%CITATION = doi:10.1088/1751-8113/48/35/355203;%%
%
%
%
%
%
%
%
}\bibitem{Sfetsos:2015nya}{K.~Sfetsos, K.~Siampos and D.~C.~Thompson,
``Generalised integrable $\lambda$- and $\eta$-deformations and their relation,''
\doilink{Nucl.\ Phys.\ B {\bf 899}, 489 (2015)}{10.1016/j.nuclphysb.2015.08.015}
\arxivlink{1506.05784}.
%%CITATION = doi:10.1016/j.nuclphysb.2015.08.015;%%
%
}\bibitem{Klimcik:2015gba}{C.~Klimcik,
``$\eta$ and $\lambda$ deformations as $\mathcal{E}$-models,''
\doilink{Nucl.\ Phys.\ B {\bf 900}, 259 (2015)}{10.1016/j.nuclphysb.2015.09.011}
\arxivlink{1508.05832}.
%%%CITATION = doi:10.1016/j.nuclphysb.2015.09.011;%%
%
}\bibitem{Hoare:2017ukq}{B.~Hoare and F.~K.~Seibold,
``Poisson-Lie duals of the $\eta$ deformed symmetric space sigma model,''
\doilink{JHEP {\bf 1711}, 014 (2017)}{10.1007/JHEP11(2017)014}
\arxivlink{1709.01448}.
%%CITATION = doi:10.1007/JHEP11(2017)014;%%
%
}\bibitem{Klimcik:1995dy}{C.~Klimcik and P.~Severa,
``Poisson-Lie T duality and loop groups of Drinfeld doubles,''
\doilink{Phys.\ Lett.\ B {\bf 372}, 65 (1996)}{10.1016/0370-2693(96)00025-1}
\arxivlink{hep-th/9512040}.
%%CITATION = doi:10.1016/0370-2693(96)00025-1;%%
%
}\bibitem{Valent:2009nv}{G.~Valent, C.~Klimcik and R.~Squellari,
``One loop renormalizability of the Poisson-Lie sigma models,''
\doilink{Phys.\ Lett.\ B {\bf 678}, 143 (2009)}{10.1016/j.physletb.2009.06.001}
\arxivlink{0902.1459}.
%%CITATION = doi:10.1016/j.physletb.2009.06.001;%%
%
}\bibitem{Sfetsos:2009dj}{K.~Sfetsos and K.~Siampos,
``Quantum equivalence in Poisson-Lie T-duality,''
\doilink{JHEP {\bf 0906}, 082 (2009)}{10.1088/1126-6708/2009/06/082}
\arxivlink{0904.4248}.
%%CITATION = doi:10.1088/1126-6708/2009/06/082;%%
%
}\bibitem{mt}{R.~R.~Metsaev and A.~A.~Tseytlin,
``Order alpha-prime (two loop) equivalence of the string equations of motion and the sigma model Weyl invariance conditions: dependence on the dilaton and the antisymmetric tensor,''
\doilink{Nucl.\ Phys.\ B {\bf 293}, 385 (1987)}{10.1016/0550-3213(87)90077-0};
%%CITATION = doi:10.1016/0550-3213(87)90077-0;%%
``Two loop beta function for the generalized bosonic sigma model,''
\doilink{Phys.\ Lett.\ B {\bf 191}, 354 (1987)}{doi:10.1016/0370-2693(87)90622-8}.
%%CITATION = doi:10.1016/0370-2693(87)90622-8;%%
%
%
%
%%%%%%%%%%%%%%%%%%%%%%%%%%%%%%% 
%%%%%%% 1910 refs:
%
%%}\bibitem{Fateev:1992tk} V.~A.~Fateev, E.~Onofri and A.~B.~Zamolodchikov,
%%``The sausage model (integrable deformations of $O(3)$ sigma model),''
%%\doilink{Nucl.\ Phys.\ B {\bf 406}, 521 (1993)}{10.1016/0550-3213(93)90001-6}.
%%%%CITATION = doi:10.1016/0550-3213(93)90001-6;%%
%%\\ V.~A.~Fateev,
%%``The sigma model (dual) representation for a two-parameter family of integrable quantum field theories,''
%%\doilink{Nucl.\ Phys.\ B {\bf 473}, 509 (1996)}{10.1016/0550-3213(96)00256-8}.
%%%%CITATION = %doi:10.1016/0550-3213(96)00256-8;%%
%%\\ S.~L.~Lukyanov,
%%``The integrable harmonic map problem versus Ricci flow,''
%%\doilink{Nucl.\ Phys.\ B {\bf 865}, 308 (2012)}{10.1016/j.nuclphysb.2012.08.002},
%%\arxivlink{1205.3201}.
%%%%CITATION = doi:10.1016/j.nuclphysb.2012.08.002;%%
%
%%}\bibitem{Fateev:2019xuq} V.~Fateev,
%%``Classical and quantum integrable sigma models. Ricci flow, ``nice duality'' and perturbed rational conformal field theories,''
%%\arxivlink{1902.02811}.
%%%%CITATION = ARXIV:1902.02811;%%
%%\\ V.~A.~Fateev and A.~V.~Litvinov,
%%``Integrability, Duality and Sigma Models,''
%%\doilink{JHEP {\bf 1811}, 204 (2018)}{10.1007/JHEP11(2018)204},
%%\arxivlink{1804.03399}.
%%%%CITATION = doi:10.1007/JHEP11(2018)204;%%
%%\\ A.~V.~Litvinov and L.~A.~Spodyneiko, %v3
%%``On dual description of the deformed $O(N)$ sigma model,''
%%\doilink{JHEP {\bf 1811}, 139 (2018)}{10.1007/JHEP11(2018)139},
%%\arxivlink{1804.07084}.
%%%%CITATION = doi:10.1007/JHEP11(2018)139;%%
%
%
%
}\bibitem{Sfetsos:1993bh}{
K.~Sfetsos,
``Effective action and exact geometry in chiral gauged WZW models,''
\arxivlink{hep-th/9305074}.\\
K.~Sfetsos and A.~A.~Tseytlin,
``Chiral gauged WZNW models and heterotic string backgrounds,''
\doilink{Nucl.\ Phys.\ B {\bf 415}, 116 (1994)}{10.1016/0550-3213(94)90069-8}
\arxivlink{hep-th/9308018}.
%%CITATION = doi:10.1016/0550-3213(94)90069-8;%%
%
%
}\bibitem{Sfetsos:19932}{K.~Sfetsos and A.~A.~Tseytlin,
``Antisymmetric tensor coupling and conformal invariance in sigma models corresponding to gauged WZNW theories,''
\doilink{Phys.\ Rev.\ D {\bf 49}, 2933 (1994)}{10.1103/PhysRevD.49.2933}
\arxivlink{hep-th/9310159}.
%%CITATION = doi:10.1103/PhysRevD.49.2933;%%
%
}\bibitem{Tseytlin:1993hm}{A.~A.~Tseytlin,
``On a `universal' class of WZW type conformal models,''
\doilink{Nucl.\ Phys.\ B {\bf 418}, 173 (1994)}{10.1016/0550-3213(94)90243-7}
\arxivlink{hep-th/9311062}.
%%CITATION = doi:10.1016/0550-3213(94)90243-7;%%
%
}\bibitem{Chung:1992mj}{S.~W.~Chung and S.~H.~H.~Tye,
``Chiral gauged WZW theories and coset models in conformal field theory,''
\doilink{Phys.\ Rev.\ D {\bf 47}, 4546 (1993)}{doi:10.1103/PhysRevD.47.4546}
\arxivlink{hep-th/9202002}.
%%CITATION = doi:10.1103/PhysRevD.47.4546;%%
%
%}\bibitem{Itsios:2014lca} G.~Itsios, K.~Sfetsos and K.~Siampos,
%``The all-loop non-abelian Thirring model and its RG flow,''
%\doilink{Phys.\ Lett.\ B {\bf 733}, 265 (2014)}{doi:10.1016/j.physletb.2014.04.061},
%\arxivlink{1404.3748}.
%%%CITATION = doi:10.1016/j.physletb.2014.04.061;%%
%
%
}\bibitem{Kutasov:1989aw}{D.~Kutasov,
``Duality off the critical point in two-dimensional systems with non-abelian symmetries,''
\doilink{Phys.\ Lett.\ B {\bf 233} (1989) 369}{doi:10.1016/0370-2693(89)91325-7}.
%%CITATION = doi:10.1016/0370-2693(89)91325-7;%%
%
%}\bibitem{n1} A.~Subbotin and I.~V.~Tyutin,
%``On the equivalence of dual theories,''
%\doilink{Int.\ J.\ Mod.\ Phys.\ A {\bf 11}, 1315 (1996)}{10.1142/S0217751X96000596},
%Erratum: [Int.\ J.\ Mod.\ Phys.\ A {\bf 11}, 2231 (1996)],
%\arxivlink{hep-th/9506132}.
%%%CITATION = %doi:10.1142/S0217751X96000596;%%
%
%L.~K.~Balazs, J.~Balog, P.~Forgacs, N.~Mohammedi, L.~Palla and J.~Schnittger,
%``Quantum equivalence of sigma models related by non-abelian duality transformations,''
%\doilink{Phys.\ Rev.\ D {\bf 57}, 3585 (1998)}{10.1103/PhysRevD.57.3585},
%\arxivlink{hep-th/9704137}.
%%%CITATION = %doi:10.1103/PhysRevD.57.3585;%%
%
%Bonneau Casteill
%
}\bibitem{Braaten:1985is}{E.~Braaten, T.~L.~Curtright and C.~K.~Zachos,
``Torsion and geometrostasis in nonlinear sigma models,''
\doilink{Nucl.\ Phys.\ B {\bf 260}, 630 (1985)}{10.1016/0550-3213(85)90053-7},
Erratum: [\doilink{Nucl.\ Phys.\ B {\bf 266}, 748 (1986)}{10.1016/0550-3213(86)90196-3}].
%%CITATION = doi:10.1016/0550-3213(86)90196-3, 10.1016/0550-3213(85)90053-7;%%
%
}\bibitem{Hull:1987pc}{C.~M.~Hull and P.~K.~Townsend,
``The two loop beta function for $\sigma$ models with torsion,''
\doilink{Phys.\ Lett.\ B {\bf 191}, 115 (1987)}{10.1016/0370-2693(87)91331-1}.\\
%%CITATION = doi:10.1016/0370-2693(87)91331-1;%%
%
D.~Zanon,
``Two loop beta functions and low-energy string effective action for the two-dimensional bosonic nonlinear $\sigma$ model with a {Wess-Zumino}-Witten  term,''
\doilink{Phys.\ Lett.\ B {\bf 191}, 363 (1987)}{10.1016/0370-2693(87)90623-X}.
%%CITATION = doi:10.1016/0370-2693(87)90623-X;%%
%
}\bibitem{adscftsol}{N.~Beisert {\it et al.}, %v3
``Review of AdS/CFT integrability: an overview,''
\doilink{Lett.\ Math.\ Phys.\ {\bf 99}, 3 (2012)}{10.1007/s11005-011-0529-2}
\arxivlink{1012.3982}.
%%CITATION = doi:10.1007/s11005-011-0529-2;%%
%
%}\bibitem{adscftsol2}
%N.~Gromov, V.~Kazakov, S.~Leurent and D.~Volin,
%``Quantum Spectral Curve for Planar $\mathcal{N} = 4$ Super-Yang-Mills Theory,''
%\doilink{Phys.\ Rev.\ Lett.\ {\bf 112}, no. 1, 011602 (2014)}{10.1103/PhysRevLett.112.011602},
%\arxivlink{1305.1939}.
%%%CITATION = doi:10.1103/PhysRevLett.112.011602;%%
%
}\bibitem{Zarembo:2010sg}{K.~Zarembo,
``Strings on semisymmetric superspaces,''
\doilink{JHEP {\bf 1005}, 002 (2010)}{doi:10.1007/JHEP05(2010)002}
\arxivlink{1003.0465}.
%%CITATION = doi:10.1007/JHEP05(2010)002;%%
%
}\bibitem{Wulff:2014kja}{L.~Wulff,
``Superisometries and integrability of superstrings,''
\doilink{JHEP {\bf 1405}, 115 (2014)}{10.1007/JHEP05(2014)115}
\arxivlink{1402.3122}.
%%CITATION = doi:10.1007/JHEP05(2014)115;%%
%
}\bibitem{betadef}{O.~Lunin and J.~M.~Maldacena,
``Deforming field theories with $U(1) \times U(1)$ global symmetry and their gravity duals,''
\doilink{JHEP {\bf 0505}, 033 (2005)}{10.1088/1126-6708/2005/05/033}
\arxivlink{hep-th/0502086}.\\
%%CITATION = doi:10.1088/1126-6708/2005/05/033;%%
%
S.~A.~Frolov, R.~Roiban and A.~A.~Tseytlin,
``Gauge-string duality for superconformal deformations of $\mathcal{N}=4$ super Yang-Mills theory,''
\doilink{JHEP {\bf 0507}, 045 (2005)}{10.1088/1126-6708/2005/07/045}
\arxivlink{hep-th/0503192}.\\
%%CITATION = doi:10.1088/1126-6708/2005/07/045;%%
%
S.~Frolov,
``Lax pair for strings in Lunin-Maldacena background,''
\doilink{JHEP {\bf 0505}, 069 (2005)}{10.1088/1126-6708/2005/05/069}
\arxivlink{hep-th/0503201}.
%%CITATION = doi:10.1088/1126-6708/2005/05/069;%%
%
}\bibitem{Hollowood:2014qma}{T.~J.~Hollowood, J.~L.~Miramontes and D.~M.~Schmidtt,
``An integrable deformation of the $AdS_5 \times S^5$ superstring,''
\doilink{J.\ Phys.\ A {\bf 47}, no. 49, 495402 (2014)}{doi:10.1088/1751-8113/47/49/495402}
\arxivlink{1409.1538}.
%%CITATION = doi:10.1088/1751-8113/47/49/495402;%%
%
}\bibitem{Delduc:2013qra}{F.~Delduc, M.~Magro and B.~Vicedo,
``An integrable deformation of the $AdS_5 \times S^5$ superstring action,''
\doilink{Phys.\ Rev.\ Lett.\ {\bf 112}, no. 5, 051601 (2014)}{10.1103/PhysRevLett.112.051601}
\arxivlink{1309.5850}.
%%CITATION = doi:10.1103/PhysRevLett.112.051601;%%
%
%}\bibitem{klimcik} \la{testc}
%C.~Klimcik,
%``Yang-Baxter sigma models and dS/AdS T duality,''
%\doilink{JHEP {\bf 0212}, 051 (2002)}{10.1088/1126-6708/2002/12/051},
%\arxivlink{hep-th/0210095}.
%%%CITATION = doi:10.1088/1126-6708/2002/12/051;%%
%
%}\bibitem{dmv} F.~Delduc, M.~Magro and B.~Vicedo,
%``On classical $q$-deformations of integrable sigma-models,''
%\doilink{JHEP {\bf 1311}, 192 (2013)}{doi:10.1007/JHEP11(2013)192},
%\arxivlink{1308.3581}.
%%%CITATION = doi:10.1007/JHEP11(2013)192;%%
%
}\bibitem{poissonlie}{C.~Klimcik and P.~Severa,
``Dual non-Abelian duality and the Drinfeld double,''
\doilink{Phys.\ Lett.\ B {\bf 351}, 455 (1995)}{10.1016/0370-2693(95)00451-P}
\arxivlink{hep-th/9502122}.\\
%%CITATION = doi:10.1016/0370-2693(95)00451-P;%%
%
C.~Klimcik,
``Poisson-Lie T-duality,''
\doilink{Nucl.\ Phys.\ Proc.\ Suppl.\ {\bf 46}, 116 (1996)}{10.1016/0920-5632(96)00013-8}
\arxivlink{hep-th/9509095}.
%%CITATION = doi:10.1016/0920-5632(96)00013-8;%%
%
}
%\bibitem{pld}{B.~Vicedo,
%``Deformed integrable $\sigma$-models, classical R-matrices and classical exchange algebra on Drinfel'd doubles,''
%\doilink{J.\ Phys.\ A {\bf 48}, no. 35, 355203 (2015)}{doi:10.1088/1751-8113/48/35/355203}
%\arxivlink{1504.06303}.\\
%%%CITATION = doi:10.1088/1751-8113/48/35/355203;%%
%%
%% K.~Sfetsos, K.~Siampos and D.~C.~Thompson,
%%``Generalised integrable $\lambda$- and $\eta$-deformations and their relation,''
%%\doilink{Nucl.\ Phys.\ B {\bf 899}, 489 (2015)}{doi:10.1016/j.nuclphysb.2015.08.015},
%%\arxivlink{1506.05784}.
%%%%CITATION = doi:10.1016/j.nuclphysb.2015.08.015;%%
%%
%C.~Klimcik,
%``$\eta$ and $\lambda$ deformations as E-models,''
%\doilink{Nucl.\ Phys.\ B {\bf 900}, 259 (2015)}{doi:10.1016/j.nuclphysb.2015.09.011}
%\arxivlink{1508.05832}.
%}%\cite{Bykov:2020tao}
\bibitem{byk3}{
D.~Bykov,
%``Quantum flag manifold $\sigma$-models and Hermitian Ricci flow,''
%\arxivlink{2006.14124};
``The $\mathsf{CP^{n-1}}$-model with fermions: a new look,''
\arxivlink{2009.04608};
``Flag manifold sigma-models and nilpotent orbits,''
\doilink{Proc. Steklov Inst. Math. \textbf{309}, 78-86 (2020)}{
doi:10.1134/S0081543820030062}
\arxivlink{1911.07768}.
}\bibitem{byk4}{
I.~Affleck, D.~Bykov and K.~Wamer,
``Flag manifold sigma models: spin chains and integrable theories,''
\arxivlink{2101.11638}.\\
%5 citations counted in INSPIRE as of 30 Jul 2021
D.~Bykov,
``Sigma models as Gross\textendash{}Neveu models,''
\doilink{Teor. Mat. Fiz. \textbf{208}, no.2, 165-179 (2021)}{
doi:10.1134/S0040577921080018}
\arxivlink{2106.15598}.
}\bibitem{Gross:1974jv}{
D.~J.~Gross and A.~Neveu,
``Dynamical symmetry breaking in asymptotically free field theories,''
\doilink{Phys. Rev. D \textbf{10}, 3235 (1974)}{
doi:10.1103/PhysRevD.10.3235}.
%1818 citations counted in INSPIRE as of 30 Jul 2021
%0 citations counted in INSPIRE as of 30 Jul 2021
%4 citations counted in INSPIRE as of 30 Jul 2021
%%CITATION = doi:10.1016/j.nuclphysb.2015.09.011;%%
%
%}\bibitem{hs} B.~Hoare and F.~K.~Seibold,
%``Poisson-Lie duals of the $\eta$ deformed symmetric space sigma model,''
%\doilink{JHEP {\bf 1711}, 014 (2017)}{doi:10.1007/JHEP11(2017)014},
%\arxivlink{1709.01448}.
%%%CITATION = doi:10.1007/JHEP11(2017)014;%%
%
}\bibitem{Klose:2010ki}{
T.~Klose,
``Review of AdS/CFT integrability, chapter IV.3: N=6 Chern-Simons and strings on AdS4xCP3,''
\doilink{Lett. Math. Phys. \textbf{99}, 401-423 (2012)}{
doi:10.1007/s11005-011-0520-y}
\arxivlink{1012.3999}.
%103 citations counted in INSPIRE as of 30 Jul 2021
}\bibitem{ABJM}{
O.~Aharony, O.~Bergman, D.~L.~Jafferis and J.~Maldacena,
``N=6 superconformal Chern-Simons-matter theories, M2-branes and their gravity duals,''
\doilink{JHEP \textbf{10}, 091 (2008)}{
doi:10.1088/1126-6708/2008/10/091}
\arxivlink{0806.1218}.
}\bibitem{sugra}{R.~Borsato and L.~Wulff,
``Target space supergeometry of $\eta$ and $\lambda$-deformed strings,''
\doilink{JHEP {\bf 1610}, 045 (2016)}{10.1007/JHEP10(2016)045}
\arxivlink{1608.03570}.
%%CITATION = doi:10.1007/JHEP10(2016)045;%%
%
}\bibitem{hsei}{
B.~Hoare and F.~K.~Seibold,
``Supergravity backgrounds of the $\eta$-deformed $AdS_2 \times S^2 \times T^6 $ and $AdS_5 \times S^5$ superstrings,''
\doilink{JHEP {\bf 1901}, 125 (2019)}{10.1007/JHEP01(2019)125}
\arxivlink{1811.07841}.
%%CITATION = doi:10.1007/JHEP01(2019)125;%%
%
%\cite{Seibold:2020ywq}
}\bibitem{seibold}{
F.~K.~Seibold, S.~J.~van Tongeren and Y.~Zimmermann, 
``The twisted story of worldsheet scattering in $\eta$-deformed $AdS_5 \times S^5$,''
\doilink{JHEP \textbf{12}, 043 (2020)}{
doi:10.1007/JHEP12(2020)043}
\arxivlink{2007.09136}.
%1 citations counted in INSPIRE as of 17 Oct 2020
%
%
}\bibitem{Polyakov:1988qz}{A.~M.~Polyakov,
``Two-dimensional quantum gravity: Superconductivity at high $T_c$,''
in: ``Fields, Strings and Critical Phenomena, Proc. of Les Houches 1988,''
eds.: E.~Brezin and J.~Zinn-Justin, North-Holland (1990).
%%CITATION = INSPIRE-273902;%%
%
}\bibitem{Guadagnini:1987ty}{E.~Guadagnini, M.~Martellini and M.~Mintchev,
``Scale invariant sigma models on homogeneous spaces,''
\doilink{Phys.\ Lett.\ B {\bf 194}, 69 (1987)}{10.1016/0370-2693(87)90771-4}.
%%CITATION = doi:10.1016/0370-2693(87)90771-4;%%
%
}\bibitem{Bardakci:1996gs}{K.~Bardakci, L.~M.~Bernardo and N.~Sochen, %v3a
``Integrable generalized Thirring model,''
\doilink{Nucl.\ Phys.\ B {\bf 487}, 513 (1997)}{doi:10.1016/S0550-3213(96)00715-8}
\arxivlink{hep-th/9607018}.
%%CITATION = doi:10.1016/S0550-3213(96)00715-8;%%
%
}\bibitem{Hull:1995gj}{O.~A.~Solovev,
``Towards conversion of the space of Thirring models into the model space for groups,''
\doilink{Phys.\ Lett.\ B {\bf 309}, 275 (1993)}{10.1016/0370-2693(93)90933-9}.\\
%%CITATION = doi:10.1016/0370-2693(93)90933-9;%%
%
C.~M.~Hull and O.~A.~Solovev,
``Conformal points and duality of non-abelian Thirring Models and interacting WZNW models,''
\doilink{Nucl.\ Phys.\ B {\bf 459}, 243 (1996)}{10.1016/0550-3213(95)00603-6}
\arxivlink{hep-th/9503021}.
%%CITATION = doi:10.1016/0550-3213(95)00603-6;%%
%
K.~Bardakci and L.~M.~Bernardo, %v3a
``The conformal points of the generalized Thirring model. 2.,''
\doilink{Nucl.\ Phys.\ B {\bf 450} 695 (1995)}{10.1016/0550-3213(95)00324-L}
\arxivlink{hep-th/9503143}.
%%CITATION = doi:10.1016/0550-3213(95)00324-L;%%
%
%}\bibitem{Georgiou:2016urf} G.~Georgiou and K.~Sfetsos,
%``A new class of integrable deformations of CFTs,''
%\doilink{JHEP {\bf 1703}, 083 (2017)}{10.1007/JHEP03(2017)083},
%\arxivlink{1612.05012}.
%%%CITATION = doi:10.1007/JHEP03(2017)083;%%
%
%
%
%
}\bibitem{Georgiou:2017aei}{G.~Georgiou, E.~Sagkrioti, K.~Sfetsos and K.~Siampos,
``Quantum aspects of doubly deformed CFTs,''
\doilink{Nucl.\ Phys.\ B {\bf 919}, 504 (2017)}{10.1016/j.nuclphysb.2017.04.004}
\arxivlink{1703.00462}.
%%CITATION = doi:10.1016/j.nuclphysb.2017.04.004;%%
%
}\bibitem{Georgiou:2017oly}{G.~Georgiou, K.~Sfetsos and K.~Siampos,
``Double and cyclic $\lambda$-deformations and their canonical equivalents,''
\doilink{Phys.\ Lett.\ B {\bf 771}, 576 (2017)}{10.1016/j.physletb.2017.06.007}
\arxivlink{1704.07834}.
%%CITATION = doi:10.1016/j.physletb.2017.06.007;%%
%
}\bibitem{Georgiou:2017jfi}{G.~Georgiou and K.~Sfetsos,
``Integrable flows between exact CFTs,''
\doilink{JHEP {\bf 1711}, 078 (2017)}{10.1007/JHEP11(2017)078}
\arxivlink{1707.05149}.
%%CITATION = doi:10.1007/JHEP11(2017)078;%%
%
}\bibitem{Gerganov:2000mt}{B.~Gerganov, A.~LeClair and M.~Moriconi,
``On the beta function for anisotropic current interactions in 2-D,''
\doilink{Phys.\ Rev.\ Lett.\ {\bf 86}, 4753 (2001)}{10.1103/PhysRevLett.86.4753}
\arxivlink{hep-th/0011189}.
%%CITATION = doi:10.1103/PhysRevLett.86.4753;%%
%
}\bibitem{LeClair:2001yp}{A.~LeClair,
``Chiral stabilization of the renormalization group for flavor and color anisotropic current interactions,''
\doilink{Phys.\ Lett.\ B {\bf 519}, 183 (2001)}{10.1016/S0370-2693(01)01089-9}
\arxivlink{hep-th/0105092}.
%%CITATION = doi:10.1016/S0370-2693(01)01089-9;%%
%
}\bibitem{Sfetsos:2014jfa}{K.~Sfetsos and K.~Siampos,
``Gauged WZW-type theories and the all-loop anisotropic non-abelian Thirring model,''
\doilink{Nucl.\ Phys.\ B {\bf 885}, 583 (2014)}{10.1016/j.nuclphysb.2014.06.012}
\arxivlink{1405.7803}.
%%CITATION = doi:10.1016/j.nuclphysb.2014.06.012;%%
%
}\bibitem{Witten:1983ar}{E.~Witten,
``Non-abelian bosonization in two dimensions,''
\doilink{Commun.\ Math.\ Phys.\ {\bf 92}, 455 (1984)}{10.1007/BF01215276}.
%%CITATION = doi:10.1007/BF01215276;%%
%
}\bibitem{Bos:1987mw}{M.~Bos,
``Dimensional regularization in the Wess-Zumino-Witten model,''
\doilink{Phys.\ Lett.\ B {\bf 189}, 435 (1987)}{10.1016/0370-2693(87)90656-3}.
%%CITATION = doi:10.1016/0370-2693(87)90656-3;%%
%
}\bibitem{McKane:1979cm}{A.~McKane and M.~Stone,
``Nonlinear sigma models: a perturbative approach to symmetry restoration,''
\doilink{Nucl.\ Phys.\ B {\bf 163} (1980) 169}{10.1016/0550-3213(80)90396-X}.
%%CITATION = doi:10.1016/0550-3213(80)90396-X;%%
%
}\bibitem{Hikami:1980hi}{S.~Hikami,
``Three loop beta-functions of nonlinear sigma models on symmetric spaces,''
\doilink{Phys.\ Lett.\ {\bf 98B}, 208 (1981)}{10.1016/0370-2693(81)90989-8}.
%%CITATION = doi:10.1016/0370-2693(81)90989-8;%%
%
}\bibitem{Brezin:1975sq}{E.~Brezin and J.~Zinn-Justin,
``Renormalization of the nonlinear sigma model in 2 + epsilon dimensions. Application to the Heisenberg ferromagnets,''
\doilink{Phys.\ Rev.\ Lett.\ {\bf 36}, 691 (1976)}{10.1103/PhysRevLett.36.691}.\\
%%CITATION = doi:10.1103/PhysRevLett.36.691;%%
%
S.~Hikami and E.~Brezin,
``Three loop calculations in the two-dimensional nonlinear sigma model,''
\doilink{J.\ Phys.\ A {\bf 11}, 1141 (1978)}{10.1088/0305-4470/11/6/015}.\\
%%CITATION = doi:10.1088/0305-4470/11/6/015;%%
%
E.~Brezin, S.~Hikami and J.~Zinn-Justin,
``Generalized nonlinear $\sigma$ models with gauge invariance,''
\doilink{Nucl.\ Phys.\ B {\bf 165}, 528 (1980)}{10.1016/0550-3213(80)90047-4}.
%%CITATION = doi:10.1016/0550-3213(80)90047-4;%%
%
}\bibitem{Balog:1993es}{%v3a
J.~Balog, P.~Forgacs, Z.~Horvath and L.~Palla,
``A new family of $SU(2)$ symmetric integrable sigma models,''
\doilink{Phys.\ Lett.\ B {\bf 324}, 403 (1994)}{10.1016/0370-2693(94)90213-5}
\arxivlink{hep-th/9307030}.\\
%%CITATION = doi:10.1016/0370-2693(94)90213-5;%%
%
J.~M.~Evans and T.~J.~Hollowood,
``Integrable theories that are asymptotically CFT,''
\doilink{ Nucl.\ Phys.\ B {\bf 438}, 469 (1995)}{10.1016/0550-3213(94)00473-R}
\arxivlink{hep-th/9407113}.
%%CITATION = doi:10.1016/0550-3213(94)00473-R;%%
%
}\bibitem{Kawaguchi:2010jg}{I.~Kawaguchi and K.~Yoshida,
``Hidden Yangian symmetry in sigma model on squashed sphere,''
\doilink{JHEP {\bf 1011}, 032 (2010)}{10.1007/JHEP11(2010)032}
\arxivlink{1008.0776}.
%%CITATION = doi:10.1007/JHEP11(2010)032;%%
%
}\bibitem{Kaloper:1997ux}{N.~Kaloper and K.~A.~Meissner,
``Duality beyond the first loop,''
\doilink{Phys.\ Rev.\ D {\bf 56}, 7940 (1997)}{10.1103/PhysRevD.56.7940}
\arxivlink{hep-th/9705193}.
%%CITATION = doi:10.1103/PhysRevD.56.7940;%%
%
}\bibitem{Parsons:1999ze}{S.~Parsons,
``T duality and conformal invariance at two loops,''
\doilink{Phys.\ Rev.\ D {\bf 61}, 086002 (2000)}{10.1103/PhysRevD.61.086002}
\arxivlink{hep-th/9912105}.\\
%%CITATION = doi:10.1103/PhysRevD.61.086002;%%
%
I.~Jack and S.~Parsons,
``$O(d,d)$ invariance at two loops and three loops,''
\doilink{Phys.\ Rev.\ D {\bf 62}, 026003 (2000)}{10.1103/PhysRevD.62.026003}
\arxivlink{hep-th/9911064}.
%%CITATION = doi:10.1103/PhysRevD.62.026003;%%
%
}\bibitem{Tseytlin:1990va}{
A.~A.~Tseytlin,
%\cite{Tseytlin:1990nb}
%}\bibitem{Tseytlin:1990nb}
%A.~A.~Tseytlin,
``Duality symmetric formulation of string world sheet dynamics,''
Phys. Lett. B \textbf{242}, 163-174 (1990)
doi:10.1016/0370-2693(90)91454-J; 
%307 citations counted in INSPIRE as of 13 Jul 2021
``Duality symmetric closed string theory and interacting chiral scalars,''
\doilink{Nucl.\ Phys.\ B {\bf 350}, 395 (1991)}{10.1016/0550-3213(91)90266-Z}.
%%CITATION = doi:10.1016/0550-3213(91)90266-Z;%%
%
%
}\bibitem{Rocek:1997hi}{M.~Rocek and A.~A.~Tseytlin,
``Partial breaking of global $D = 4$ supersymmetry, constrained superfields, and three-brane actions,''
\doilink{Phys.\ Rev.\ D {\bf 59}, 106001 (1999)}{10.1103/PhysRevD.59.106001}
\arxivlink{hep-th/9811232}.
%%CITATION = doi:10.1103/PhysRevD.59.106001;%%
%
}\bibitem{Abdalla:1980jt}{E.~Abdalla, M.~C.~B.~Abdalla and M.~Gomes,
``Anomaly in the nonlocal quantum charge of the $\mathbb{C}\mathbf{P}^{n-1}$ model,''
\doilink{Phys.\ Rev.\ D {\bf 23}, 1080 (1981)}{10.1103/PhysRevD.23.1800}.\\
%%CITATION = doi:10.1103/PhysRevD.23.1800;%%
%
E.~Abdalla, M.~Forger and M.~Gomes,
``On the origin of anomalies in the quantum nonlocal charge for the generalized nonlinear $\sigma$ models,''
\doilink{Nucl.\ Phys.\ B {\bf 210}, 181 (1982)}{doi:10.1016/0550-3213(82)90238-3}.
%%CITATION = doi:10.1016/0550-3213(82)90238-3;%%
%
}\bibitem{ey1}{J.~M.~Evans, D.~Kagan and C.~A.~S.~Young,
``Nonlocal charges and quantum integrability of sigma models on the symmetric spaces $SO(2n)/SO(n)\times SO(n)$ and $Sp(2n) / Sp(n) \times Sp(n)$,''
\doilink{Phys.\ Lett.\ B {\bf 597}, 112 (2004)}{doi:10.1016/j.physletb.2004.04.042}
\arxivlink{hep-th/0404003}.\\
%%CITATION = doi:10.1016/j.physletb.2004.04.042;%%
J.~M.~Evans, D.~Kagan, N.~J.~MacKay and C.~A.~S.~Young,
``Quantum, higher-spin, local charges in symmetric space sigma models,''
\doilink{JHEP {\bf 0501}, 020 (2005)}{doi:10.1088/1126-6708/2005/01/020}
\arxivlink{hep-th/0408244}.
%%CITATION = doi:10.1088/1126-6708/2005/01/020;%%
%
}\bibitem{cpn1}{A.~V.~Litvinov,
``Integrable $\mathfrak{gl}(n|n)$ Toda field theory and its sigma-model dual,''
\doilink{Pisma Zh. Eksp. Teor. Fiz. \textbf{110}, no.11, 723-726 (2019)}{
doi:10.1134/S0021364019230048}
\arxivlink{1901.04799}.
%%CITATION = ARXIV:1901.04799;%%
%
}\bibitem{bbr}{D.~Bykov,
``The worldsheet low-energy limit of the $AdS_4 \times \mathbb{C}\mathbf{P}^3$ superstring,''
\doilink{Nucl.\ Phys.\ B {\bf 838} (2010) 47}{doi:10.1016/j.nuclphysb.2010.05.013}
\arxivlink{1003.2199}.\\
%%CITATION = doi:10.1016/j.nuclphysb.2010.05.013;%%
%
B.~Basso and A.~Rej,
``On the integrability of two-dimensional models with $U(1) \times SU(N)$ symmetry,''
\doilink{Nucl.\ Phys.\ B {\bf 866} (2013) 337}{doi:10.1016/j.nuclphysb.2012.09.003}
\arxivlink{1207.0413}.
%
}\bibitem{Gomes:1982qh}{M.~Gomes, E.~Abdalla and M.~C.~B.~Abdalla,
``On the nonlocal charge of the {CP}$^{(N-1)}$ model and its supersymmetric extension to all orders,''
\doilink{Phys.\ Rev.\ D {\bf 27}, 825 (1983)}{10.1103/PhysRevD.27.825}.
%%CITATION = doi:10.1103/PhysRevD.27.825;%%
%
}\bibitem{Squellari:2014jfa}{R.~Squellari, %v3a
``Yang-Baxter $\sigma$ model: Quantum aspects,''
\doilink{Nucl.\ Phys.\ B {\bf 881}, 502 (2014)}{10.1016/j.nuclphysb.2014.02.009}
\arxivlink{1401.3197}.
%%CITATION = doi:10.1016/j.nuclphysb.2014.02.009;%%
%
}\bibitem{oneloopeta}{%v3a
C.~Klimcik and G.~Valent,
``One loop renormalizability of all 2-D dimensional Poisson-Lie sigma models,''
\doilink{Phys.\ Lett.\ B {\bf 565}, 237 (2003)}{10.1016/S0370-2693(03)00764-0}
\arxivlink{hep-th/0304053}.
%%CITATION = doi:10.1016/S0370-2693(03)00764-0;%%
%
%G.~Valent, C.~Klimcik and R.~Squellari,
%``One loop renormalizability of the Poisson-Lie sigma models,''
%\doilink{Phys.\ Lett.\ B {\bf 678}, 143 (2009)}{10.1016/j.physletb.2009.06.001},
%\arxivlink{0902.1459}.
%%%CITATION = doi:10.1016/j.physletb.2009.06.001;%%
%
}\bibitem{Kagan:2005wt}{D.~Kagan and C.~A.~S.~Young,
``Conformal sigma-models on supercoset targets,''
\doilink{Nucl.\ Phys.\ B {\bf 745}, 109 (2006)}{10.1016/j.nuclphysb.2006.02.027}
\arxivlink{hep-th/0512250}.
%%CITATION = doi:10.1016/j.nuclphysb.2006.02.027;%%
%
}\bibitem{Babichenko:2006uc}{A.~Babichenko,
``Conformal invariance and quantum integrability of sigma models on symmetric superspaces,''
\doilink{Phys.\ Lett.\ B {\bf 648}, 254 (2007)}{doi:10.1016/j.physletb.2007.03.003}
\arxivlink{hep-th/0611214}.
%%CITATION = doi:10.1016/j.physletb.2007.03.003;%%
%
%}\bibitem{Castellani:1999fz}{R. Gilmore,
%``Lie algebras and some of their applications,'' (Dover, 2005).\\
%%
%A.~Salam and J.~A.~Strathdee,
%``On Kaluza-Klein Theory,''
%\doilink{Annals Phys.\ {\bf 141}, 316 (1982)}{10.1016/0003-4916(82)90291-3}.\\
%%%CITATION = doi:10.1016/0003-4916(82)90291-3;%%
%%
%L.~Castellani,
%``On $G/H$ geometry and its use in M theory compactifications,''
%\doilink{Annals Phys.\ {\bf 287}, 1 (2001)}{10.1006/aphy.2000.6097}
%\arxivlink{hep-th/9912277}.
%%%CITATION = doi:10.1006/aphy.2000.6097;%%
%%
}\bibitem{bonneau}{G.~Bonneau, G.~Valent and F.~Delduc,
``Renormalization properties of bosonic nonlinear sigma models built on compact homogeneous Kahler manifolds,''
\doilink{Phys.\ Lett.\ B {\bf 196}, 456 (1987)}{10.1016/0370-2693(87)90801-X}.\\
%%CITATION = doi:10.1016/0370-2693(87)90801-X;%%
%
C.~Becchi, A.~Blasi, G.~Bonneau, R.~Collina and F.~Delduc,
``Renormalizability and infrared finiteness of nonlinear $\sigma$ models: a regularization independent analysis for compact coset spaces,''
\doilink{Commun.\ Math.\ Phys.\ {\bf 120}, 121 (1988)}{10.1007/BF01223209}.\\
%%CITATION = doi:10.1007/BF01223209;%%
%
A.~V.~Bratchikov,
``Renormalization properties of two-dimensional homogeneous symplectic sigma models,''
\doilink{Mod.\ Phys.\ Lett.\ A {\bf 7}, 2229 (1992)}{10.1142/S0217732392001993}.
%%%CITATION = doi:10.1142/S0217732392001993;%%
%
%%%%%%%%%%%%%%%%%
%% 2103 refs:
%
%%}\bibitem{intRG} 
%%V.~A.~Fateev, E.~Onofri and A.~B.~Zamolodchikov,
%%``Integrable deformations of the $O(3)$ sigma model. The sausage model,''
%%\doilink{Nucl.\ Phys.\ B {\bf 406}, 521 (1993)}{10.1016/0550-3213(93)90001-6}.
%%
%%V.~A.~Fateev,
%%``Classical and Quantum Integrable Sigma Models. Ricci Flow, ``Nice Duality'' and Perturbed Rational Conformal Field Theories,''
%%\doilink{J. Exp. Theor. Phys. \textbf{129}, no.4, 566-590 (2019)}{10.1134/S1063776119100042}
%%\arxivlink{1902.02811}.
%%
%%S.~L.~Lukyanov,
%%``The integrable harmonic map problem versus Ricci flow,''
%%\doilink{Nucl.\ Phys.\ B {\bf 865}, 308 (2012)}{10.1016/j.nuclphysb.2012.08.002}
%%\arxivlink{1205.3201}.
% 
%
}\bibitem{DLMV}{
F.~Delduc, S.~Lacroix, M.~Magro and B.~Vicedo,
``Integrable coupled $\sigma$ models,''
\doilink{Phys. Rev. Lett. \textbf{122}, no.4, 041601 (2019)}{doi:10.1103/PhysRevLett.122.041601}
\arxivlink{1811.12316};
``Assembling integrable $\sigma$-models as affine Gaudin models,''
\doilink{JHEP \textbf{06}, 017 (2019)}{doi:10.1007/JHEP06(2019)017}
\arxivlink{1903.00368}.
%
}\bibitem{ABL}{
G.~Arutyunov, C.~Bassi and S.~Lacroix,
``New integrable coset sigma models,''
\doilink{JHEP {\bf 2103}, 062 (2021)}{doi:10.1007/JHEP03(2021)062}
\arxivlink{2010.05573}.
%  
%}\bibitem{klim} 
%C.~Klimcik,
%``Yang-Baxter sigma models and dS/AdS T-duality,''
%\doilink{JHEP \textbf{12}, 051 (2002)}{doi:10.1088/1126-6708/2002/12/051}
%\arxivlink{hep-th/0210095};
%``On integrability of the Yang-Baxter sigma-model,''
%\doilink{J. Math. Phys. \textbf{50}, 043508 (2009)}{doi:10.1063/1.3116242}
%\arxivlink{0802.3518}.
%
%}\bibitem{balog}
%J.~Balog, P.~Forgacs, Z.~Horvath and L.~Palla,
%``A New family of SU(2) symmetric integrable sigma models,''
%\doilink{Phys.\ Lett.\ B {\bf 324}, 403 (1994)}{doi:10.1016/0370-2693(94)90213-5}
%\arxivlink{hep-th/9307030}.
%
}\bibitem{DMV}{
F.~Delduc, M.~Magro and B.~Vicedo,
%``On classical $q$-deformations of integrable sigma-models,''
%\doilink{JHEP \textbf{11}, 192 (2013)}{doi:10.1007/JHEP11(2013)192}
%\arxivlink{1308.3581};
``Integrable double deformation of the principal chiral model,''
\doilink{Nucl. Phys. B \textbf{891}, 312-321 (2015)}{doi:10.1016/j.nuclphysb.2014.12.018}
\arxivlink{1410.8066}.
%
}\bibitem{today}{
F.~Delduc, S.~Lacroix, K.~Sfetsos and K.~Siampos,
``RG flows of integrable $\sigma$-models and the twist function,''
\doilink{JHEP {\bf 2102}, 065 (2021)}{doi:10.1007/JHEP02(2021)065}
\arxivlink{2010.07879}.
%  
%%  
%%}\bibitem{mt} R.~R.~Metsaev and A.~A.~Tseytlin, 
%%``Order alpha-prime (two loop) equivalence of the string equations of motion and the sigma model Weyl invariance conditions: dependence on the dilaton and the antisymmetric tensor,''
%%\doilink{Nucl.\ Phys.\ B {\bf 293}, 385 (1987)}{10.1016/0550-3213(87)90077-0};
%%``Two loop $\beta$-function for the generalized bosonic sigma model,''
%%\doilink{Phys.\ Lett.\ B {\bf 191}, 354 (1987)}{doi:10.1016/0370-2693(87)90622-8}.
%
%%}\bibitem{Bos}
%%M.~Bos,
%%``Dimensional Regularization in the Wess-Zumino-Witten Model,''
%%\doilink{Phys. Lett. B \textbf{189}, 435-441 (1987)}{doi:10.1016/0370-2693(87)90656-3}.
%
}\bibitem{Ketov}{
%C.~M.~Hull and P.~K.~Townsend,
%``The Two Loop Beta Function for $\sigma$ Models With Torsion,''
%\doilink{Phys.\ Lett.\ B {\bf 191}, 115 (1987)}{doi:10.1016/0370-2693(87)91331-1}.
%
S.~V.~Ketov,
``Two loop calculations in $\sigma$ model with torsion,''
\doilink{Nucl.\ Phys.\ B {\bf 294}, 813 (1987)}{doi:10.1016/0550-3213(87)90609-2}.
%
%D.~Zanon,
%``Two Loop Beta Functions and Low-energy String Effective Action for the Two-dimensional Bosonic Nonlinear $\sigma$ Model With a {Wess-Zumino}-witten Term,''
%\doilink{Phys.\ Lett.\ B {\bf 191}, 363 (1987)}{doi:10.1016/0370-2693(87)90623-X}.
%
}\bibitem{Bap}{
M.~Bos,
``An example of dimensional regularization with antisymmetric tensors,''
\doilink{Annals Phys. \textbf{181}, 177 (1988)}{doi:10.1016/0003-4916(88)90164-9}.
%
}\bibitem{KZ}{
V.~G.~Knizhnik and A.~B.~Zamolodchikov,
``Current algebra and Wess-Zumino model in two-dimensions,''
\doilink{Nucl. Phys. B \textbf{247}, 83 (1984)}{doi:10.1016/0550-3213(84)90374-2}.
%
}\bibitem{shifman}{
D.~Schubring and M.~Shifman,
``Sigma model on a squashed sphere with a Wess-Zumino term,''
\doilink{Phys.\ Rev.\ D {\bf 103}, no. 2, 025016 (2021)}{doi:10.1103/PhysRevD.103.025016}
\arxivlink{2002.04696}.
%
}\bibitem{romans}{
D.~N.~Page and C.~N.~Pope,
``Which compactifications of $D=11$ supergravity are stable?,''
\doilink{Phys.\ Lett.\  {\bf 144B}, 346 (1984)}{doi:10.1016/0370-2693(84)91275-9}.\\
%
L.~J.~Romans,
``New compactifications of chiral N=2,  d=10 supergravity,''
\doilink{Phys. Lett. B \textbf{153}, 392 (1985)}{doi:10.1016/0370-2693(85)90479-4}.\\
%
P.~Candelas and X.~C.~de la Ossa,
``Comments on conifolds,''
\doilink{Nucl. Phys. B \textbf{342}, 246 (1990)}{doi:10.1016/0550-3213(90)90577-Z}.\\
%
I.~R.~Klebanov and E.~Witten,
``Superconformal field theory on three-branes at a Calabi-Yau singularity,''
\doilink{Nucl. Phys. B \textbf{536}, 199 (1998)}{doi:10.1016/S0550-3213(98)00654-3}
\arxivlink{hep-th/9807080}.
%
}\bibitem{GMM}{
E.~Guadagnini, M.~Martellini and M.~Mintchev,
``Scale invariance of sigma models  on homogeneous spaces,''
\doilink{Phys. Lett. B \textbf{194}, 69 (1987)}{doi:10.1016/0370-2693(87)90771-4}.\\
%
E.~Guadagnini,
``Current algebra in $\sigma$ models on homogeneous spaces,''
\doilink{Nucl. Phys. B \textbf{290}, 417 (1987)}{doi:10.1016/0550-3213(87)90195-7}.
%
%
}\bibitem{PZT}{
L.~A.~Pando Zayas and A.~A.~Tseytlin,
``Conformal sigma models for a class of T(p,q) spaces,''
\doilink{Class. Quant. Grav. \textbf{17}, 5125-5131 (2000)}{doi:10.1088/0264-9381/17/24/312}
\arxivlink{hep-th/0007086}.
%
}\bibitem{maillet}{
J.~M.~Maillet,
``Kac-Moody algebra and extended Yang-Baxter relations in the O($N$) nonlinear $\sigma$ model,''
\doilink{Phys. Lett. B \textbf{162}, 137 (1985)}{doi:10.1016/0370-2693(85)91075-5};
``New integrable canonical structures in two-dimensional models,''
\doilink{Nucl. Phys. B \textbf{269}, 54 (1986)}{doi:10.1016/0550-3213(86)90365-2}.
%
}\bibitem{LMV}{
S.~Lacroix, M.~Magro and B.~Vicedo,
``Local charges in involution and hierarchies in integrable sigma-models,''
\doilink{JHEP \textbf{09}, 117 (2017)}{doi:10.1007/JHEP09(2017)117}
\arxivlink{1703.01951}.
%
%}\bibitem{Sfetsos:2013wia} 
%K.~Sfetsos,
%``Integrable interpolations: From exact CFTs to non-Abelian T-duals,''
%\doilink{Nucl.\ Phys.\ B {\bf 880}, 225 (2014)}{doi:10.1016/j.nuclphysb.2014.01.004}
%\arxivlink{1312.4560}.
%
%T.~J.~Hollowood, J.~L.~Miramontes and D.~M.~Schmidtt,
%``Integrable Deformations of Strings on Symmetric Spaces,''
%\doilink{JHEP \textbf{11}, 009 (2014)}{doi:10.1007/JHEP11(2014)009}
%\arxivlink{1407.2840}.
%
}\bibitem{sfetsos}{G.~Georgiou and K.~Sfetsos,
``A new class of integrable deformations of CFTs,''
\doilink{JHEP {\bf 1703}, 083 (2017)}{10.1007/JHEP03(2017)083}
\arxivlink{1612.05012}.
%``Integrable flows between exact CFTs,''
%\doilink{JHEP {\bf 1711}, 078 (2017)}{10.1007/JHEP11(2017)078}
%\arxivlink{1707.05149}.
%
%G.~Georgiou, E.~Sagkrioti, K.~Sfetsos and K.~Siampos,
%``Quantum aspects of doubly deformed CFTs,''
%\doilink{Nucl.\ Phys.\ B {\bf 919}, 504 (2017)}{10.1016/j.nuclphysb.2017.04.004}
%\arxivlink{1703.00462}.
%
%G.~Georgiou, K.~Sfetsos and K.~Siampos,
%``Double and cyclic $\lambda$-deformations and their canonical equivalents,''
%\doilink{Phys.\ Lett.\ B {\bf 771}, 576 (2017)}{10.1016/j.physletb.2017.06.007}
%\arxivlink{1704.07834}.
%
}\bibitem{s19}{
G.~Georgiou, E.~Sagkrioti, K.~Sfetsos and K.~Siampos,
``An exact symmetry in $\lambda$-deformed CFTs,''
\doilink{JHEP \textbf{01}, 083 (2020)}{doi:10.1007/JHEP01(2020)083}
\arxivlink{1911.02027}.
%
%}\bibitem{ss}
%K.~Sfetsos and K.~Siampos,
%``Gauged WZW-type theories and the all-loop anisotropic non-abelian Thirring model,''
%\doilink{Nucl.\ Phys.\ B {\bf 885}, 583 (2014)}{10.1016/j.nuclphysb.2014.06.012},
%\arxivlink{1405.7803}.
%
%}\bibitem{ah}
%C.~Appadu and T.~J.~Hollowood,
%``Beta function of $k$ deformed $AdS_5 \times S^5$ string theory,''
%\doilink{JHEP {\bf 1511}, 095 (2015)}{10.1007/JHEP11(2015)095},
%\arxivlink{1507.05420}.
%
}\bibitem{HS}{
C.~M.~Hull and B.~J.~Spence,
``The gauged nonlinear $\sigma$ model with {Wess-Zumino} term,''
\doilink{Phys. Lett. B \textbf{232}, 204-210 (1989)}{doi:10.1016/0370-2693(89)91688-2}.
%
}\bibitem{witten}{
E.~Witten,
``On holomorphic factorization of WZW and coset models,''
\doilink{Commun. Math. Phys. \textbf{144}, 189-212 (1992)}{doi:10.1007/BF02099196}.
%
}\bibitem{belo}{
V.~V.~Belokurov and P.~M.~de Barrush Pasheku Seara de Sa,
``Ultraviolet finiteness of the Wess-Zumino-Witten gauge model on homogeneous manifolds,''
Moscow Univ.\ Phys.\ Bull.\  {\bf 45N3}, 13 (1990)
[Vestn.\ Mosk.\ Univ.\ Fiz.\ Astron.\  {\bf 31N3}, 13 (1990)].
%
%}\bibitem{BBS}
%K.~Bardakci, L.~M.~Bernardo and N.~Sochen,
%``Integrable generalized Thirring model,''
%\doilink{Nucl. Phys. B \textbf{487}, 513-525 (1997)}{doi:10.1016/S0550-3213(96)00715-8}
%\arxivlink{hep-th/9607018}.
%
}\bibitem{bpz}{
P.~Basu and L.~A.~Pando Zayas,
``Chaos rules out integrability of strings on AdS$_5 \times T^{1,1}$,''
\doilink{Phys. Lett. B \textbf{700}, 243-248 (2011)}{doi:10.1016/j.physletb.2011.04.063}
\arxivlink{1103.4107}.
%
}\bibitem{Basu:2011fw}{
P.~Basu and L.~A.~Pando Zayas,
``Analytic non-integrability in string theory,''
\doilink{Phys. Rev. D \textbf{84}, 046006 (2011)}{10.1103/PhysRevD.84.046006}
\arxivlink{1105.2540}.
%
}\bibitem{yosh}{
I.~Kawaguchi, D.~Orlando and K.~Yoshida,
``Yangian symmetry in deformed WZNW models on squashed spheres,''
\doilink{Phys.\ Lett.\ B {\bf 701}, 475 (2011)}{doi:10.1016/j.physletb.2011.06.007}
\arxivlink{1104.0738}.\\
%  
I.~Kawaguchi and K.~Yoshida,
``A deformation of quantum affine algebra in squashed Wess-Zumino-Novikov-Witten models,''
\doilink{J.\ Math.\ Phys.\  {\bf 55}, 062302 (2014)}{doi:10.1063/1.4880341}
\arxivlink{1311.4696}.
%
%S.~Demulder, S.~Driezen, A.~Sevrin and D.~C.~Thompson,
%``Classical and Quantum Aspects of Yang-Baxter Wess-Zumino Models,''
%\doilink{JHEP {\bf 1803}, 041 (2018)}{doi:10.1007/JHEP03(2018)041}
%\arxivlink{1711.00084}.
%
%}\bibitem{Tseytlin:1991wr} 
%A.~A.~Tseytlin,
%``Duality and dilaton,''
%\doilink{Mod.\ Phys.\ Lett.\ A {\bf 6}, 1721 (1991)}{10.1142/S021773239100186X}, \\
%{\small \url{https://www.academia.edu/40796716/Tseytlin1991_duality_dilaton_MPLA} }
%
%P.~E.~Haagensen and K.~Olsen,
%``T duality and two loop renormalization flows,''
%\doilink{Nucl.\ Phys.\ B {\bf 504}, 326 (1997)}{doi:10.1016/S0550-3213(97)00496-3}
%\arxivlink{hep-th/9704157}.
%
%N.~Kaloper and K.~A.~Meissner,
%``Duality beyond the first loop,''
%\doilink{Phys. Rev. D \textbf{56}, 7940-7953 (1997)}{doi:10.1103/PhysRevD.56.7940}
%\arxivlink{hep-th/9705193}.
%
%}\bibitem{bu}
%T.~H.~Buscher,
%``Path Integral Derivation of Quantum Duality in Nonlinear Sigma Models,''
%\doilink{Phys.\ Lett.\ B {\bf 201}, 466 (1988)}{doi:10.1016/0370-2693(88)90602-8}.
%
%A.~S.~Schwarz and A.~A.~Tseytlin,
%``Dilaton shift under duality and torsion of elliptic complex,''
%\doilink{Nucl.\ Phys.\ B {\bf 399}, 691 (1993)}{doi:10.1016/0550-3213(93)90514-P}
%\arxivlink{hep-th/9210015}.
%  
}\bibitem{hassler}{
F.~Hassler and T.~B.~Rochais,
``O($D$,$D$)-covariant two-loop $\beta$-functions and Poisson-Lie T-duality,''
\arxivlink{2011.15130}.\\
%
F.~Hassler,
``RG flow of integrable $\mathcal{E}$-models,''
\doilink{Phys. Lett. B \textbf{818}, 136367 (2021)}{
doi:10.1016/j.physletb.2021.136367}
\arxivlink{2012.10451}.
%
}\bibitem{Tseytlin:1992pq}{A.~A.~Tseytlin,
``String vacuum backgrounds with covariantly constant null Killing vector and 2-d quantum gravity,''
\doilink{Nucl. Phys. B \textbf{390}, 153-172 (1993)}{10.1016/0550-3213(93)90389-7}
\arxivlink{hep-th/9209023};
``Finite $\s$-models and exact string solutions with Minkowski signature metric,''
\doilink{Phys.\ Rev.\ D {\bf 47}, 3421 (1993)}{10.1103/PhysRevD.47.3421}
\arxivlink{hep-th/9211061}.
%
%%}\bibitem{HLT3}
%%B.~Hoare, N.~Levine and A.~A.~Tseytlin,
%%``Sigma models with local couplings: a new integrability -- RG flow connection,''
%%\doilink{JHEP \textbf{11}, 020 (2020)}{10.1007/JHEP11(2020)020}
%%\arxivlink{2008.01112}.
%
%
%%%%%%%%%%%%%%%%%%%%
%%%%%% 2008 refs
%
%%%Fateev:1992tk,Fateev:2019xuq,lukyanov
%%}\bibitem{Fateev:2019xuq} V.~A.~Fateev, E.~Onofri and A.~B.~Zamolodchikov,
%%``Integrable deformations of the $O(3)$ sigma model. The sausage model,''
%%\doilink{Nucl.\ Phys.\ B {\bf 406}, 521 (1993)}{10.1016/0550-3213(93)90001-6}.
%%%%CITATION = doi:10.1016/0550-3213(93)90001-6;%%
%%
%%V.~A.~Fateev,
%%``Classical and Quantum Integrable Sigma Models. Ricci Flow, ``Nice Duality'' and Perturbed Rational Conformal Field Theories,''
%%\doilink{J. Exp. Theor. Phys. \textbf{129}, no.4, 566-590 (2019)}{10.1134/S1063776119100042}
%%\arxivlink{1902.02811}.
%%
%%S.~L.~Lukyanov,
%%``The integrable harmonic map problem versus Ricci flow,''
%%\doilink{Nucl.\ Phys.\ B {\bf 865}, 308 (2012)}{10.1016/j.nuclphysb.2012.08.002}
%%\arxivlink{1205.3201}.
%%%%CITATION = doi:10.1016/j.nuclphysb.2012.08.002;%%
%
%%% HLT1,HLT2
%%}\bibitem{HLT} B.~Hoare, N.~Levine and A.~A.~Tseytlin,
%%``Integrable 2d sigma models: quantum corrections to geometry from RG flow,''
%%\doilink{Nucl. Phys. B \textbf{949}, 114798 (2019)}{10.1016/j.nuclphysb.2019.114798}
%%\arxivlink{1907.04737}.
%%
%%B.~Hoare, N.~Levine and A.~A.~Tseytlin,
%%``Integrable sigma models and 2-loop RG flow,''
%%\doilink{JHEP {\bf 1912}, 146 (2019)}{doi:10.1007/JHEP12(2019)146}
%%\arxivlink{1910.00397}.
%%%%CITATION = doi:10.1007/JHEP12(2019)146;%%
%
%%}\bibitem{Klimcik:2002zj} C.~Klim\v{c}\'{i}k,
%%``Yang-Baxter sigma models and dS/AdS T duality,''
%%\doilink{JHEP \textbf{12}, 051 (2002)}{10.1088/1126-6708/2002/12/051}
%%\arxivlink{hep-th/0210095}.
%
%%}\bibitem{Delduc:2013fga} F.~Delduc, M.~Magro and B.~Vicedo,
%%``On classical $q$-deformations of integrable sigma-models,''
%%\doilink{JHEP {\bf 1311}, 192 (2013)}{10.1007/JHEP11(2013)192}
%%\arxivlink{1308.3581}.
%%%%CITATION = doi:10.1007/JHEP11(2013)192;%%
%
%%}\bibitem{Sfetsos:2013wia} K.~Sfetsos,
%%``Integrable interpolations: From exact CFTs to non-Abelian T-duals,''
%%\doilink{Nucl.\ Phys.\ B {\bf 880}, 225 (2014)}{10.1016/j.nuclphysb.2014.01.004}
%%\arxivlink{1312.4560}.
%%%%CITATION = doi:10.1016/j.nuclphysb.2014.01.004;%%
%
%%}\bibitem{Hollowood:2014rla} T.~J.~Hollowood, J.~L.~Miramontes and D.~M.~Schmidtt,
%%``Integrable Deformations of Strings on Symmetric Spaces,''
%%\doilink{JHEP {\bf 1411}, 009 (2014)}{10.1007/JHEP11(2014)009}
%%\arxivlink{1407.2840}.
%%%%CITATION = doi:10.1007/JHEP11(2014)009;%%
%
%%}\bibitem{Valent:2009nv} G.~Valent, C.~Klim\v{c}\'{i}k and R.~Squellari,
%%``One loop renormalizability of the Poisson-Lie sigma models,''
%%\doilink{Phys. Lett. B \textbf{678}, 143-148 (2009)}{10.1016/j.physletb.2009.06.001}
%%\arxivlink{0902.1459}.
%
%%%Itsios:2014lca,Sfetsos:2014jfa
%%}\bibitem{Sfetsos:2014jfa}
%%%v2
%%G.~Itsios, K.~Sfetsos and K.~Siampos,
%%``The all-loop non-Abelian Thirring model and its RG flow,''
%%\doilink{Phys. Lett. B \textbf{733}, 265-269 (2014)}{10.1016/j.physletb.2014.04.061}
%%\arxivlink{1404.3748}.
%%
%%K.~Sfetsos and K.~Siampos,
%%``Gauged WZW-type theories and the all-loop anisotropic non-abelian Thirring model,''
%%\doilink{Nucl.\ Phys.\ B {\bf 885}, 583 (2014)}{10.1016/j.nuclphysb.2014.06.012}
%%\arxivlink{1405.7803}.
%%%%CITATION = doi:10.1016/j.nuclphysb.2014.06.012;%%
%
%%%ah
%%}\bibitem{Appadu:2015nfa} C.~Appadu and T.~J.~Hollowood,
%%``Beta function of $k$ deformed $AdS_5 \times S^5$ string theory,''
%%\doilink{JHEP {\bf 1511}, 095 (2015)}{10.1007/JHEP11(2015)095}
%%\arxivlink{1507.05420}.
%%%%CITATION = doi:10.1007/JHEP11(2015)095;%%
%
}\bibitem{t-dep-1d}{S. Bouquet and A. Bourdier,
``Notion of integrability for time-dependent Hamiltonian systems:
Illustrations from the relativistic motion of a charged particle,''
\doilink{Phys.\ Rev.\ E \textbf{57}, 1273 (1998)}{10.1103/PhysRevE.57.1273}.\\
%
M. V. Bartuccelli and G. Gentile,
``On a class of integrable time-dependent dynamical systems,''
\doilink{Phys.\ Lett.\ A \textbf{307}, 274-280 (2003)}{10.1016/S0375-9601(02)01731-0}.\\
%
R. M. Angelo, E. I. Duzzioni and A. D. Ribeiro,
``Integrability in time-dependent systems with one degree of freedom,''
\doilink{J. Phys. A \textbf{45}, 5 (2012)}{10.1088/1751-8113/45/5/055101}
\arxivlink{1106.6034}.
%
}\bibitem{Papadopoulos:2002bg}{G.~Papadopoulos, J.~Russo and A.~A.~Tseytlin,
``Solvable model of strings in a time dependent plane wave background,''
\doilink{Class. Quant. Grav. \textbf{20}, 969-1016 (2003)}{10.1088/0264-9381/20/5/313}
\arxivlink{hep-th/0211289}.\\
%
M.~Blau, M.~O'Loughlin, G.~Papadopoulos and A.~A.~Tseytlin,
``Solvable models of strings in homogeneous plane wave backgrounds,''
\doilink{Nucl.\ Phys.\ B {\bf 673}, 57 (2003)}{10.1016/j.nuclphysb.2003.09.018}
\arxivlink{hep-th/0304198}.
%%CITATION = doi:10.1016/j.nuclphysb.2003.09.018;%%
%
}\bibitem{yoshida}{A.~Borowiec, H.~Kyono, J.~Lukierski, J.~I.~Sakamoto and K.~Yoshida,
``Yang-Baxter sigma models and Lax pairs arising from $\kappa$-Poincar\'e $r$-matrices,''
\doilink{JHEP {\bf 1604}, 079 (2016)}{10.1007/JHEP04(2016)079}
\arxivlink{1510.03083}.\\
%%CITATION = doi:10.1007/JHEP04(2016)079;%%
%
H.~Kyono, J.~I.~Sakamoto and K.~Yoshida,
``Lax pairs for deformed Minkowski space-times,''
\doilink{JHEP {\bf 1601}, 143 (2016)}{10.1007/JHEP01(2016)143}
\arxivlink{1512.00208}.
%%CITATION = doi:10.1007/JHEP01(2016)143;%%
%
}\bibitem{Maldacena:2002fy}{J.~M.~Maldacena and L.~Maoz,
``Strings on pp waves and massive two-dimensional field theories,''
\doilink{JHEP \textbf{12}, 046 (2002)}{10.1088/1126-6708/2002/12/046}
\arxivlink{hep-th/0207284}.
%
}\bibitem{Russo:2002qj}{J.~G.~Russo and A.~A.~Tseytlin,
``A class of exact pp wave string models with interacting light cone gauge actions,''
\doilink{JHEP \textbf{09}, 035 (2002)}{10.1088/1126-6708/2002/09/035}
\arxivlink{hep-th/0208114}.
%
}\bibitem{Bakas:2002kt}{I.~Bakas and J.~Sonnenschein,
``On Integrable models from pp wave string backgrounds,''
\doilink{JHEP \textbf{12}, 049 (2002)}{10.1088/1126-6708/2002/12/049}
\arxivlink{hep-th/0211257}.
%
%}\bibitem{Tseytlin:1992pq} A.~A.~Tseytlin,
%``String vacuum backgrounds with covariantly constant null Killing vector and 2-d quantum gravity,''
%\doilink{Nucl. Phys. B \textbf{390}, 153-172 (1993)}{10.1016/0550-3213(93)90389-7}
%\arxivlink{hep-th/9209023};
%%v2
%``A class of finite two-dimensional sigma models and string vacua,''
%\doilink{Phys. Lett. B \textbf{288}, 279 (1992)}{doi:10.1016/0370-2693(92)91104-H}
%\arxivlink{hep-th/9205058}.
%%%CITATION = doi:10.1016/0370-2693(92)91104-H;%%
%
}\bibitem{Tseytlin:1992ee}{A.~A.~Tseytlin,
``Finite $\s$-models and exact string solutions with Minkowski signature metric,''
\doilink{Phys.\ Rev.\ D {\bf 47}, 3421 (1993)}{10.1103/PhysRevD.47.3421}
\arxivlink{hep-th/9211061}.
%%CITATION = doi:10.1103/PhysRevD.47.3421;%%
%
}\bibitem{Schmidhuber:1994bv}{C.~Schmidhuber and A.~A.~Tseytlin,
``On string cosmology and the RG flow in 2-d field theory,''
\doilink{Nucl.\ Phys.\ B {\bf 426}, 187 (1994)}{10.1016/0550-3213(94)90131-7}
\arxivlink{hep-th/9404180}.
%%CITATION = doi:10.1016/0550-3213(94)90131-7;%%
%
%v2
}\bibitem{Belinsky:1971nt}{
V.~A.~Belinsky and V.~E.~Zakharov,
``Integration of the Einstein Equations by the Inverse Scattering Problem Technique and the Calculation of the Exact Soliton Solutions,''
Sov.\ Phys.\ JETP {\bf 48}, 985 (1978)
[Zh.\ Eksp.\ Teor.\ Fiz.\ {\bf 75}, 1953 (1978)];
``Stationary Gravitational Solitons with Axial Symmetry,''
Sov.\ Phys.\ JETP {\bf 50}, 1 (1979)
[Zh.\ Eksp.\ Teor.\ Fiz.\ {\bf 77}, 3 (1979)].\\
%%CITATION = SPHJA,50,1;%%
%
%%CITATION = SPHJA,48,985;%%
D.~Maison,
``Are the stationary, axially symmetric Einstein equations completely integrable?,''
\doilink{Phys.\ Rev.\ Lett.\ {\bf 41}, 521 (1978)}{doi:10.1103/PhysRevLett.41.521};
%%CITATION = doi:10.1103/PhysRevLett.41.521;%%
``Stationary, Axially Symmetric Einstein Spaces: A Completely Integrable Hamiltonian System?,''
\doilink{J.\ Math.\ Phys.\ {\bf 20}, 871 (1979)}{doi:10.1063/1.524134}.
%%CITATION = doi:10.1063/1.524134;%%
%
}\bibitem{Breitenlohner:1986um}{
P.~Breitenlohner and D.~Maison,
``On the Geroch group,''
Ann.\ Inst.\ H.\ Poincare Phys.\ Theor.\ {\bf 46}, 215 (1987).
%
}\bibitem{Nicolai:1991tt}{
H.~Nicolai,
``Two-dimensional gravities and supergravities as integrable system,''
\doilink{Lect.\ Notes Phys.\ {\bf 396}, 231 (1991)}{doi:10.1007/3-540-54978-1_12}.\\
%%CITATION = doi:10.1007/3-540-54978-1_12;%%
%
H.~Nicolai, D.~Korotkin and H.~Samtleben,
``Integrable classical and quantum gravity,''
NATO Advanced Study Institute on Quantum Fields and Quantum Space Time,
22 July--3 August 1996. Cargese, France, pp. 203--243.
\arxivlink{hep-th/9612065}.
%%CITATION = HEP-TH/9612065;%%
%
%%%
%
}\bibitem{Arutyunov:2003za}{G.~Arutyunov, J.~Russo and A.~A.~Tseytlin,
``Spinning strings in $AdS_5 \times S^5$: New integrable system relations,''
\doilink{Phys.\ Rev.\ D {\bf 69} (2004) 086009}{10.1103/PhysRevD.69.086009}
\arxivlink{hep-th/0311004}.
%%CITATION = doi:10.1103/PhysRevD.69.086009;%%
%
}\bibitem{Goldschmidt:1980wq}{Y.~Y.~Goldschmidt and E.~Witten,
``Conservation laws in some two-dimensional models,''
\doilink{Phys. Lett. B \textbf{91}, 392-396 (1980)}{10.1016/0370-2693(80)91004-7}.
%
}\bibitem{Evans:1999mj}{
J.~M.~Evans, N.~J.~MacKay and M.~Hassan,
``Conserved charges and supersymmetry in principal chiral models,''
\arxivlink{hep-th/9711140}.\\
%13 citations counted in INSPIRE as of 06 Jul 2021
%
J.~M.~Evans, M.~Hassan, N.~J.~MacKay and A.~J.~Mountain,
``Local conserved charges in principal chiral models,''
\doilink{Nucl. Phys. B \textbf{561}, 385-412 (1999)}{10.1016/S0550-3213(99)00489-7}
\arxivlink{hep-th/9902008}.
%\cite{Evans:1997xu}
%}\bibitem{Evans:1997xu}
%
%
}\bibitem{Evans:2000qx}{J.~M.~Evans and A.~J.~Mountain,
``Commuting charges and symmetric spaces,''
\doilink{Phys. Lett. B \textbf{483}, 290-298 (2000)}{10.1016/S0370-2693(00)00566-9}
\arxivlink{hep-th/0003264}.
%
%LMV
}\bibitem{Lacroix:2017isl}{S.~Lacroix, M.~Magro and B.~Vicedo,
``Local charges in involution and hierarchies in integrable sigma-models,''
\doilink{JHEP \textbf{09}, 117 (2017)}{10.1007/JHEP09(2017)117}
\arxivlink{1703.01951}.
%
%}\bibitem{Faddeev:1985qu}{L.~D.~Faddeev and N.~Y.~Reshetikhin, %removed
%``Integrability of the principal chiral field model in (1+1)-dimension,''
%\doilink{Annals Phys. \textbf{167}, 227 (1986)}{10.1016/0003-4916(86)90201-0}.
%
% -> DLMV
%}\bibitem{Delduc:2019bcl} F.~Delduc, S.~Lacroix, M.~Magro and B.~Vicedo,
%``Assembling integrable $\sigma$-models as affine Gaudin models,''
%\doilink{JHEP \textbf{06}, 017 (2019)}{10.1007/JHEP06(2019)017}
%\arxivlink{1903.00368}.
%
%% -> maillet
%}\bibitem{Maillet:1985ek} J.~M.~Maillet,
%``New Integrable Canonical Structures in Two-dimensional Models,''
%\doilink{Nucl. Phys. B \textbf{269}, 54-76 (1986)}{10.1016/0550-3213(86)90365-2}
%
%}\bibitem{Sevostyanov:1995hd}{A.~Sevostyanov, %removed
%``The classical R matrix method for nonlinear sigma model,''
%\doilink{Int. J. Mod. Phys. A \textbf{11}, 4241-4254 (1996)}{10.1142/S0217751X96001978}
%\arxivlink{hep-th/9509030}.
%
%}\bibitem{Basu:2011fw} P.~Basu and L.~A.~Pando Zayas,
%``Analytic Non-integrability in String Theory,''
%\doilink{Phys. Rev. D \textbf{84}, 046006 (2011)}{10.1103/PhysRevD.84.046006}
%\arxivlink{1105.2540}.
%
}\bibitem{Stepanchuk:2012xi}{A.~Stepanchuk and A.~A.~Tseytlin,
``On (non)integrability of classical strings in p-brane backgrounds,''
\doilink{J. Phys. A \textbf{46}, 125401 (2013)}{10.1088/1751-8113/46/12/125401}
\arxivlink{1211.3727}.
%
}\bibitem{Osborn:1987au}{H.~Osborn,
``Renormalization and composite operators in nonlinear $\sigma$ models,''
\doilink{Nucl.\ Phys.\ B {\bf 294}, 595 (1987)}{10.1016/0550-3213(87)90599-2};
%%CITATION = doi:10.1016/0550-3213(87)90599-2;%%
``Weyl consistency conditions and a local renormalization group equation for general renormalizable field theories,''
\doilink{Nucl.\ Phys.\ B {\bf 363}, 486 (1991)}{10.1016/0550-3213(91)80030-P}.
%%CITATION = doi:10.1016/0550-3213(91)80030-P;%%
%
%v3
}\bi{byk}{
D.~Bykov and D.~Lust,
``Deformed $\sigma$-models, Ricci flow and Toda field theories,''
\arxivlink{2005.01812}.
%%CITATION = ARXIV:2005.01812;%%
%
%v3
}\bi{byk2}{
D.~Bykov,
``Quantum flag manifold $\sigma$-models and Hermitian Ricci flow,''
\arxivlink{2006.14124}.
%%CITATION = ARXIV:2006.14124;%%
%
%
%
%%--> Young:2005jv
%C.~A.~S.~Young,
%``Non-local charges, Z(m) gradings and coset space actions,''
%\doilink{Phys. Lett. B \textbf{632}, 559-565 (2006)}{doi:10.1016/j.physletb.2005.10.090},
%\arxivlink{hep-th/0503008}.
%
}\bibitem{yo}{
D.~Kagan and C.~A.~S.~Young,
``Conformal sigma-models on supercoset targets,''
\doilink{Nucl. Phys. B \textbf{745}, 109-122 (2006)}{doi:10.1016/j.nuclphysb.2006.02.027}
\arxivlink{hep-th/0512250}.
%
%\cite{Levkov:2020lfa}
}\bibitem{Levkov:2020lfa}{D.~G.~Levkov, V.~E.~Maslov and E.~Y.~Nugaev,
``Chaotic solitons in driven sine-Gordon model,''
\doilink{Fractals {\bf 139}, 110079 (2020)}{doi:10.1016/j.chaos.2020.110079}
\arxivlink{2004.13052}.
%%CITATION = doi:10.1016/j.chaos.2020.110079;%%
}\bibitem{Tseytlin:1986ws}{
A.~A.~Tseytlin,
``Conformal anomaly in two-dimensional sigma model on curved background and strings,''
\doilink{Phys. Lett. B \textbf{178}, 34 (1986)}{
doi:10.1016/0370-2693(86)90465-X};
``$\sigma$ model Weyl invariance conditions and string equations of motion,''
\doilink{Nucl. Phys. B \textbf{294}, 383-411 (1987)}{
doi:10.1016/0550-3213(87)90588-8}.
}\bibitem{Ishii:2021asw}{
T.~Ishii, S.~Kushiro and K.~Yoshida,
``Chaotic string dynamics in deformed $T^{1,1}$,''
\doilink{JHEP \textbf{05}, 158 (2021)}{doi:10.1007/JHEP05(2021)158}
\arxivlink{2103.12416}.
%2 citations counted in INSPIRE as of 31 Jul 2021
}\bibitem{Costello:2019tri}{K.~Costello, E.~Witten and M.~Yamazaki,
``Gauge theory and integrability, I,''
\doilink{ICCM Not. 6, 46-191 (2018)}{10.4310/ICCM.2018.v6.n1.a6}
\arxivlink{1709.09993}.\\
%%CITATION = doi:10.4310/ICCM.2018.v6.n1.a6;%%
%
K.~Costello and M.~Yamazaki,
``Gauge theory and integrability, III''
\arxivlink{1908.02289}.
%
}\bibitem{Delduc:2019whp}{F.~Delduc, S.~Lacroix, M.~Magro and B.~Vicedo,
``A unifying 2d action for integrable $\sigma$-models from 4d Chern-Simons theory,''
\doilink{Lett. Math. Phys. {\bf 110}, 1645 (2020)}{10.1007/s11005-020-01268-y}
\arxivlink{1909.13824}.
%%CITATION = doi:10.1007/s11005-020-01268-y;%%
%
}\bibitem{Lax:1968fm}{L.~D.~Faddeev and L.~A.~Takhtajan,
``Hamiltonian methods in the theory of solitons,''
Berlin, Springer (1987).
%
%\cite{Borsato:2020wwk}
}\bibitem{BW}{
R.~Borsato and L.~Wulff,
``Quantum correction to generalized $T$ dualities,''
\doilink{Phys. Rev. Lett. \textbf{125}, no.20, 201603 (2020)}{
doi:10.1103/PhysRevLett.125.201603}
\arxivlink{2007.07902}.\\
%19 citations counted in INSPIRE as of 28 Jun 2021
%\cite{Codina:2020yma}
%
%\cite{Hassler:2020tvz}
F.~Hassler and T.~Rochais,
``$\alpha'$-corrected Poisson-Lie T-duality,''
\doilink{Fortsch. Phys. \textbf{68}, no.9, 2000063 (2020)}{
doi:10.1002/prop.202000063}
\arxivlink{2007.07897}.\\
%15 citations counted in INSPIRE as of 28 Jun 2021
%
T.~Codina and D.~Marques,
``Generalized dualities and higher derivatives,''
\doilink{JHEP \textbf{10}, 002 (2020)}{
doi:10.1007/JHEP10(2020)002}
\arxivlink{2007.09494}.
}
%19 citations counted in INSPIRE as of 28 Jun 2021
%
\end{thebibliography}
